\documentclass{aa}

\usepackage{graphicx}
\usepackage{txfonts}
\usepackage{lscape}
\begin{document}

   \title{Characterization of Omega-WINGS galaxy clusters}

   \subtitle{I. Stellar light and mass profiles}

   \author{S.~Cariddi\inst{1} \and M.~D'Onofrio\inst{1,2} \and G.~Fasano\inst{2} \and B.M.~Poggianti\inst{2} \and A.~Moretti\inst{2} \and M.~Gullieuszik\inst{2} \and D.~Bettoni\inst{2} \and M.~Sciarratta\inst{1}}

   \institute{
                Dipartimento di Fisica e Astronomia ``Galileo Galilei'', Universit\`a degli Studi di Padova, Vicolo dell'Osservatorio 3, 35122 Padova, Italy
                \and
                INAF-OAPD, Vicolo dell'Osservatorio 5, 35122 Padova, Italy}

   \date{Received July 19, 2017; Accepted September 21, 2017}

% \abstract{}{}{}{}{} 
% 5 {} token are mandatory
 
  \abstract
  % context heading (optional)
  % {} leave it empty if necessary  
   {Galaxy clusters are the largest virialized structures in the observable Universe.  Knowledge of their properties provides many useful astrophysical and cosmological information.}
  % aims heading (mandatory)
   {Our aim is to derive the luminosity and stellar mass profiles of the nearby galaxy clusters of the Omega-WINGS survey and to study the main scaling relations valid for such systems.}
  % methods heading (mandatory)
   {We merged data from the WINGS and Omega-WINGS databases, sorted the sources according to the distance from the brightest cluster galaxy (BCG), and calculated the integrated luminosity profiles in the $B$ and $V$ bands, taking into account extinction, photometric and spatial completeness, K correction, and background contribution. Then, by exploiting the spectroscopic sample we derived the stellar mass profiles of the clusters.}
  % results heading (mandatory).
   {We obtained the luminosity profiles of 46 galaxy clusters, reaching $r_{200}$ in 30 cases, and the stellar mass profiles of 42 of our objects. We successfully fitted all the integrated luminosity growth profiles with one or two embedded S\'ersic components, deriving the main clusters parameters.
   Finally, we checked the main scaling relation among the clusters parameters in comparison with those obtained for a selected sample of early-type galaxies (ETGs) of the same clusters.}
  % conclusions heading (optional), leave it empty if necessary 
   {We found that the nearby galaxy clusters are non-homologous structures such as ETGs and exhibit a color-magnitude (CM) red-sequence relation very similar to that observed for galaxies in clusters.  These properties  are not expected in the current cluster formation scenarios. In particular the existence of a CM relation for clusters, shown here for the first time, suggests that the baryonic structures grow and evolve in a similar way at all scales.}


   \maketitle
%
%-------------------------------------------------------------------

\section{Introduction}

Galaxy clusters are the largest virialized structures that we observe in the Universe. Their study offers the possibility of significantly improving our understanding of many astrophysical and cosmological problems (e.g., \citealt{all11}). For example, the determination of their masses and density profiles is of fundamental importance for determining the dark matter (DM) content and distribution in the galaxy halos, mechanisms underlying the formation and evolution of the structure, and fraction of baryons inside clusters. Furthermore, each of these astrophysical questions is linked to many others. For example, knowledge of the baryon fraction is crucial for understanding baryonic physics and correctly calibrating cosmological simulations. It follows that much of our present understanding of the Universe is based on accurate measurements of galaxy cluster properties. 

Everyone working in this field knows how difficult it is to determine the luminosity profiles of galaxy clusters because of galactic extinction, background galaxies contamination, completeness of the data, and membership uncertainty. These difficulties are the reason for the relatively small number of clusters luminosity determinations (see, e.g., \citealt{oem74}; \citealt{dre78}; \citealt{ada98}; \citealt{car96}; \citealt{gir0}). Even more difficult is the estimate of clusters masses and mass profiles that are biased by various effects depending on the applied methods. The masses inferred from either X-ray (e.g., \citealt{ett13}; \citealt{mau16}) or optical data are, for example, based on the assumption of dynamical equilibrium, while those obtained by gravitational lensing (e.g., \citealt{ume14}; \citealt{mer15}) require a good knowledge of the geometry of the potential well. Discrepancies by a factor of 2-3 between the masses obtained by various methods have been reported (e.g., \citealt{wu97}). 

Today, thanks to the large field of view of many optical cluster surveys, such as the Sloan Digital Sky Survey (e.g.,~\citealt{aba3}) and the Canada-France-Hawaii Telescope Legacy Survey (e.g.,~\citealt{hud12}), the idea of reconstructing the stellar mass profiles of galaxy clusters starting from their integrated luminosity profiles has become possible. The optical data of modern surveys have drastically reduced the problems mentioned above affecting the precision of light profile measurements. In particular, some of the techniques already used to derive the surface brightness distribution of ETGs have been now adapted to the case of galaxy clusters. These systems are already known for sharing with ETGs many scaling relations (see, e.g., \citealt{sch93}; \citealt{ada98}; \citealt{ann94}; \citealt{fuj99}; \citealt{fri99}; \citealt{mil99}) that might provide useful insights into formation mechanisms and evolutionary processes. Their existence is expected on the basis of simple models of structure formation, such as the gravitational collapse of density fluctuations of collisionless DM halos. The \cite{gun72} model for example predicts that all the existing collapsed DM halos are virialized and characterized by a constant mean density, depending by the critical density of the Universe at that redshift and the adopted cosmology (\citealt{pee80}; \citealt{eke96}). If DM halos are structurally homologous systems with similar velocity dispersion profiles, as cosmological simulations predict (\citealt{col96}; \citealt{nav97}), and if the light profiles of the clusters trace the DM potential, then looking at the projected properties of galaxy clusters we expect to find many of the scaling relations observed in ETGs. This is the case, for example, for the fundamental plane relation, i.e., the relation involving the effective surface brightness $I_{\it eff}$, effective radius $r_{\it eff}$, and  velocity dispersion $\sigma$ (\citealt{djo87}; \citealt{dre87}), which appears to originate from a common physical mechanism valid both for ETGs and clusters (\citealt{cap94}; \citealt{don17}). The combined analysis of the scaling relations of ETGs and galaxy clusters could provide important information concerning the mass assembly at different scales.

With this paper we start a series of works aimed at addressing such issue. The first work is dedicated to the problem of the accurate determination of stellar light and stellar mass profiles of galaxy clusters. We exploit for this goal the large optical and spectroscopic database provided by the WIde-field Nearby Galaxy-cluster Survey (WINGS; \citealt{fas6}; \citealt{var9}; \citealt{cav9}) and its Omega-WINGS extension (\citealt{gul15}; \citealt{mor17}), which is available for nearby galaxy clusters ($0.04 \lesssim z \lesssim 0.07$). The first step carried out here provides the integrated luminosity (and stellar mass) profiles for 46 (42) nearby clusters.

The paper is organized as follows: In Section \ref{sec:data} we present the characteristics of our spectro-photometric data sample; in Section \ref{sec:photometry} we merge the data of the two photometric surveys to maximize the spatial coverage and we derive the integrated luminosity profiles, color profiles, and surface brightness profiles; in this analysis we take into account the completeness effects, K correction, and background subtraction. Finally, we calculate also the total flux coming from the faint end of the cluster luminosity function. In Section \ref{sec:mass} we derive the stellar mass profiles starting from previously derived luminosity profiles and spectroscopic data of the surveys, which were already used to get the cluster membership and spectrophotometric masses (\citealt{cav9}; \citealt{fri7}, \citealt{fri11}). Finally, we successfully reconstruct the integrated stellar mass profiles up to $r=r_{200}$ for 30 of our 46 clusters, and up to $r = 2\, r_{200}$ for 3 of these clusters. In Section \ref{sec:res} we discuss the fitting procedure used to reproduce the integrated luminosity profiles of our clusters through a S\'ersic (\citeyear{ser63}, \citeyear{ser68}) law. From these fits we obtain the main photometric parameters ($n$, $r_{\it eff}$, $I_{\it eff}$) and we use these parameters to construct the main scaling relations of clusters that are compared with those already found for ETGs (Section~\ref{sec:relations}). In Section \ref{sec:conclusions} we present a general discussion and our conclusions.

Throughout the paper we have assumed a $\Lambda$-CDM universe with $H_0 = 70$ km s$^{-1}$ Mpc$^{-1}$ and $\Omega_m = 0.3$.

\section{Data sample}
\label{sec:data}

The WINGS survey is a spectrophotometric wide-field survey of 76 galaxy clusters selected from the ROSAT X-ray-brightest Abell-type Cluster Sample Survey \citep{ebe96} and its extensions (\citealt{ebe98}, \citeyear{ebe00}). It consists of $B$- and $V$-band observations  ($34' \times 34'$ field of view; FoV) obtained with the Wide Field Camera on the INT and the 2.2~m MPG/ESO telescopes (\citealt{fas6}; \citealt{var9}), $J$- and $K$-band images obtained with the Wide Field CAMera at UKIRT \citep{val9}, and $U$-band observations performed with the INT, LBT, and BOK telescopes \citep{omi14}. Spectroscopic observations were performed by \cite{cav9} for a subsample of 6137 galaxies using 2dF-AAT and WYFFOS-WHT.

The unicity of WINGS lies in the combination of the size of the sample with the depth of the observations. However, the original project only covered the cluster cores up to $\sim0.6 \, r_{200}$ in almost all cases. Thus, WINGS alone does not allow a proper study of the transition regions between the cluster cores and the field.
This is a severe limitation. In fact, several studies have proved that many galaxy properties are a function of the clustercentric distance (e.g., \citealt{lew2}; \citealt{gom3}). In particular, \cite{fas15} found that the morphology-density relation \citep{dre80} holds only in the cores of WINGS clusters.

\begin{figure*}
        \includegraphics[width=\textwidth]{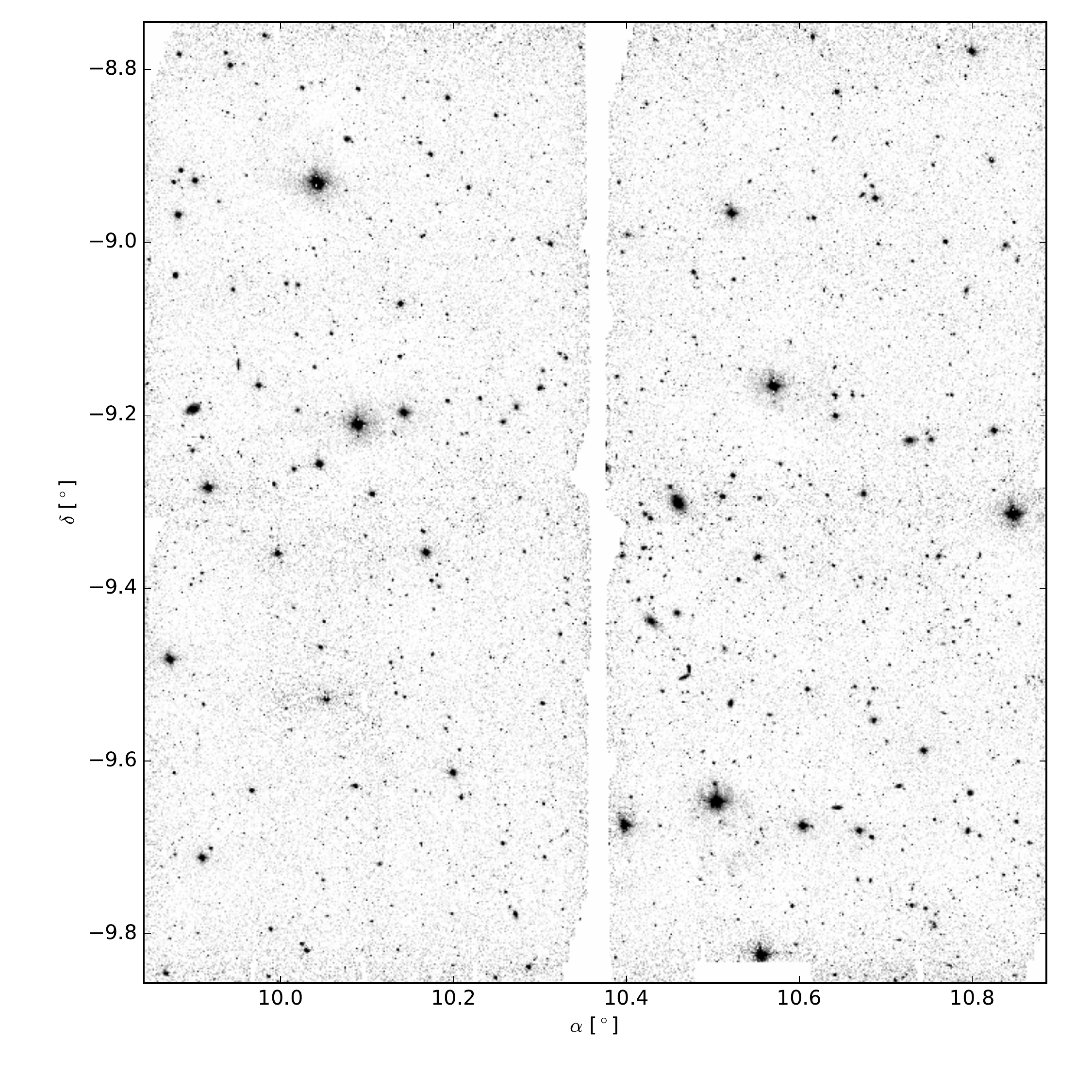}
    \caption{Omega-WINGS V-band image of A85.}
    \label{fig:a85}
\end{figure*}

In order to overcome the limited FoV problem, an extension of the original survey was performed with OmegaCAM at VST. $B$- and $V$-band data for 46 WINGS clusters were obtained by \citealt{gul15} (see, e.g., Figure~\ref{fig:a85}); these data cover an area of $1^\circ \times 1^\circ$. A spectroscopic follow up of 17985 galaxies in 34 clusters with the AAOmega spectrograph at AAT was also performed by \citealt{mor17}. Table \ref{tab:clusters} in Appendix shows a recap of the observations carried out in the WINGS and  Omega-WINGS surveys.

The photometric and spectroscopic WINGS/Omega-WINGS catalogs are now available on the Virtual Observatory (\citealt{mor14}; \citealt{gul15}). The WINGS database includes not only the magnitudes of the galaxies in the field, but also important quantities derived from the photometric and spectroscopic analyses, such as effective radii and surface brightness, flattening, masses, light indexes, and velocity dispersions. In this work we used the following data sets: WINGS photometric $B$- and $V$-band data, WINGS spectroscopic data,
Omega-WINGS $B$- and $V$-band data, and Omega-WINGS spectroscopic data.

The $V$- and $B$-band magnitudes used in this work are the SExtractor AUTO magnitudes (see \citealt{Bertin} for further details), whose $V$-band completeness was calculated by \cite{var9} for the WINGS sample (90\% completeness threshold at $m_V \sim 21.7$ mag), and by \cite{gul15} for the Omega-WINGS sample (90\% completeness threshold at $m_V \sim 21.2$ mag).

The objects in the catalogs are divided into three different categories: stars, galaxies, and unknown, according to their SExtractor stellarity index and a number of diagnostic diagrams used to improve the classifications (details in \citealt{var9} and \citealt{gul15}). We rejected the stars and focused on galaxies and unknown objects.

%In total we got the photometry of $\sim 865\,000$ galaxies and unknown objects into 46 different clusters, and the spectroscopy of $\sim 22\,674$ galaxies in 42 clusters.

The spectroscopic redshifts of our galaxies were measured using a semi-automatic method and the mean redshift of the clusters and rest-frame velocity dispersions \citep{pac16}. The latter were derived using
a clipping procedure. The galaxies members were those laying within 3 root-mean-squares (RMS) from the cluster redshift. The $r_{200}$ radius was computed as in 
\cite{Poggiantietal2006} and used to scale the distances from the BCG. A correction for both geometrical and magnitude incompleteness was applied to
the spectroscopic catalog, using the ratio between the number of spectra yielding a redshift to the total number of galaxies in the parent photometric catalog, calculated as a function of both the $V$-band magnitude and the projected radial distance from the BCG \citep[see][]{pac16}.

Owing to the limits of the spectroscopic survey our stellar mass profile analysis is restricted here only to 42 out of 46 clusters.

\section{Photometric profiles of clusters}
\label{sec:photometry}

To build the photometric profiles of each cluster we performed the following steps: the WINGS and Omega-WINGS catalogs were merged, a cross-match between the photometric and spectroscopic catalogs was performed, and all the galaxies classified as ``non members'' in the latter were removed from the main source catalog and saved into a rejected objects catalog. For each object in the main catalog, the integrated intensity within a circular area centered on the BCG was calculated, taking into account the magnitude $m_i$, position $r_i$, and completeness $cc_i$ (see below). The field intensity per square degree $I_{\it field}$ was calculated starting from the \cite{ber6} number counts, taking into account the already rejected objects. After sorting the galaxies for increasing cluster-centric distance, the final integrated intensity growth curve of each cluster was derived according to the following formula:

\begin{equation}
I(r_n) = \sum_{i=1}^{n} ({\it cc_i} \cdot 10^{-0.4 m_i}) - \pi r_n^2 \cdot I_{\it field} {\rm ,}
\label{eq:i}
\end{equation}

\noindent where $i$ is the index associated with each catalog object and $r_n$ the distance of the $n$-th object from the BCG. The intensity profiles were transformed into integrated magnitude, $(B-V)$ color, and surface brightness profiles. The K correction was applied to each radial bin, according to the color index of the galaxy population in the bin and the mean redshift of the cluster.

We now analyze the previous points in detail. In particular Section~\ref{sec:preliminary} focuses on the preliminary work carried out on the two catalogs, Section~\ref{sec:cc} on the completeness correction calculation, Section~\ref{sec:field} on the determination of the field galaxy contribution, and Section~\ref{sec:profiles} on the photometric profiles construction. Finally, Section~\ref{sec:faint-obj-corr} deals with the calculation of the faint objects correction that is later used for correctly deriving the stellar mass profiles.

\subsection{Preliminary work}
\label{sec:preliminary}

The Omega-WINGS images are typically four times larger than the WINGS images. These images also have broader gaps between the CCDs with the central cluster regions usually laying out of the FoV (see, e.g., Figure~\ref{fig:a85}).

In order to combine the larger Omega-WINGS FoV with the WINGS information about the central regions, we merged the two catalogs. For the objects in common between the two catalogs, we decided to use the WINGS original magnitude since its photometry is the most precise. Then we rejected all the sources classified as ``stars'', keeping only galaxies and unknown objects.

Finally,  we removed all galaxies classified by \cite{mor17} as ``non members'' via a cross-check with the spectroscopic catalog.  These objects were moved to a rejected sources catalog that was later used for improving the statistical field subtraction, as described in Section~\ref{sec:field}.

\subsection{Completeness correction}
\label{sec:cc}

The detection rate is the probability to observe a source as a function of a series of parameters, the most important of which, in our case, are its magnitude and position.

Starting from this definition, we got the completeness correction $cc_i$ of the $i$-th object as the inverse of the detection rate of an object of magnitude $m_i$ and distance from the BCG $r_i$ times the probability that this object is a galaxy. We can compute our completeness correction as the product of three terms:  first, the photometric completeness correction $c_{\it ph}(m_i)$, i.e., the inverse of the probability to observe an object of magnitude $m_i$;
second, the areal correction ${\it ac}(r_i)$, i.e., the inverse of the probability that an object with distance $r_i$ from the BCG lies inside the FoV;
and third, the probability that the considered object is a galaxy $c_{\it un}(m_i)$, that is 1 for objects classified as ``galaxies'' and a function of $m_i$ for unknown objects.

Summarizing,

\begin{equation}
{\it cc}_i = c_{\it ph} (m_i) \cdot {\it ac} (r_i) \cdot c_{\it un} (m_i) {\rm .}
\label{eq:compl}
\end{equation}

Each term is now be analyzed in detail. We start from $c_{\it ph}$.

The function $c_{\it ph} (m_i)$ for the $V$-band WINGS data was calculated by \cite{var9} through the detection rate of artificial stars randomly added to the WINGS images. Likely, the detection rate of galaxies follows a slightly different trend as they are not point sources and their detection probability is not only a function of their magnitude, but also depends on a series of parameters (e.g., morphology, compactness, and inclination) whose simulated distribution should correctly match the observed distribution. This introduces an uncertainty in any measurement of the photometric completeness correction that we wanted to avoid. For this reason we decided to introduce a photometric cut at $m_V = 20$ mag, which is the limit within which the spectroscopic sample is representative of the photometric sample. At this magnitude the detection rate found by \cite{var9} is equal to 97\% and galaxies are easily distinguished from stars. This allows us to assume that second-order dependences, connected to the above-mentioned galaxy parameters, cannot significantly modify the artificially-calculated completeness.

The Omega-WINGS $V$-band completeness, instead, was calculated by \cite{gul15} as a function of the WINGS $V$-band completeness by comparing the number of objects in each magnitude bin, after matching the total number of sources in the magnitude range 16 mag < $m_V$ < 21 mag to account for the different sky coverage. These authors found that if we limit our analysis to objects brighter than $m_V = 20$ mag, the detection rate of the two surveys is the same. All these considerations allowed us to safely assume a $V$-band photometric correction $c_{\it ph} = 1$ for every object and discard galaxies fainter than $m_V=20$ mag. We see the consequences of this choice below.

The $B$-band completeness correction was not evaluated for the two surveys. As a consequence, we chose to characterize our clusters using only the $V$-band-limited sample of objects with $m_V \leq 20$ mag.

In order to understand  whether, by assuming $c_{\it ph} = 1$ also for the $B$ band, we were introducing a systematic bias, we considered the following facts: first of all, the equal number of sources in the two bands rules out the possibility of a drastically different $c_{\it ph}$ value between the two bands in the considered photometric range; moreover, the integrated $B$-band intensity of all the sources of our catalog with $m_B \leq 20$ mag is almost 10 times larger than the combined intensity of the sources with $B$-band magnitude in the range 20 mag < $m_B \leq$ 22 mag; finally, the integrated color indexes of the clusters are always close to $(B-V)\sim1$, no matter what cut in $B$ or $V$ magnitude we introduce.

These considerations made us confident that, by assuming a $B$-band completeness of $c_{\it ph} = 1$ as well, we were not introducing any systematic error.

The second term on the right side of Equation~\ref{eq:compl} concerns the probability that an object distant $r_i$ from the BCG is observed. This probability might be defined by the fraction of the circle centered on the BCG (with radius $r_i$ and thickness of 1 pixel) that resides in the FoV. The areal correction ${\it ac}(r_i)$ is defined as the inverse of this probability.

The third term of Equation~\ref{eq:compl} is $c_{\it un}(m_i)$. Under the reasonable assumption that no object identified as a galaxy was misclassified, this term is different from 1 only for the unknown-type objects and corresponds to the probability that the considered unknown object is a galaxy. This probability has also been calculated by \cite{var9}.

To summarize, we can approximate the completeness correction with the following formula:

\begin{equation}
{\it cc}_i \simeq \left\{\begin{array}{ll}
{\it ac} (r_i) & {\rm if \ the \ object \ is \ a \ galaxy,}\\
{\it ac} (r_i) \cdot c_{\it un} (m_i) & {\rm if \ the \ object \ is \ an \ unknown.}\\
\end{array}\right.
\label{eq:compl_approx}
\end{equation}

\subsection{Field subtraction}
\label{sec:field}

To calculate the intensity per square degree emitted by the field galaxies we used the galaxy number counts measured by \cite{ber6}. These authors give the galaxy number counts normalized to an area of 1 square degree for the $B$ and $V$ bands, from magnitude 16 to 28, with bins of width 0.5 mag.

In our case the flux emitted by the statistically measured field galaxies is given by

\begin{equation}
I_{\it Berta} = \sum_{j=1}^{p} N_j \cdot f_j(m_V \leq 20) \cdot 10^{-0.4 m_j} {\rm ,}
\label{eq:berta_flux}
\end{equation}

\noindent where $j$ is the index of each magnitude bin, $p$ the total number of bins, $N_j$ the number of counts in the $j$-th bin, $m_j$ the average magnitude of the galaxies in the $j$-th bin, and $f_j(m_V \leq 20)$ the fraction of objects in the considered magnitude bin with a $V$-band magnitude lower than 20.

For the $V$-band data $f_j(m_V \leq 20)$ is a step function equal to 1 for magnitudes brighter than 20 and equal to 0 for fainter magnitudes. 

\begin{figure}
        \includegraphics[width=0.45\textwidth]{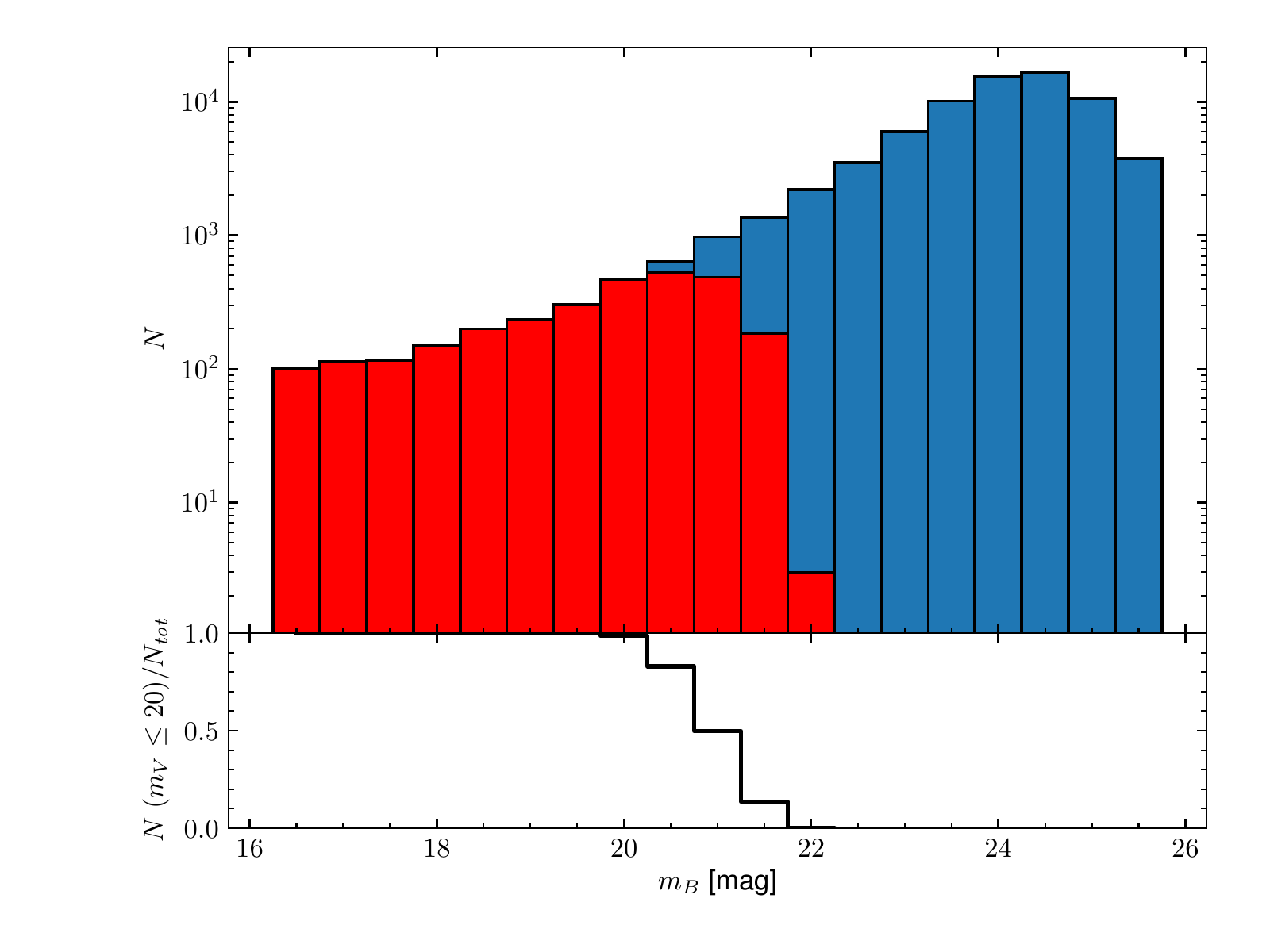}
    \caption{Upper panel: Total number of galaxies in each $B$-band magnitude interval detected by \cite{ber6} (in blue) and fraction of these galaxies with $m_V \leq 20$ mag (in red). Lower panel: Fraction $f_j$ of galaxies with $V$-band magnitude brighter than 20 as a function of their $B$-band magnitude is shown.}
    \label{fig:ber}
\end{figure}

For the $B$ band, since we are working with a $V$-band-limited sample of galaxies, we need to subtract the field contribution given only by objects with $m_V \leq 20$ mag. 
To achieve this, we downloaded the original photometric data by \cite{ber6} and we rebuilt the histogram of galaxy counts. The trend observed is visible in Figure~\ref{fig:ber}.
The figure shows in red the number of galaxies with magnitude $m_V \leq 20$ mag. The values of $f_j$ considered in this case are those plotted in the bottom panel of the figure. 

Since we had already removed a certain amount of field objects on the basis of the spectroscopic information (that excluded their membership), we risked overcorrecting the field subtraction. In order to avoid this, we calculated the intensity per square degree associated with all the sources in the rejected-objects catalog ($I_{\it rej}$), i.e.,

\begin{equation}
I_{\it rej} = \sum_{k=1}^{q} \frac{10^{-0.4 m_k}}{A} {\rm ,}
\label{eq:rejected}
\end{equation}

\noindent where $k$ is the index of the considered rejected object, $q$ the total number of the rejected galaxies, $m_k$ the magnitude of the $k$-th object, and $A$ the FoV area in units of square degrees.
The lower magnitude limit corresponds to the lower limit of the tabulated Berta number counts, and the upper magnitude limit to our photometric cut.

We therefore calculated the field galaxies intensity per square degree as

\begin{equation}
I_{\it field} = I_{\it Berta} - I_{\it rej} {\rm .}
\label{eq:field}
\end{equation}

\subsection{Photometric profiles}
\label{sec:profiles}

Equation~\ref{eq:i} allowed us to calculate the integrated intensity at the distance of every source from the BCG, but to have equally spaced points we rebinned the data through a weighted least squares (WLS) interpolation with a sampling of $0.05 \ r_{200}$. Then we converted the intensity profiles into integrated absolute magnitude profiles ($M_B(\leq r)$ and $M_V(\leq r)$) and we derived the integrated color index profiles

%\begin{equation}
%\begin{array}{lll}
%M_B(\leq r) = -2.5 \, {\rm log}_{10} \left( I_B(\leq r) \right) - 5 \, {\rm log}_{10} (D) + 5 {\rm ,} \\
%M_V(\leq r) = -2.5 \, {\rm log}_{10} \left( I_V(\leq r) \right) - 5 \, {\rm log}_{10} (D) + 5 {\rm ,} \\
%\end{array}
%\label{eq:mag}
%\end{equation}

%\noindent where $I_B(r)$ and $I_V(r)$ are the WLS interpolated intensity within the radius $r$ and $D$ is the luminosity distance of the cluster in parsec derived from the distance modulus provided by \cite{var9};

\begin{equation}
(B-V)(\leq r) = M_B(\leq r) - M_V(\leq r) {\rm ,}
\label{eq:int_color}
\end{equation}

\noindent and the local color index profiles

\begin{equation}
(B-V)(r) = M_B(r) - M_V(r) {\rm ,}
\label{eq:diff_color}
\end{equation}

\noindent where $M_B(r)$ and $M_V(r)$ are the local values at each radius $r$ obtained by differentiating the integrated values.

Finally, we applied the K correction following \cite{Poggianti1997} using the local color index in each bin radius and the mean redshift of the clusters. If $(B-V)\geq0.8$ we applied the mean correction valid for early-type systems, if $0.8<(B-V)\leq0.5$ we used the typical correction of Sa galaxies, and if $(B-V)<0.5$ we adopted the correction valid for Sc and Irregular galaxies. This is clearly an approximation, as the K correction should be applied to the magnitude of each galaxy by a precise knowledge of its morphological type and redshift, which are available only for the spectroscopic sample. However, as long as our procedure is correct the mean K correction of the galaxy population within each radial bin can reliably be calculated by introducing errors not larger than 0.05 mag.

Finally we got the surface brightness profiles

\begin{equation}
\begin{array}{lll}
\mu_B (r) = -2.5 \, {\rm log} \left( I_B(r)/A_{\it ring} \right) {\rm ,} \\
\mu_V (r) = -2.5 \, {\rm log} \left( I_V(r)/A_{\it ring} \right) {\rm ,} \\
\end{array}
\label{eq:sb}
\end{equation}

\noindent where $I_B(r)$ and $I_V(r)$ are the local K-corrected values of the intensity measured in a ring of area
$A_{\it ring} = \pi \left( (r + 0.025 \, r_{200})^2 - (r - 0.025 \, r_{200})^2 \right)$ at each position $r$.

\begin{figure*}[t]
   \centering
        \includegraphics[width=0.45\textwidth]{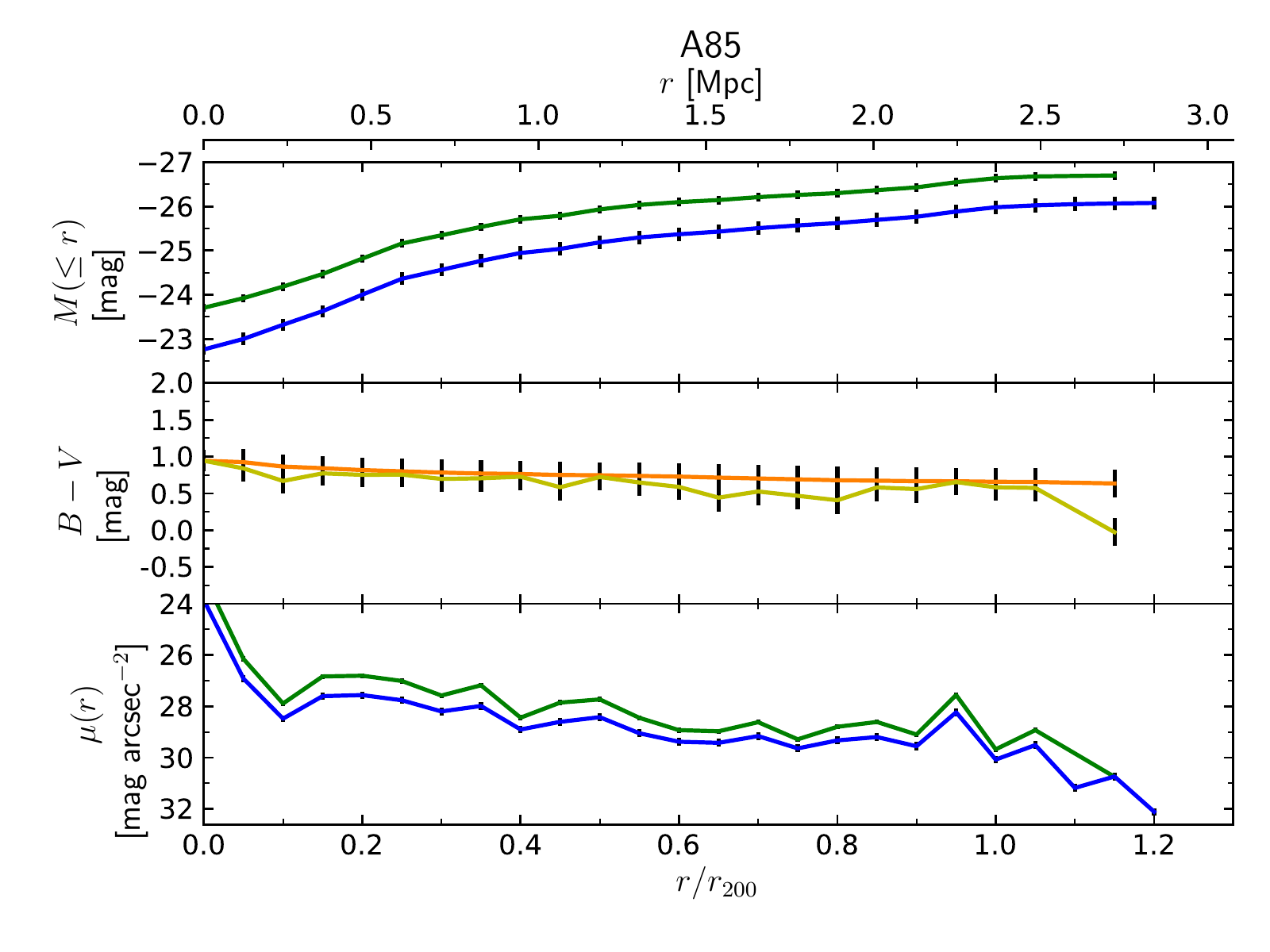}
        \includegraphics[width=0.45\textwidth]{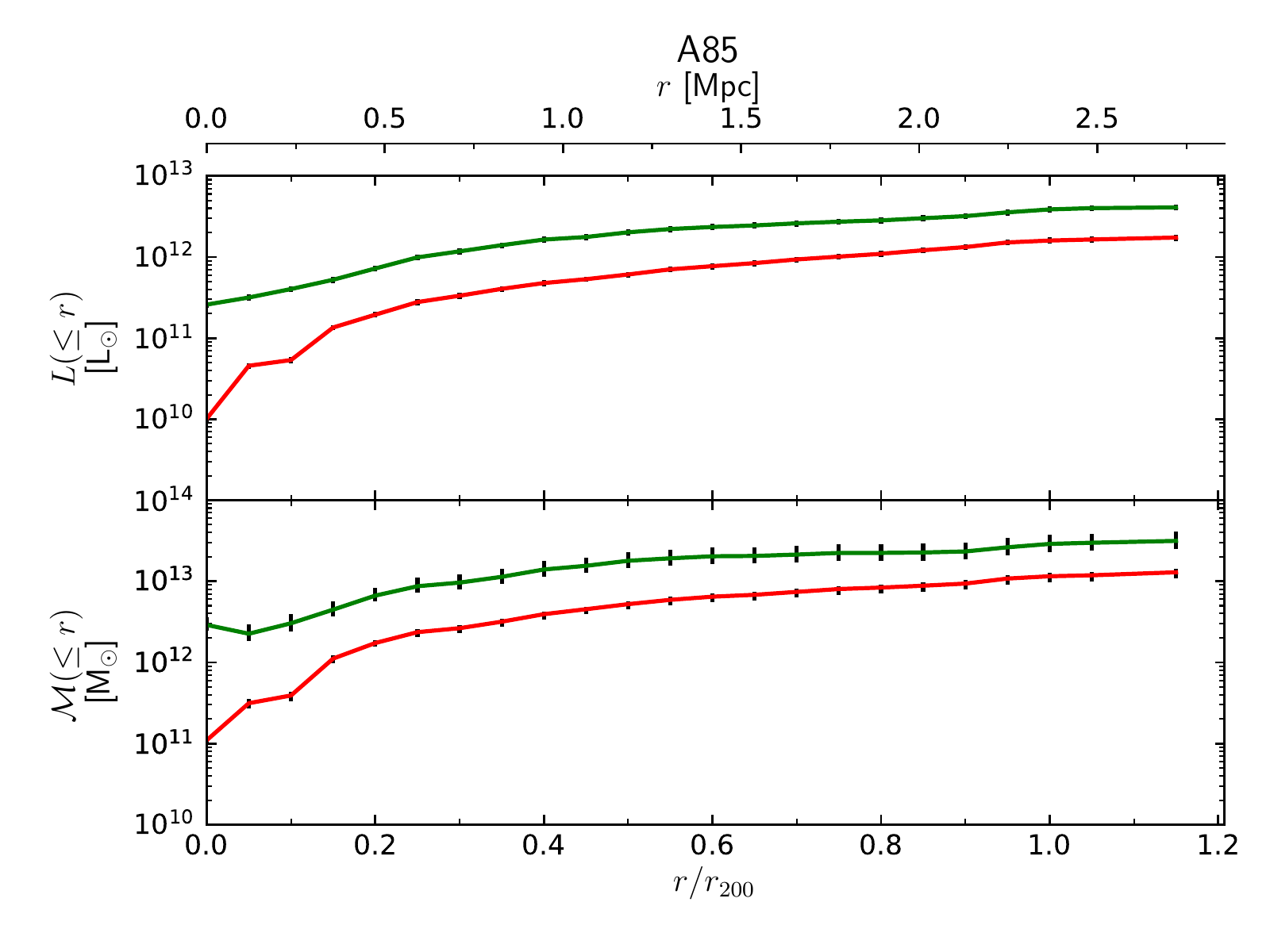}
    \caption{Photometric and stellar mass profiles of the cluster A85. Left panel, upper plot: Integrated magnitude profiles, where the blue lines correspond to the $B$-band data and green lines to the $V$-band data, are shown. Left panel, central plot: The color index profile, where the orange lines correspond to the integrated color index $(B-V)(\leq r)$ and yellow lines to the local color index $(B-V)(r)  $ are shown. Left panel, lower plot: Surface brightness profiles, with the same color code of the upper plot, are shown. Right panel, upper plot: $V$-band integrated luminosity profile for the objects in the photometric (green line) and spectroscopic (red line) samples are shown. Right panel, lower plot: The integrated mass profile for the objects in the photometric and spectroscopic samples, with the same color code as the upper plot, is shown.}
    \label{fig:plots-example}
\end{figure*}

\subsection{Correction for faint objects}
\label{sec:faint-obj-corr}

What is lacking in our luminosity profiles is a quantification of the total light coming from the sources fainter than our magnitude cut.
In order to get such contribution, we used the parametrization of the luminosity function (LF) provided by \cite{mor15} for the $V$-band data of the stacked WINGS sample. It is based on a double Schechter function, i.e.,

\begin{equation}
\phi(L) = \left( \frac{\phi_V^b}{L_V} \right) \cdot \left( \frac{L}{L_V^b} \right)^{\alpha_b} \cdot e^{-L/L_V^b} + \left( \frac{\phi_V^f}{L_V} \right) \cdot \left( \frac{L}{L_V^f} \right)^{\alpha_f} \cdot e^{-L/L_V^f} {\rm ,}
\label{eq:dsf}
\end{equation}

\noindent where $\phi_V^b$ and $\phi_V^b$ are normalization constants, $L_V^b$ is the luminosity associated with $M_V^b = -21.25$ mag, and $\alpha^b = -1.10$, $L_V^f$ is the luminosity associated with  $M_V^f = -16.25$ mag, and $\alpha^f = -1.5$.

This allowed us to calculate an approximated LF correction $c_{LF}$, which is valid under the implicit assumptions that all the clusters have a similar LF and that $\phi(L)$ does not depend on $r$ as the ratio between the total expected $V$-band cluster intensity $I_{\it tot}$ and the observed intensity $I_{\it obs}$,

\begin{equation}
c_{\rm LF} = \frac{I_{\it tot}}{I_{\it obs}} = \frac{\int_{0}^{L_{\rm BCG}} L \cdot \phi(L)dL}{\int_{L_{\it V,cut}}^{L_{\rm BCG}} L \cdot \phi(L)dL} {\rm ,}
\label{eq:lfc}
\end{equation}

\noindent where $L_{\rm BCG}$ is the $V$-band luminosity of the BCG, and $L_{\it V,cut}$ is the $V$-band luminosity at magnitude 20. The two integrals represent the luminosity density associated with a distribution of objects with LF $\phi(L)$ and luminosity within the integration interval. Both the integrals can be solved through the incomplete gamma function and lead to a correction on the order of, at most, $5\%$.

Since it is a very small value and the $B$-band LF was not derived, we decided to not apply such a correction to our photometric profiles.

\section{Stellar mass profiles of clusters}
\label{sec:mass}

The spectrophotometric masses of all the galaxies in the spectroscopic sample are already public (\citealt{fri11}) or have been measured by Moretti et al. (private communication) with the same spectral energy distribution (SED)-fitting procedure described in \cite{fri7}. Because the memberships of these objects are known on the basis of redshift measurements, we could proceed to calculate the stellar light profiles by repeating the same procedure described in Section~\ref{sec:photometry} with two fundamental differences. First, the photometric completeness correction $c_{\it ph}$ in Equation~\ref{eq:compl} is now significantly larger than 1. In fact, the spectroscopic sample at $m_V=20$ is more than 80\% complete (\citealt{mor17}). However, in this case we can get a more precise measurement of $c_{\it ph}$ because the photometric sample is approximately 100\% complete. 
Second, the statistical field subtraction is not needed, as the membership of each object is known.

The integrated spectroscopic stellar mass profiles can be calculated according to the following formula:

\begin{equation}
\mathcal{M}_{\it sp}(\leq r_n) = \sum_{i=1}^{n} c_{\it ph}(m_i) \cdot {\it ac}(r_i) \cdot \mathcal{M}_i {\rm ,}
\label{eq:M}
\end{equation}

\noindent where $\mathcal{M}_i$ is the mass of the $i$-th galaxy, $c_{\it ph}(m_i)$
is the ratio between the total number of objects in a given bin of magnitude in the photometric and spectroscopic samples,
and the last term has already been defined in Section~\ref{sec:photometry}. As for the photometric profiles, the stellar mass profiles were also rebinned to have equally spaced points every 0.05 $r_{200}$.

The total photometric stellar mass profiles of the clusters can be finally obtained through the relation 

\begin{equation}
\mathcal{M}_{\it ph} (\leq r) = c_{\rm LF} \cdot \mathcal{M}_{\it sp} (\leq r) \cdot \frac{L_{\it ph} (\leq r)}{L_{\it sp} (\leq r)} {\rm ,}
\label{eq:mph}
\end{equation}

\noindent where $L_{\it ph} (\leq r)$ and $L_{\it sp} (\leq r)$ are the integrated luminosity within the radius $r$ of the photometric and spectroscopic samples, respectively, and $c_{LF}$ is the above-mentioned correction for faint objects. Behind this relation there is the implicit assumption that the measured mass-to-light ratio at each radius is representative of the true mass-to-light
ratio in the cluster. This is a valid assumption since the spectroscopic sample well represents the photometric sample within $m_V = 20$ mag. 

\section{Final light and mass profiles of clusters}
\label{sec:res}

In this section we discuss the properties of the light and stellar mass profiles of our clusters created through the aforementioned procedure. In total we got the stellar mass profiles of 42 of our 46 clusters, reaching $r_{200}$ in 30 cases and exceeding $2 \, r_{200}$ in 3 of them. For the mass values at various radii, see Table~\ref{tab:phot_and_mass} in Appendix.

The left panel of Figure~\ref{fig:plots-example} shows the photometric profiles of the cluster A85; all the others can be found in Appendix. In the upper plot we see the integrated magnitude profiles, in the central plot the integrated and local $(B-V)$ color profiles, and in the lower plot the surface brightness profiles. Blue and green lines are $B$- and $V$-band data, respectively while yellow and  orange lines are the local and integrated colors. In the right part of Figure~\ref{fig:plots-example} we present the light and mass profiles of A85 for the photometric and spectroscopic samples. The upper panel shows the integrated $V$-band luminosity of the photometric (green line) and spectroscopic (red line) samples, while the lower panel shows the associated stellar mass profiles.

A number of considerations emerge from these plots: Most of the growth curves seem to be still increasing at the maximum photometric radius $r_{\it max,ph}$. The $(B-V)$ integrated color is that typical of an evolved stellar population ($\overline{(B-V)} \sim 1$) and usually shows a gradient between the central and outer regions, which are redder in the center and bluer in the outskirts; in the most extreme case (i.e.,~IIZW108) it is equal to $\Delta (B-V)/\Delta r = 0.36$ mag $r_{200}^{-1}$. The local colors are, generally, more noisy than the integrated colors because of the lack of sources in some radial bins. The surface brightness profiles, although dominated by random fluctuation at the adopted spatial binning, show a clear cusp in the central region and very different gradients when the profiles are plotted in units of $r_{200}$. The spectroscopic and photometric light and stellar mass profiles are very different from cluster to cluster.

Concerning the last point we believe that the origin of the systematic difference between the photometric and spectroscopic profiles is due to the observational difficulty of
positioning the multi-object spectrograph fibers to get a simultaneous coverage of the whole cluster region, particularly in the dense core of the clusters.
In most cases the spectra of the most luminous galaxies in the cluster center have not been obtained and sometimes even the BCG spectrum is missing. 
The consequence is that the completeness correction $c_{\it ph}(m_i)$  could not be calculated in certain magnitude bins, i.e.,
the ratio between the total number of objects in a given magnitude bin from
the photometric sample and corresponding number from the spectroscopic sample. Hence, the lost flux could not be redistributed between the observed sources (e.g., in Equation~\ref{eq:M}), resulting in a net displacement of the two curves.
The effect appears to be larger in the center and smaller in the outer regions (e.g., Figure~\ref{fig:plots-example}), thus supporting our explanation.
Clearly, the correct mass and light profiles are those based on the most complete photometric sample.

\subsection{Profile fitting and effective parameters calculation}
\label{sec:eff}

In order to obtain the main structural parameters and the asymptotic luminosity of our clusters we decided to fit the growth curves with some empirical models. In choosing a model we made the following considerations. The surface brightness profiles display a central cusp followed by a steady decrease, as in many ETGs following the S\'ersic profile or when the bulge and disk components are both visible in late-type galaxies. \cite{mar4} preliminarily attempted the fit of the WINGS cluster profiles using the King (\citeyear{king62}) and the De Vaucouleurs (\citeyear{dev48}) laws, and the King model was also used by, for example, \cite{ada98} to fit the number density of clusters. In case of hydrostatic equilibrium and isothermality of the intracluster medium (ICM), the ICM intensity profile has traditionally been reproduced by the standard $\beta$-model (\citealt{cav76}; \citealt{for84}). In principle this model should be able to reproduce correctly the stellar light profiles of our clusters as well because clusters are thought to be scale invariant (see, e.g., \citealt{kai86}, or \citealt{nav97}), and because both the ICM and the stellar light distribution are tracers of the same DM potential well.

Consequently, we decided to fit the integrated luminosity profiles of our clusters by using the same empirical laws used for ETGs, i.e., the King, the S\'ersic (\citeyear{ser63}, \citeyear{ser68}), and with the standard $\beta$-models.

The integrated light for a King profile is given by the following expression:

\begin{equation}
L(\leq r) = \int 2 \pi r \, k \left( 1/\sqrt{r^2/r_c^2+1} - 1/\sqrt{r_t^2/r_c^2+1} \right)^2 dr + L_{\rm ZP}  {\rm ,}
\label{eq:king}
\end{equation}

\noindent where $k$ is a scale factor, $r$ is the radius, $r_c$ is the core radius, $r_t$ the tidal radius, and $L_{\rm ZP}$ the zero-point luminosity (i.e., the luminosity of the BCG).

\begin{figure*}[t]
        \includegraphics[width=\textwidth]{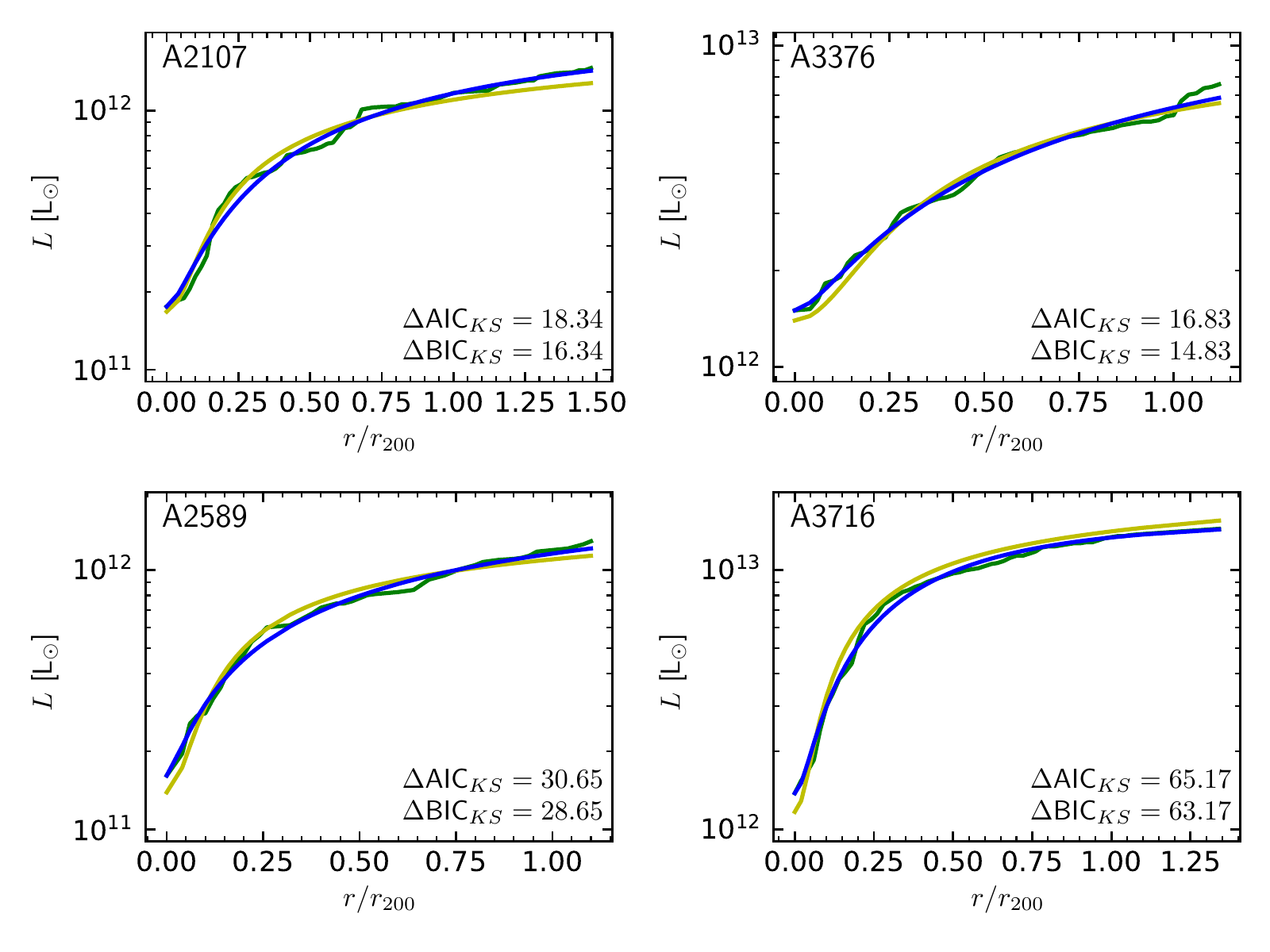}
        \caption{Growth curves of the 4 clusters (green lines) best fitted with the King models (yellow lines) and corresponding S\'ersic fits (blue lines). The cluster name is shown in the upper left corner of each panel, while in the lower right corner we plotted the discrepancy between the two models quantified with the two criteria defined in Section~\ref{sec:eff}.}
        \label{fig:king-sersic}
\end{figure*}

The integral can be solved as

\begin{equation}
\begin{array}{lll}
L(\leq r) & = & \pi k \left\lbrace r_c^2 \, {\rm log}\left( r^2/r_c^2 + 1\right) + 1/\sqrt{r_t^2/r_c^2+1} \, \cdot \right. \\
& \cdot & \left. \left[ r^2/\sqrt{r_t^2/r_c^2+1} + r_c^2 \, \cdot \right. \right. \\
& \cdot & \left. \left. \left( 1/\sqrt{r_t^2/r_c^2+1} - 4\sqrt{r^2/r_c^2 +1} \right)\right]\right\rbrace + L_{\rm ZP} {\rm .}\\
\end{array}
\label{eq:king_i}
\end{equation}

The S\'ersic law is now widely adopted to fit the profiles of ETGs (see, e.g., \citealt{cao93}; \citealt{don94}).
The integrated profile is given by

\begin{equation}
L(\leq r) = \int 2 \pi r \, I_{\it eff} e^{-b_n\left[ (r/r_{\it eff})^{1/n} -1 \right]} dr + L_{\rm ZP} {\rm ,}
\label{eq:sersic}
\end{equation}

\noindent where $r_{\it eff}$ is the effective radius (i.e., the radius containing half of the total luminosity), $I_{\it eff}$ the effective intensity (i.e., the intensity at $r = r_{\it eff}$), $n$ the S\'ersic index, and $b_n$ a function of $n$ that can only be numerically derived; for $n \leq 0.5,$ we used the \citealt{mac3} approximation and for $n > 0.5$ we used the \citealt{pru97} approximation. This leads to

\begin{equation}
L(\leq r) = 2 \pi n \, r_{\it eff}^2 \, I_{\it eff}  \, \frac{e^{b_n}}{b_n^{2n}} \, \gamma \left( 2n,b_n(r/r_{\it eff})^{1/n} \right) + L_{\rm ZP} {\rm ,}
\label{eq:sersic_i}
\end{equation}

\noindent where $\gamma \left( 2n,b_n(r/r_{\it eff})^{1/n} \right)$ is the incomplete gamma function. 

For more reliable fits, especially in the central regions, where the integrated luminosity steeply rises, we rebinned the profiles with a radial spacing of $0.02 \, r_{200}$ and then we applied the models to the data and we realized that the integrated King profiles systematically fail to reproduce the data, when compared to the S\'ersic profiles. In Figure~\ref{fig:king-sersic} we show the four best King fits that we produced and the adopted goodness-of-fit criteria values, which are explained in Section~\ref{sec:bm}, used to discriminate between the two models. In all cases, the goodness-of-fit criteria strongly point toward the S\'ersic model, and this evidence is even stronger for the remaining 42 clusters. In addition, to correctly reproduce the bulk of the points, the King profiles cannot match the central luminosity values, which are known for their high precision. As a consequence, we decided to reject the King model.

\begin{figure*}[t]
        \includegraphics[width=\textwidth]{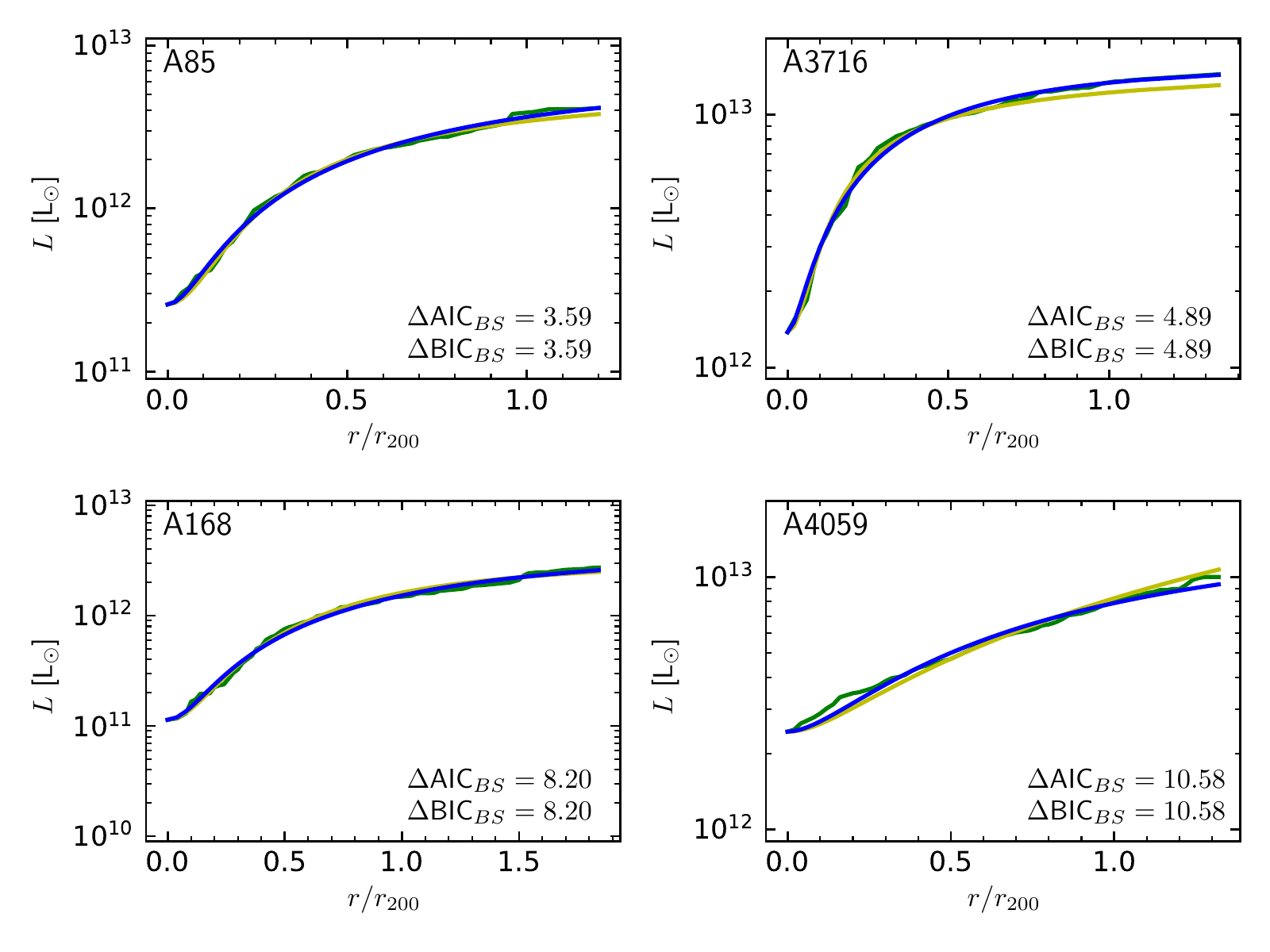}
        \caption{Light profiles of the 4 clusters (green lines) best fitted with the $\beta$-models (yellow lines) and the corresponding S\'ersic models (blue lines). The cluster name is shown in the upper left corner of each panel, while in the lower right corner we plot the discrepancy between the two models quantified with the two criteria defined in Section~\ref{sec:eff}. Since the number of free parameters and data points are equal in the case of the two models, $\Delta {\rm BIC}_{\rm BS} = {\rm BIC}_{\rm Beta} - {\rm BIC}_{\rm Sersic} = \Delta {\rm AIC}_{\rm BS} = {\rm AIC}_{\rm Beta} - {\rm AIC}_{\rm Sersic} = \Delta \chi^2$.}
        \label{fig:beta-sersic}
\end{figure*}

The same conclusion was reached using the standard $\beta$-models, whose luminosity profile can be calculated as

\begin{equation}
L(\leq r) = \int 2 \pi r \, I_0 \, \left[ 1 + \left(\frac{r}{r_c}\right)^2 \right]^{0.5 - 3 \, \beta} dr + L_{\rm ZP} {\rm ,}
\label{eq:beta}
\end{equation}

\noindent where $I_0$ is the central intensity, $r_c$ the core radius, and $\beta$ the ratio of the specific energy of the galaxies to the specific energy of the gas. The integral can be easily solved as

\begin{equation}
L(\leq r) = \frac{2 \pi r_c^2 \, I_0}{3 - 6 \beta} \, \left[ 1 + \left( \frac{r}{r_c} \right) \right]^{1.5 - 3 \beta} + L_{\rm ZP} {\rm .}
\label{eq:beta_i}
\end{equation}

\begin{figure}[t]
   \centering
        \includegraphics[width=0.45\textwidth]{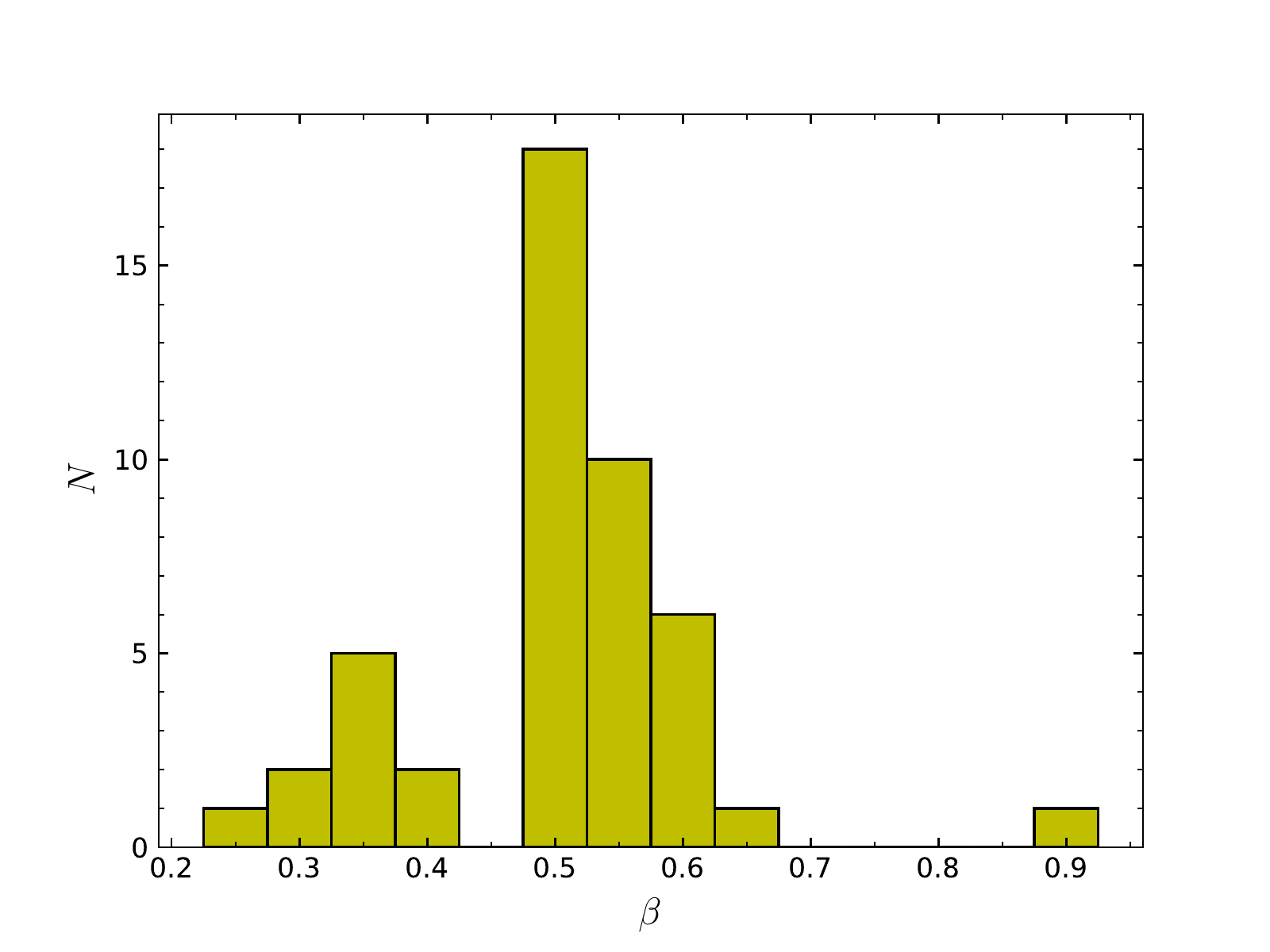}
    \caption{Histogram of all the $\beta$ parameters values derived starting from the growth curve fitting.}
    \label{fig:hist_beta}
\end{figure}

Overall, the standard beta-model provides a worse representation of the integrated light profiles than the S\'ersic model. In Figure~\ref{fig:beta-sersic} there is a visual representation of four clusters best fitted by the $\beta$-models in comparison with the  S\'ersic fits. In this case too the goodness-of-fit criteria point toward the choice of the S\'ersic model with only a few borderline cases and more than 40 out of 46 cases strongly in favor of the latter. However,  most of the derived $\beta$ parameter values are compatible with those typical of the ICM in galaxy groups (see, e.g., \citealt{pon93}) and clusters (e.g., \citealt{moh97}), which span the range from $\beta \sim 0.5$ to $\beta \sim 0.65$ (Figure~\ref{fig:hist_beta}).

\begin{figure*}[t]
   \centering
        \includegraphics[width=0.45\textwidth]{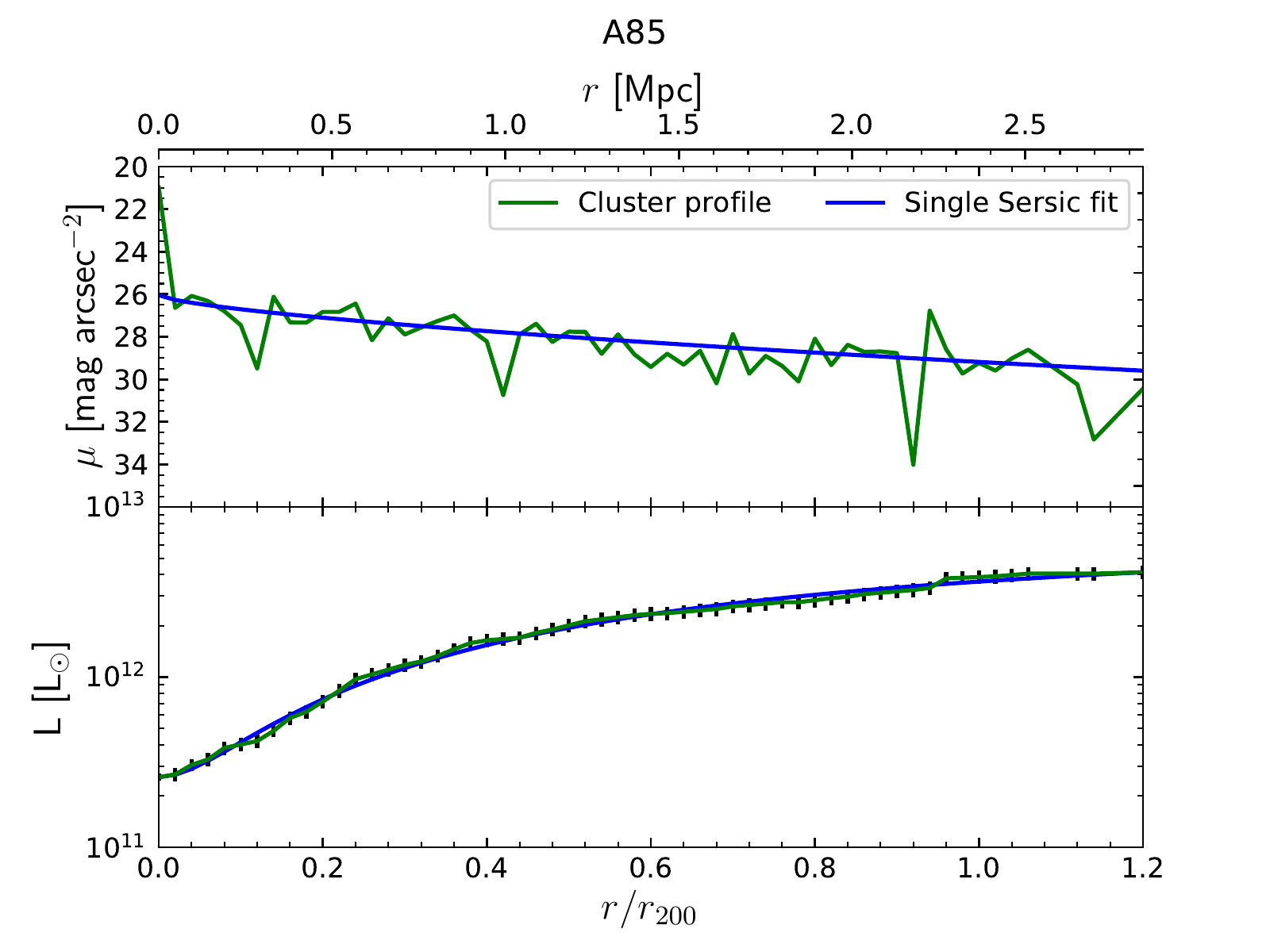}\includegraphics[width=0.45\textwidth]{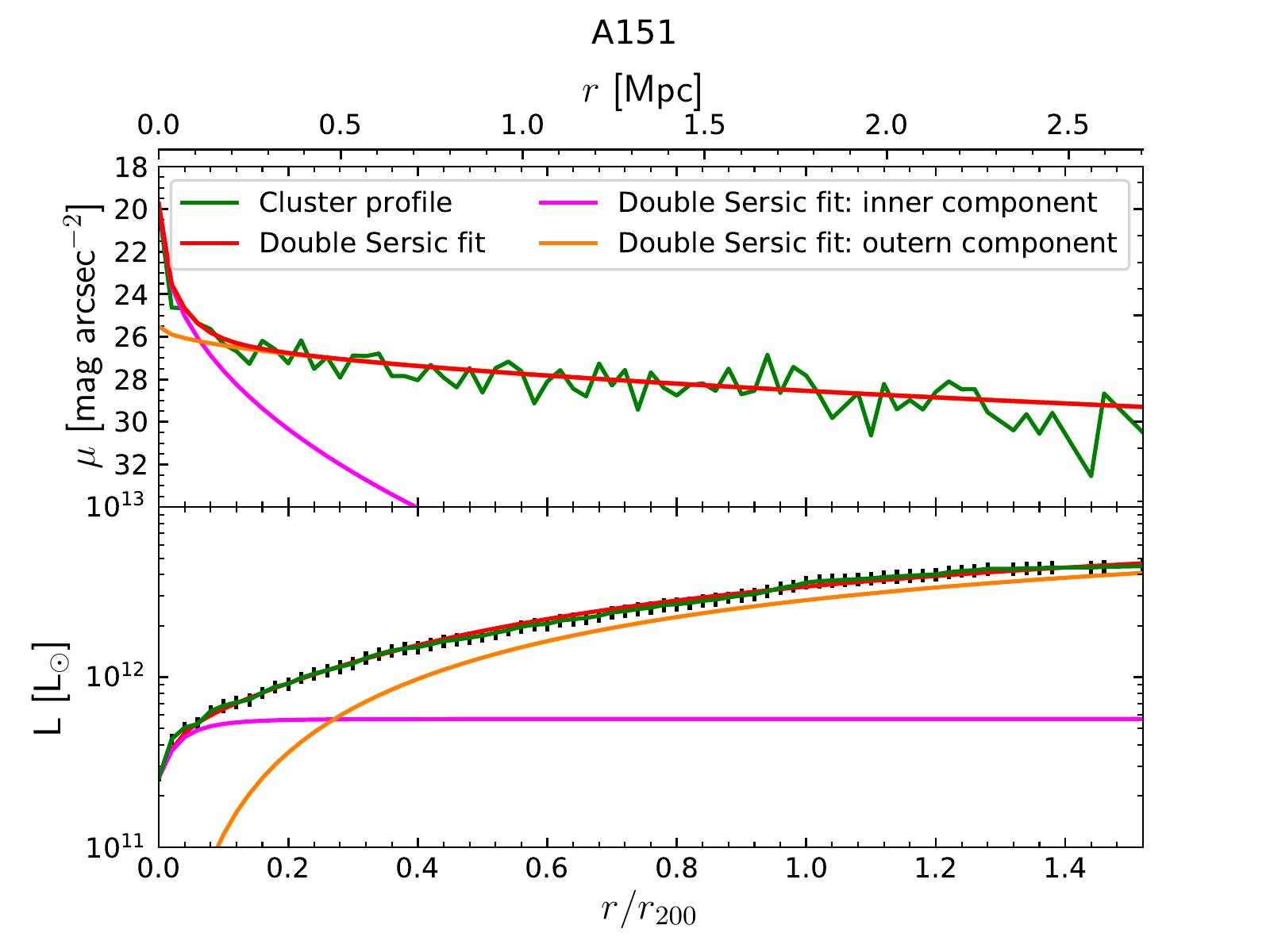}
    \caption{Left panel: $V$-band surface brightness and integrated luminosity profiles of the cluster A85 (green line) with the best single S\'ersic fit (in blue). Right panel: $V$-band surface brightness and integrated luminosity profiles of the cluster A151 (green line) with the best double S\'ersic fit (in red) are shown; the pink and orange lines indicate the inner and outer S\'ersic components, respectively.}
    \label{fig:fits-es}
\end{figure*}

\begin{figure*}
        \includegraphics[width=0.45\textwidth]{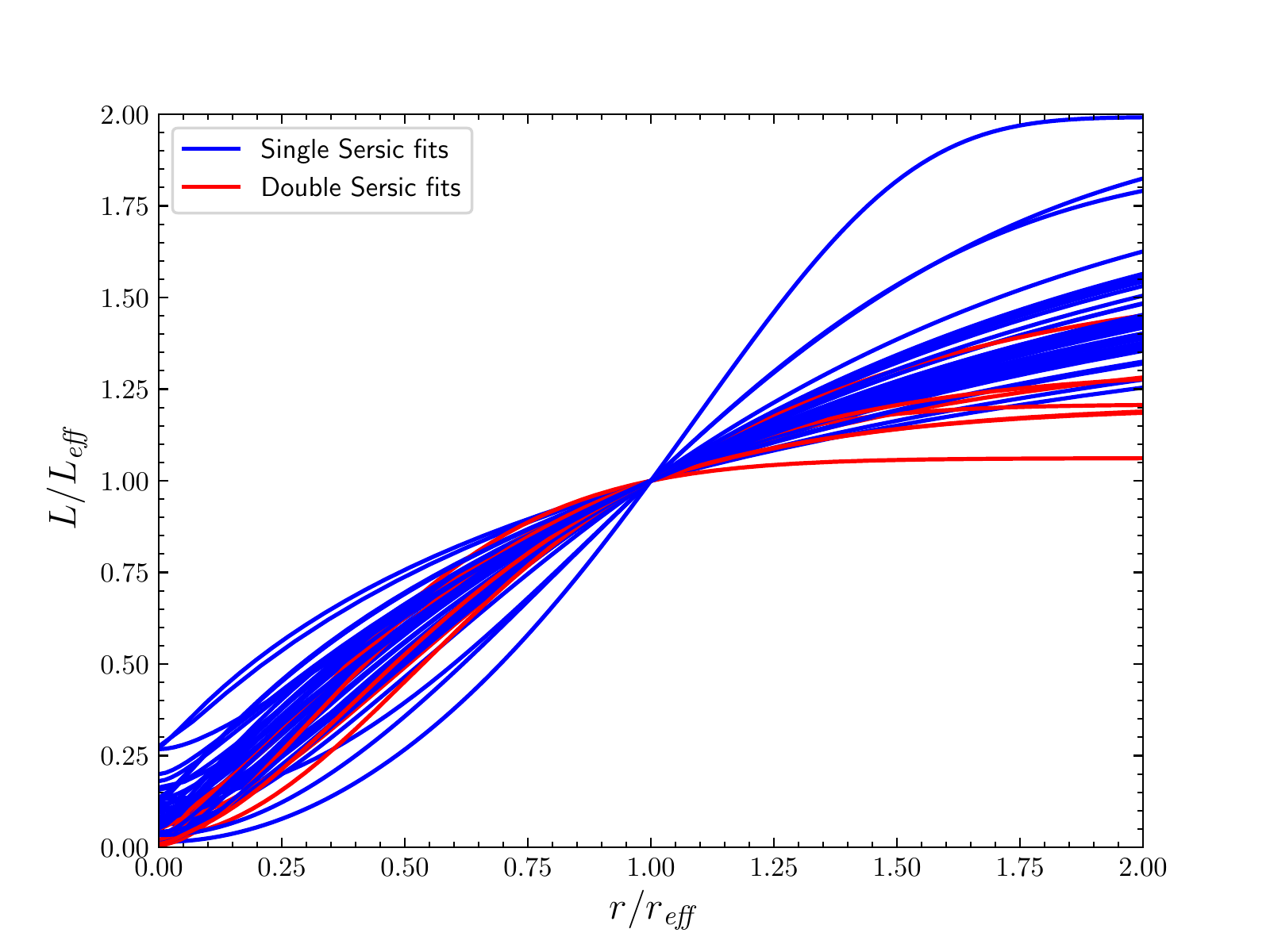}\includegraphics[width=0.45\textwidth]{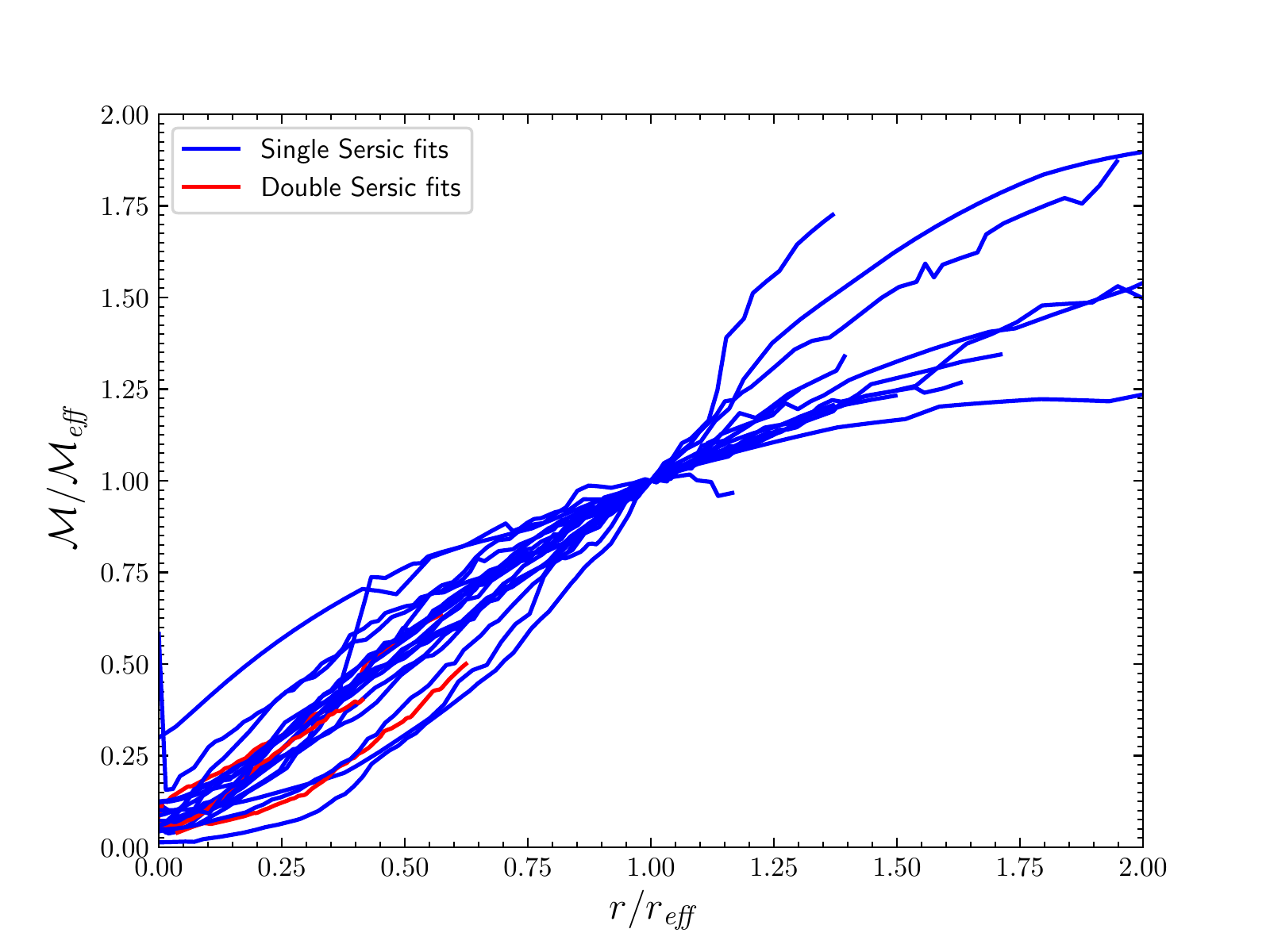}
        \includegraphics[width=0.45\textwidth]{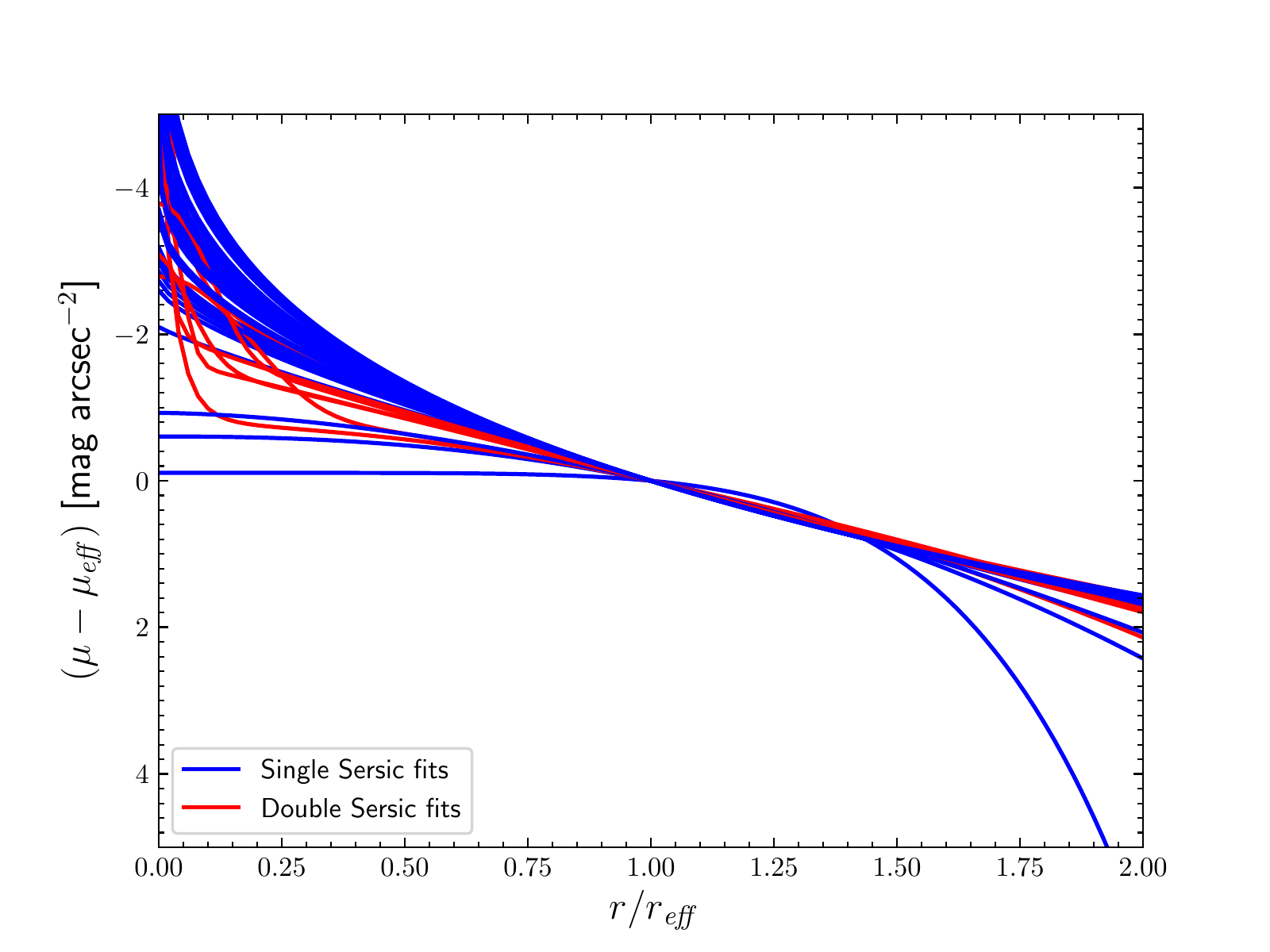}\includegraphics[width=0.45\textwidth]{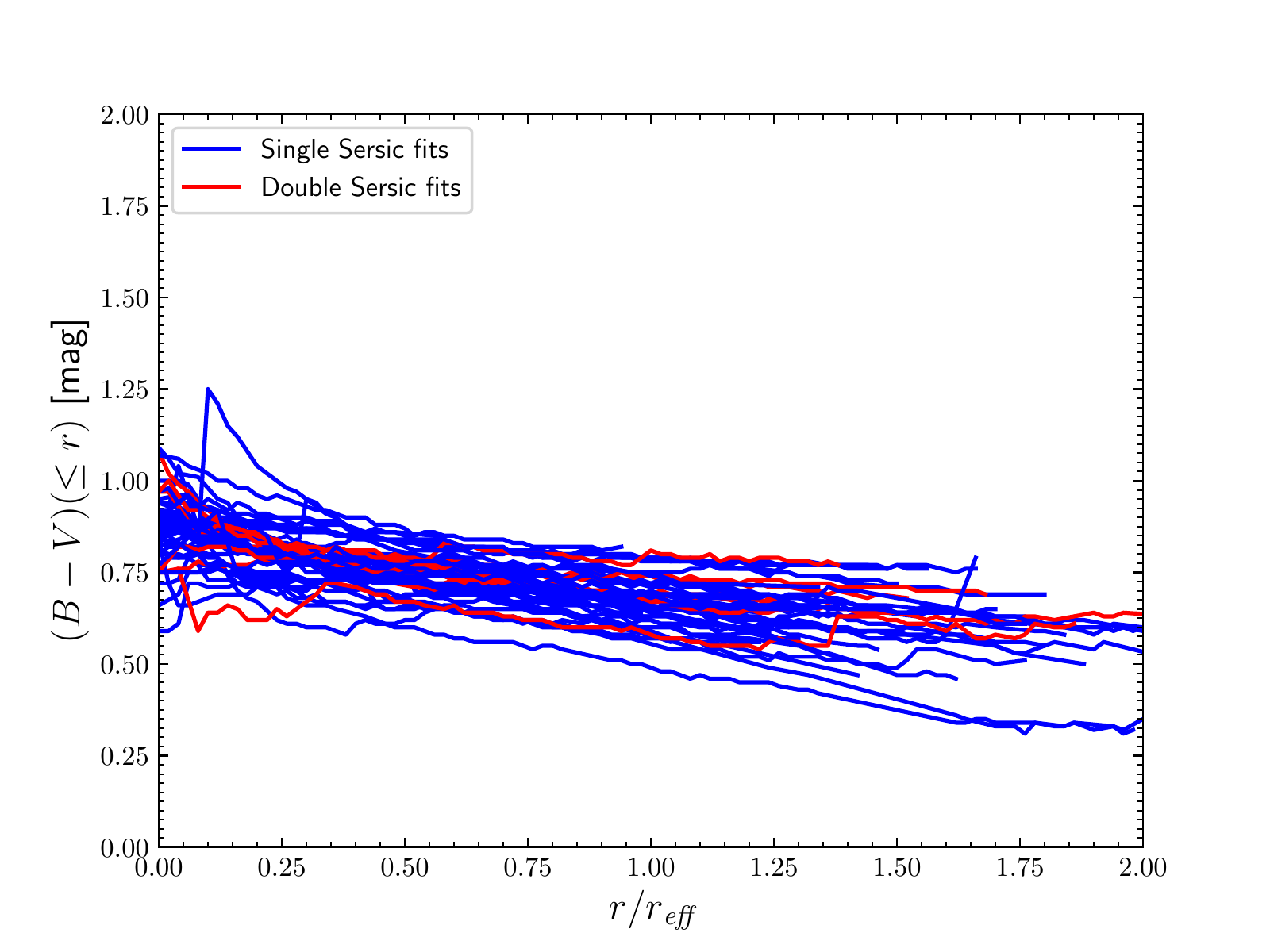}
        \caption{Left panels: Superposition of the luminosity and surface brightness profiles associated with all the best fit models, normalized to the effective parameters. Right panels: The mass and color profiles with the same normalization are superimposed. Blue and red lines are associated with the single and double S\'ersic fits used to reproduce the luminosity profiles (see Section~\ref{sec:eff} for details).}
        \label{fig:stack}
\end{figure*}

The integrated S\'ersic profile was able to correctly reproduce most of the observed profiles, but the presence of some luminosity bumps in the profiles of some clusters resulted in some poor fits.
In these cases (see plots in Appendix) an integrated double S\'ersic profile (i.e., the superposition of two S\'ersic profiles) was used to better reproduce the luminosity profiles.

\subsection{Best model selection}
\label{sec:bm}

Since by increasing the number of free parameters the $\chi^2$ obviously decreases, we tested the goodness of our fits 
with the single and double S\'esic laws through the following criteria:

\begin{itemize}

\item Akaike (\citeyear{aka73}) information criterion (AIC),

\begin{equation}
{\rm AIC} = \chi^2 + 2k + \frac{2k \, (k+1)}{N-k-1} {\rm ,}
\label{eq:aic}
\end{equation}

\noindent where $k$ is the number of free parameters and $N$ the number of data points to be fitted, and the

\item Bayesian information criterion (BIC, \citealt{sch78}), where

\begin{equation}
{\rm BIC} = \chi^2 + k \, {\rm ln} \left( N \right) {\rm .}
\label{eq:bic}
\end{equation}

\end{itemize}

The rule for both criteria is to choose the model able to get the minimum AIC or BIC.

A comparison between these criteria can be found in \cite{bur2}, according to which both the criteria can be obtained, by changing the prior, in the same Bayesian context. These authors identified two main theoretical advantages of the AIC over the BIC. First, the AIC is derived from the principles of information, while the BIC is not; and second, the BIC prior is not sensible in the information theory context. Moreover, as a result of simulations, the authors also concluded that the AIC is less biased than the BIC.

Despite this, we chose to favor the BIC over the AIC for two main reasons. First, the BIC is built starting from a vague or uniform prior \citep{bur2}, which is a good assumption in our context, in which we have no theoretical justification to privilege one category of models with respect to the other. Second, the BIC penalizes more strongly models with a higher number of free parameters \citep{kr95}, thus it reduces the risk of adopting overcomplicated models.

In order to understand how strongly one model is favored in comparison with another, we used the criterion defined by \cite{kr95}, according to which if we call $\Delta {\rm BIC}$ the difference between the BICs of the two models:

\begin{itemize}

\item $0 \leq \Delta {\rm BIC} < 2$ is not worth more than a bare mention
\item $2 \leq \Delta {\rm BIC} < 6$ indicates positive evidence toward the lowest BIC model
\item $6 \leq \Delta {\rm BIC} < 10$ indicates strong evidence toward the lowest BIC model
\item $\Delta {\rm BIC} \geq 10$ indicates very strong evidence toward the lowest BIC model.

\end{itemize}

To further reduce the risk of adopting overcomplicated models, we chose favor the single S\'ersic models unless strong or very strong evidence in support of the double S\'ersic models was present. This resulted in 39 single and 7 double S\'ersic fits (e.g., Figure~\ref{fig:fits-es} for the clusters A85 and A151), whose effective parameters are tabulated in Table~\ref{tab:eff}. For all the double S\'ersic profiles, the inner component is always smaller and fainter, but with a higher value of surface brightness than the outer component.

For a recap of the fitting parameters see Tables~\ref{tab:single_fits} and \ref{tab:double_fits}, while the fits are shown in the figures in Appendix.

The confidence intervals around the best-fit parameters were calculated with the following procedure. The $\chi^2$ was recomputed giving to all the points a constant weight calculated in such a way to have a final $\chi^2 = N-k,$ where $N-k$ is the number of degrees of freedom. Then, each parameter was individually modified until the $\chi^2$ reached the value $N-k+9,$ i.e., the interval containing a probability of 99.73\% of finding the true parameter value.

In the double S\'ersic fits sometimes the threshold $N-k+9$ could not be reached and the limit was marked as ``undefined''. This happens in two possible cases: first, when the S\'ersic index $n_{\it in}$ cannot be safely constrained owing to a very limited number of data points in which the inner component dominates; second, when inside the central region the inner component is not significantly brighter than the outer component and the outer region displays a very disturbed profile, no significant increase in the $\chi^2$ is possible by increasing the inner-component effective radius value $r_{\it eff,in}$.

The confidence limits derived in this way are likely an overestimation of the true errors in the structural parameters because they were calculated by ignoring the mutual correlations that may exist between the parameters. All of these limits are on the order of few percent (see Table \ref{tab:single_fits}).

Almost all the profiles seem to well represent both the luminosity and surface brightness of our clusters, however at least in one case (i.e., A1631a) our best model selection criterion preferred a single S\'ersic fit where a human analysis of the surface brightness profile would suggest a two-component model.

In a few cases the total asymptotic luminosity may have been overestimated. In fact, in the case of A151, A754, and maybe also A3560, the fitted profile intercepts with an increasing trend the very upper edge of the growth curve, which instead appears to flatten. A possible way to improve the quality of these fits could be the implementation of a simultaneous minimization of the residuals of both the integrated luminosity and surface brightness profiles. Instead, the profiles of A1991, A2415, and A2657 display some significant sudden increases of the luminosity profile that could be due either to some ongoing major merger or the presence of important background structures.

No fit displays a central surface brightness higher than the observed surface brightness (within the error); in fact, to avoid unphysical divergencies at small radii all the models with higher values were immediately rejected, even if their reduced $\chi^2$ and their AIC or BIC parameter were preferable. The same was not done in case of much lower values because the central surface brightness has almost certainly been overestimated; in fact, the central point displays the luminosity of the BCG and of all the nearby galaxies entirely collapsed to a single point with no information concerning its real distribution. As a consequence, we chose not to limit our best model selection for matching a likely unrealistic value.

In a few cases (i.e., A147, A1631a, A1991, A2657, A2717, A3128, A3532, A3556, and IIZW108) the fitted profile is unable to correctly reproduce the fluctuations of the luminosity profile at small radii, however this is not, generally, a problem. In fact, in case of infalling structures larger than the adopted spatial scale we expect to see such fluctuations. Both A1991 and A2657 display important fluctuations at larger radii too. As a consequence, we can assume that these two systems either are experiencing some relevant merging event or are made of very strongly bound substructures.

Finally, the presence/absence of a cool core (see, e.g., \citealt{hen9}) does not seem to influence the number of components used to fit the luminosity profiles. In fact six of the seven clusters in our sample analyzed by \cite{hen9} can be parametrized with a single S\'ersic profile, even though two of these (i.e., A119, and A4059) have a cool-core and the remaining four (i.e., A85, A3266, A3558, and A3667) do not. The only exception is A3158, which has no cool core and a double-S\'ersic parametrization. No connection seems to exist either between the cool core presence/absence and the best-fit parameters values.

\section{Stacked profiles and main relations among structural parameters}
\label{sec:relations}

\subsection{Profiles analysis}
\label{sec:prof}

The various panels of Figure~\ref{fig:stack} show the whole set of luminosity, surface brightness, mass, and color profiles of the clusters stacked in four different plots and normalized to the effective structural parameters. Both in the central and  outer regions the cluster profiles show very different behaviors. The central surface brightness spans a range of $\sim6$ mag arcsec$^{-2}$, while the amount of light and mass within and beyond $r_{\it eff}$ appears to differ  up to a factor of $\sim2$ in units of $L_{\it eff}$ and $\mathcal{M}_{\it eff}$.

This is clearly an evidence of a marked difference in the global structure of clusters. Galaxy clusters do not seem to share
a common light and mass distribution.
The only similar behavior is visible in the stacked color profiles, showing that all the clusters have similar $(B-V)(\leq r)$ color profiles dominated by an old stellar population in the center and by a bluer population in the outer parts. Despite the large spread observed (around 0.3 mag), all the measured profiles are compatible with an old average stellar population.

As in the case of galaxies, the best way to establish the main cluster properties is to study the relations among the structural parameters. This is the aim of this section.

When available, we compared the measured parameters with those from a sample of 261 ETGs studied in our previous works (e.g., \citealt{don8}; \citealt{don14}), whose structural parameters (e.g., effective radius and luminosity, mass, velocity dispersion, and S\'ersic index) are available from the WINGS database. The idea is to quantify the correspondences and differences between the structural parameters of clusters and ETGs.

\begin{figure*}[t]
        \includegraphics[width=0.45\textwidth]{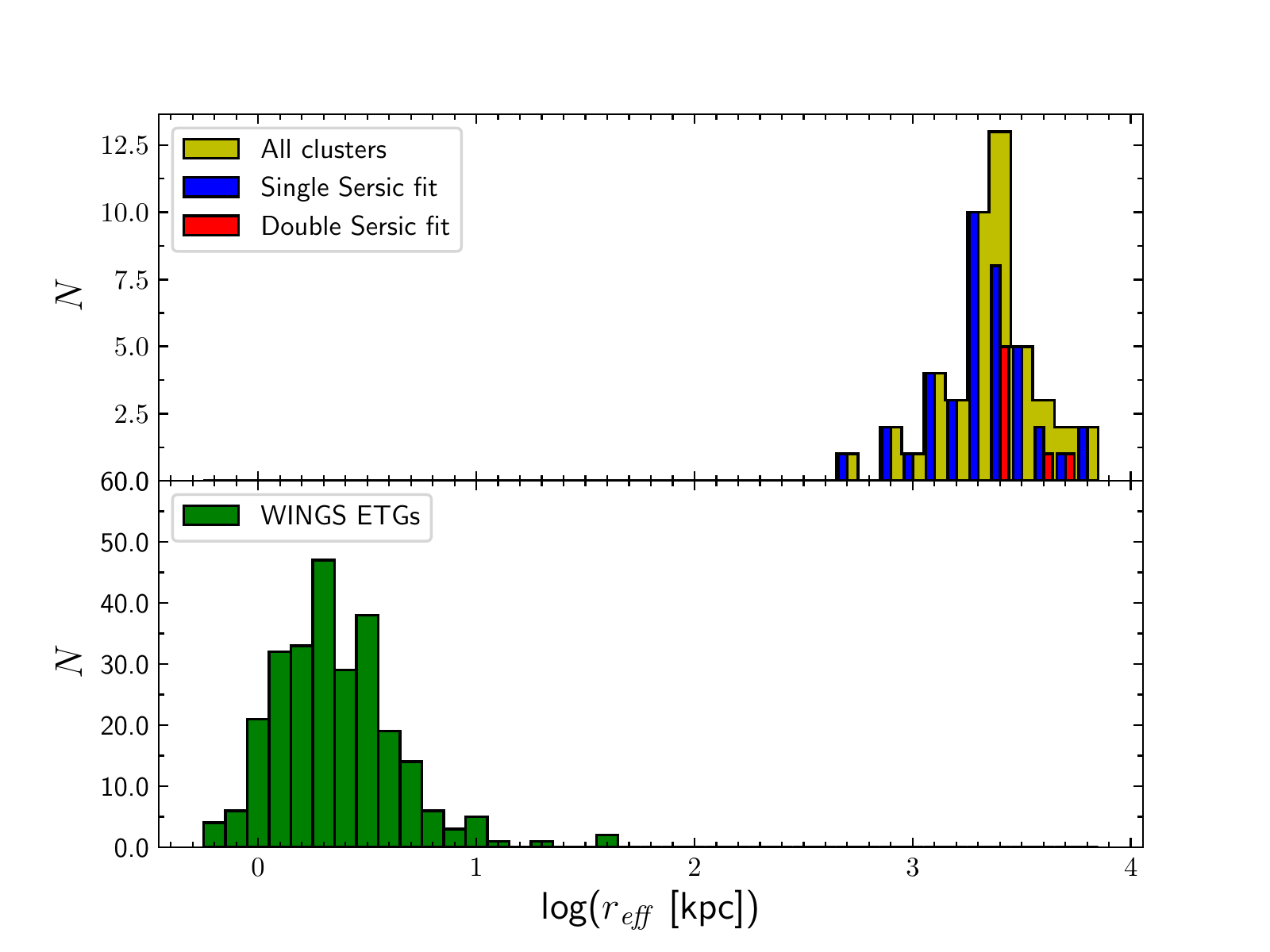}\includegraphics[width=0.45\textwidth]{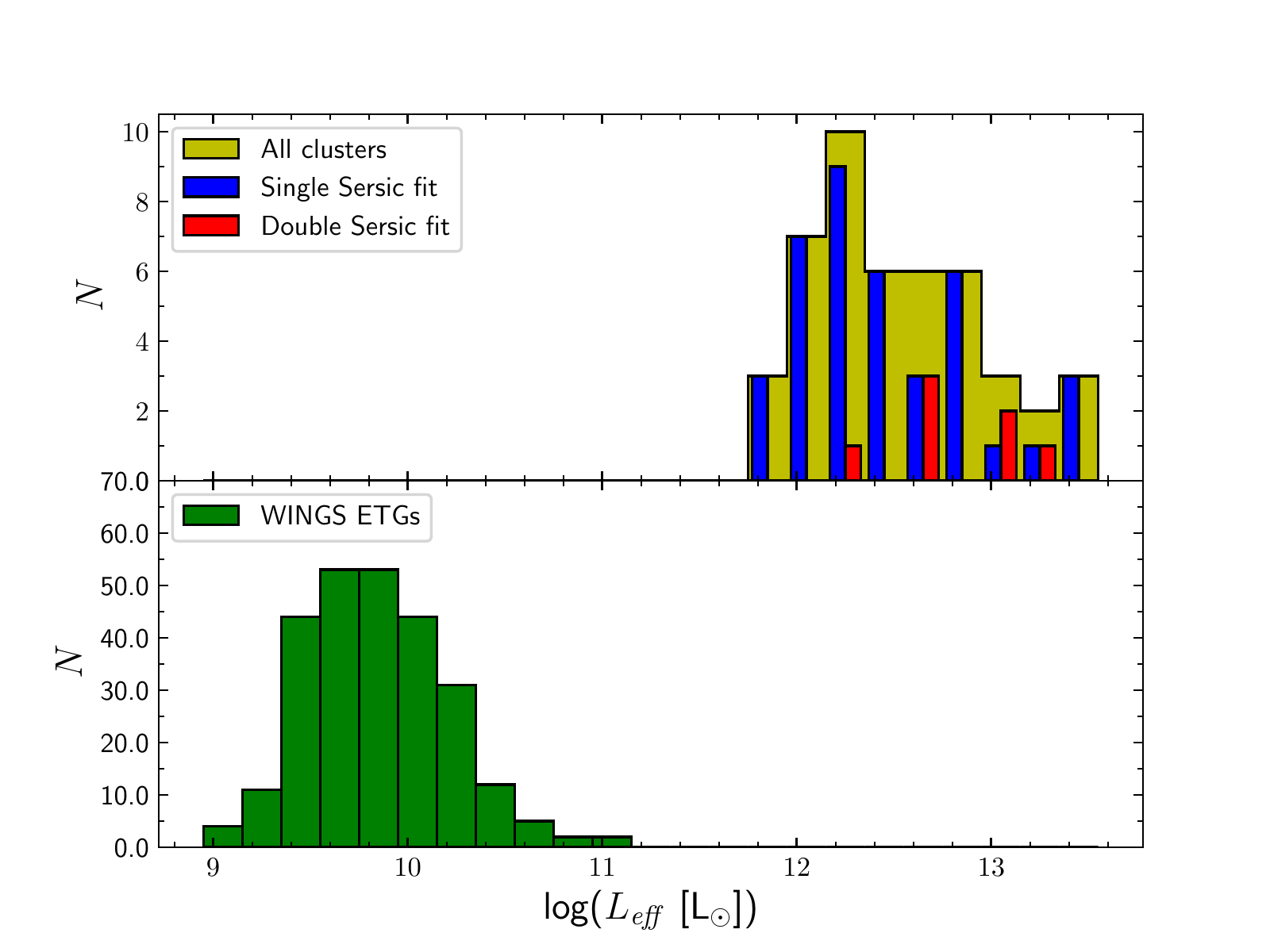}
        \includegraphics[width=0.45\textwidth]{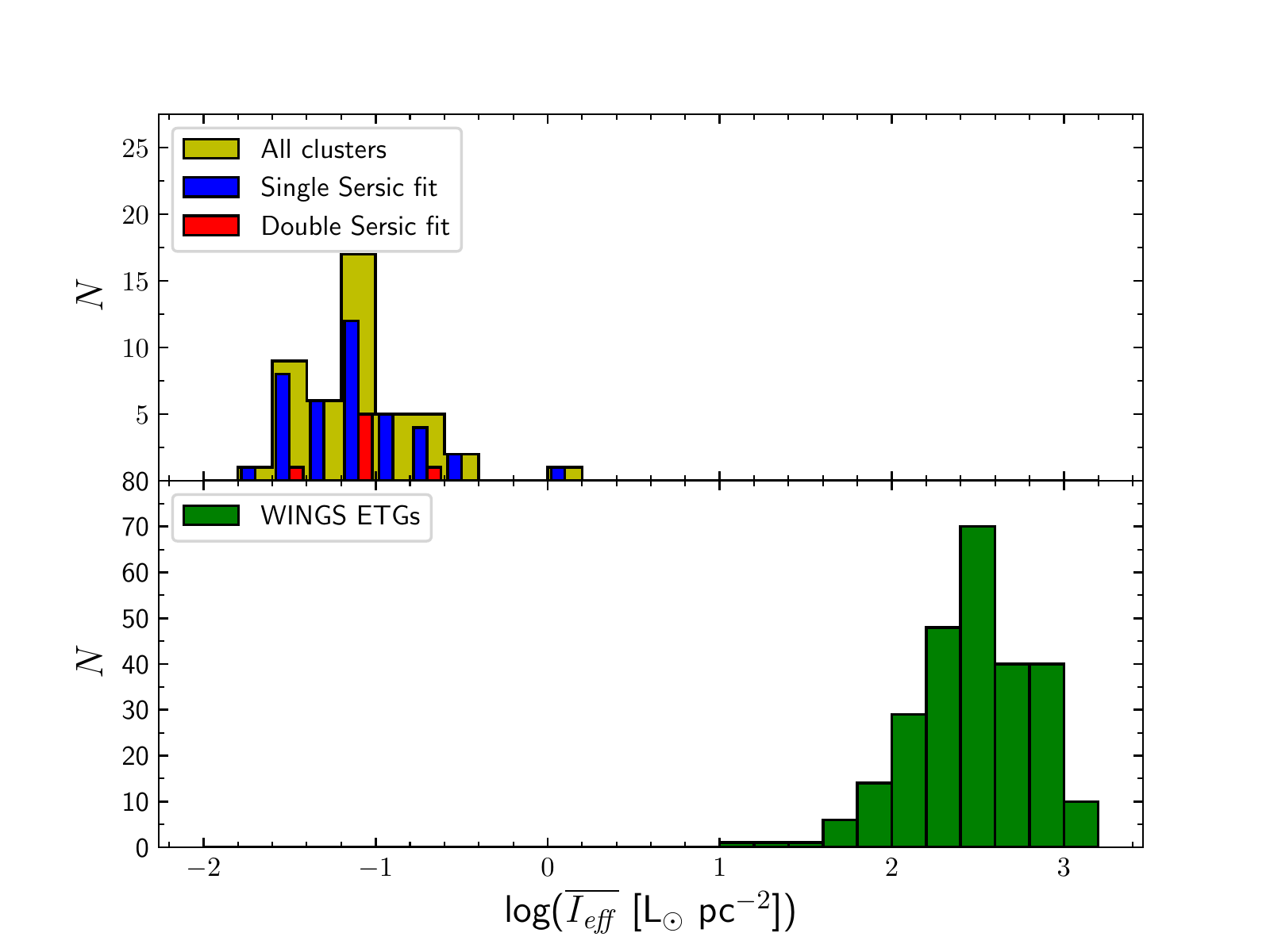}\includegraphics[width=0.45\textwidth]{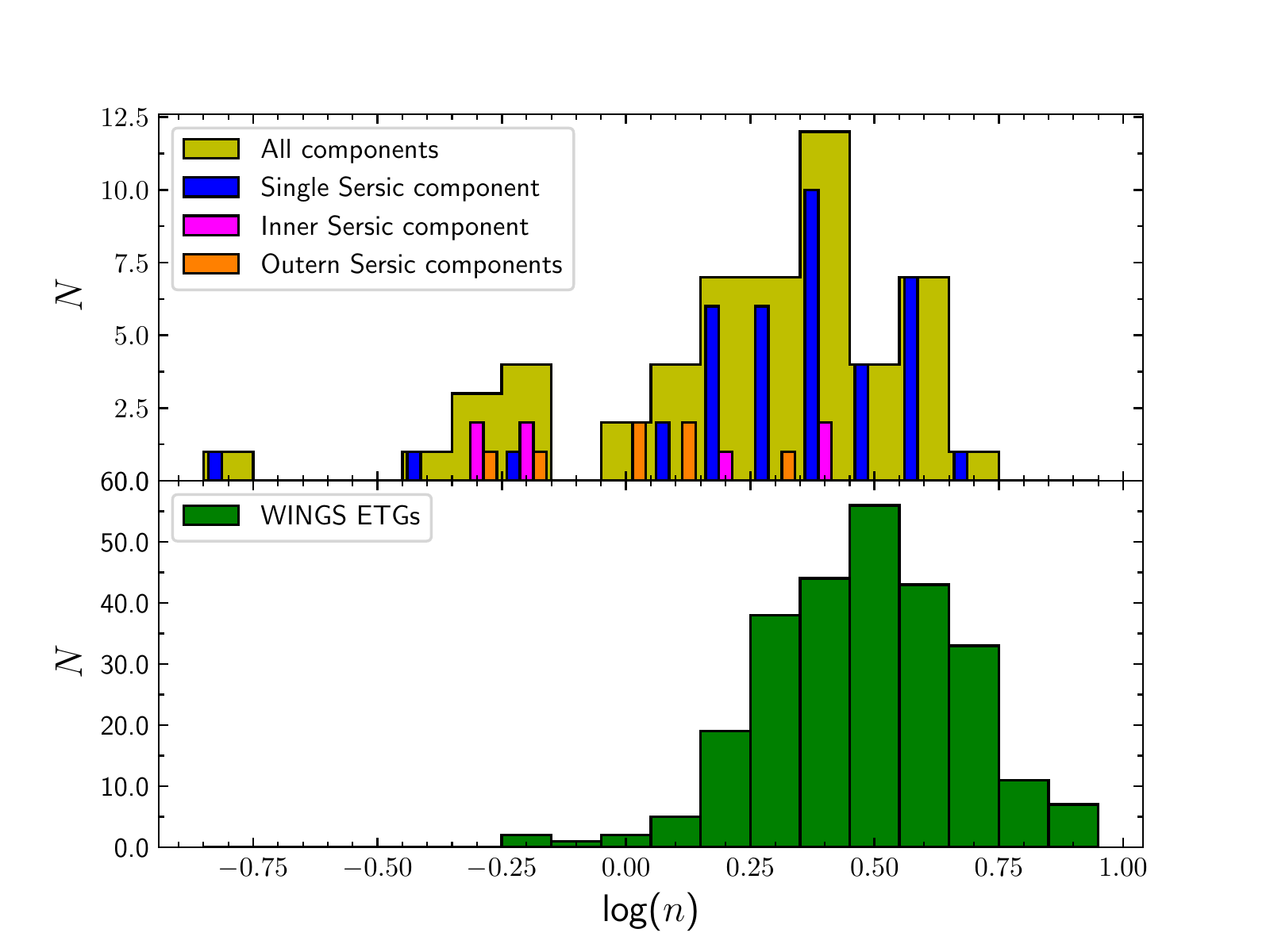}
        \caption{Histograms of the effective parameters (i.e., radius, average intensity, and luminosity) and S\'ersic index distribution. Upper panels: Shown are single S\'ersic fits data in blue, double S\'ersic fits data in red (plus inner S\'ersic in pink and outer S\'ersic in orange), and all of these combined in yellow. Lower panels: ETG data are shown in green.}
        \label{fig:hist}
\end{figure*}

Figure~\ref{fig:hist} provides the histograms of the observed distributions of the structural parameters. Almost all the components follow some Gaussian-like distribution. The range of values spanned by each parameter, although at different scales, is comparable. For example, the effective radii span in both samples a factor of $\sim 50$, while luminosities and average effective intensities span a larger interval, i.e., up to a factor of $\sim 100$ for galaxies and around 30 for clusters.

Once we observed that clusters and ETGs have a similar distribution of the photometric structural parameters at different scales, we decided
to investigate the main relations known to be valid for ETGs.

\subsection{Color-magnitude diagram of galaxy clusters}
\label{sec:CM}

\begin{figure*}
        \includegraphics[width=\textwidth]{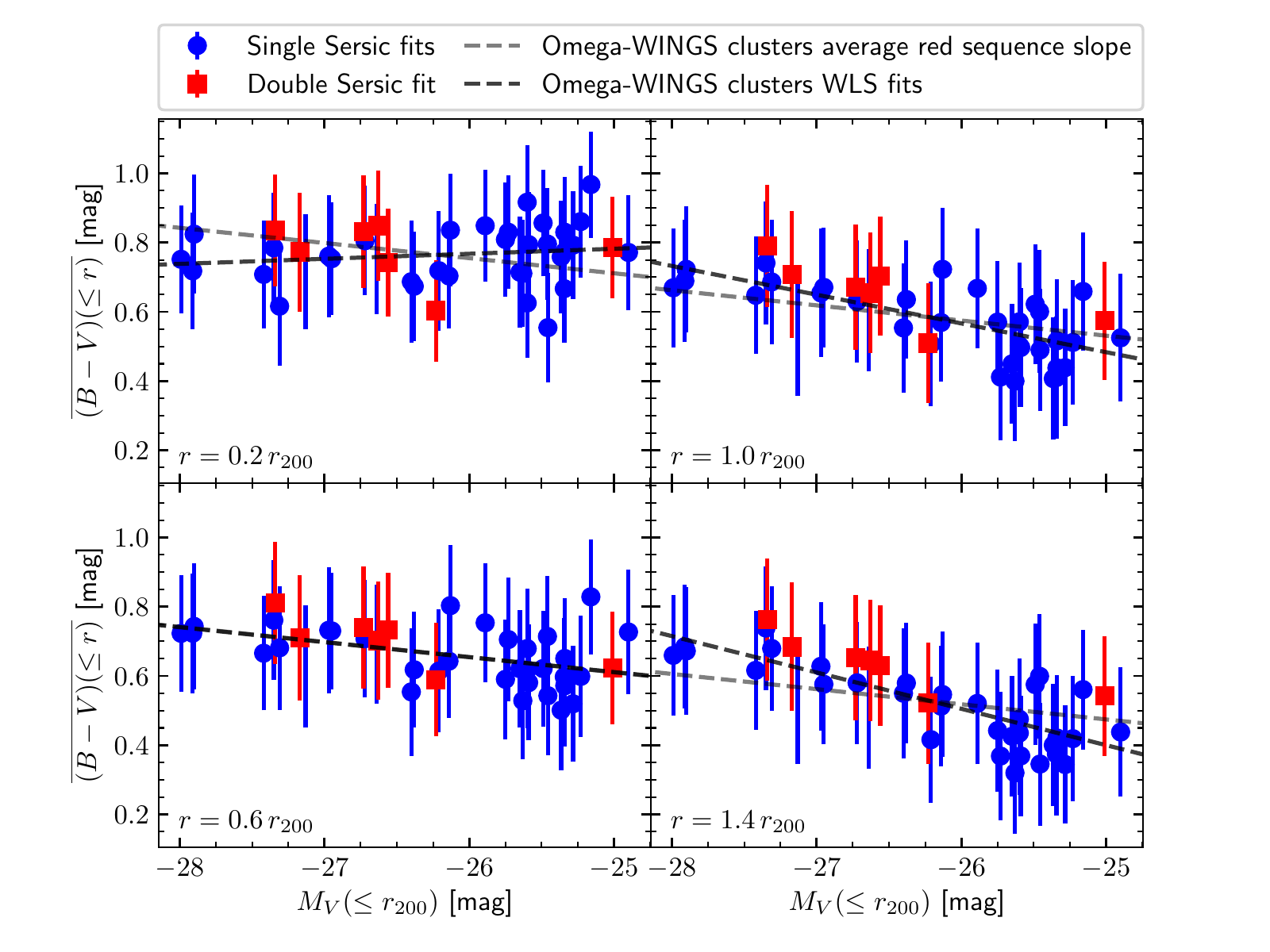}
        \caption{Color-magnitude relation for the whole sample of clusters considering different cluster areas. On the $x$-axis the integrated $V$-band absolute magnitude within $r_{200}$ is indicated, and on the $y$-axis the average of all the integrated colors within the considered radius, plotted in the lower left corner of the panel, is indicated. The gray dashed line gives the average red sequence slope observed for the CM relations of the galaxies of the WINGS clusters derived by \cite{val11}. The average red sequence slope is plotted by matching the intercept of the line to the mean color of the distribution. The bold dashed line is the WLS fit of the observed distribution.}
        \label{fig:CM}
\end{figure*}

We started by comparing the color-magnitude (CM) relation $(\overline{B-V})(\leq r)-M_V$ of clusters with the average red sequence slope found by \cite{val11} for the WINGS galaxies in the CM diagrams of the single clusters. In Figure~\ref{fig:CM} each cluster is represented by a dot with a $(\overline{B-V})(\leq r)$ color that is the average integrated color index measured within various fractions of $r_{200}$ and a total magnitude $M_V$ that corresponds to the total magnitude of the cluster within $r_{200}$. The WLS fit of the clusters is only steeper than the average red sequence slope of the galaxies in clusters (i.e., $-0.04$) when the mapped region of the clusters is larger $0.6 \, r_{200}$. The plots clearly indicate that the most massive clusters are, on average, the reddest, while
the less luminous clusters are the bluest;  this is also observed in the red sequence of ETGs. Also, in the central region all the clusters seem to have approximately the same color.

The WLS fits of our data (bold dashed lines) provide the following relations, which are valid for various fractions of the cluster areas (in $r_{200}$ units):

\begin{equation}
\begin{array}{lll}
\overline{(B-V)}(\leq 0.2 \, r_{200}) & \!\!\!\!= & \!\!\!\!+1.15\pm0.39 + 0.01\pm0.02\,M_V(r_{200}),\\      
\overline{(B-V)}(\leq 0.6 \, r_{200}) & \!\!\!\!= & \!\!\!\!-0.47\pm0.35 - 0.04\pm0.01\,M_V(r_{200}) {\rm ,}\\
\overline{(B-V)}(\leq 1.0 \, r_{200}) & \!\!\!\!= & \!\!\!\!-1.59\pm0.36 - 0.08\pm0.01\,M_V(r_{200}) {\rm ,}\\
\overline{(B-V)}(\leq 1.4 \, r_{200}) & \!\!\!\!= & \!\!\!\!-2.23\pm0.35 - 0.11\pm0.01\,M_V(r_{200}) {\rm ,}\\
\end{array}
\end{equation}

\noindent while the corresponding average red sequence slopes for the galaxies of the WINGS clusters in the same areas (light gray dashed lines) are

\begin{equation}
\begin{array}{lll}
\overline{(B-V)}(\leq 0.2 \, r_{200}) & \!\!\!\!= & \!\!\!\!-0.38\pm0.16 - 0.04\pm0.01\,M_V(r_{200}) {\rm ,}\\
\overline{(B-V)}(\leq 0.6 \, r_{200}) & \!\!\!\!= & \!\!\!\!-0.48\pm0.17 - 0.04\pm0.01\,M_V(r_{200}) {\rm ,}\\
\overline{(B-V)}(\leq 1.0 \, r_{200}) & \!\!\!\!= & \!\!\!\!-0.56\pm0.17 - 0.04\pm0.01\,M_V(r_{200}) {\rm ,}\\
\overline{(B-V)}(\leq 1.4 \, r_{200}) & \!\!\!\!= & \!\!\!\!-0.62\pm0.18 - 0.04\pm0.01\,M_V(r_{200}) {\rm .}\\
\end{array}
\end{equation}

The CM relation of galaxy clusters is found here for the first time. An explanation of its existence should be found in the current models of cluster formation and evolution. We will dedicate a future work to a possible theoretical interpretation of what is observed.

\begin{figure}
        \includegraphics[width=0.45\textwidth]{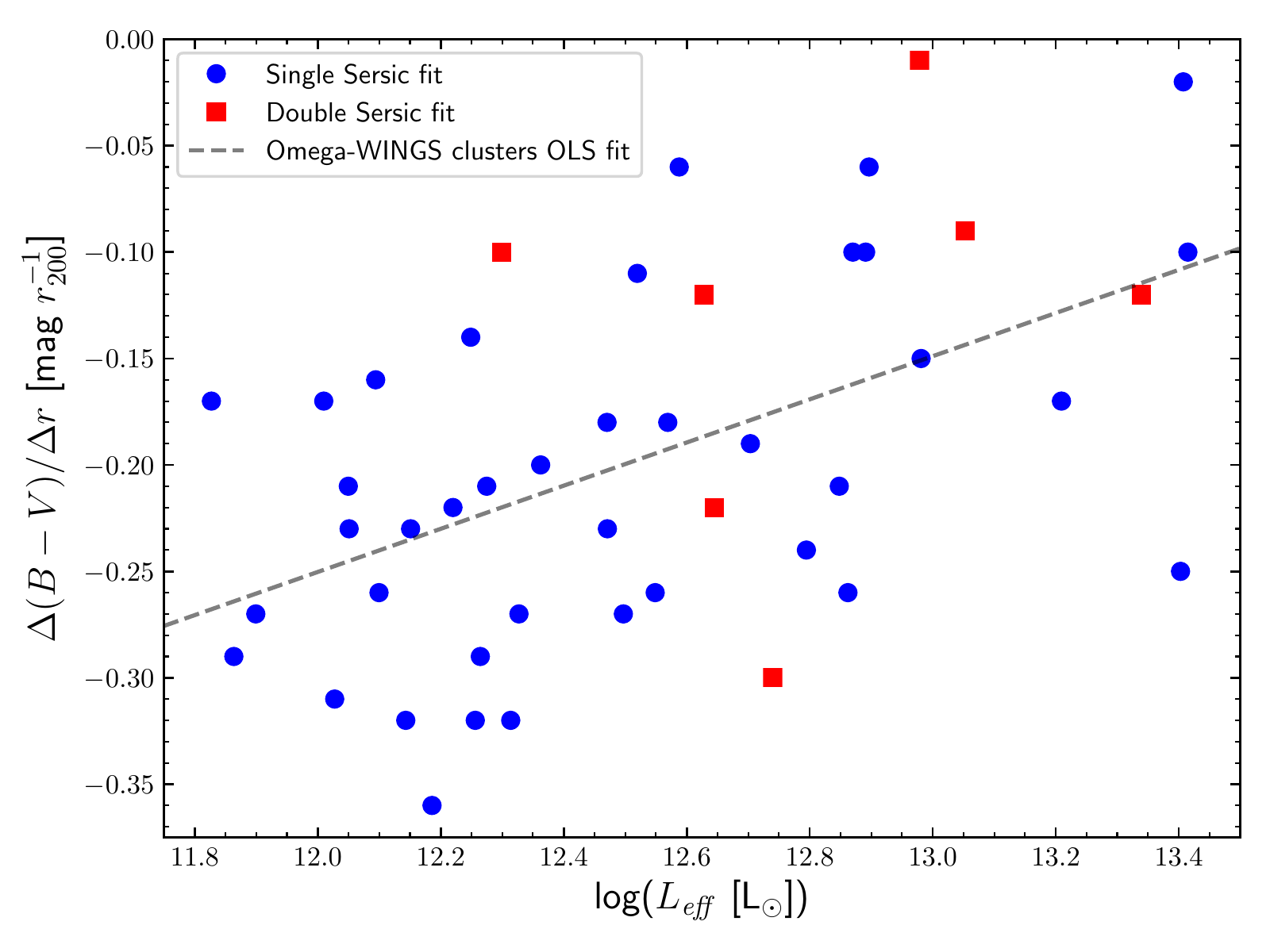}
        \caption{Color gradient as a function of the effective luminosity of the clusters. Color code as in the previous figures.}
        \label{fig:gradiente}
\end{figure}

In addition to the CM relation a correlation between the mean effective luminosity of clusters $L_{\it eff}$ and the color gradient $\Delta (B-V)/\Delta r$ is significant in our data (see Figure~\ref{fig:gradiente} and Table \ref{tab:phot_and_mass}), i.e.,

\begin{equation}
\Delta(B-V)/\Delta r = -1.47\pm0.32 + 0.10\pm0.03 \, {\rm log}(L_{\it eff}) {\rm .}
\end{equation}

The color gradient $\Delta (B-V)/\Delta r$ is negative because clusters, such as ETGs, are redder in the center and bluer in the outskirts, and this gradient appears to be larger in fainter clusters. This is at variance with the case of ETGs, where the optical gradient does not seem to correlate with the galaxy luminosity (see, e.g., \citealt{lab10}).

\subsection{Main scaling relations of galaxy clusters}
\label{sec:stat}

Table \ref{tab:correlations} presents the data of the mutual correlations among the structural parameters of galaxies and clusters. The following figures show the most famous ETGs parameters correlations extended to the domain of galaxy clusters.

\begin{table*}
        \caption{Summary of all the discussed OLS/WLS relations: $y$- and $x$-axis (Column~1), sample used to compute the relation (Column~2, where g = galaxies and c = clusters), sample size (Column~3), zero-order polynomial term (Column~4), first-order polynomial term (Column~5), RMS (Column~6), Pearson correlation coefficient PCC (Column~7), Pearson p-value in logarithmic units (Column~8, where the unreported values indicate a probability smaller than $3 \times 10^{-7}$, which comes from the 5 RMS criterion), Spearman's rank correlation $\rho$ (Column~9), Spearman p-value in logarithmic units (Column~10, where the unreported values again indicate a probability smaller than $3 \times 10^{-7}$).}
        \centering
        \label{tab:correlations}
        \begin{tabular}{lccccccccc}
                \hline
                $y-x$ & Sample & $N$ & $c_0$ & $c_1$ & RMS & PCC & ${\rm log} (p_{\rm P})$ & $\rho$ & ${\rm log} (p_{\rm S})$\\
                \hline
                $\overline{(B-V)}(\leq 0.2 \, r_{200})-M_V(r_{200})$ & c & 46 & $1.15\pm0.39$ & $-0.15\pm0.15$ & 0.08 & 0.15 & $-0.48$ & 0.21 & $-0.73$\\
                $\overline{(B-V)}(\leq 0.6 \, r_{200})-M_V(r_{200})$ & c & 46 & $-0.47\pm0.35$ & $-0.04\pm0.01$ & 0.07 & $-0.46$ & $-2.66$ & $-0.46$ & $-2.67$\\
                $\overline{(B-V)}(\leq 1.0 \, r_{200})-M_V(r_{200})$ & c & 46 & $-1.59\pm0.36$ & $-0.08\pm0.01$ & 0.07 & $-0.69$ & $-6.41$ & +0.67 & $-5.95$\\
                $\overline{(B-V)}(\leq 1.4 \, r_{200})-M_V(r_{200})$ & c & 46 & $-2.23\pm0.35$ & $-0.11\pm0.01$ & 0.07 & $-0.78$ & - & $-0.75$ & -\\
                $\Delta(B-V)/\Delta r - {\rm log}(L_{\it eff}$) & c & 46 & $-1.47\pm0.32$ & $0.10\pm0.03$ & $0.07$ & $0.51$ & $3.59$ & $0.50$ & $-3.37$\\
                log($\overline{I_{\it eff}}) - {\rm log}(r_{\it eff}$) & g & 261 & $2.82\pm0.02$ & $-0.99\pm0.05$ & $0.21$ & $-0.79$ & - & $-0.75$ & -\\
                log($\overline{I_{\it eff}}) - {\rm log}(r_{\it eff}$) & c & 53 & $2.75\pm0.29$ & $-1.02\pm0.09$ & $0.31$ & $-0.84$ & - & $-0.57$ & $-5.09$\\
                log($L_{\it eff}) - {\rm log}(r_{\it eff}$) & g & 261 & $9.48\pm0.02$ & $1.03\pm0.05$ & $0.40$ & $0.80$ & - & $0.75$ & -\\
                log($L_{\it eff}) - {\rm log}(r_{\it eff}$) & c & 53 & $9.34\pm0.29$ & $0.97\pm0.10$ & $0.33$ & $0.70$ & - & $0.64$ & $-5.94$\\
                log($\mathcal{M}_{\it eff}) - {\rm log}(r_{\it eff}$) & g & 261 & $10.10\pm0.03$ & $0.91\pm0.06$ & $0.31$ & $0.64$ & - & $0.59$ & -\\
                log($\mathcal{M}_{\it eff}) - {\rm log}(r_{\it eff}$) & c & 20 & $10.10\pm0.03$ & $0.91\pm0.06$ & $0.26$ & $0.54$ & $-1.94$ & $0.40$ & $-1.16$\\     
                log($L_{\it eff}) - {\rm log}(\sigma$) & g & 261 & $6.09\pm0.21$ & $1.73\pm0.10$ & $0.23$ & $0.75$ & - & $0.78$ & -\\
                log($L_{\it eff}) - {\rm log}(\sigma$) & c & 41 & $7.90\pm1.59$ & $1.63\pm0.56$ & $0.39$ & $0.42$ & $-2.25$ & $0.38$ & $-1.88$\\
                log($r_{\it eff}) - {\rm log}(\sigma$) & c & 41 & $1.93\pm0.33$ & $1.56\pm0.68$ & $0.47$ & $0.34$ & - & $0.35$ & -\\
                log($r_{\it eff}) - {\rm log}(n$) & g & 261 & $-0.28\pm0.04$ & $1.33\pm0.08$ & $0.26$ & $0.49$ & - & $0.49$ & -\\
                log($r_{\it eff}) - {\rm log}(n$) & c & 38 & $2.97\pm0.07$ & $1.02\pm0.12$ & $0.24$ & $0.33$ & $-1.35$ & $0.40$ & $-1.91$\\
                log($L_{\it eff}) - {\rm log}(n$) & g & 261 & $9.09\pm0.05$ & $1.58\pm0.09$ & $0.32$ & $0.44$ & - & $0.47$ & -\\
                log($L_{\it eff}) - {\rm log}(n$) & c & 38 & $12.08\pm0.09$ & $1.08\pm0.22$ & $0.48$ & $0.05$ & $-0.12$ & $-0.12$ & $-0.31$\\
                log($\sigma) - {\rm log}(L_{\it sub}/L_{\it main}$) & c & 22 & $2.92\pm0.03$ & $0.12\pm0.06$ & $0.08$ & $0.43$ & $-1.36$ & $0.49$ & $-1.68$\\
                \hline
        \end{tabular}
\end{table*}

\begin{figure}
        \includegraphics[width=0.45\textwidth]{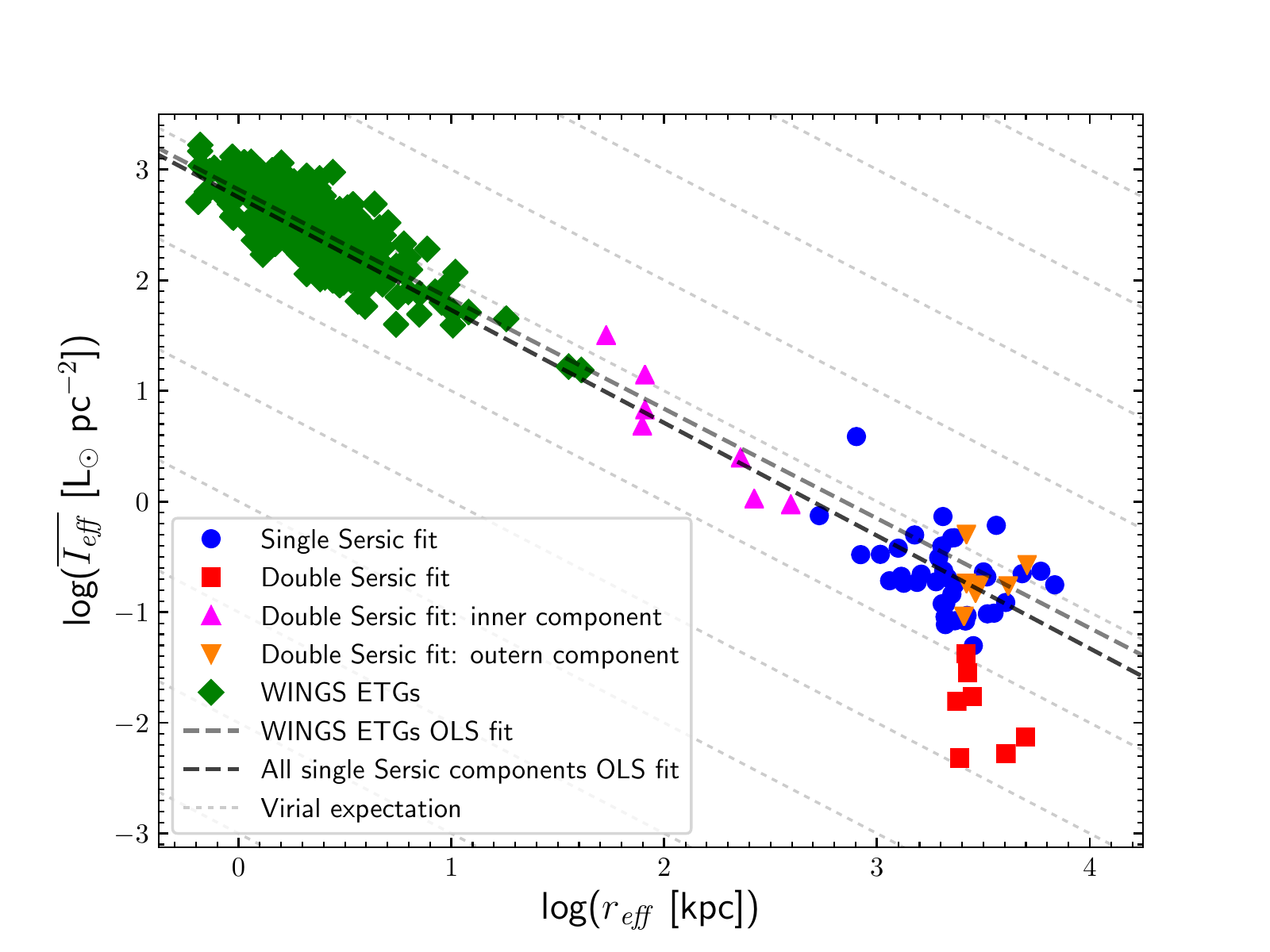}
    \caption{Kormendy $\overline{I_{\it eff}}$-$r_{\it eff}$ relation for ETGs and galaxy clusters. Green diamonds indicate ETGs. The blue circles are associated with single S\'ersic fits of galaxy clusters, red squares to double S\'ersic fits, pink triangles to the inner components of double S\'ersic fits, and reversed orange tringles to the outer components of the double S\'ersic fits.}
    \label{fig:kor}
\end{figure}

Figure~\ref{fig:kor} compares the Kormendy relation (\citealt{kor77}; \citealt{ham87}; \citealt{cap92}; \citealt{don17}) of ETGs with that of our clusters. The green diamonds correspond to the WINGS ETGs, blue dots to the single S\'ersic fits, red squares to the general parameters of the double S\'ersic profiles, pink triangles to the inner components of the double S\'ersic fits, and orange reversed triangles to the outer components of the double S\'ersic profiles. 

The effective parameters of galaxy clusters follow the same relation previously found for ETGs. Clusters reside along the high-radii tail of galaxies and share the same zone-of-esclusion (ZoE; details in \citealt{ben92}) of ETGs. The ordinary least squares (OLS) linear interpolation of both the samples provides the same slope within the errors (see Table~\ref{tab:correlations}), compatible with that expected by the scalar virial theorem (light gray dashed lines) when a constant mass-to-light ratio is assumed. For a better explanation of the observed distribution in the $\overline{I_{\it eff}} - r_{\it eff }$ plane, see \cite{don17}.

%The relations parameters are summarized in Table~\ref{tab:correlations}, where $c_0$ and $c_1$ are the zero- and first-order polynomial coefficients of the fit (i.e., $y = c_0 + c_1 \, x$), and PCC and $\rho$ are the Pearson Correlation Coefficient and the Spearman's correlation rank, which measure, respectively, the linear and monotonic relationship between parameters.

\begin{figure*}
        \includegraphics[width=0.45\textwidth]{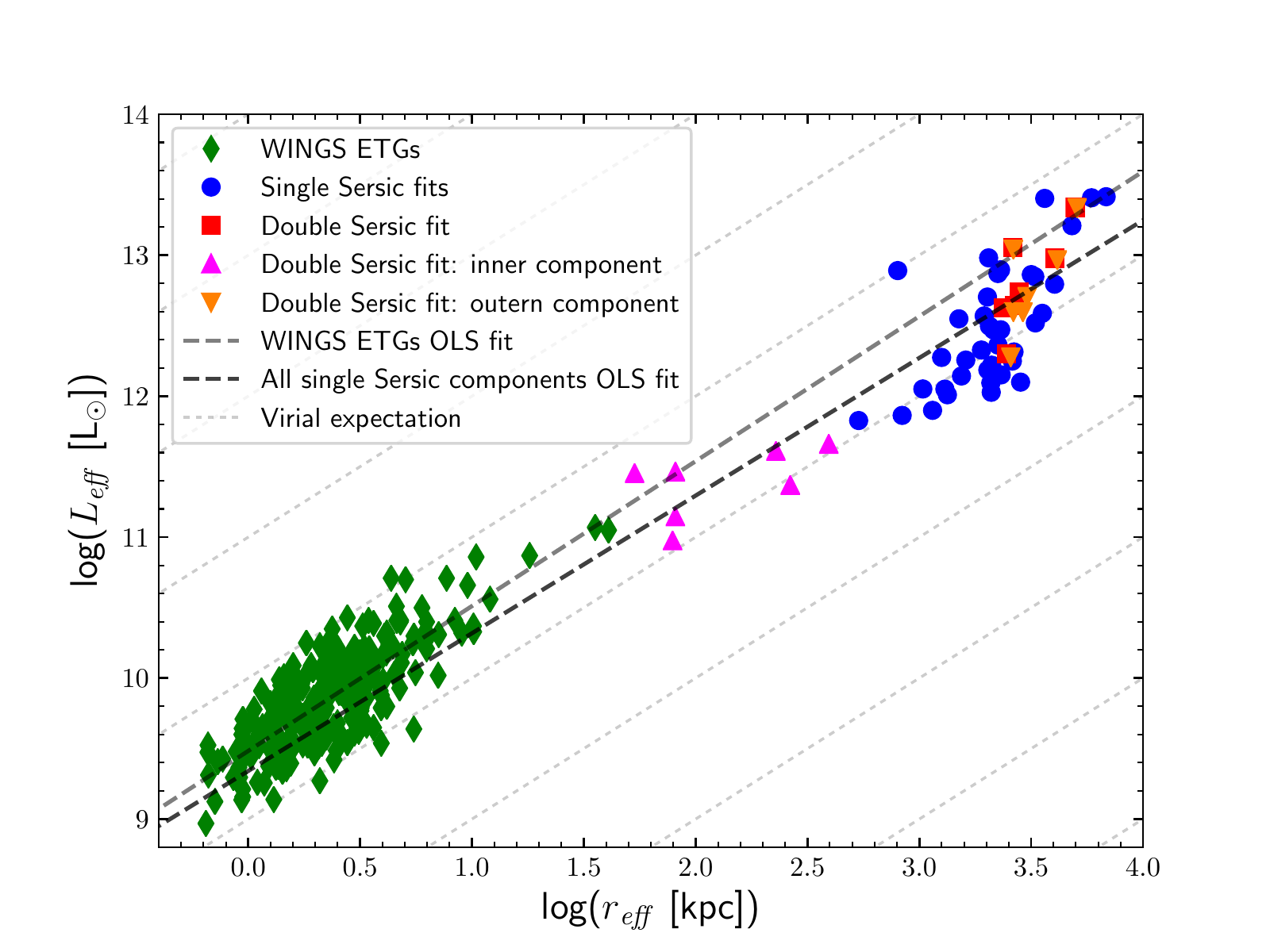}\includegraphics[width=0.45\textwidth]{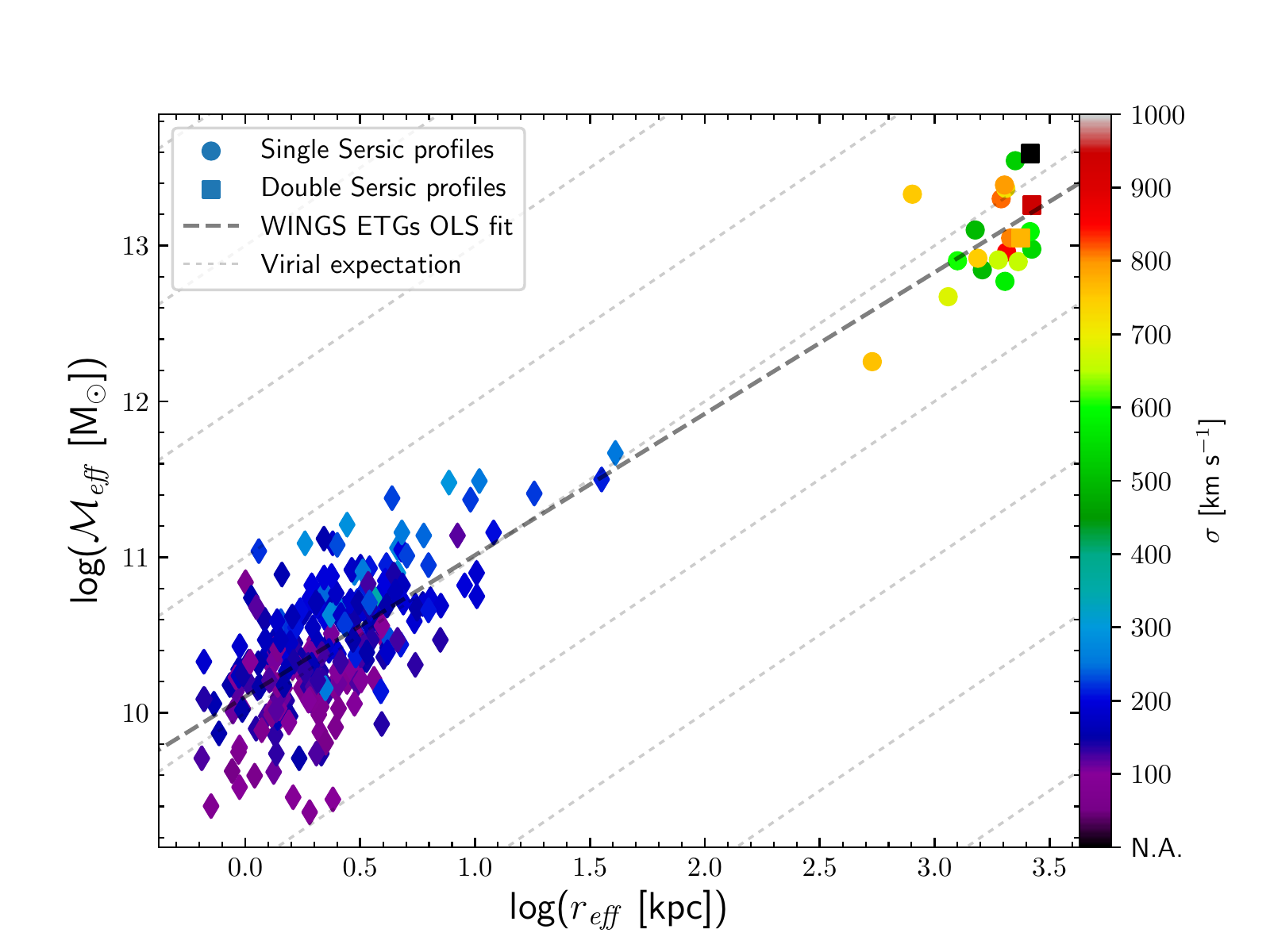}
    \caption{Left plot: $L_{\it eff}$-$r_{\it eff}$ relation for ETGs and clusters. Right plot: $\mathcal{M}_{\it eff}$-$r_{\it eff}$ relation for the same samples is shown. The symbols are as those in the previous figure, the colors represent the velocity dispersion of each point (color bar on the right).}
    \label{fig:r-LM}
\end{figure*}

Figure~\ref{fig:r-LM} shows the $L_{\it eff}-r_{\it eff}$ and the $\mathcal{M}_{\it eff}-r_{\it eff}$ relations between the effective luminosity/mass and the effective radius. Again we see that the distribution of ETGs and clusters follows the expected behavior on the basis of the Virial theorem (see Table~\ref{tab:correlations}). The position of each object in these planes depends on the zero point of the virial relation, which is different for each system. To see such an effect note how the position in the $\mathcal{M}_{\it eff}-r_{\it eff}$ depends on the central velocity dispersion $\sigma$ (color scale on the right plot). The velocity dispersion values on the plot are those tabulated by \citealt{don8} for all the galaxies and were provided us by \cite{Biv17} for the clusters.

\begin{figure}[t]
        \includegraphics[width=0.45\textwidth]{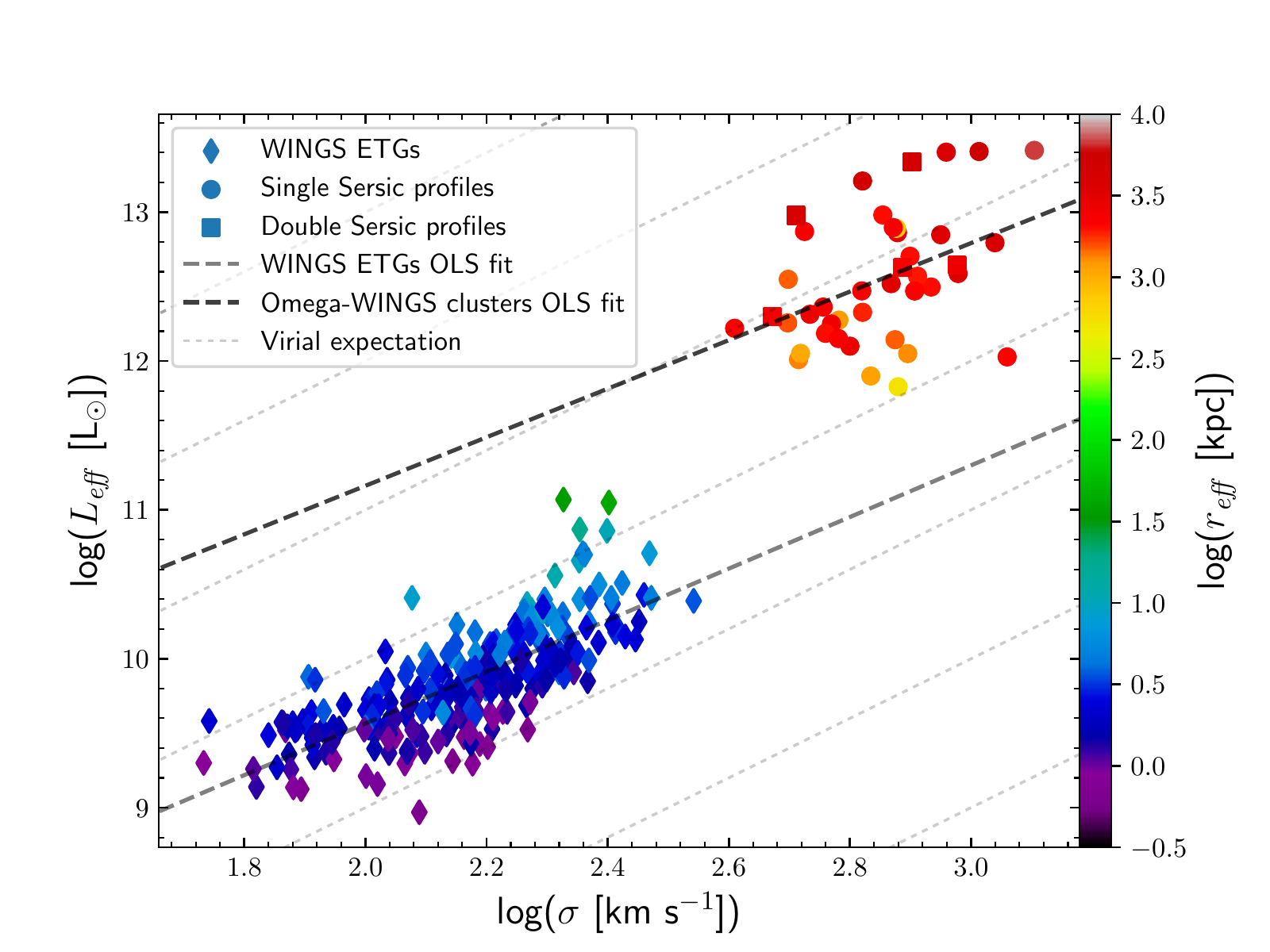}
    \caption{$L_{\it eff}-\sigma$ relation for the WINGS ETGs and our galaxy clusters. The symbols are as those in the previous figures, the colors represent the effective radius (color bar on the right).}
    \label{fig:L-sigma}
\end{figure}

To complete the series of plots dedicated to the virial equilibrium of our clusters, we also show the $L_{\it eff}-\sigma$ relation (\citealt{fj76}). A plot similar to Figure~\ref{fig:L-sigma} has been shown by \cite{don17} with a possible explanation of the observed distribution. According to the authors the position of each point in the diagram is given by its virial equilibrium and by a variable zero point that turns out to depend on the effective radius $r_{\it eff}$ and mean $\mathcal{M}/L$ ratio. The slope different from 2 often claimed for this relation arises when structures with different zero points are mixed together.

%The fitted distributions for ETGs and clusters are given in Tab. \ref{tab:correlations}.

This set of figures clearly shows that ETGs and clusters share the same virial relations. The occasional deviations come from the different zero points of the different systems in each diagram.

%All the previous findings seem to suggest a deeper-than-expected link between the processes ruling the galaxy formation and the galaxy clusters formation. In order to test this hypothesis, we investigated for more analogies between the two classes of objects.
%
%An important relation between the galaxy parameters is the Faber-Jackson (\citeyear{fj76}) relation, which links the effective luminosity to the velocity dispersion $\sigma$. This relation arises from the virial theorem, which connects the velocity dispersion of a system with its mass, and from the fact that the light distribution in galaxies is a good tracer of the mass distribution. We investigated whether this relation is valid also for galaxy clusters (Figure~\ref{fig:L-sigma}). 

%The OLS fits of the two samples show similar slopes:
%
%\begin{equation}
%{\rm log}(L_{\it eff}) = 6.10\pm0.21 + 1.73\pm0.10 \, {\rm log}(\sigma) {\rm \ for\ galaxies,}
%\end{equation}
%\begin{equation}
%{\rm log}(L_{\it eff}) = 1.93\pm0.33 + 1.56\pm0.68 \, {\rm log}(\sigma) {\rm \ for\ clusters.}
%\end{equation}
%
%\noindent and both the PCC and $\rho$ parameter identified a possible weak correlation.

Now we discuss the non-homology of clusters. First, we remember that a virialized structure is not necessarily a homologous structure (i.e., a structure with scale free properties). In the case of ETGs this has been proved by several works (\citealt{mic85}; \citealt{sch86}; \citealt{cap87}; \citealt{dec88}; \citealt{cap89}; \citealt{bur93}; \citealt{cao93}; \citealt{you94}; \citealt{pru97}) by showing that the light profiles of these galaxies are best fitted by the S\'ersic law and that the S\'ersic index $n$ correlates with the luminosity, mass, and radius of the galaxies themselves.

%As widely demonstrated by the observations of other authors (\citealt{mic85}; \citealt{sch86}; \citealt{cap87}; \citealt{dec88}; \citealt{cap89}; \citealt{bur93}; \citealt{cao93}; \citealt{you94}; \citealt{pru97}), galaxies are non-homologous systems. In fact, a scattered-yet-well-defined correlation between their S\'ersic index and their effective radius or luminosity is visible, which means that their light distribution is affected by their size. By 

\begin{figure*}
        \includegraphics[width=0.45\textwidth]{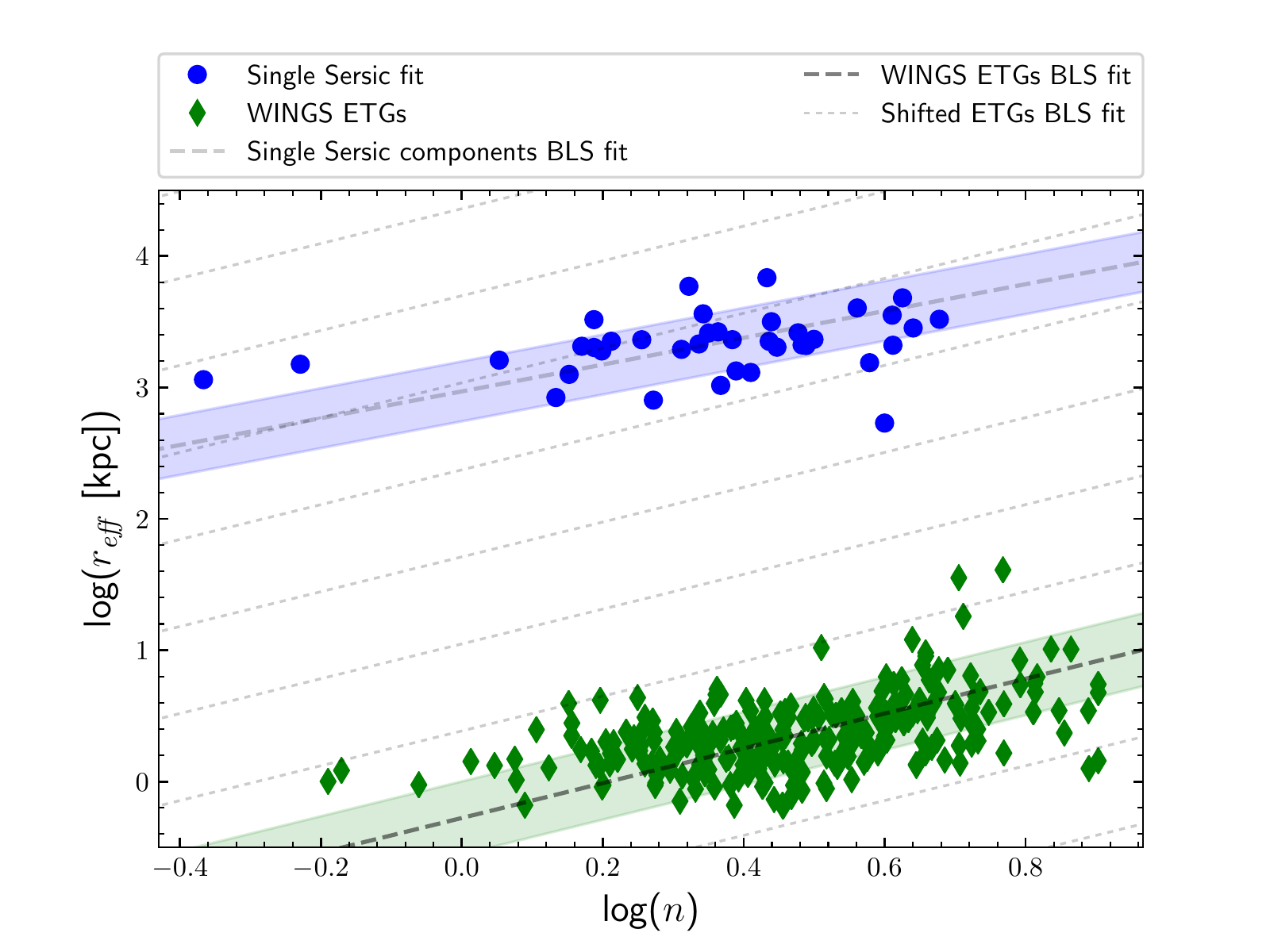} \includegraphics[width=0.45\textwidth]{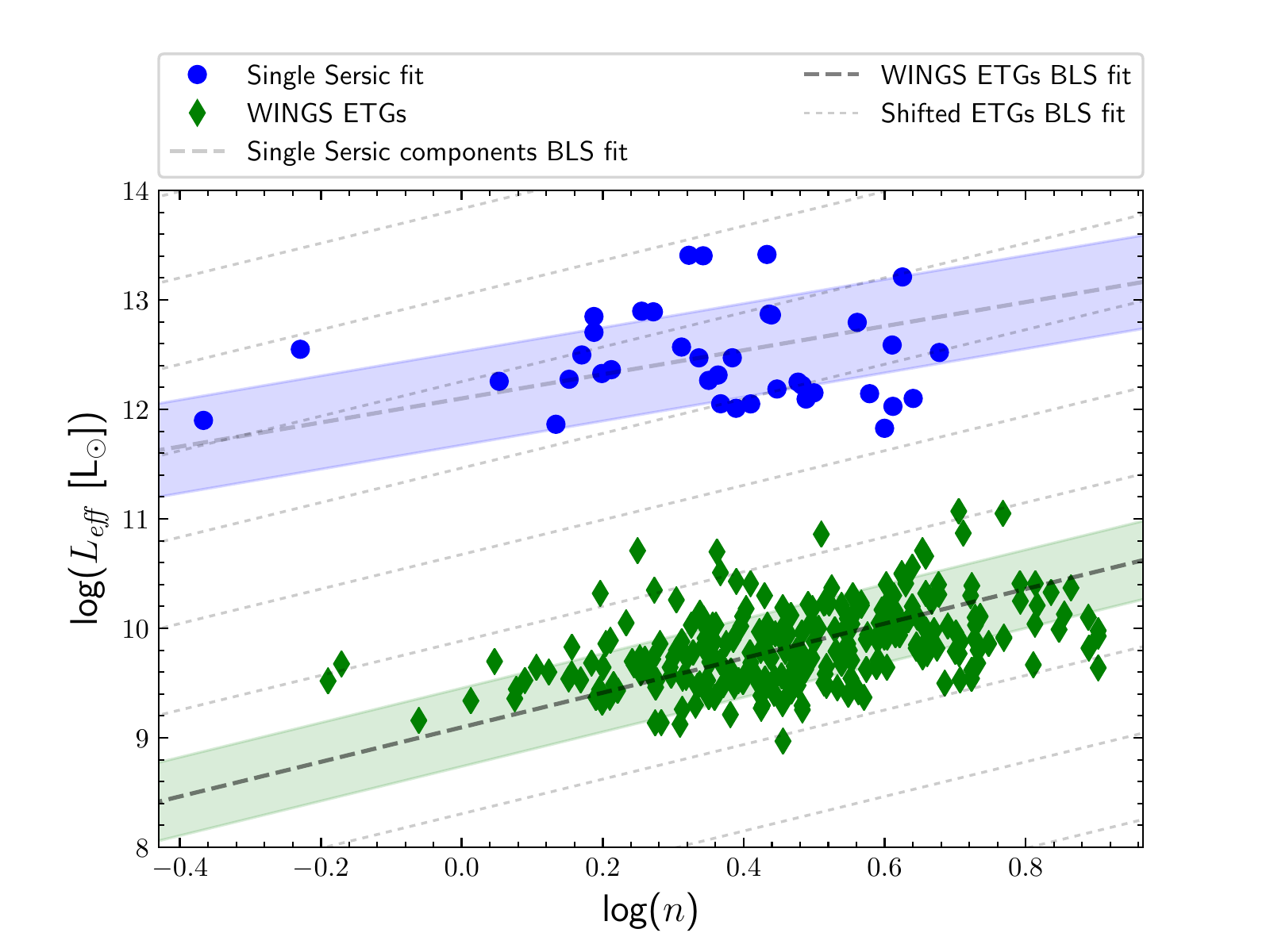}                                         
        \caption{Correlations of $r_{\it eff}-n$ and $L_{\it eff}-n$ in log units for ETGs (green dots) and clusters (blu dots). The gray dashed lines are parallel to the BLS fit of the WINGS ETGs. The bold gray dashed line indicates the BLS fit of our clusters. The filled areas  identify the limits of the RMS standard deviation of the observed distributions with the same color code of the sample to which they refer.}
    \label{fig:n-rL}
\end{figure*}

Figure~\ref{fig:n-rL} shows the distribution of galaxy clusters with respect to the ETGs in the $r_{\it eff} - n$ and $L_{\it eff} - n$ diagrams in logarithmic units. The dashed lines gives the bi-weighted least square (BLS) fit of the two distributions. This kind of fit was applied because we do not know a priori which variable drives the correlation. We removed from the plot the cluster A1631a because a visual inspection of the surface brightness profile suggests the presence of a second cluster component in the same area.

%\begin{figure}
%       \includegraphics[width=0.45\textwidth]{./stacked_profiles.pdf}
%       \caption{Superimposition of the surface brightness profiles associated to all the best fit models, scaled to take into account the different effective parameters.}
%    \label{fig:stack}
%\end{figure}

Two considerations emerge from these plots: first, both the classes of objects span the same range of $n$; second, the slope found for the ETGs also seems to be a plausible slope for galaxy clusters. 
Considering that almost all the clusters are in the luminosity range $\sim 10^{12} - 10^{13} L_\odot$, while the galaxies span the range $\sim 10^{9} - 10^{11} L_\odot$, it is not surprising to see that the $L_{\it eff}-n$ correlation, well visible in galaxies, is almost absent in clusters.
The $r_{\it eff}-n$ correlation is, on the other hand, well visible for both the types of structures. This means that clusters with the same luminosity can have very different structures with different values of $r_{\it eff}$ and, of course, $n$. In other words, clusters are likely non-homologous systems, such as ETGs.

The idea that clusters are non-homologous systems is not predicted by cluster simulations. The DM halos emerging from numerical simulations are structurally homologous systems with similar velocity dispersion profiles (\citealt{col96}; \citealt{nav97}). We do not have enough data at the moment to check the consistency of the DM profiles with the observed stellar light profiles, so we will addressed this problem in a future work.

%The $\mathcal{M}/L$ ratio display a lack of correlation with $L$ (Figure~\ref{fig:ML-L}), and a positive correlation with $\mathcal{M}$ (Figure~\ref{fig:ML-M}) both in the case of galaxies and in the case of galaxy clusters. This relation appears to be very scattered in the case of ETGs, however it is impressive to note that the OLS fits of the two samples display a perfectly parallel trend:

%\begin{equation}
%{\rm log}(\mathcal{M}/L) = -0.73\pm0.27 + 0.12\pm0.03\,{\rm log}(\mathcal{M}) {\rm \ for\ galaxies,}
%\end{equation}
%\begin{equation}
%{\rm log}(\mathcal{M}/L) = -0.75\pm0.46 + 0.12\pm0.03\,{\rm log}(\mathcal{M}) {\rm \ for\ clusters.}
%\end{equation}

%Figure~\ref{fig:ML-M} suggests that the galaxy clusters $\mathcal{M}/L$ ratio is generally compatible with that of the most massive ETGs, which could be linked to similar stellar populations. In order to test our hypothesis, we built the color-magnitude diagram of our clusters (Figure~\ref{fig:col-mag}) with the average color index $\overline{(B-V)}$ plotted as a function of the asymptotic $V$-band absolute magnitude, calculated from the fitting parameters. 

\begin{figure}[t]
        \includegraphics[width=0.45\textwidth]{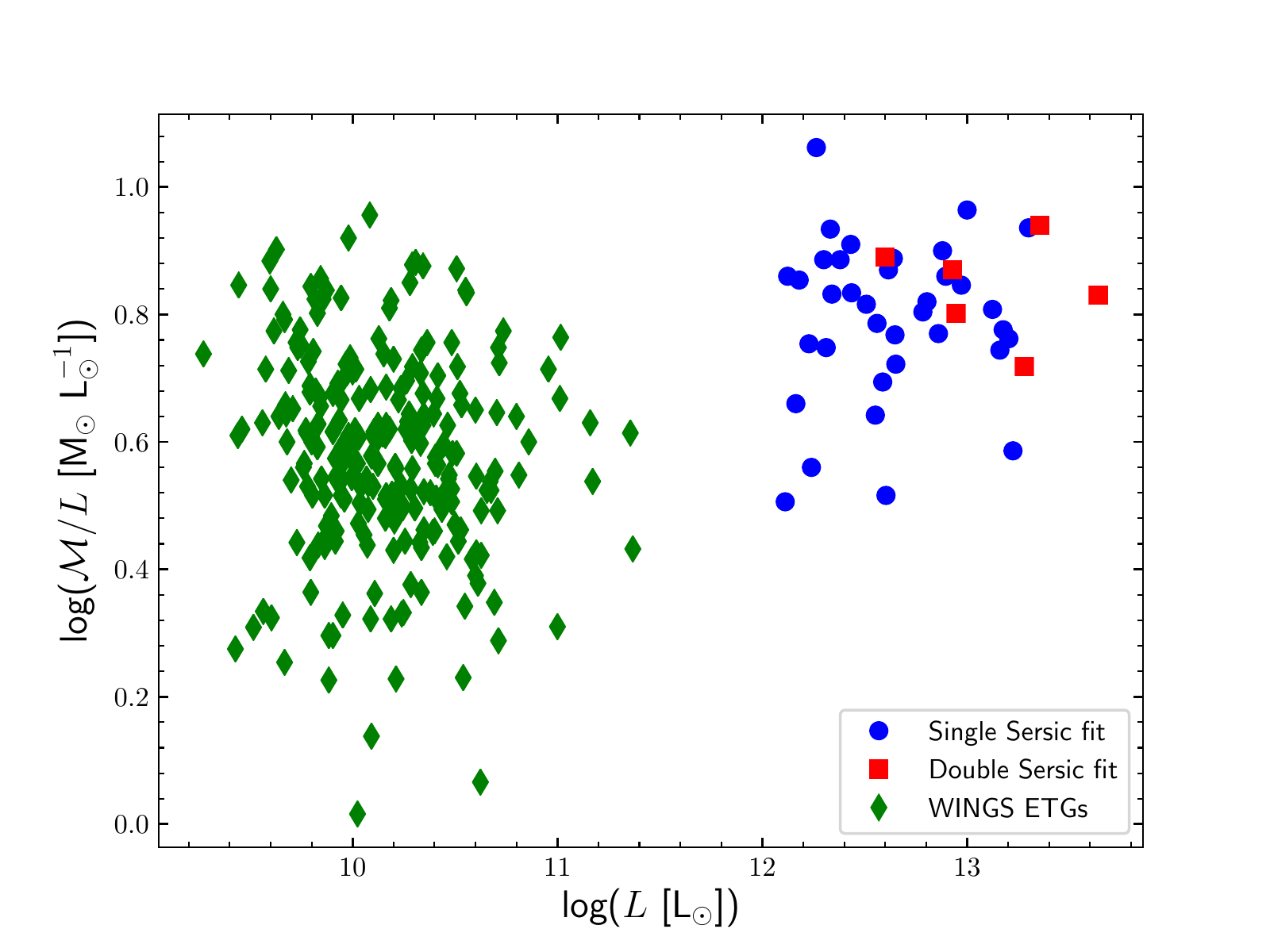}
    \caption{Distribution of the $\mathcal{M}/L$ ratio as a function of the total luminosity $L$ for both the samples. Same symbols and color code as in the previous figures.}
    \label{fig:ML-L}
\end{figure}

Finally, we analyze the stellar mass-to-light ratios of ETGs and clusters. Figure~\ref{fig:ML-L} shows the distribution of the $\mathcal{M}/L$ ratios as a function of the total luminosities. We see that ETGs span a factor of 10 in $\mathcal{M}/L$ and that the stellar mass-to-light ratio does not correlate with the luminosity. This seems to be in contradiction with the claimed relation between the dynamical $\mathcal{M}/L$ ratio and the luminosity (see, e.g., \citealt{cap6}, or \citealt{cap8}). The mean stellar $\mathcal{M}/L$ ratio of galaxy clusters spans a much smaller interval of values and again no correlation is seen with the luminosity. The combination of the two samples seems to suggest a trend with $L$ of the mass-to-light ratio, but this is a misleading conclusion originating from the absence of clusters with low $\mathcal{M}/L$ values.  

\begin{figure}
        \includegraphics[width=0.45\textwidth]{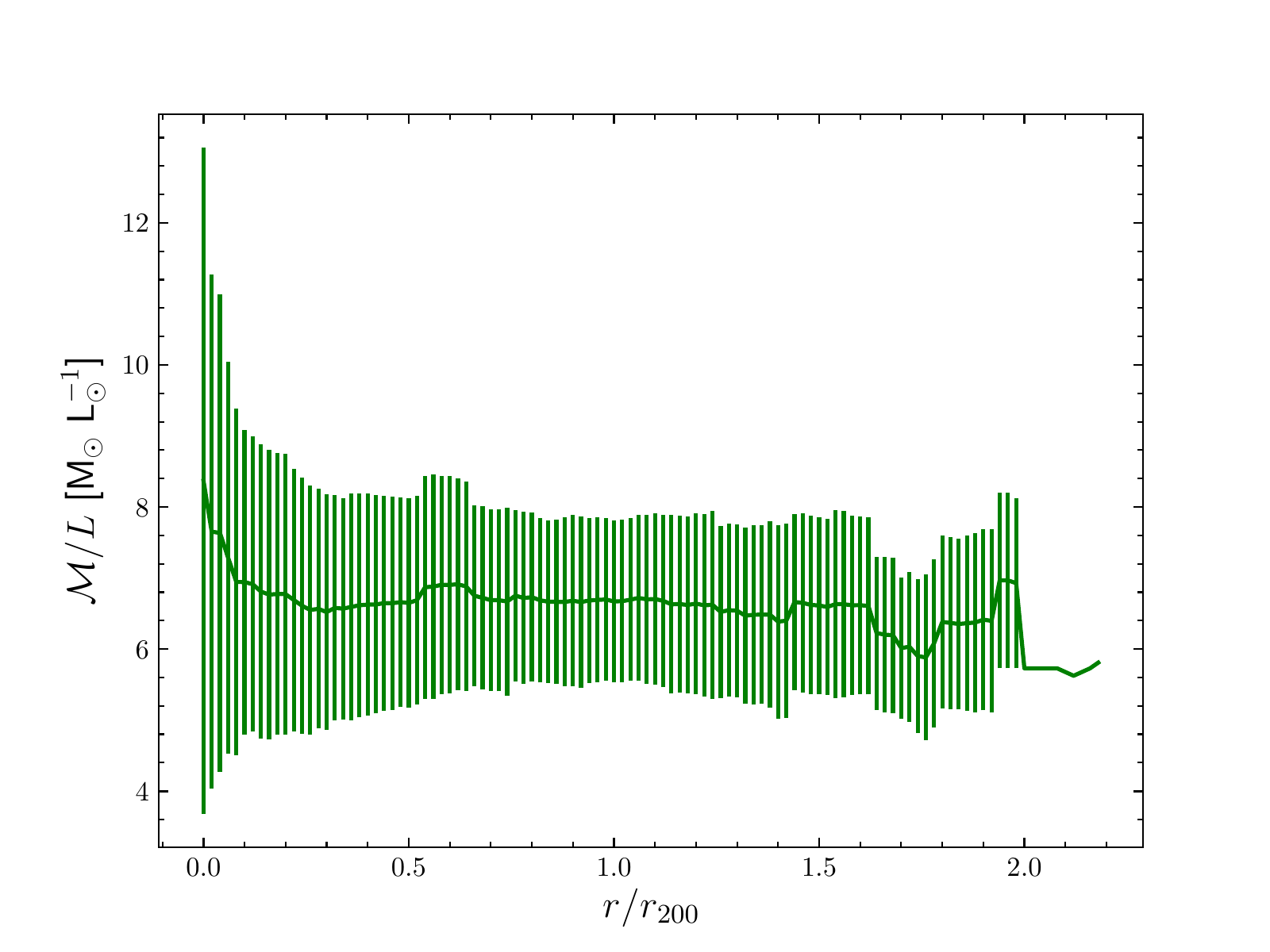}
        \caption{Average value and dispersion of the mass-to-light ratio of all the clusters at various radii in units of $r_{200}$.}
        \label{fig:stack2}
\end{figure}

Figure~\ref{fig:stack2} shows the stellar $\mathcal{M}/L$ ratio as a function of radius in units of $r_{200}$ for all our clusters. The constant value of $\mathcal{M}/L$ with the small spread at medium and large radii and the increase of such a spread in the inner region.

These figures provide further evidence that nearby clusters are dominated by an old stellar population almost over the whole extension of their profiles, which is an observational fact that must be reproduced by models of cluster formation and evolution.

\begin{figure}[t]
        \includegraphics[width=0.45\textwidth]{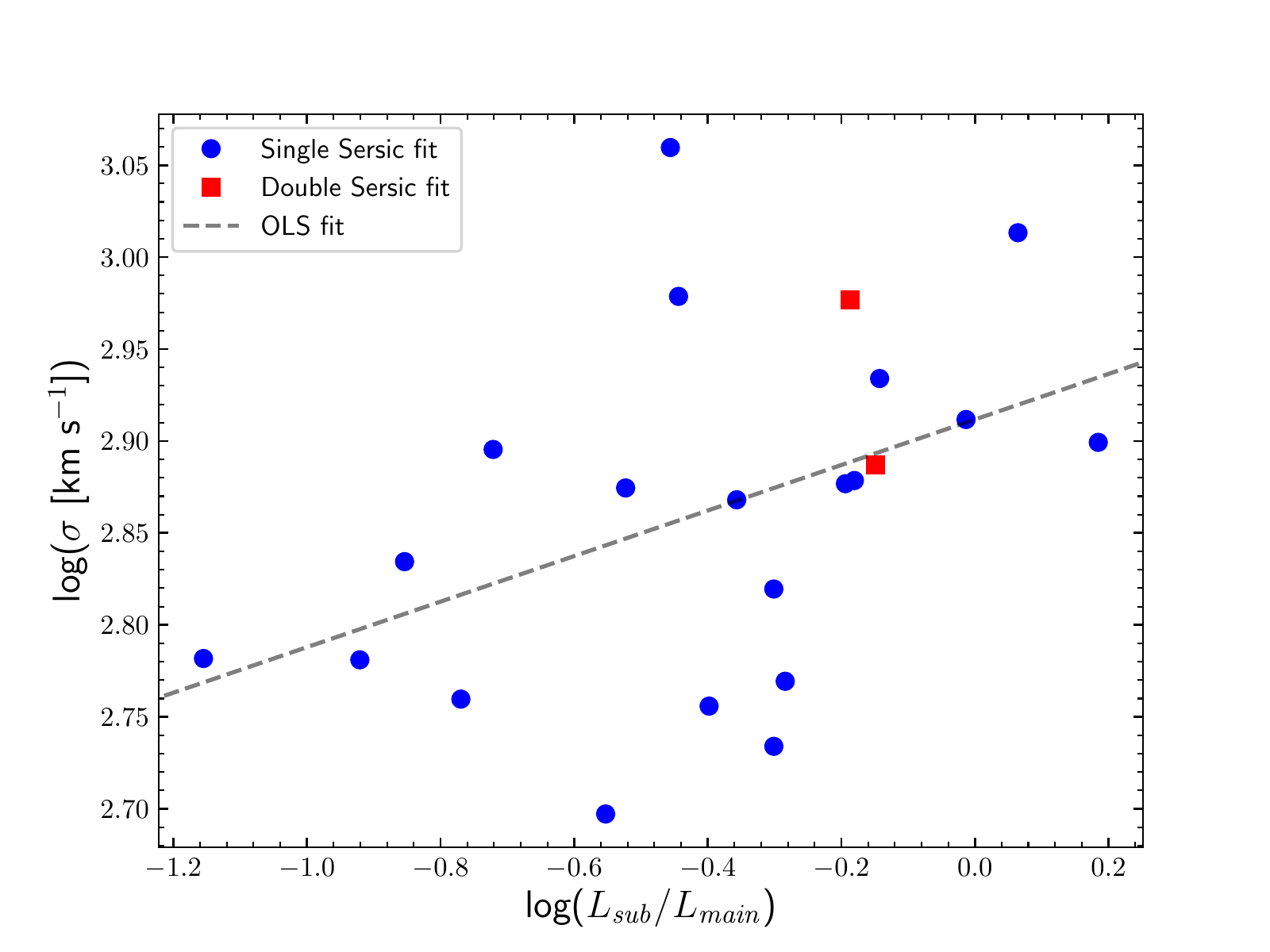}
    \caption{Correlation between the cluster velocity dispersion calculated by \cite{Biv17} and the amount of cluster light inside the subtructures identified by \cite{ram7}.}
    \label{fig:subs}
\end{figure}

We conclude by observing that the ratio between the luminosity of cluster substructures and the main cluster component measured by \cite{ram7} mildly correlates with the cluster velocity dispersion (see Figure~\ref{fig:subs} and Table~\ref{tab:correlations}). This is a somewhat expected result, as the velocity dispersion should increase when a merging occurs.

%\begin{figure}[t]
%       \includegraphics[width=0.45\textwidth]{./sigma-r200.pdf}
%    \caption{Correlation between the virial radius and the velocity dispersion. The colors are a function of the cluster mass.}
%    \label{fig:r200-sigma}
%\end{figure}

\section{Summary and conclusions}
\label{sec:conclusions}

We have produced the stellar light (and mass) profiles of 46 (42) nearby galaxy clusters observed by the WINGS and Omega-WINGS surveys. The best fit of the growth curves was obtained with the S\'ersic law, which was compared with the King and the $\beta$ models.
We derived from the analysis of the light profiles the main cluster parameters, i.e., effective radius, total luminosity and mass, effective surface brightness, $(B-V)$ colors, and S\'ersic index. Then we used such parameters in combination with the measured velocity dispersion of the clusters to test the main scaling relations already analyzed in the past for ETGs (see, e.g., \citealt{don17}).

When fitting the light profiles we found that 7 out of 46 clusters are best fitted by a double S\'ersic profile (an inner bright structure plus an outer faint structure). The presence of multiple components seems disconnected from other cluster properties such as the number of substructures visible in the optical images or a difference in the stellar populations content. This presence also does not seem to be linked to the presence of the BCG in the center of the clusters; the BCG effects will be investigated in a forthcoming work.

All the analyzed relations confirm that the clusters of our sample are well-virialized structures. Notably, the same relations that are valid for the ETGs are visible for clusters at different scales, providing a clear indication that a scale-free phenomenon of mass accretion regulated by gravitation is at work. Like ETGs, clusters also exhibit a degree of non-homology (varying values of the S\'ersic index even for clusters with the same luminosity) in their visible light and stellar mass profiles, and a very robust correlation of the S\'ersic index with the effective radius. This is a somewhat unexpected property on the basis of numerical simulations (see, e.g, \citealt{col96}; \citealt{nav97}), which predict self-similar DM halos with Navarro-Frenk-White profiles. 

The most interesting and new relation found here for the first time is the existence of a color-magnitude relation for clusters. When this relation is calculated considering only the galaxies within an area of $0.6 \, r_{200}$, the CM slope perfectly matches the average red sequence slope found for the galaxies in the Omega-WINGS clusters. The CM cluster relation appears even more clearly when the analysis of the cluster properties is pushed beyond $0.6 \, r_{200}$. The blue (red) clusters are the faint (bright) clusters.
The existence of such relations must find an explanation in the current paradigm of galaxy and cluster formation. In fact, it is not easy to understand why the most massive structures preferentially host the older and redder galaxies, while the less massive clusters host the younger and bluer galaxies. In fact, the hierarchical accretion scenario predicts that the first structures to form are the smallest structures, while the biggest structures are the latest to form.

The questions of to what extent the CM cluster relation is a cluster environment effect and what model of cluster formation and evolution is consistent with such data are left to future investigations.

Finally we observed that the cluster luminosity correlates with the intrinsic $(B-V)$ color gradient measured within $r_{200}$. We see that the faintest clusters show the largest color gradients. This behavior is not observed in ETGs, where the optical color gradient appears uncorrelated with the galaxy luminosity \citep{lab10}.

In forthcoming papers we will investigate how the BCG and cluster properties are connected, which models of clusters and galaxies can explain the observed CM relation,
what are the similarities and differences between the stellar mass/light profiles and the hydrostatic/dynamic mass profiles, and what is the baryon fraction in the local universe.

\begin{acknowledgements}
The authors thank the anonymous referee for the valuable suggestions that helped to improve the overall quality of this research, Mr. Davide Bonanno Consiglio for his contribution in the statistical analysis, and Dr. Andrea Biviano for his precious comments and suggestions.
\end{acknowledgements}

\begin{appendix}
\section{Tables and figures}
\label{sec:fig}

In the following Appendix we gathered all the tables and figures omitted from the main article.

In Table~\ref{tab:clusters} we summarize the data sets available for each WINGS galaxy cluster.

In Table~\ref{tab:phot_and_mass} we compared the main parameters of our clusters.

Table~\ref{tab:eff} presents a summary of the main effective parameters of each cluster. In case of the double S\'ersic fits, the average effective intensity is given at the cluster effective radius, which does not correspond to either of the effective radii of the two different components.

In Tables~\ref{tab:single_fits} and \ref{tab:double_fits} we tabulated the best single and double S\'ersic fit parameters, plus the reduced $\chi^2$, AIC, and BIC values of all the best-fit models.

\begin{figure*}[t]
   \centering
        \includegraphics[width=0.45\textwidth]{./A85_plots.pdf}        \includegraphics[width=0.45\textwidth]{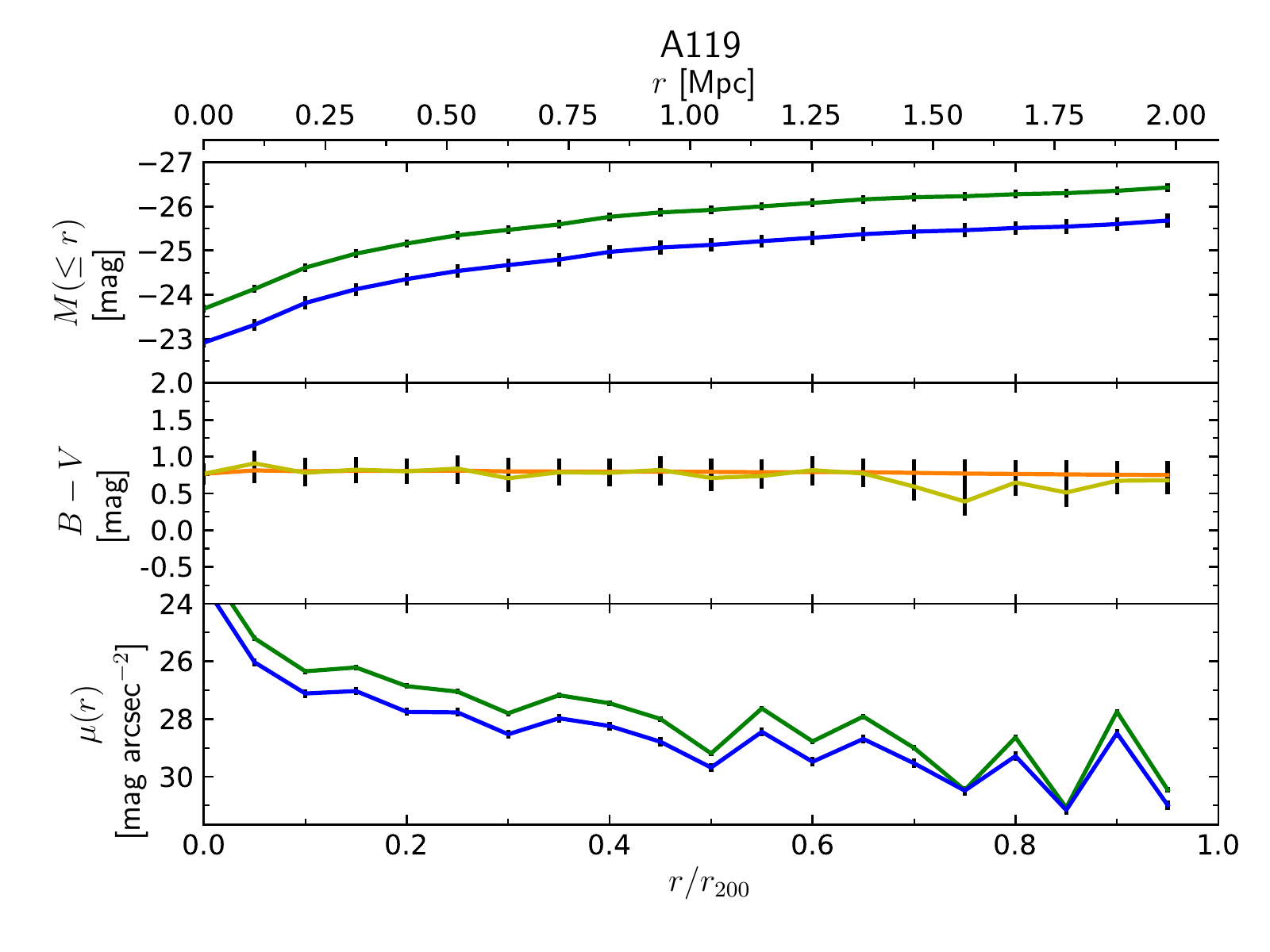}
        \includegraphics[width=0.45\textwidth]{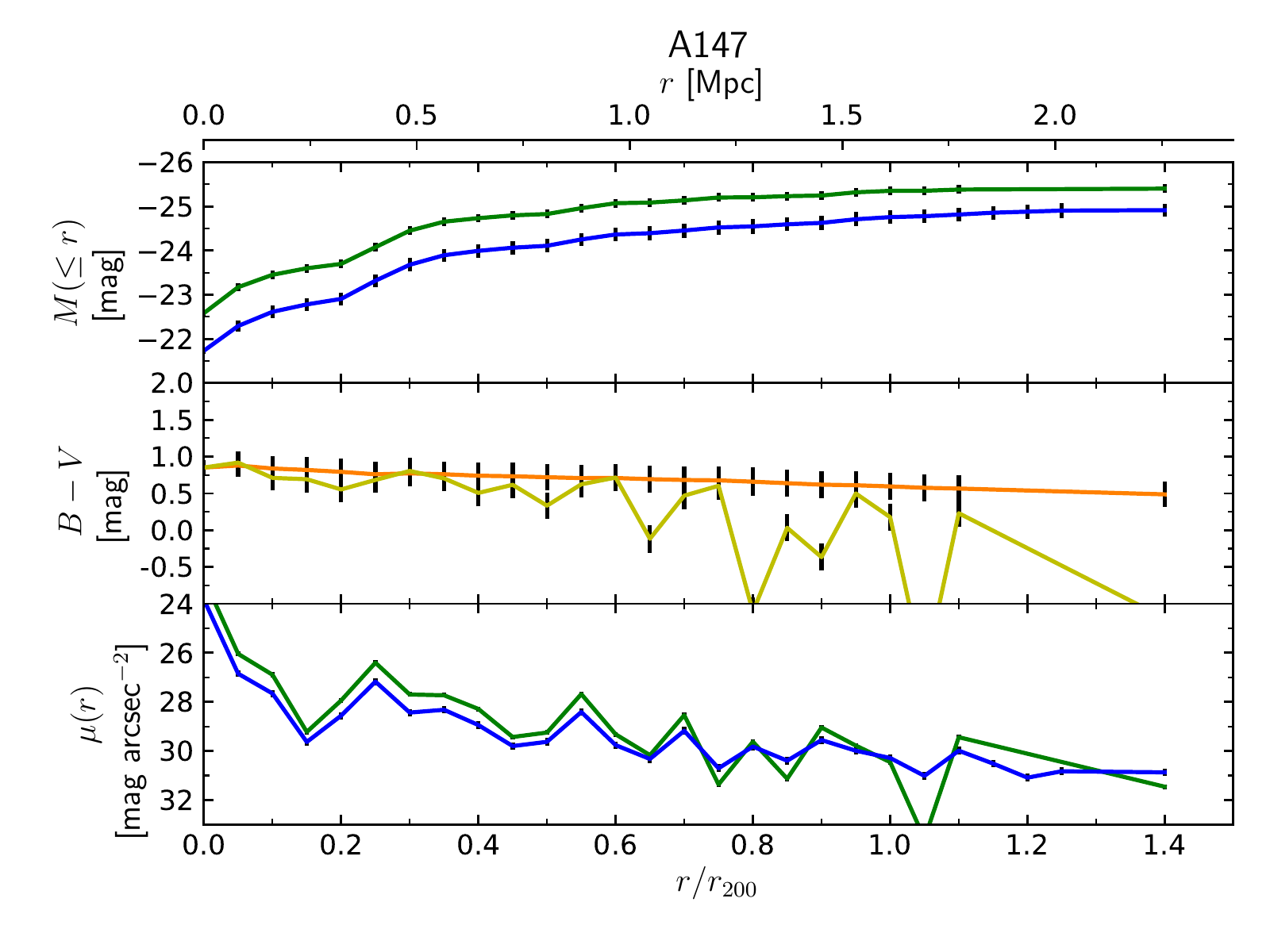}       \includegraphics[width=0.45\textwidth]{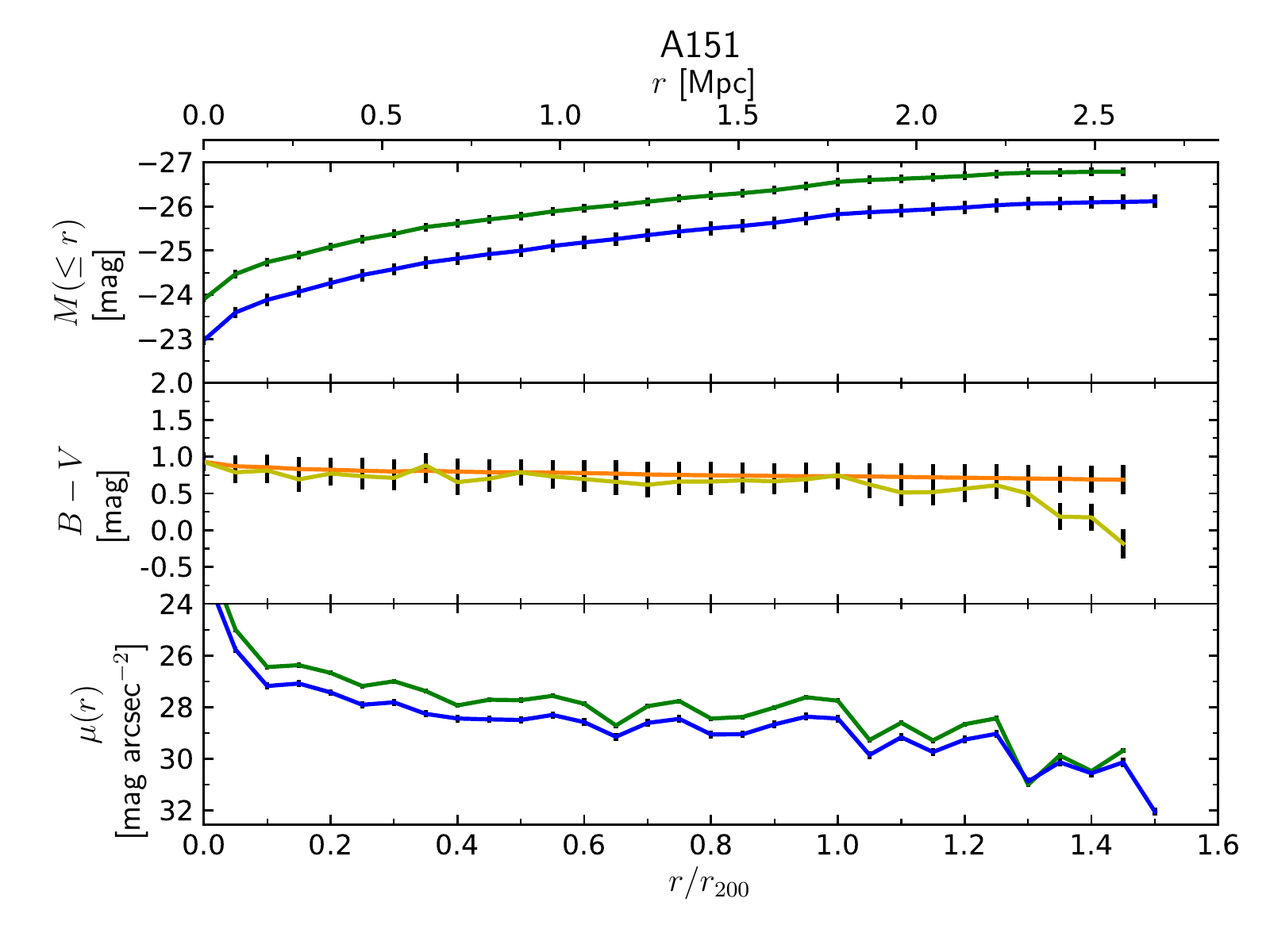}
        \includegraphics[width=0.45\textwidth]{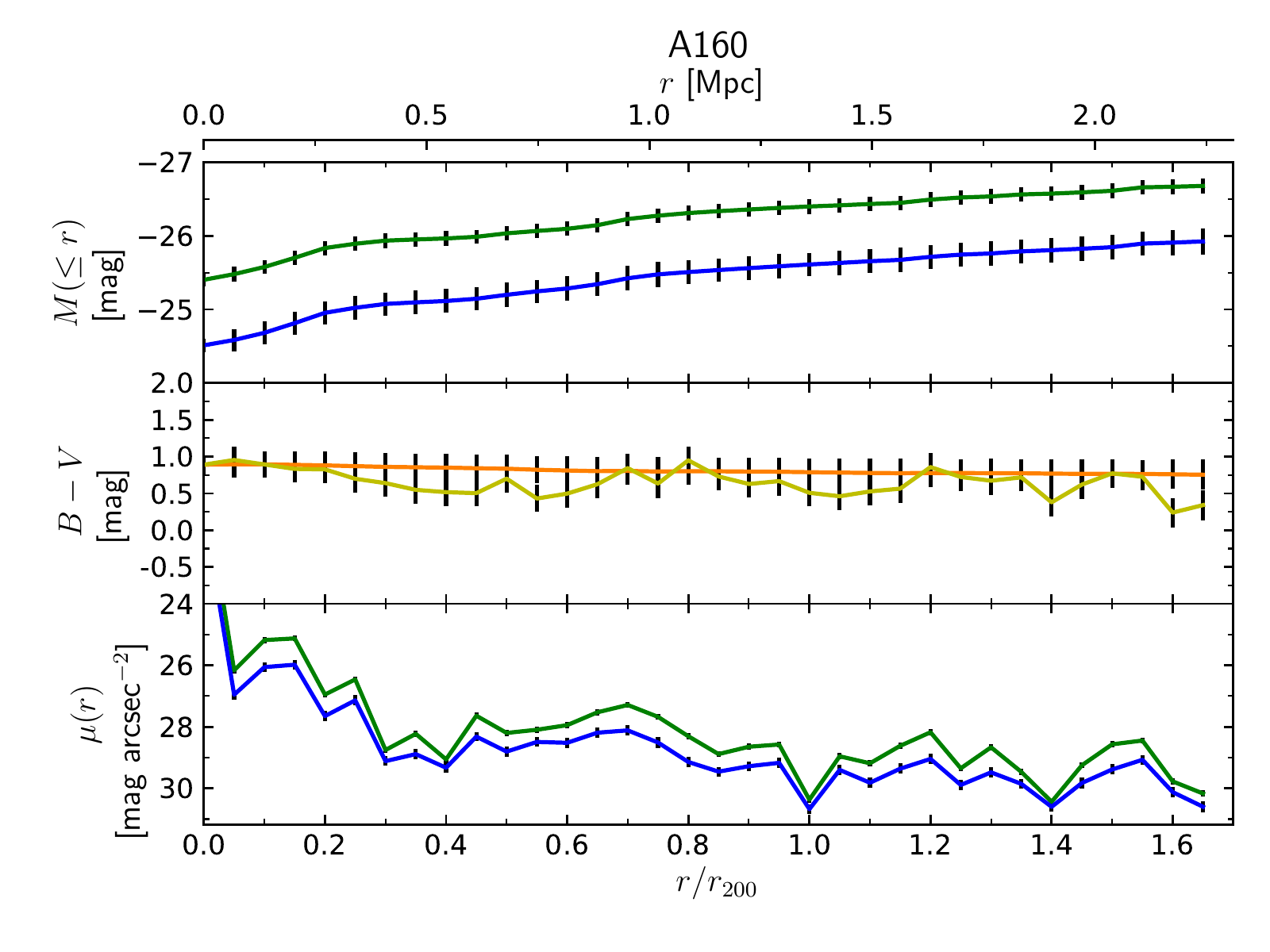}       \includegraphics[width=0.45\textwidth]{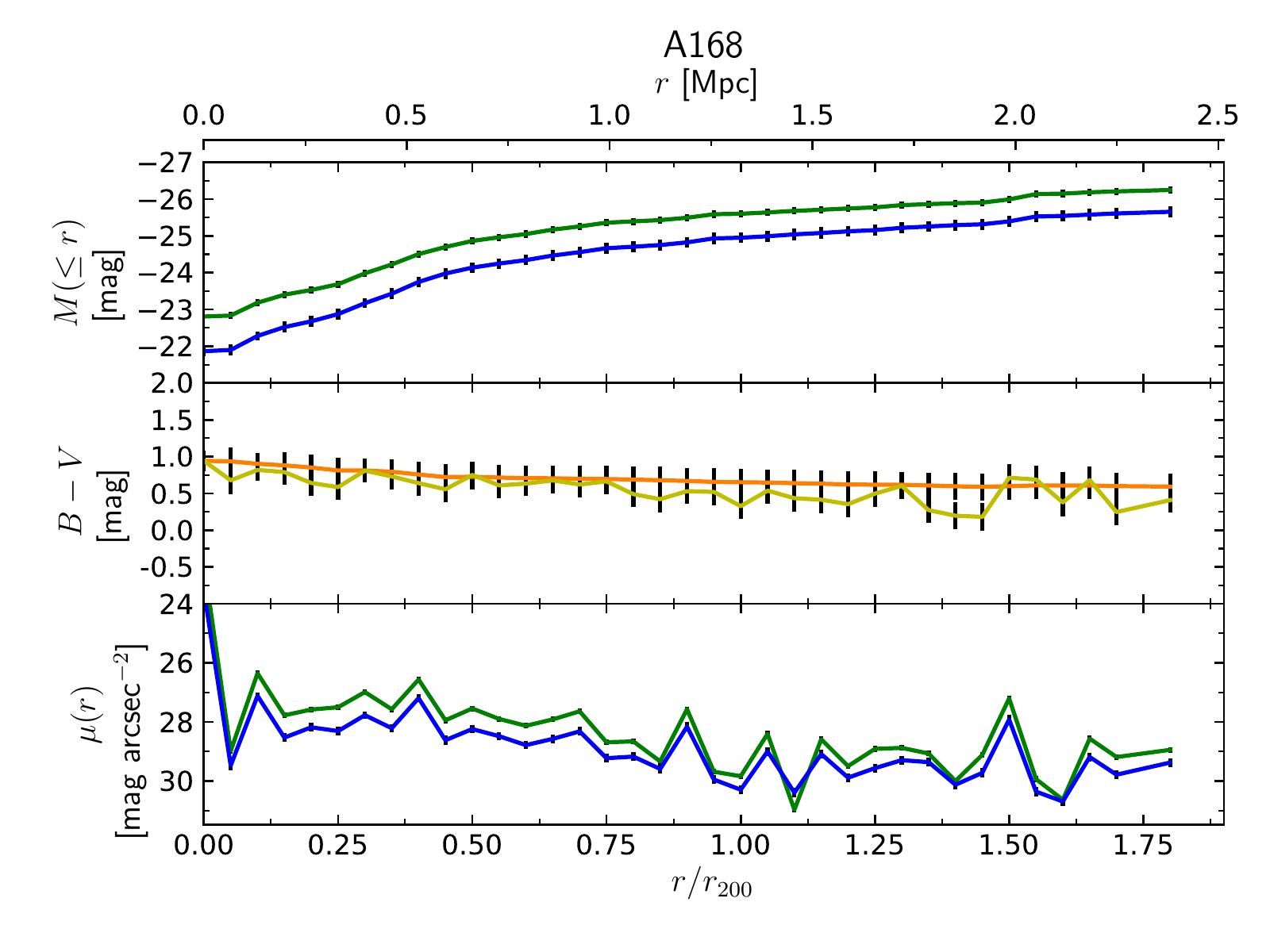}
    \caption{Photometric profiles of Omega-WINGS galaxy clusters. The color code is the same as in Figure~\ref{fig:plots-example}, left panel.}
    \label{fig:plots-begin}
\end{figure*}

\begin{figure*}[t]
   \centering
        \includegraphics[width=0.45\textwidth]{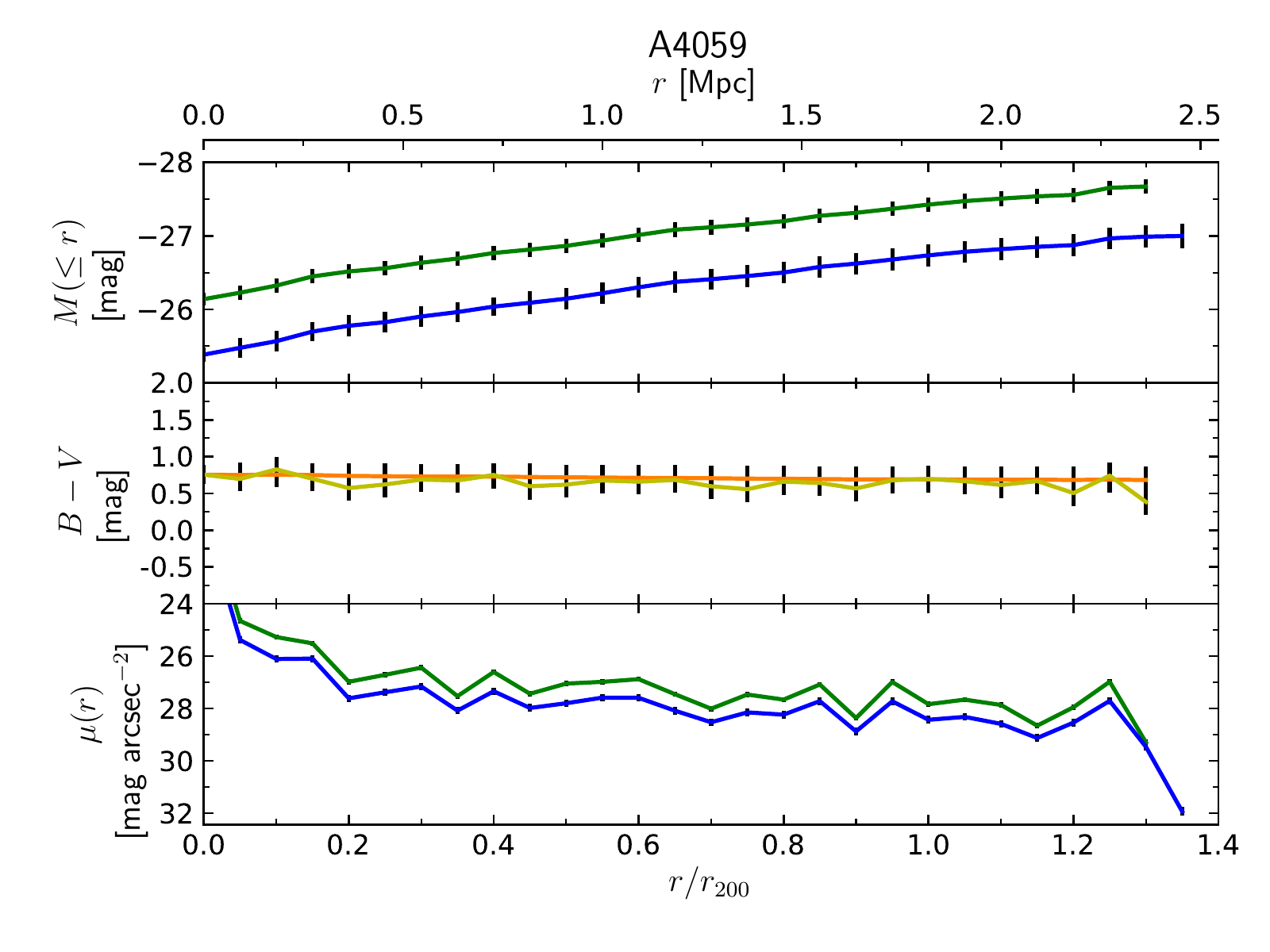}      \includegraphics[width=0.45\textwidth]{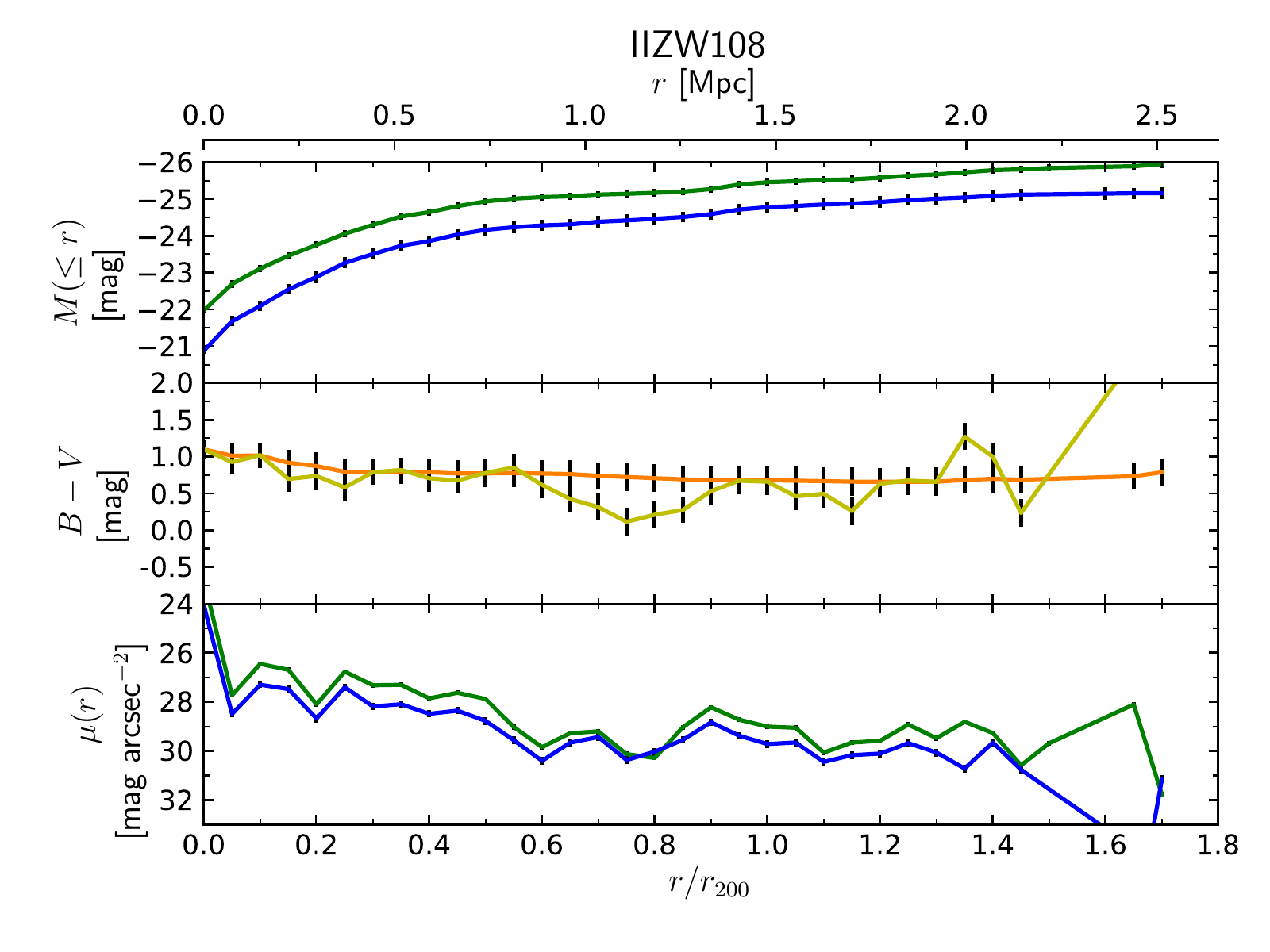}
        \includegraphics[width=0.45\textwidth]{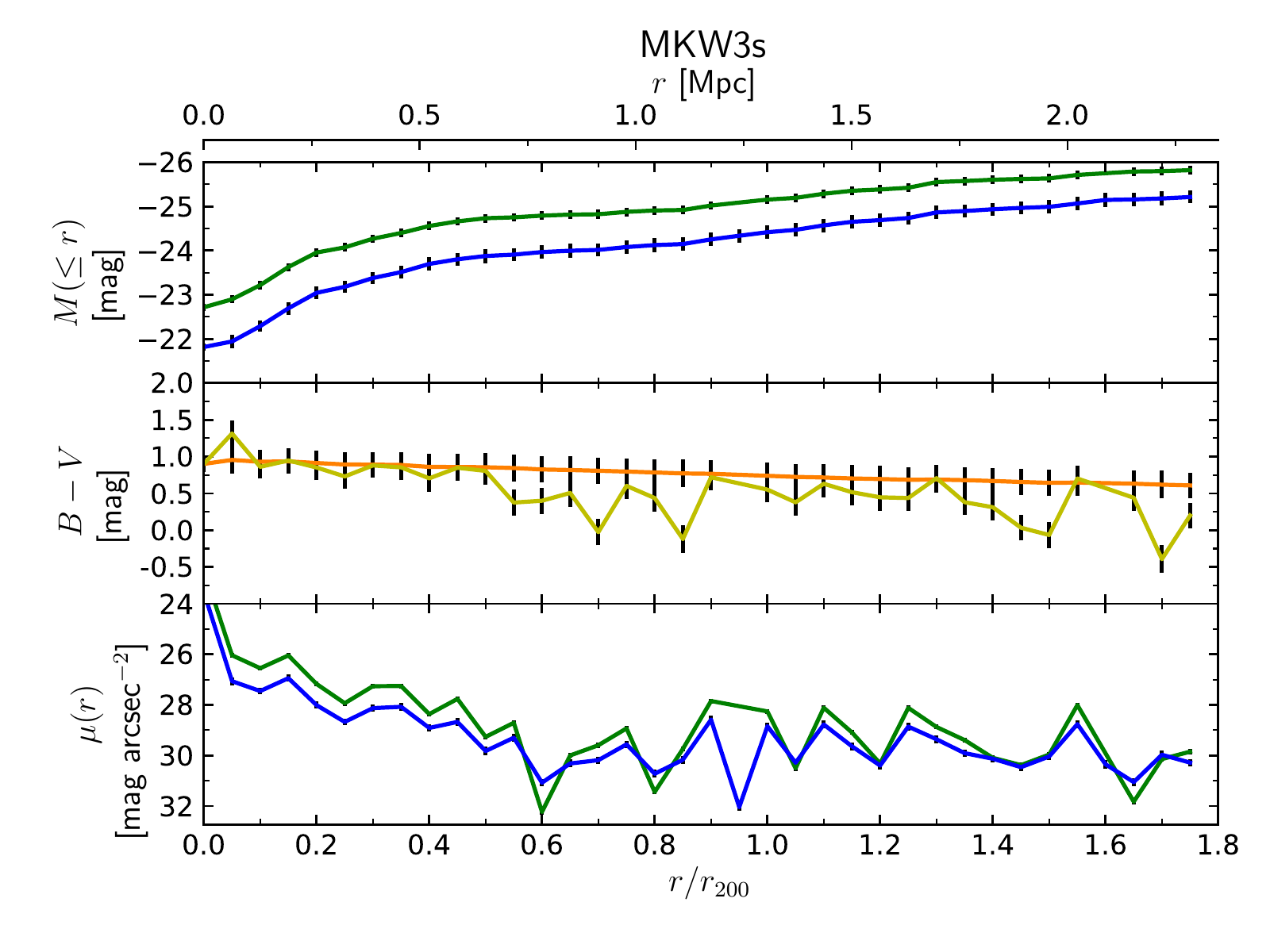}      \includegraphics[width=0.45\textwidth]{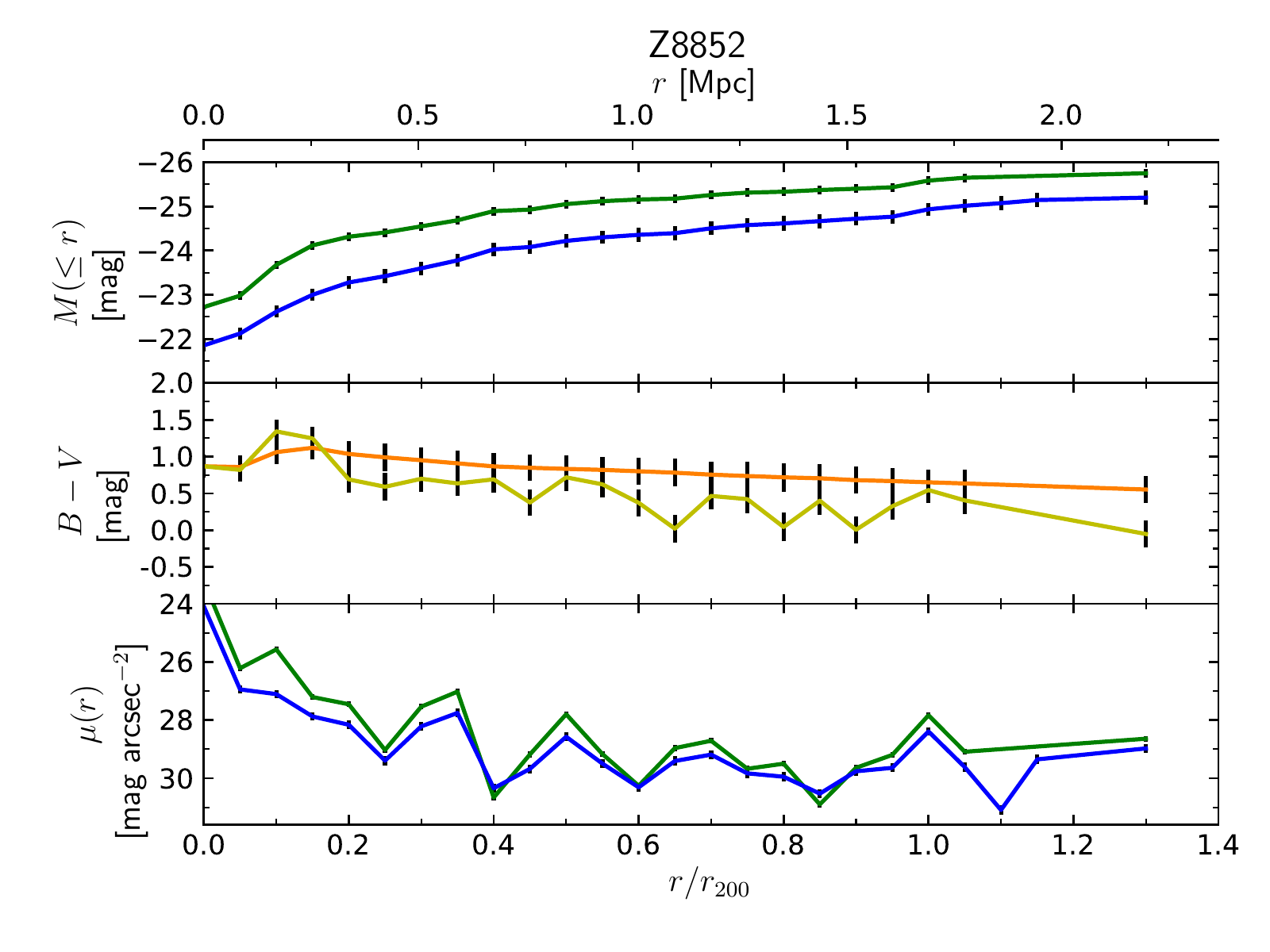}
    \caption{Photometric profiles of Omega-WINGS galaxy clusters, continued.}
    \label{fig:plots-end}
\end{figure*}

\begin{figure*}[t]
   \centering
        \includegraphics[width=0.45\textwidth]{./mass_A85_plots.pdf}   \includegraphics[width=0.45\textwidth]{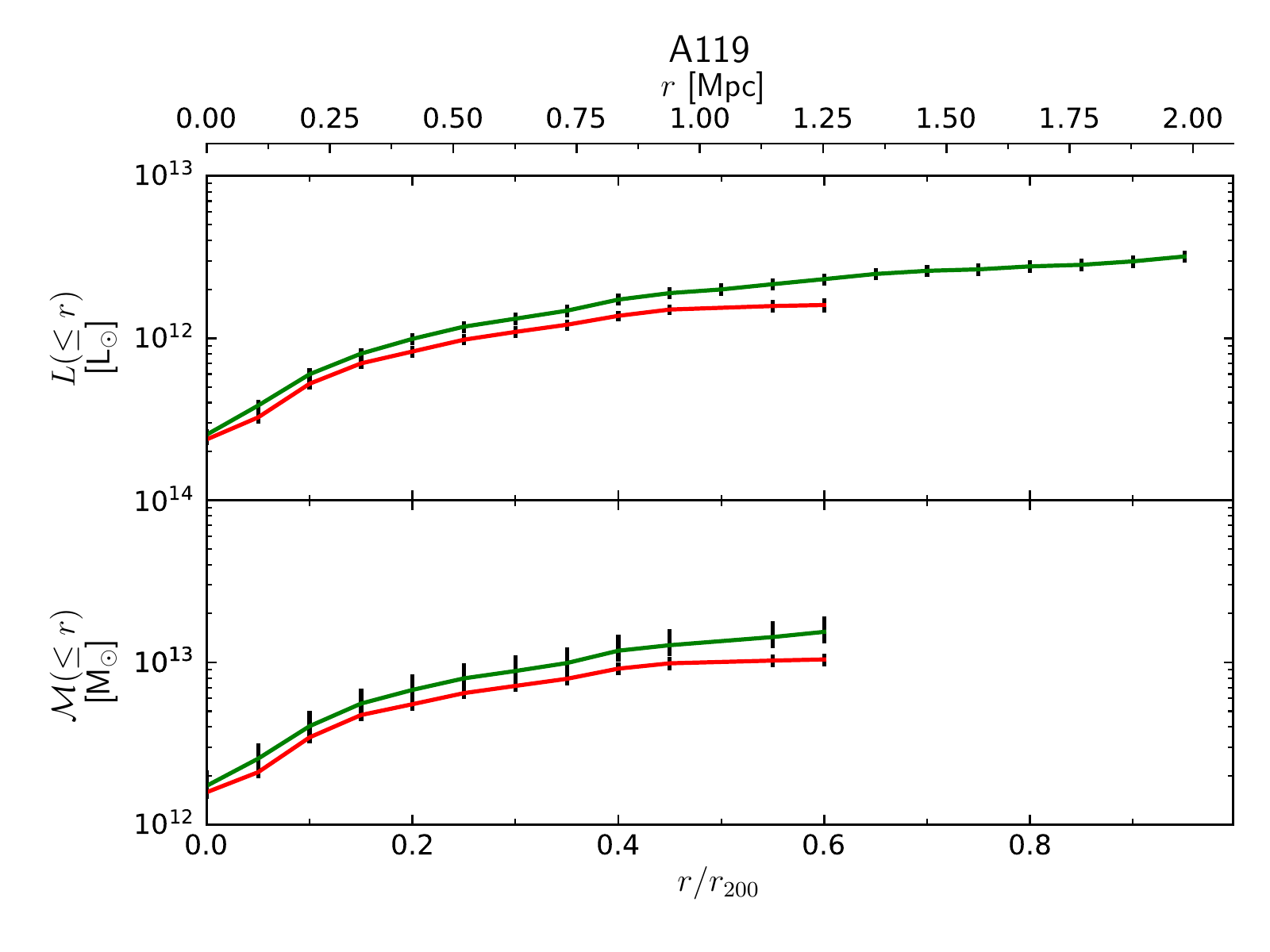}
        \includegraphics[width=0.45\textwidth]{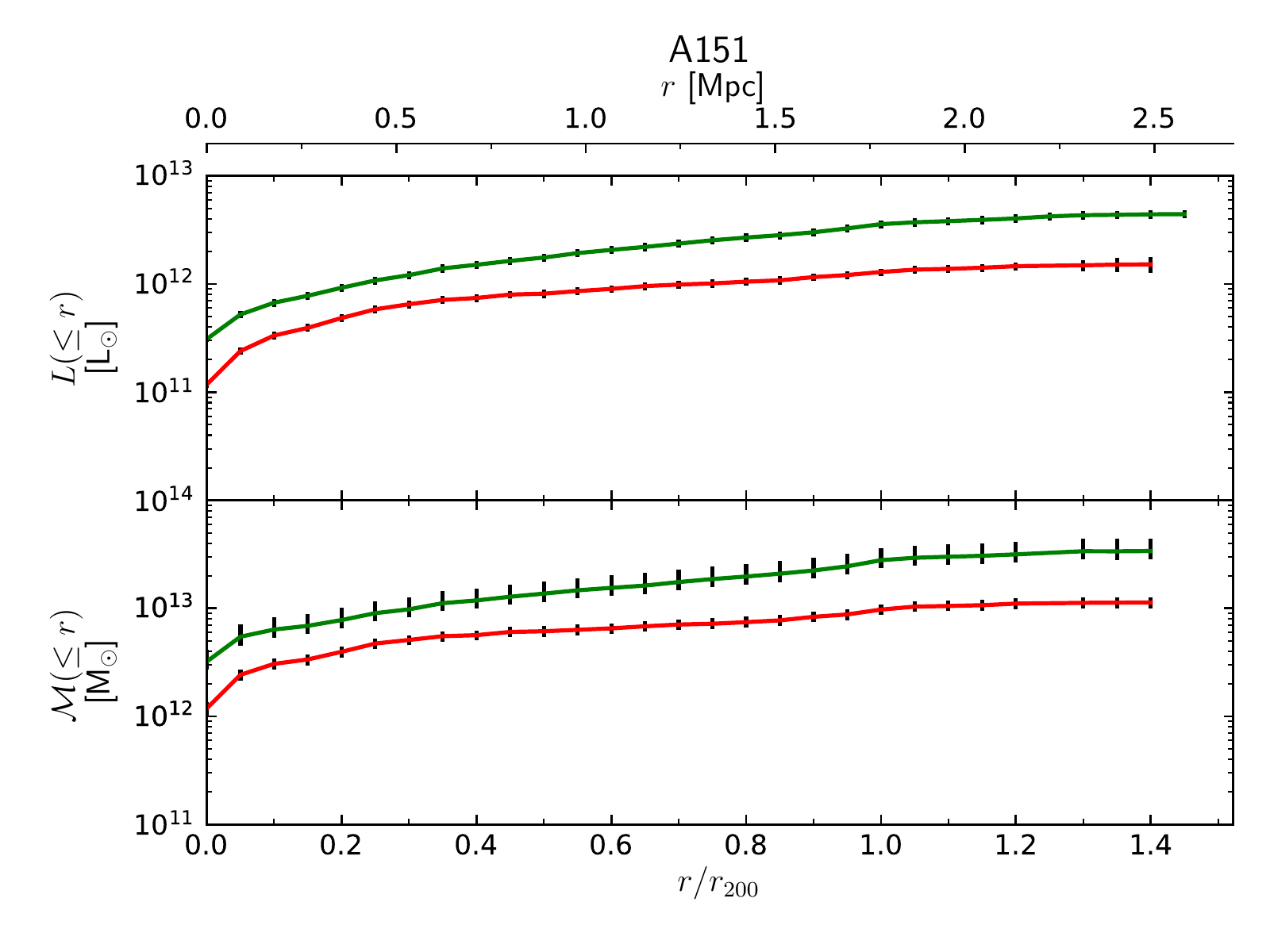}  \includegraphics[width=0.45\textwidth]{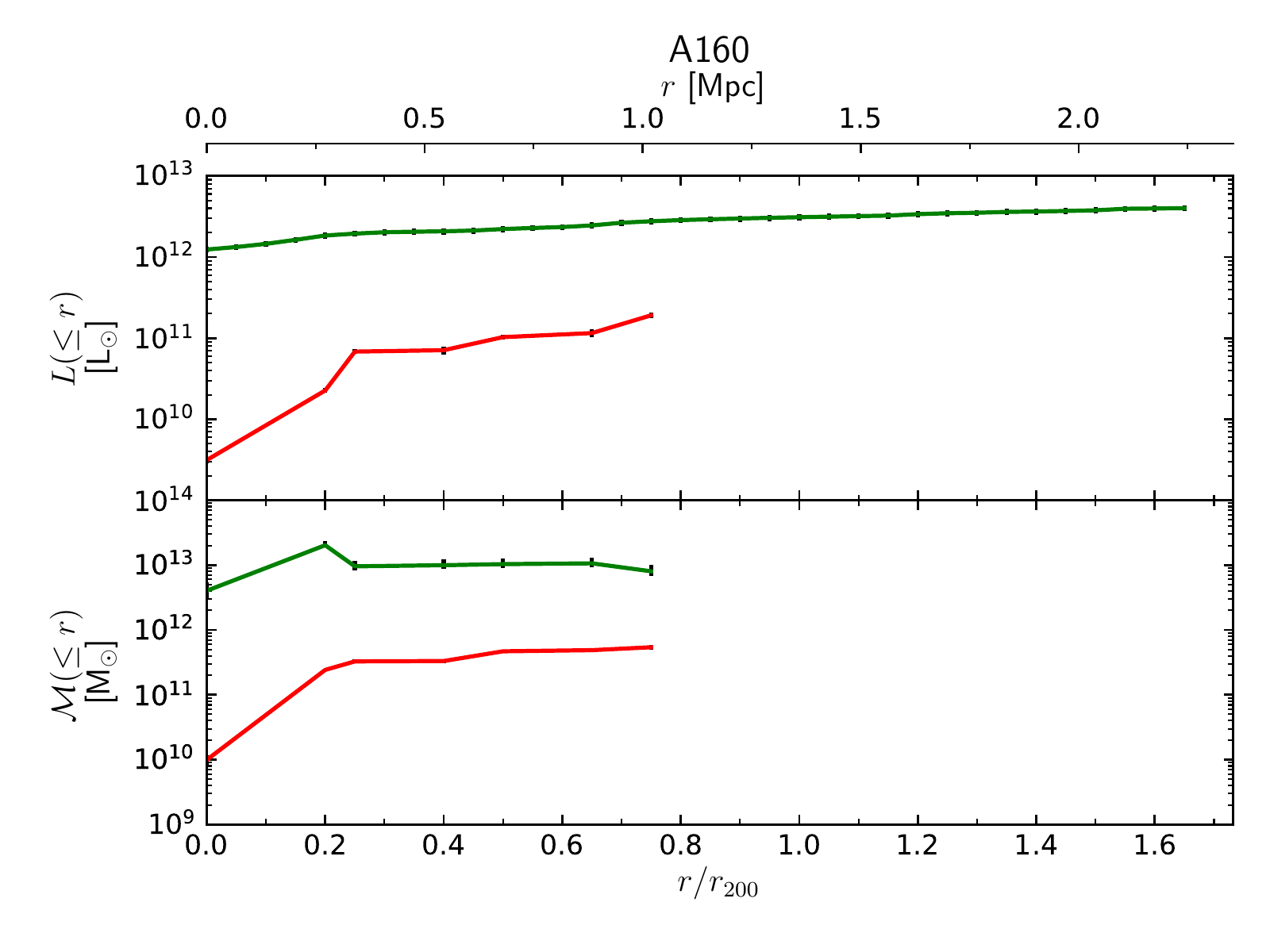}
        \includegraphics[width=0.45\textwidth]{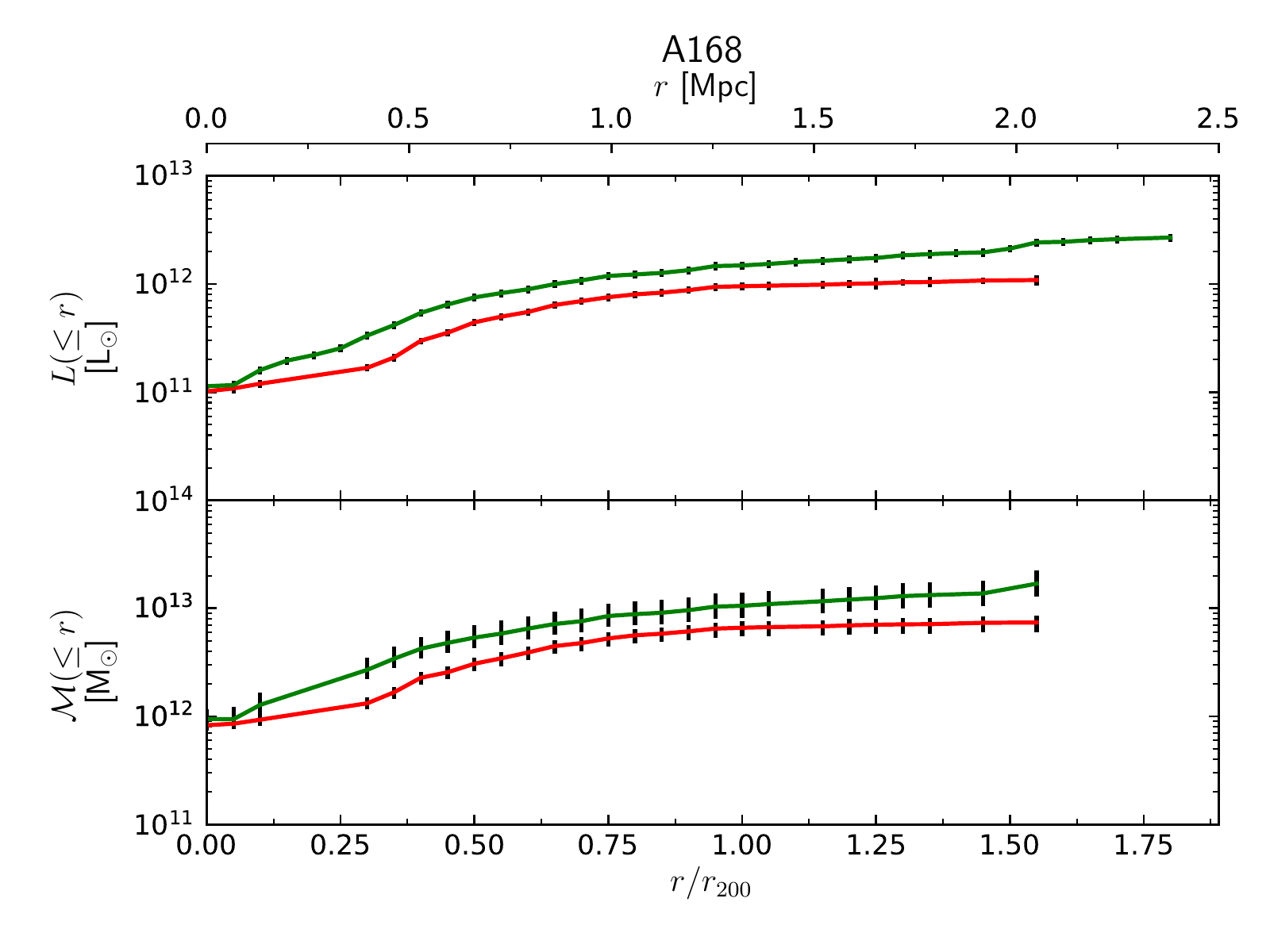}\includegraphics[width=0.45\textwidth]{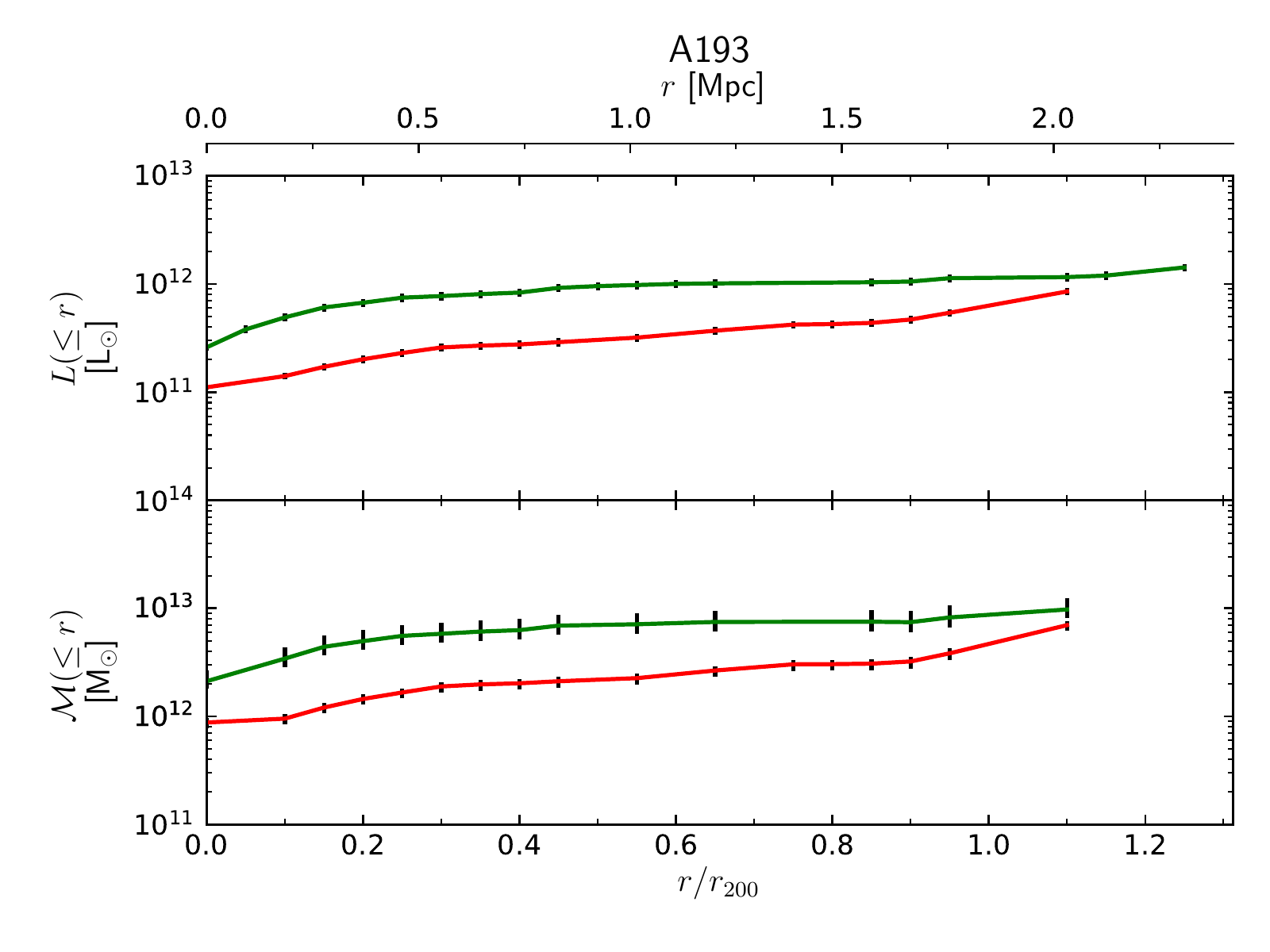}
    \caption{Mass profiles of Omega-WINGS galaxy clusters. The color code is the same as in Figure~\ref{fig:plots-example}, right panel.}
    \label{fig:mass_plots-begin}
\end{figure*}

\begin{figure*}[t]
        \includegraphics[width=0.45\textwidth]{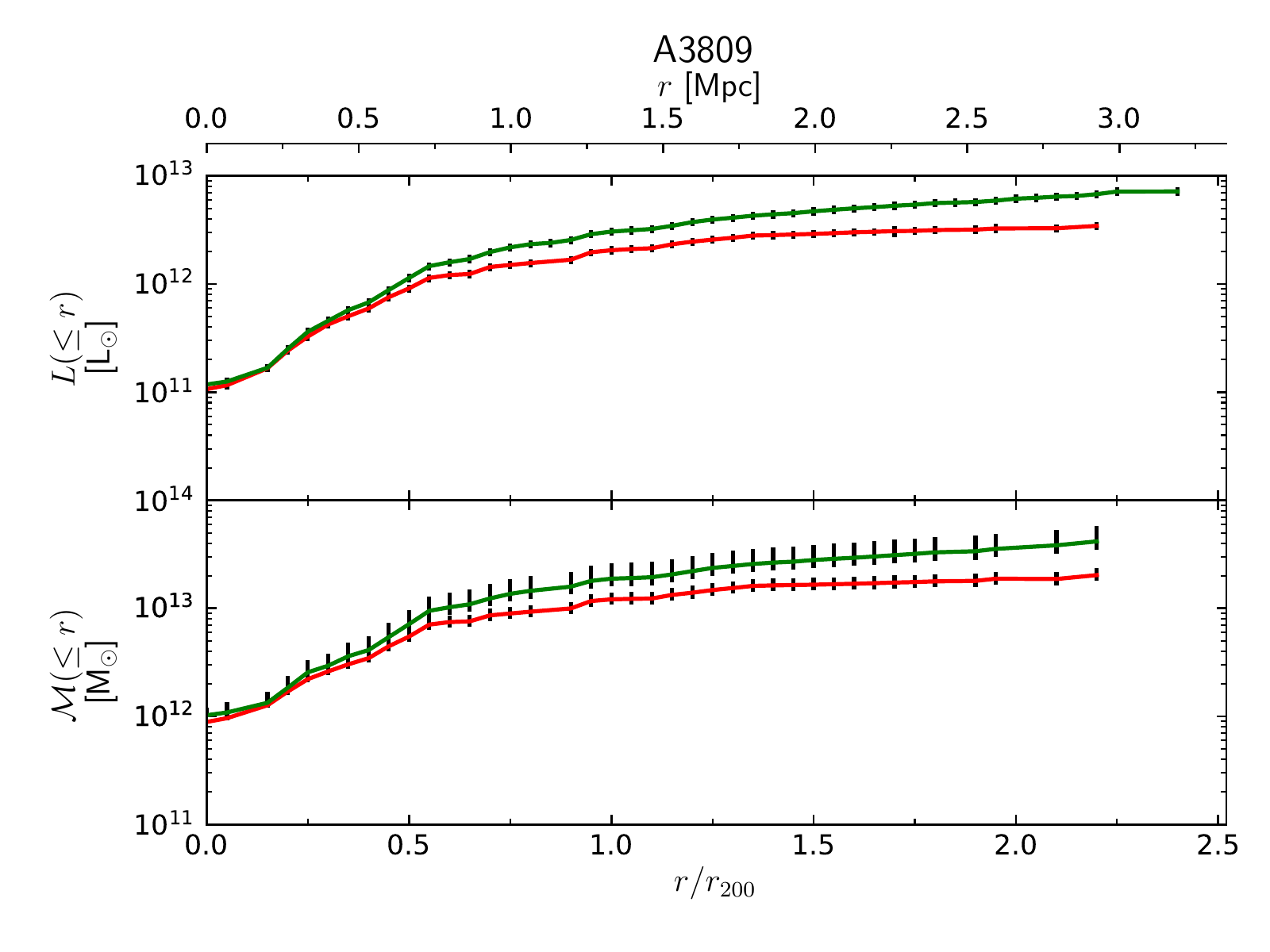}\includegraphics[width=0.45\textwidth]{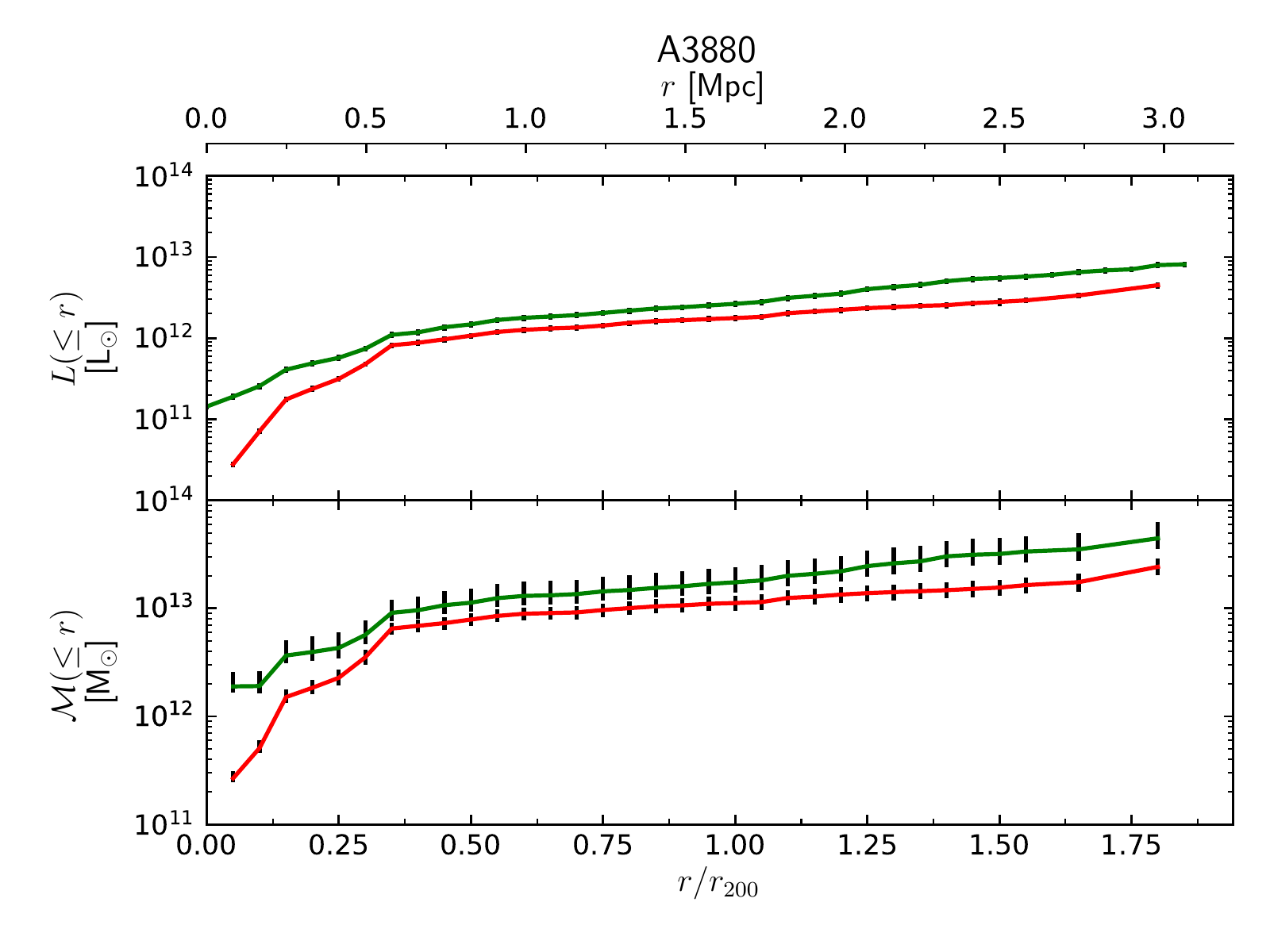}
        \includegraphics[width=0.45\textwidth]{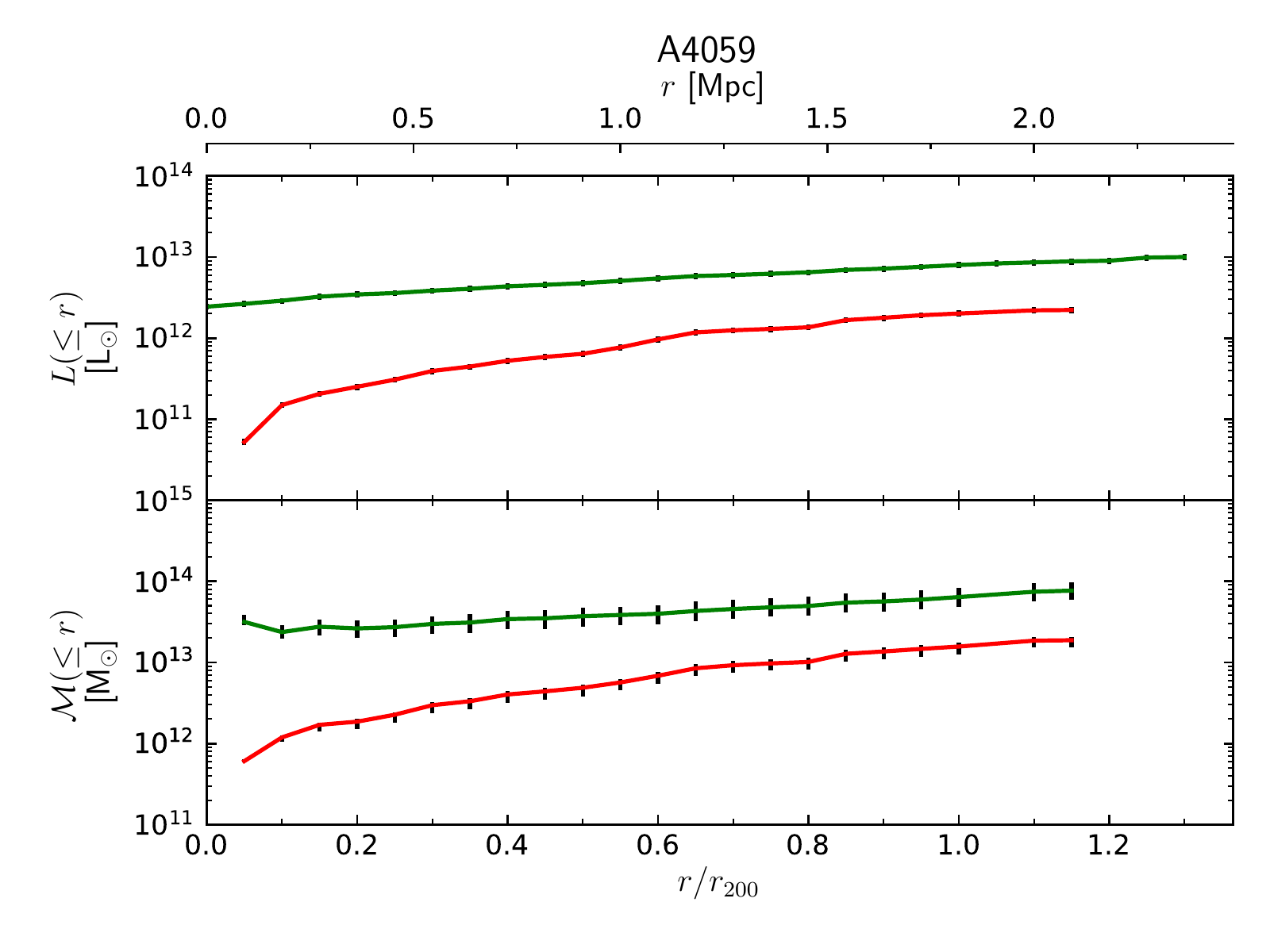}\includegraphics[width=0.45\textwidth]{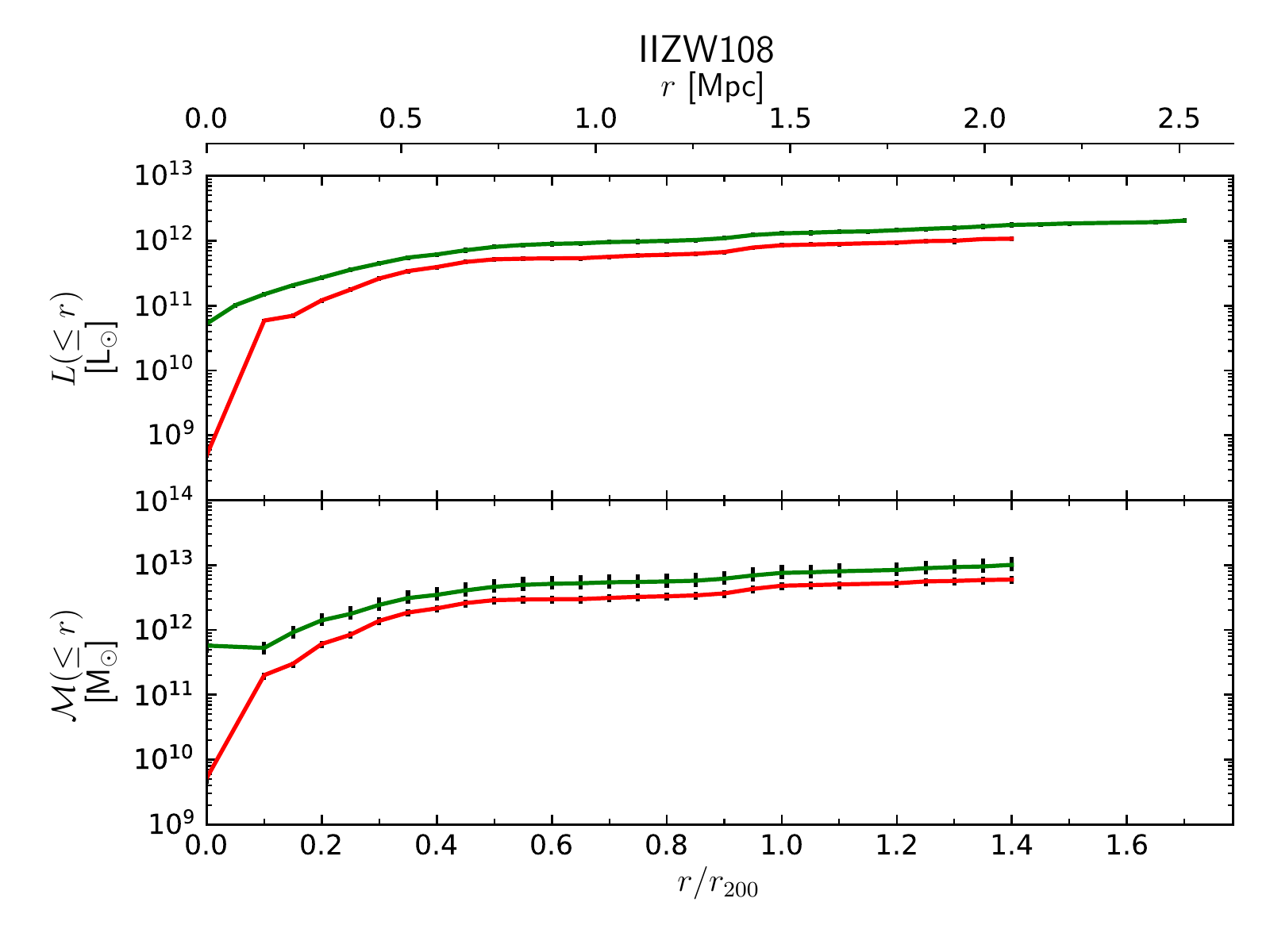}
        \includegraphics[width=0.45\textwidth]{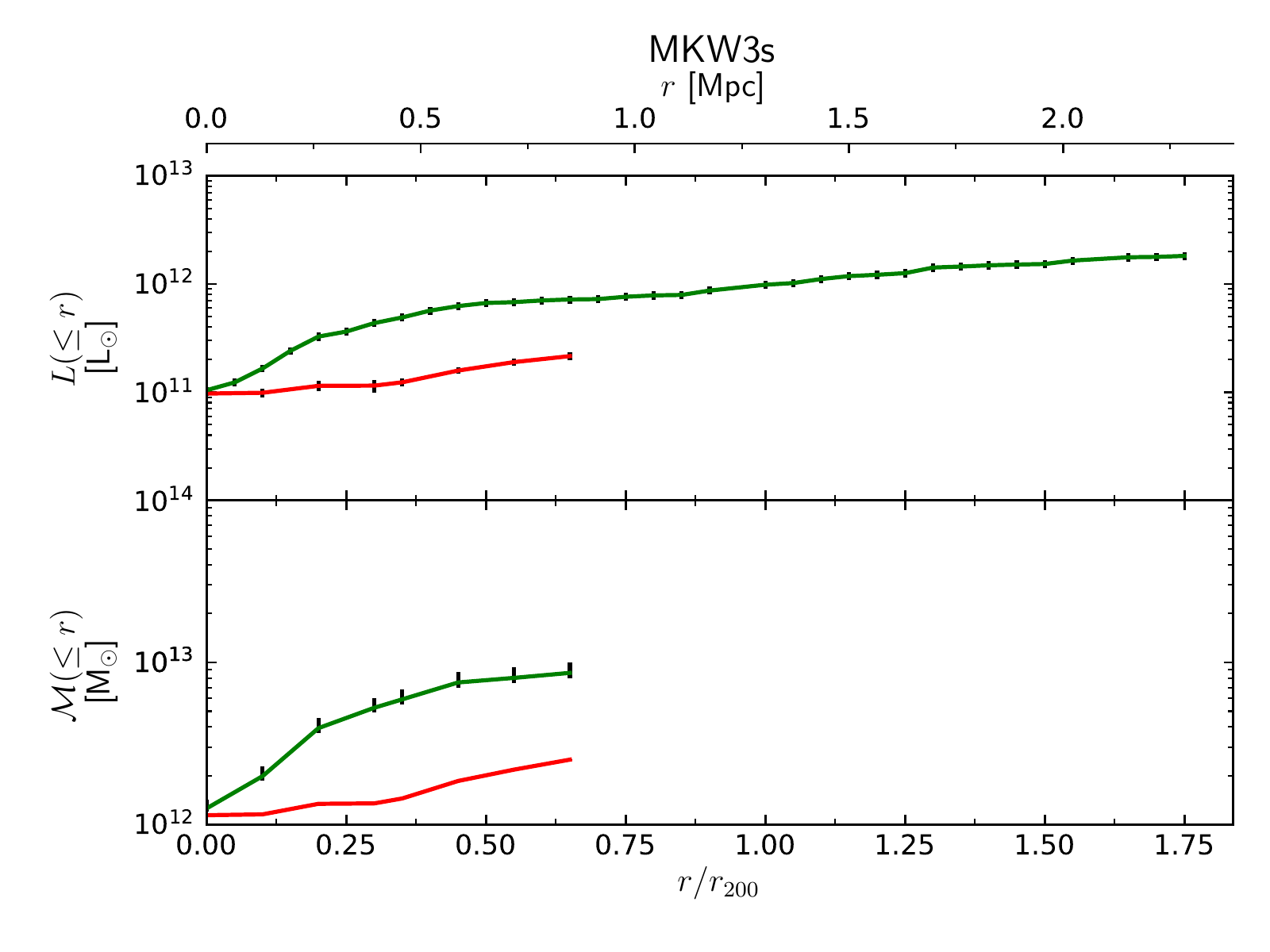} \includegraphics[width=0.45\textwidth]{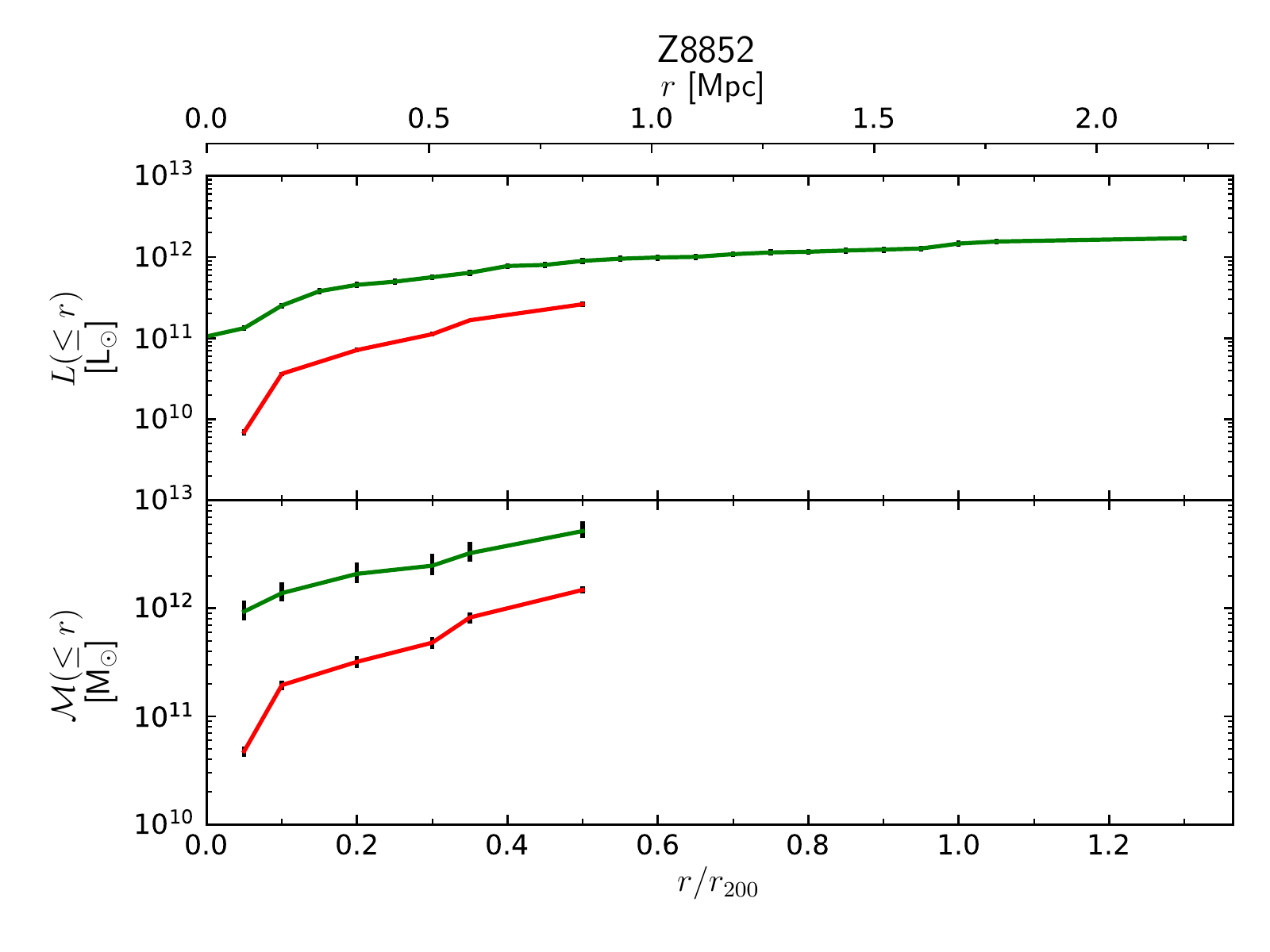}
    \caption{Mass profiles of Omega-WINGS galaxy clusters, continued.}
    \label{fig:mass_plots-end}
\end{figure*}

\begin{figure*}[t]
   \centering
        \includegraphics[width=0.45\textwidth]{./A85_fit.pdf}  \includegraphics[width=0.45\textwidth]{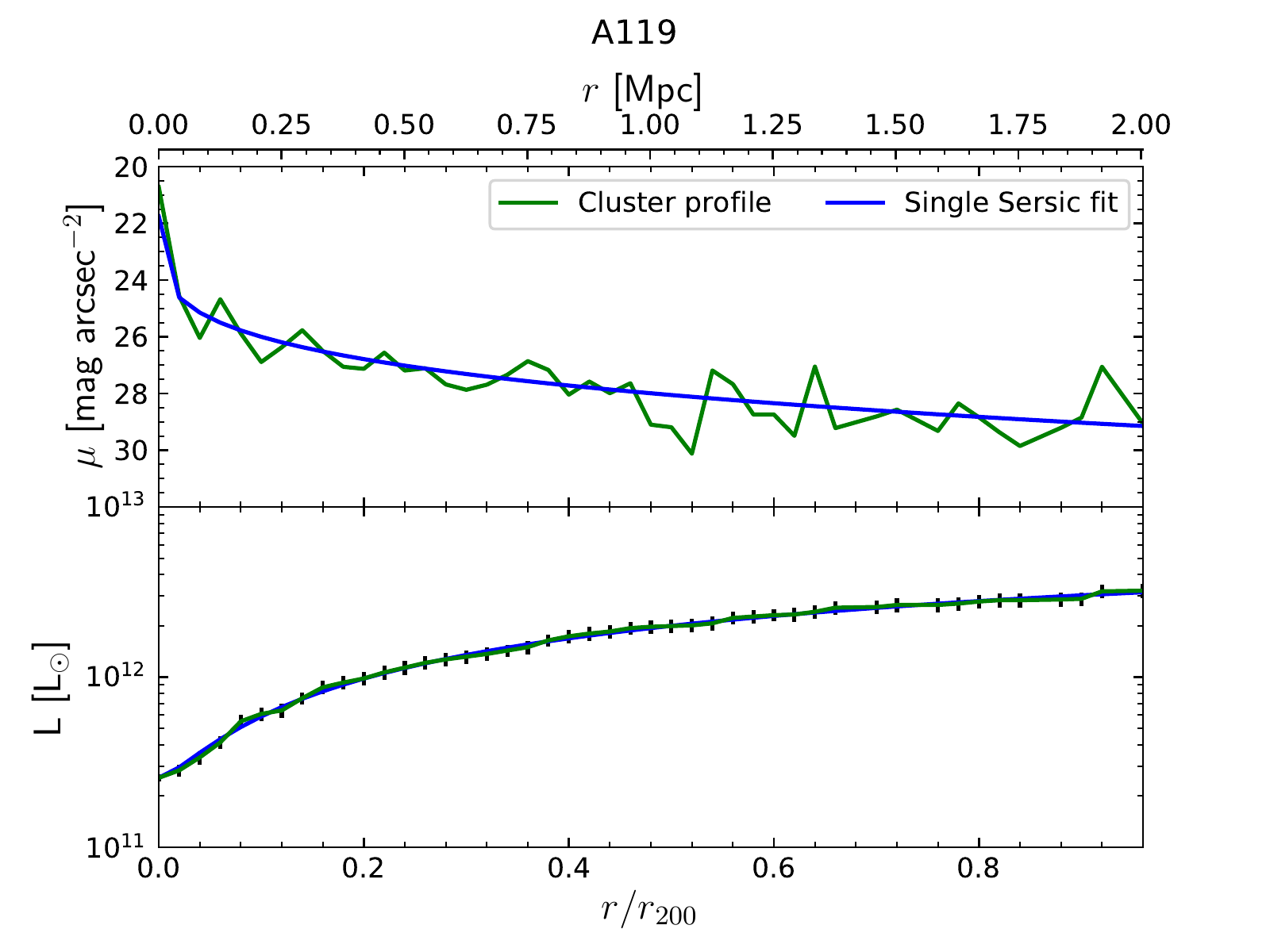}
        \includegraphics[width=0.45\textwidth]{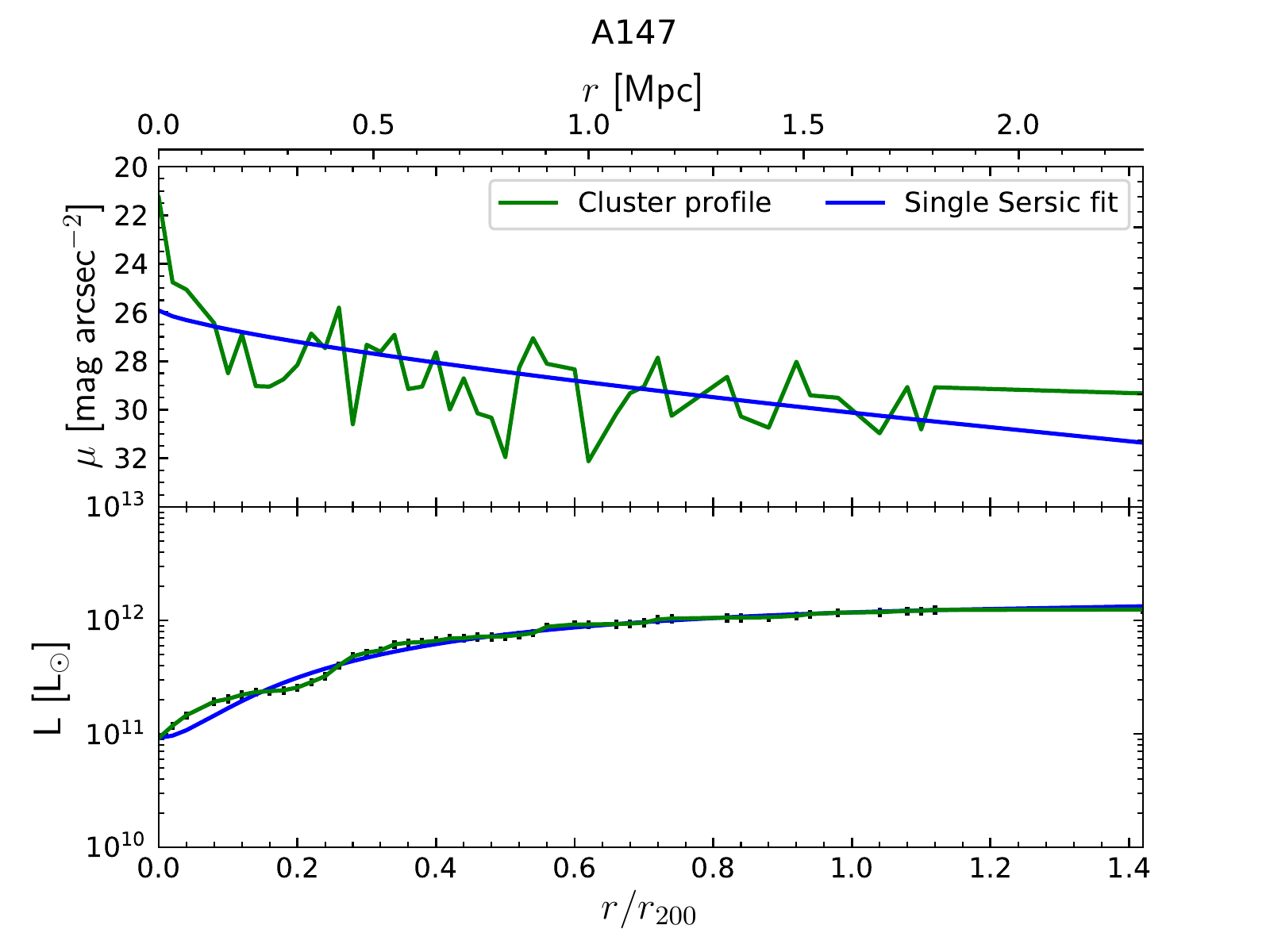} \includegraphics[width=0.45\textwidth]{./A151_fit.pdf}
        \includegraphics[width=0.45\textwidth]{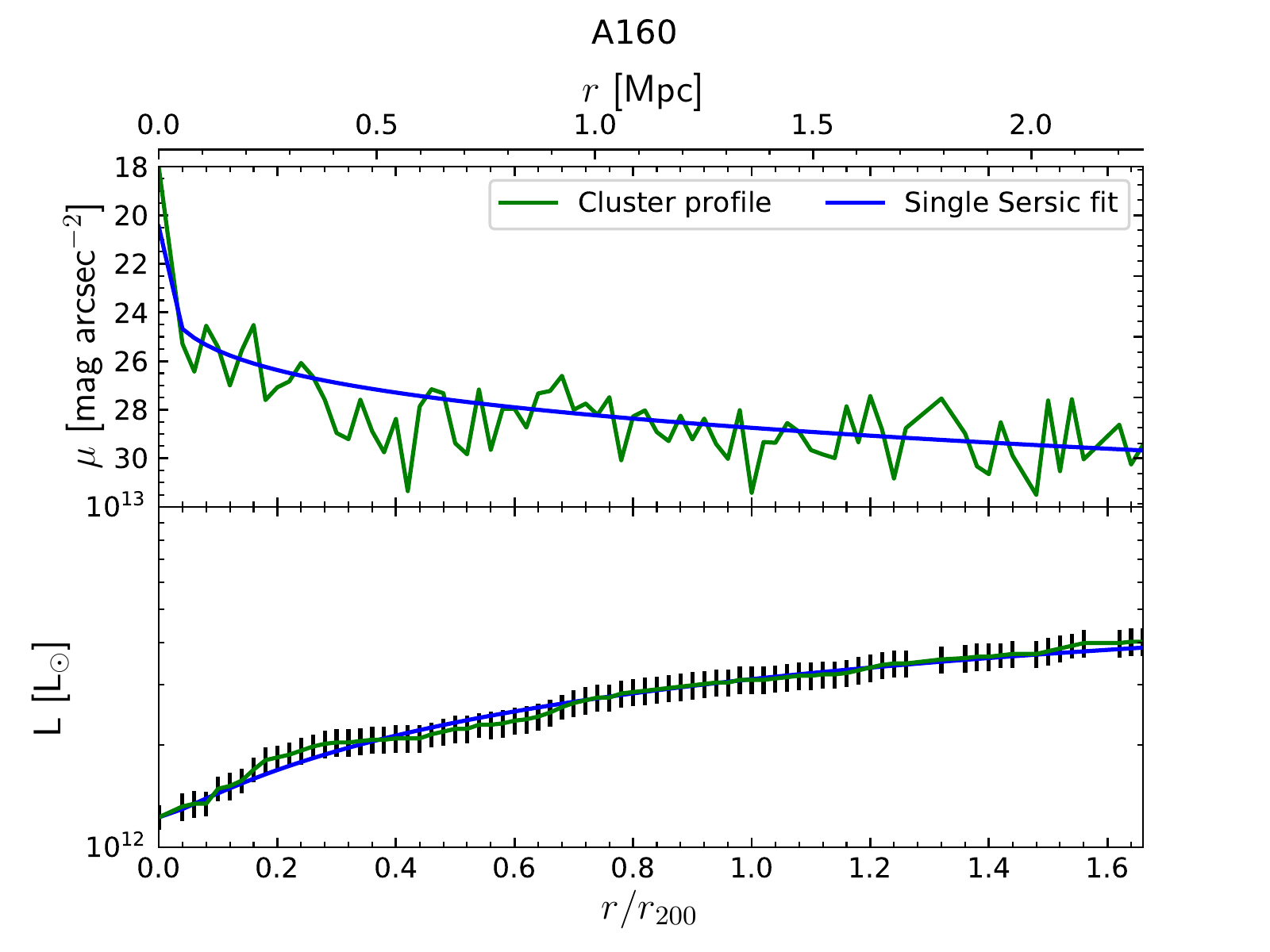} \includegraphics[width=0.45\textwidth]{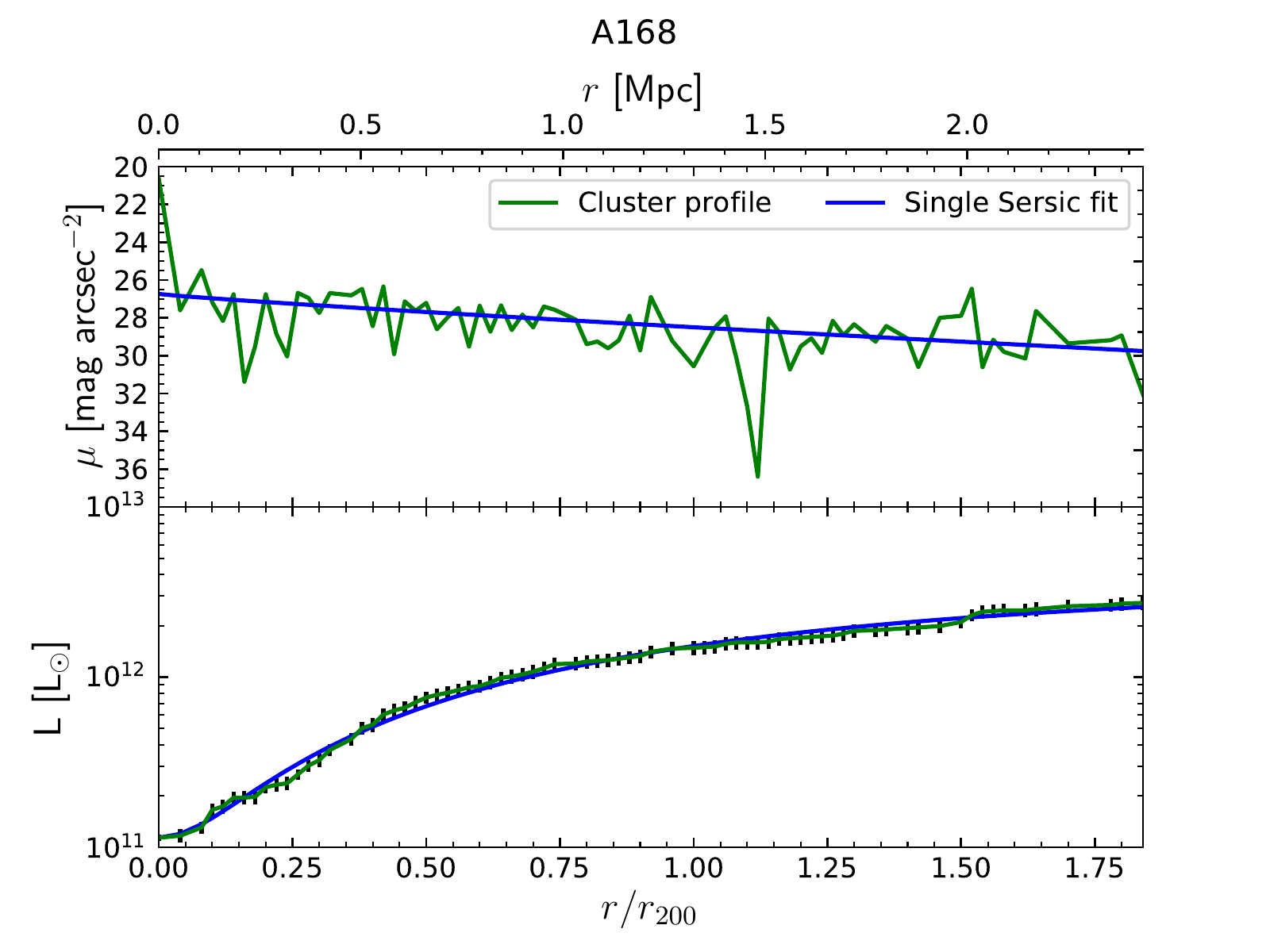}
    \caption{Photometric decomposition of Omega-WINGS galaxy clusters luminosity profiles. The color code is the same as in Figure~\ref{fig:fits-es}.}
    \label{fig:fits-begin}
\end{figure*}

\begin{figure*}[t]
   \centering
        \includegraphics[width=0.45\textwidth]{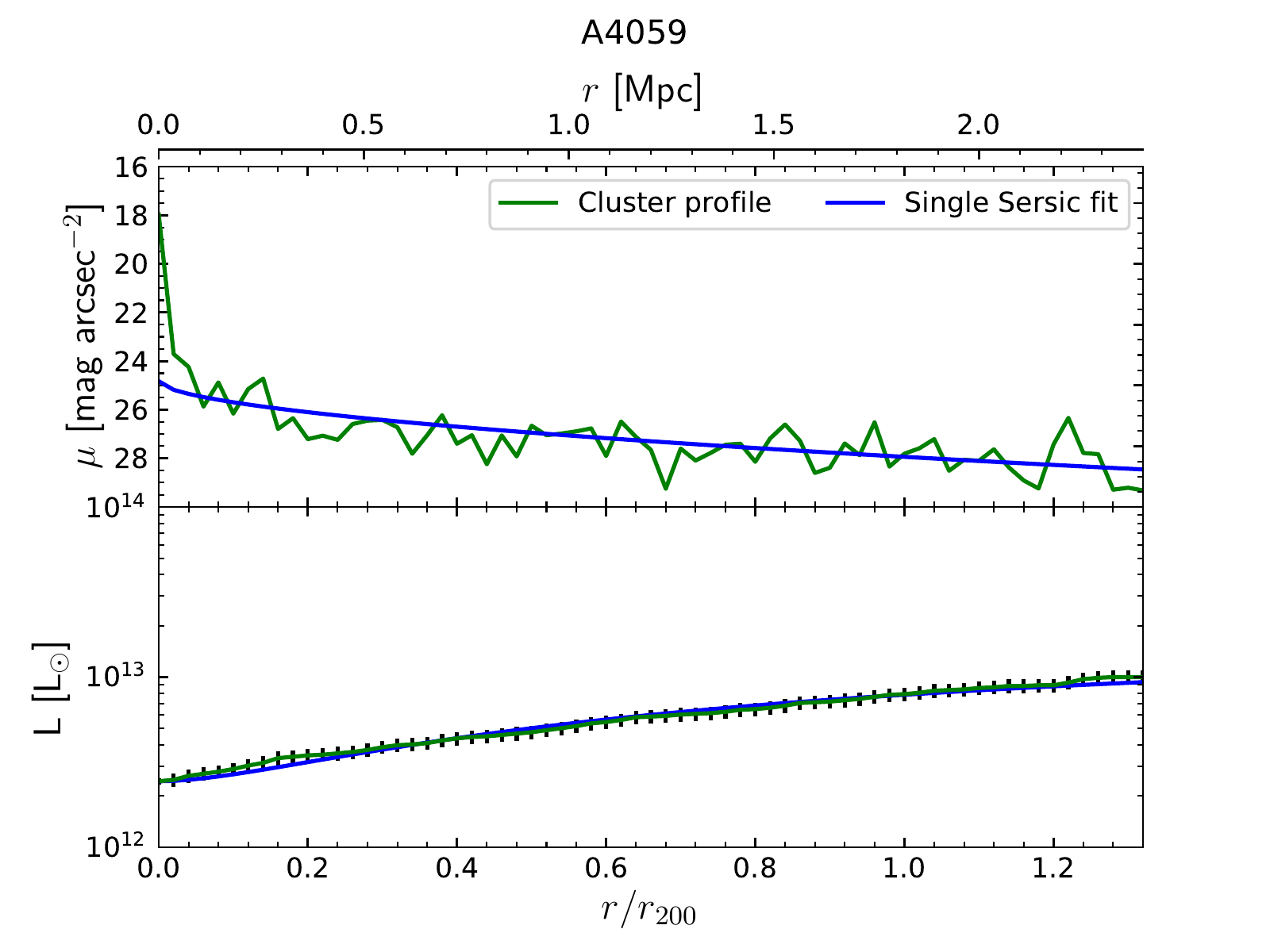}        \includegraphics[width=0.45\textwidth]{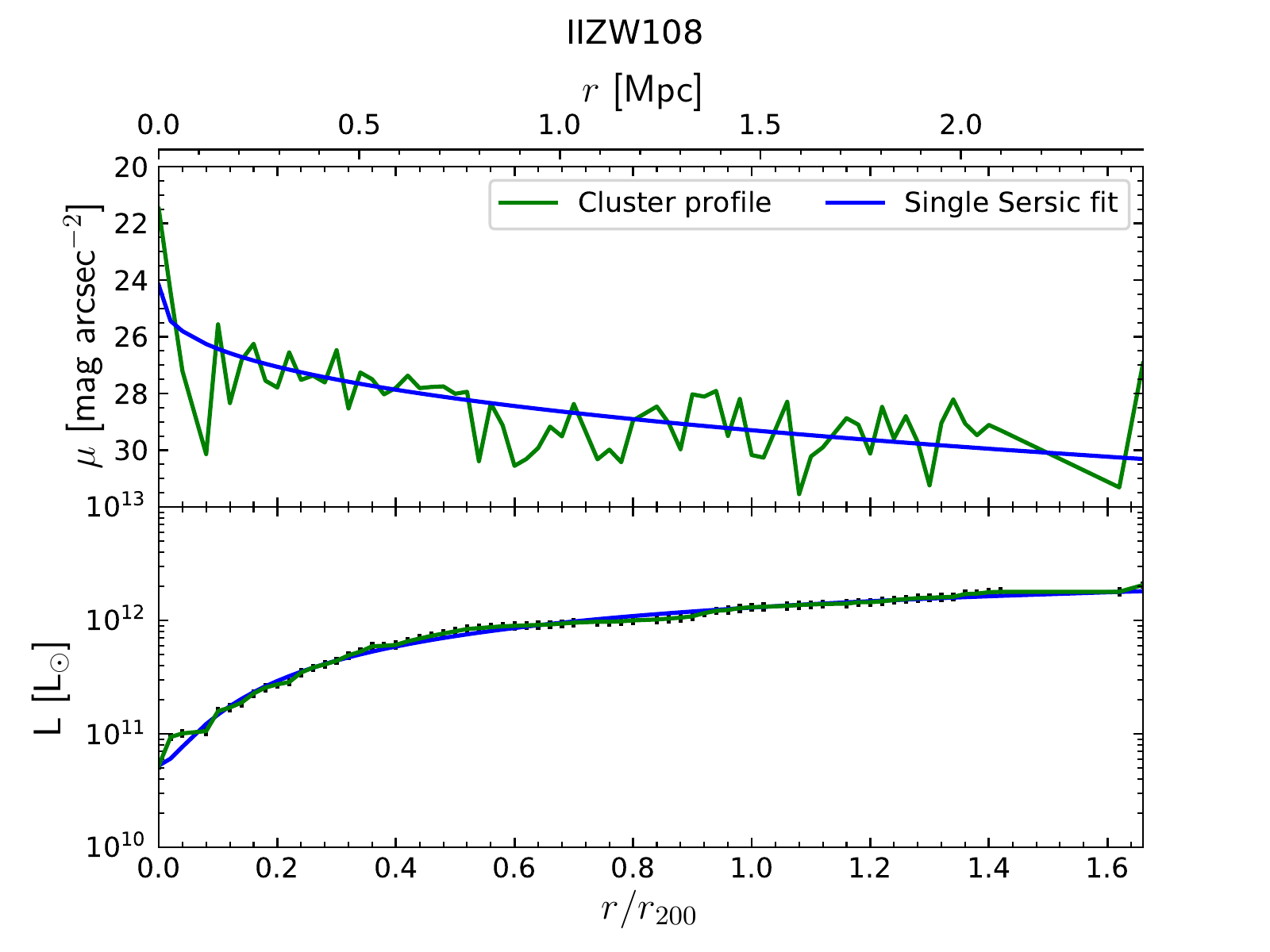}
        \includegraphics[width=0.45\textwidth]{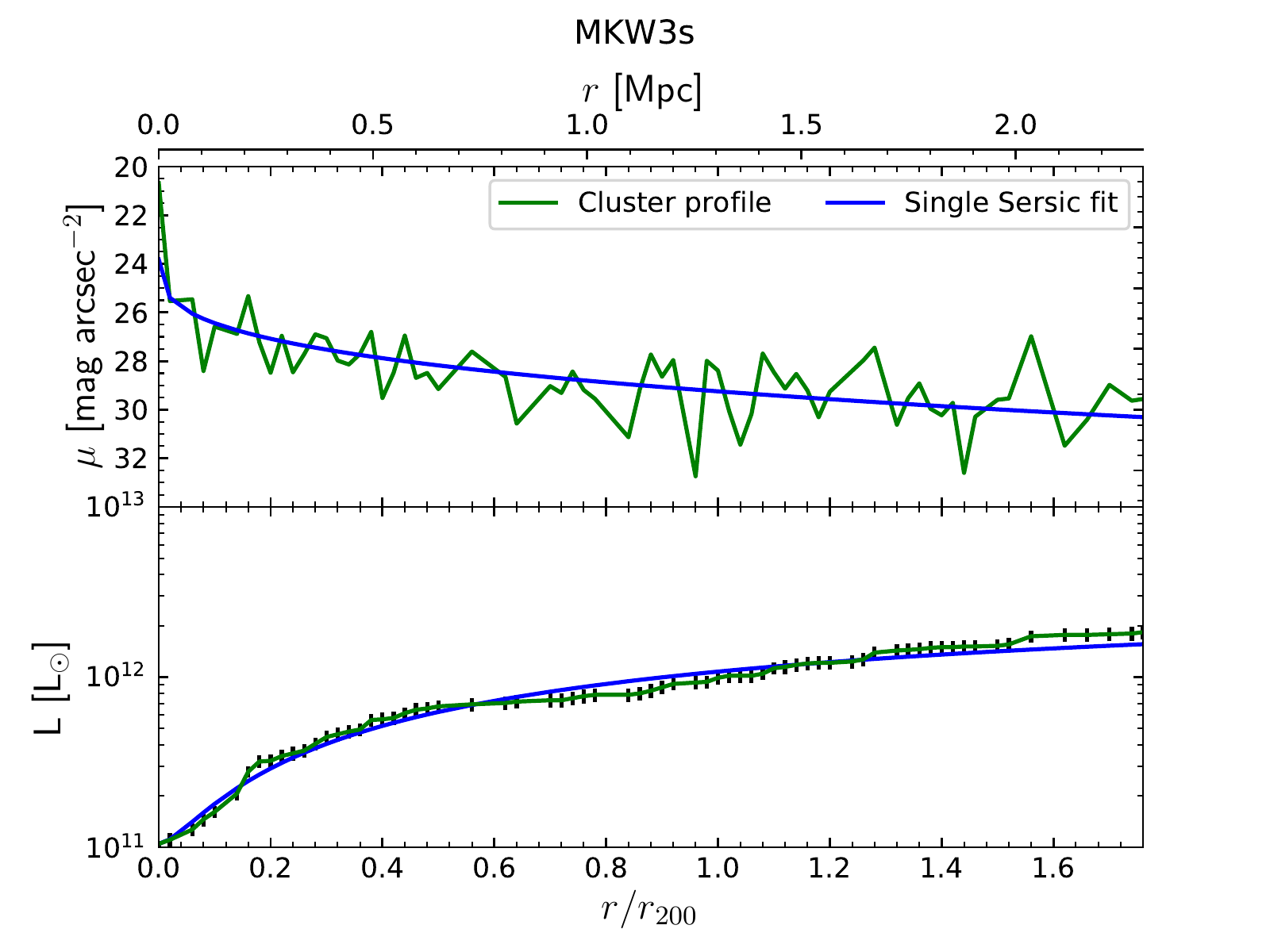}        \includegraphics[width=0.45\textwidth]{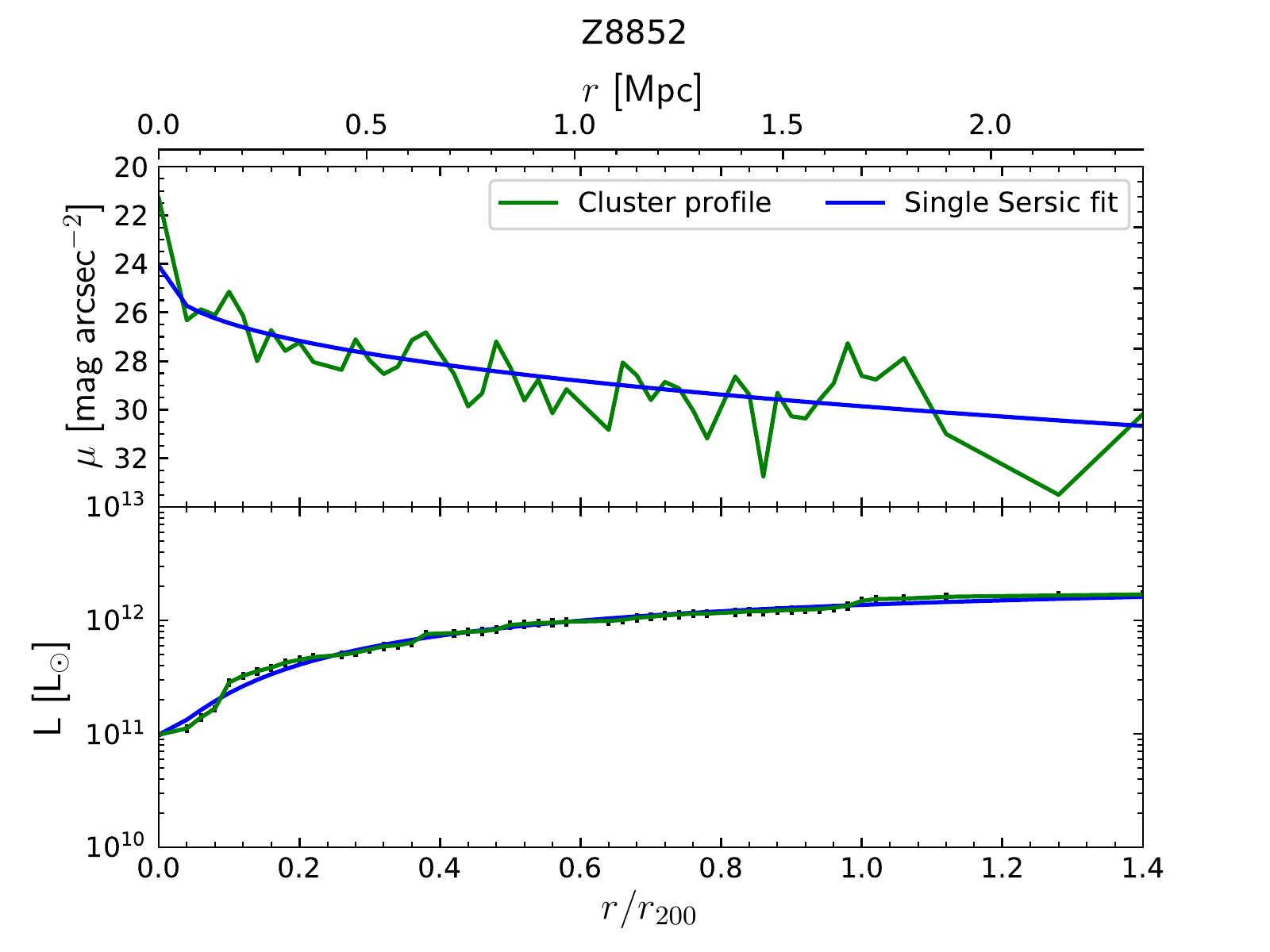}
    \caption{Photometric decomposition of Omega-WINGS galaxy clusters luminosity profiles, continued.}
    \label{fig:fits-end}
\end{figure*}

In Figures~\ref{fig:plots-begin}-\ref{fig:plots-end} we plotted the photometric profiles of our clusters, in Figures~\ref{fig:mass_plots-begin}-\ref{fig:mass_plots-end} their stellar mass profiles, and in Figures~\ref{fig:fits-begin}-\ref{fig:fits-end} the best fits to their luminosity profiles.

\newpage

\begin{table*}
\caption{Recap of all the WINGS and Omega-WINGS observations.}
\centering
\label{tab:clusters}
\begin{tabular}{lccclccc}
\hline
WINGS   &       Omega-WINGS     &       WINGS   & Omega-WINGS   &       WINGS   &       Omega-WINGS     &       WINGS   & Omega-WINGS\\
cluster &       photometry      &       spectroscopy    & spectroscopy  &       cluster &       photometry      &       spectroscopy    & spectroscopy\\
\hline
A85     &       yes     &       no      &       yes     &       A2589   &       yes     &       yes     &       no\\
A119    &       yes     &       yes     &       no      &       A2593   &       yes     &       yes     &       no\\
A133    &       no      &       no      &       no      &       A2622   &       no      &       yes     &       no\\
A147    &       yes     &       no      &       no      &       A2626   &       no      &       yes     &       no\\
A151    &       yes     &       yes     &       no      &       A2657   &       yes     &       no      &       no\\
A160    &       yes     &       yes     &       no      &       A2665   &       yes     &       no      &       no\\
A168    &       yes     &       no      &       yes     &       A2717   &       yes     &       no      &       yes\\
A193    &       yes     &       yes     &       yes     &       A2734   &       yes     &       no      &       yes\\
A311    &       no      &       no      &       no      &       A3128   &       yes     &       yes     &       yes\\
A376    &       no      &       yes     &       no      &       A3158   &       yes     &       yes     &       yes\\
A500    &       yes     &       yes     &       yes     &       A3164   &       no      &       no      &       no\\
A548b   &       no      &       no      &       no      &       A3266   &       yes     &       yes     &       yes\\
A602    &       no      &       no      &       no      &       A3376   &       yes     &       yes     &       yes\\
A671    &       no      &       yes     &       no      &       A3395   &       yes     &       yes     &       yes\\
A754    &       yes     &       yes     &       yes     &       A3395   &       yes     &       yes     &       yes\\
A780    &       no      &       no      &       no      &       A3490   &       no      &       yes     &       no\\
A957    &       yes     &       yes     &       yes     &       A3497   &       no      &       yes     &       no\\
A970    &       yes     &       yes     &       yes     &       A3528a  &       yes     &       no      &       yes\\
A1069   &       yes     &       yes     &       yes     &       A3528b  &       yes     &       no      &       yes\\
A1291   &       no      &       yes     &       no      &       A3530   &       yes     &       no      &       yes\\
A1631a  &       yes     &       yes     &       yes     &       A3532   &       yes     &       no      &       yes\\
A1644   &       no      &       yes     &       no      &       A3556   &       yes     &       yes     &       yes\\
A1688   &       no      &       no      &       no      &       A3558   &       yes     &       no      &       yes\\
A1736   &       no      &       no      &       no      &       A3560   &       yes     &       yes     &       yes\\
A1795   &       no      &       yes     &       no      &       A3562   &       no      &       no      &       no\\
A1831   &       no      &       yes     &       no      &       A3667   &       yes     &       no      &       yes\\
A1983   &       yes     &       yes     &       no      &       A3716   &       yes     &       no      &       yes\\
A1991   &       yes     &       yes     &       no      &       A3809   &       yes     &       yes     &       yes\\
A2107   &       yes     &       yes     &       no      &       A3880   &       yes     &       no      &       yes\\
A2124   &       no      &       yes     &       no      &       A4059   &       yes     &       no      &       yes\\
A2149   &       no      &       no      &       no      &       IIZW108 &       yes     &       yes     &       yes\\
A2169   &       no      &       yes     &       no      &       RX0058  &       no      &       yes     &       no\\
A2256   &       no      &       no      &       no      &       RX1022  &       no      &       yes     &       no\\
A2271   &       no      &       no      &       no      &       RX1740  &       no      &       yes     &       no\\
A2382   &       yes     &       yes     &       yes     &       MKW3s   &       yes     &       yes     &       no\\
A2399   &       yes     &       yes     &       yes     &       Z1261   &       no      &       no      &       no\\
A2415   &       yes     &       yes     &       yes     &       Z2844   &       no      &       yes     &       no\\
A2457   &       yes     &       yes     &       yes     &       Z8338   &       no      &       yes     &       no\\
A2572a  &       no      &       yes     &       no      &       Z8852   &       yes     &       yes     &       no\\
\hline
\end{tabular}
\end{table*}

\newpage

\begin{table*}
\caption{Recap of the main cluster parameters: Cluster name (Column~1), LF correction (Column~2), observed over total expected cluster luminosity (Column~3), color gradient (Column~4), number of substructures from \cite{ram7} (Column~5), total luminosity of the substructures over the main cluster component luminosity (Column~6), effective radius in units of $r_{200}$ (Column~7), effective LF-corrected mass (Column~8), $r_{200}$ in units of kpc (Column~9), and integrated mass within $r_{200}$ (Column~10).}
\centering
\label{tab:phot_and_mass}
\begin{tabular}{lccccccccc}
\hline
Cluster & LF corr. & $L_{\it obs}/L_{\it exp}$ & $\Delta(B-V)/\Delta R$ & $N_{\it sub}$ & $L_{\it sub}/L_{\it main}$ & $r_{\it eff}/r_{200}$ & $\mathcal{M}_{\it eff}$ & $r_{200}$ & $\mathcal{M}_{200}$\\
 & & & mag $r_{200}^{-1}$ & & & & $10^{13}$ M$_\odot$ & kpc & $10^{13}$ M$_\odot$ \\
\hline
A85 & 1.03 & 0.66 & $-$0.27 & 2 & 0.72 & 0.87 & 0.91 & 2366 & 1.15\\
A119 & 1.02 & 0.42 & $-$0.06 & 2 & 0.36 & - & - & 2087 & -\\
A147 & 1.03 & 0.85 & $-$0.29 & 2 & 0.34 & - & - & 1612 & -\\
A151 & 1.03 & 0.52 & $-$0.12 & 2 & 0.71 & 1.33 & 1.12 & 1779 & 1.00\\
A160 & 1.03 & 0.61 & $-$0.11 & 3 & 0.44 & - & - & 1359 & -\\
A168 & 1.03 & 0.75 & $-$0.32 & 1 & 0.28 & 1.22 & 0.70 & 1323 & 0.67\\
A193 & 1.03 & 0.93 & $-$0.17 & 0 & - & 0.29 & 0.18 & 1847 & 0.40\\
A500 & 1.05 & 0.75 & $-$0.23 & 1 & 0.50 & 1.22 & 0.79 & 1895 & 0.71\\
A754 & 1.03 & 0.58 & $-$0.18 & 2 & 0.97 & 0.88 & 2.00 & 2216 & 2.14\\
A957 & 1.02 & 0.60 & $-$0.26 & 0 & - & 1.83 & - & 1550 & 0.41\\
A970 & 1.04 & 0.72 & $-$0.32 & 1 & 0.30 & 0.76 & 0.83 & 2031 & 1.10\\
A1069 & 1.05 & 0.66 & $-$0.32 & 1 & 0.50 & 1.59 & 0.95 & 1667 & 0.78\\
A1631a & 1.03 & 0.84 & $-$0.15 & 0 & - & 1.11 & 2.33 & 1839 & 1.78\\
A1983 & 1.03 & 0.63 & $-$0.22 & - & - & - & - & 1276 & -\\
A1991 & 1.04 & 0.77 & $-$0.20 & 1 & 0.40 & - & - & 1441 & -\\
A2107 & 1.02 & 0.72 & $-$0.17 & 0 & - & - & - & 1436 & -\\
A2382 & 1.04 & 0.76 & $-$0.18 & 0 & - & 1.28 & 1.12 & 1675 & 0.78\\
A2399 & 1.04 & 0.85 & $-$0.27 & - & - & 1.08 & 0.81 & 1755 & 0.79\\
A2415 & 1.04 & 1.38 & $-$0.27 & 1 & 0.14 & 0.69 & 0.47 & 1661 & 0.76\\
A2457 & 1.04 & 0.97 & $-$0.21 & 1 & 0.07 & 0.77 & 0.80 & 1634 & 0.95\\
A2589 & 1.02 & 0.59 & $-$0.31 & 1 & 0.35 & - & - & 1978 & -\\
A2593 & 1.02 & 0.74 & $-$0.23 & - & - & - & - & 1700 & -\\
A2657 & 1.02 & 0.76 & $-$0.16 & 1 & 0.77 & - & - & 924 & -\\
A2665 & 1.04 & 0.57 & $-$0.29 & 1 & 0.03 & - & - & 1500 & -\\
A2717 & 1.03 & 0.62 & $-$0.10 & - & - & 1.87 & - & 1315 & 0.27\\
A2734 & 1.04 & 0.68 & $-$0.14 & 3 & 0.52 & 1.39 & 1.23 & 1875 & 1.00\\
A3128 & 1.04 & 0.63 & $-$0.19 & 3 & 1.53 & 1.00 & 2.45 & 2016 & 2.45\\
A3158 & 1.04 & 0.61 & $-$0.22 & 1 & 0.65 & 1.08 & 1.82 & 2461 & 1.76\\
A3266 & 1.04 & 0.47 & $-$0.24 & 0 & - & - & - & 3170 & -\\
A3376 & 1.03 & 0.52 & $-$0.26 & 2 & 0.66 & 1.55 & - & 2043 & 3.43\\
A3395 & 1.03 & 0.25 & $-$0.10 & - & - & - & - & 2912 & -\\
A3528a & 1.03 & 0.55 & $-$0.21 & - & - & 1.34 & - & 2450 & 1.15\\
A3528b & 1.03 & 0.49 & $-$0.30 & 1 & 0.25 & - & - & 2078 & -\\
A3530 & 1.03 & 0.52 & $-$0.09 & 0 & - & 1.61 & 3.91 & 1625 & 1.91\\
A3532 & 1.04 & 0.52 & $-$0.17 & 0 & - & 2.48 & - & 1940 & 0.07\\
A3556 & 1.03 & 0.63 & $-$0.10 & - & - & 1.39 & 3.51 & 1616 & 2.51\\
A3558 & 1.03 & 0.38 & $-$0.25 & 0 & - & - & - & 2424 & -\\
A3560 & 1.03 & 0.27 & $-$0.12 & - & - & 2.45 & - & 2030 & 2.23\\
A3667 & 1.04 & 0.29 & $-$0.02 & 3 & 1.16 & 2.42 & - & 2434 & 2.45\\
A3716 & 1.03 & 0.89 & $-$0.10 & 1 & 0.64 & 0.39 & 2.14 & 2053 & 3.80\\
A3809 & 1.04 & 1.01 & $-$0.26 & - & - & 1.13 & 1.26 & 1330 & 1.20\\
A3880 & 1.04 & 0.42 & $-$0.01 & 0 & - & 2.44 & - & 1656 & 1.12\\
A4059 & 1.03 & 0.63 & $-$0.06 & - & - & 1.27 & - & 1818 & 1.57\\
IIZW108 & 1.03 & 0.67 & $-$0.36 & 1 & 0.17 & 1.37 & 0.59 & 1478 & 0.48\\
MKW3s & 1.03 & 0.63 & $-$0.23 & 1 & 0.12 & - & - & 1305 & -\\
Z8852 & 1.02 & 0.74 & $-$0.21 & 2 & 0.19 & - & - & 1690 & -\\
\hline
\end{tabular}
\end{table*}

\newpage

\begin{table*}
        \caption{Effective parameters in phisical units of all the clusters measured through the integrated light profiles fitting: Cluster name (Column~1), effective radius (Columns~2), effective luminosity (Columns~3) and average intensity within the effective radius (Columns~4).}
        \centering
        \label{tab:eff}
        \begin{tabular}{lccc}
                \hline
                Cluster & $r_{\it eff}$ &       $L_{\it eff}$   &       $\overline{I_{\it eff}}$\\
                &       kpc     &       10$^{12}$ L$_\odot$     &       L$_\odot$ pc$^{-2}$\\
                \hline
                A85 & 2058 & 3.14 & 0.24\\
                A119 & 3548 & 3.87 & 0.10\\
                A147 & 838 & 0.73 & 0.33\\
                A151 & 2373 & 4.25 & 0.24\\
                A160 & 3302 & 3.31 & 0.10\\
                A168 & 1614 & 1.80 & 0.22\\
                A193 & 536 & 0.67 & 0.75\\
                A500 & 2312 & 2.96 & 0.18\\
                A754 & 1950 & 3.71 & 0.31\\
                A957 & 2837 & 1.26 & 0.05\\
                A970 & 1544 & 1.39 & 0.19\\
                A1069 & 2651 & 2.06 & 0.09\\
                A1631a & 2041 & 9.57 & 0.73\\
                A1983 & 2105 & 1.66 & 0.12\\
                A1991 & 2248 & 2.30 & 0.15\\
                A2107 & 1335 & 1.02 & 0.18\\
                A2382 & 2144 & 2.95 & 0.20\\
                A2399 & 1895 & 2.12 & 0.19\\
                A2415 & 1146 & 0.79 & 0.19\\
                A2457 & 1258 & 1.88 & 0.38\\
                A2589 & 2097 & 1.07 & 0.08\\
                A2593 & 1037 & 1.12 & 0.33\\
                A2657 & 2088 & 1.24 & 0.09\\
                A2665 & 2595 & 1.84 & 0.09\\
                A2717 & 2453 & 1.99 & 0.11\\
                A2734 & 2606 & 1.77 & 0.08\\
                A3128 & 2016 & 5.05 & 0.40\\
                A3158 & 2658 & 4.41 & 0.20\\
                A3266 & 4026 & 6.23 & 0.12\\
                A3376 & 3167 & 7.28 & 0.23\\
                A3395 & 6843 & 25.99 & 0.18\\
                A3528a & 3283 & 7.05 & 0.21\\
                A3528b & 2800 & 5.49 & 0.22\\
                A3530 & 2618 & 11.30 & 0.52\\
                A3532 & 4811 & 16.19 & 0.22\\
                A3556 & 2246 & 7.42 & 0.47\\
                A3558 & 3636 & 25.29 & 0.61\\
                A3560 & 4978 & 21.84 & 0.28\\
                A3667 & 5890 & 25.55 & 0.23\\
                A3716 & 801 & 7.78 & 3.86\\
                A3809 & 1503 & 3.54 & 0.50\\
                A3880 & 4037 & 9.52 & 0.19\\
                A4059 & 2309 & 7.88 & 0.47\\
                IIZW108 & 2025 & 1.53 & 0.12\\
                MKW3s & 2323 & 1.42 & 0.08\\
                Z8852 & 1301 & 1.12 & 0.21\\
                \hline
        \end{tabular}
\end{table*}

\begin{table*}
\caption{Best fit parameters for all clusters reproduced by a single S\'ersic model: Cluster name (Column~1), parameters of the best single S\'ersic decomposition (Columns~2$-$4), normalized $\chi^2$ value associated with the best single S\'ersic decomposition (Column~5), AIC value associated with the best single S\'ersic decomposition (Column~6), and BIC value associated with the best single S\'ersic decomposition (Column~7).}
\centering
\label{tab:single_fits}
\begin{tabular}{lcccccc}
\hline
Cluster & $n$ & $r_{\it eff}/r_{200}$ & log ($I_{\it eff}$ [$L_\odot$ pc$^{-2}$]) & $\chi_{\rm BM}^2$ & AIC$_{\rm BM}$ & BIC$_{\rm BM}$\\
\hline
A85 & $1.48^{+0.07}_{-0.07}$ & $0.87^{+0.03}_{-0.02}$ & $-1.00^{+0.01}_{-0.01}$ & 0.33 & 23.93 & 30.06\\
A119 & $4.08^{+0.08}_{-0.07}$ & $1.70^{+0.03}_{-0.03}$ & $-1.54^{+0.01}_{-0.01}$ & 0.12 & 11.01 & 16.43\\
A147 & $1.36^{+0.13}_{-0.07}$ & $0.52^{+0.01}_{-0.02}$ & $-0.85^{+0.01}_{-0.01}$ & 1.44 & 67.71 & 73.20\\
A160 & $4.76^{+0.15}_{-0.10}$ & $2.43^{+0.07}_{-0.02}$ & $-1.62^{+0.01}_{-0.01}$ & 0.18 & 19.26 & 26.29\\
A168 & $1.13^{+0.06}_{-0.07}$ & $1.22^{+0.04}_{-0.04}$ & $-0.97^{+0.01}_{-0.01}$ & 0.66 & 54.49 & 61.48\\
A193 & $3.98^{+0.32}_{-0.25}$ & $0.29^{+0.01}_{-0.01}$ & $-0.76^{+0.01}_{-0.01}$ & 0.34 & 18.86 & 24.00\\
A500 & $2.42^{+0.14}_{-0.09}$ & $1.22^{+0.04}_{-0.02}$ & $-1.21^{+0.02}_{-0.01}$ & 0.30 & 29.60 & 36.78\\
A754 & $2.05^{+0.06}_{-0.05}$ & $0.88^{+0.01}_{-0.02}$ & $-0.94^{+0.01}_{-0.01}$ & 0.22 & 17.16 & 23.07\\
A957 & $4.37^{+0.29}_{-0.16}$ & $1.83^{+0.09}_{-0.04}$ & $-1.86^{+0.03}_{-0.01}$ & 0.37 & 27.61 & 33.95\\
A970 & $3.79^{+0.13}_{-0.12}$ & $0.76^{+0.02}_{-0.01}$ & $-1.28^{+0.01}_{-0.01}$ & 0.47 & 29.98 & 35.95\\
A1069 & $2.31^{+0.08}_{-0.07}$ & $1.59^{+0.03}_{-0.03}$ & $-1.48^{+0.01}_{-0.01}$ & 0.21 & 20.78 & 27.70\\
A1631a & $0.17^{+0.03}_{-0.02}$ & $1.11^{+0.06}_{-0.03}$ & $-0.17^{+0.01}_{-0.01}$ & 1.39 & 108.58 & 115.61\\
A1983 & $3.04^{+0.39}_{-0.08}$ & $1.65^{+0.18}_{-0.02}$ & $-1.41^{+0.05}_{-0.01}$ & 1.76 & 136.50 & 143.53\\
A1991 & $1.63^{+0.15}_{-0.09}$ & $1.56^{+0.09}_{-0.03}$ & $-1.22^{+0.03}_{-0.01}$ & 0.94 & 83.90 & 91.26\\
A2107 & $2.45^{+0.11}_{-0.09}$ & $0.93^{+0.02}_{-0.02}$ & $-1.23^{+0.01}_{-0.01}$ & 0.59 & 40.34 & 46.67\\
A2382 & $2.17^{+0.06}_{-0.06}$ & $1.28^{+0.03}_{-0.03}$ & $-1.12^{+0.01}_{-0.01}$ & 0.25 & 26.82 & 34.22\\
A2399 & $1.58^{+0.11}_{-0.10}$ & $1.08^{+0.03}_{-0.02}$ & $-1.10^{+0.02}_{-0.01}$ & 0.60 & 51.65 & 58.76\\
A2415 & $0.43^{+0.24}_{-0.05}$ & $0.69^{+0.06}_{-0.02}$ & $-0.88^{+0.06}_{-0.02}$ & 1.65 & 95.13 & 101.26\\
A2457 & $1.42^{+0.11}_{-0.05}$ & $0.77^{+0.02}_{-0.01}$ & $-0.79^{+0.01}_{-0.01}$ & 0.28 & 25.35 & 32.14\\
A2589 & $4.09^{+0.18}_{-0.13}$ & $1.06^{+0.03}_{-0.02}$ & $-1.69^{+0.01}_{-0.01}$ & 0.28 & 18.21 & 23.70\\
A2593 & $2.33^{+0.13}_{-0.09}$ & $0.61^{+0.01}_{-0.01}$ & $-0.94^{+0.01}_{-0.01}$ & 0.50 & 31.21 & 37.12\\
A2657 & $3.08^{+0.56}_{-0.19}$ & $2.26^{+0.25}_{-0.07}$ & $-1.52^{+0.05}_{-0.02}$ & 2.10 & 148.95 & 155.74\\
A2665 & $2.24^{+0.17}_{-0.07}$ & $1.73^{+0.10}_{-0.03}$ & $-1.51^{+0.03}_{-0.01}$ & 0.67 & 51.47 & 58.26\\
A2734 & $3.00^{+0.19}_{-0.12}$ & $1.39^{+0.06}_{-0.03}$ & $-1.59^{+0.02}_{-0.01}$ & 0.45 & 31.41 & 37.69\\
A3128 & $1.54^{+0.07}_{-0.04}$ & $1.00^{+0.02}_{-0.02}$ & $-0.77^{+0.01}_{-0.01}$ & 1.22 & 86.56 & 93.26\\
A3266 & $3.64^{+0.10}_{-0.09}$ & $1.27^{+0.03}_{-0.03}$ & $-1.44^{+0.01}_{-0.01}$ & 0.22 & 16.18 & 21.86\\
A3376 & $2.75^{+0.13}_{-0.09}$ & $1.55^{+0.06}_{-0.03}$ & $-1.14^{+0.02}_{-0.01}$ & 0.20 & 16.79 & 22.87\\
A3395 & $2.71^{+0.11}_{-0.04}$ & $2.35^{+0.14}_{-0.05}$ & $-1.19^{+0.02}_{-0.01}$ & 0.52 & 28.33 & 33.81\\
A3528a & $1.54^{+0.08}_{-0.05}$ & $1.34^{+0.07}_{-0.03}$ & $-1.05^{+0.02}_{-0.01}$ & 0.47 & 36.20 & 42.82\\
A3532 & $4.22^{+0.25}_{-0.09}$ & $2.48^{+0.15}_{-0.02}$ & $-1.18^{+0.03}_{-0.01}$ & 0.69 & 56.01 & 62.96\\
A3556 & $2.73^{+0.30}_{-0.10}$ & $1.39^{+0.13}_{-0.03}$ & $-0.85^{+0.04}_{-0.01}$ & 0.87 & 64.57 & 71.32\\
A3558 & $2.20^{+0.05}_{-0.03}$ & $1.50^{+0.03}_{-0.02}$ & $-0.67^{+0.01}_{-0.01}$ & 0.10 & 10.51 & 16.00\\
A3667 & $2.10^{+0.08}_{-0.02}$ & $2.42^{+0.12}_{-0.02}$ & $-1.03^{+0.02}_{-0.01}$ & 0.48 & 33.99 & 40.32\\
A3716 & $1.87^{+0.06}_{-0.12}$ & $0.39^{+0.01}_{-0.01}$ & $0.15^{+0.01}_{-0.02}$ & 0.30 & 21.98 & 28.06\\
A3809 & $0.59^{+0.04}_{-0.03}$ & $1.13^{+0.02}_{-0.01}$ & $-0.50^{+0.01}_{-0.01}$ & 0.54 & 61.97 & 69.99\\
A4059 & $1.80^{+0.13}_{-0.06}$ & $1.27^{+0.09}_{-0.03}$ & $-0.79^{+0.03}_{-0.01}$ & 0.28 & 23.60 & 30.22\\
IIZW108 & $2.80^{+0.15}_{-0.08}$ & $1.37^{+0.04}_{-0.03}$ & $-1.40^{+0.02}_{-0.01}$ & 0.86 & 62.84 & 69.54\\
MKW3s & $3.16^{+0.42}_{-0.13}$ & $1.78^{+0.18}_{-0.04}$ & $-1.57^{+0.05}_{-0.01}$ & 1.35 & 95.40 & 102.10\\
Z8852 & $2.57^{+0.19}_{-0.12}$ & $0.77^{+0.03}_{-0.02}$ & $-1.16^{+0.02}_{-0.01}$ & 0.89 & 47.72 & 53.46\\
\hline
\end{tabular}
\end{table*}

\newpage

\begin{table*}
\caption{Best fit parameters for all clusters reproduced by a double S\'ersic model: Cluster name (Column~1), parameters of the best single S\'ersic decomposition (Columns~2$-$7), normalized $\chi^2$ value associated with the best double S\'ersic decomposition (Column~8), AIC value associated with the best double S\'ersic decomposition (Column~9), and BIC value associated with the best double S\'ersic decomposition (Column~10).}
\centering
\label{tab:double_fits}
\begin{tabular}{lcccccccccc}
\hline
Cluster & $n_{\it in}$ & $r_{\it eff,in}/r_{200}$ & log($I_{\it eff,in}$)  & $n_{\it out}$ & $r_{\it eff,out}/r_{200}$ & log($I_{\it eff,out}$) & $\chi_{\rm BM}^2$ & AIC$_{\rm BM}$ & BIC$_{\rm BM}$\\
 &  & & [$L_\odot$ pc$^{-2}$] & & & [$L_\odot$ pc$^{-2}$] & & & \\
\hline
A151 & $2.31^{undef}_{-1.62}$ & $0.03^{+0.02}_{-0.01}$ & $0.80^{+0.39}_{-0.19}$ & $1.88^{+0.81}_{-0.03}$ & $1.48^{+0.36}_{-0.01}$ & $-1.14^{+0.10}_{-0.01}$ & 0.26 & 30.10 & 42.76\\
A2717 & $0.62^{undef}_{undef}$ & $0.06^{+0.12}_{-0.03}$ & $-0.04^{+0.48}_{-0.43}$ & $0.97^{+0.35}_{-0.03}$ & $1.95^{+0.78}_{-0.04}$ & $-1.31^{+0.09}_{-0.01}$ & 2.04 & 174.27 & 187.92\\
A3158 & $0.51^{+2.83}_{-0.35}$ & $0.16^{undef}_{-0.02}$ & $-0.26^{+0.33}_{-0.09}$ & $0.69^{+1.01}_{-0.04}$ & $1.18^{+6.31}_{-0.07}$ & $-1.04^{+0.21}_{-0.02}$ & 0.18 & 22.34 & 33.71\\
A3528b & $0.51^{+2.95}_{-0.37}$ & $0.11^{+0.27}_{-0.02}$ & $0.11^{+0.32}_{-0.10}$ & $1.16^{+0.97}_{-0.04}$ & $1.46^{+1.21}_{-0.06}$ & $-1.06^{+0.17}_{-0.01}$ & 0.19 & 23.37 & 35.13\\
A3530 & $2.70^{undef}_{-1.79}$ & $0.05^{+0.04}_{-0.01}$ & $0.41^{+0.39}_{-0.16}$ & $0.52^{+0.17}_{-0.01}$ & $1.62^{+0.49}_{-0.03}$ & $-0.47^{+0.06}_{-0.01}$ & 0.42 & 45.60 & 59.18\\
A3560 & $1.68^{undef}_{-1.65}$ & $0.04^{+0.04}_{-0.02}$ & $0.35^{+0.46}_{-0.37}$ & $1.13^{+0.22}_{-0.02}$ & $2.50^{+1.18}_{-0.05}$ & $-0.87^{+0.08}_{-0.01}$ & 0.68 & 55.71 & 68.11\\
A3880 & $0.67^{+4.66}_{-0.64}$ & $0.16^{undef}_{-0.05}$ & $-0.34^{+0.43}_{-0.17}$ & $0.93^{+0.31}_{-0.02}$ & $2.50^{+1.75}_{-0.05}$ & $-1.03^{+0.10}_{-0.01}$ & 1.51 & 134.11 & 147.83\\
\hline
\end{tabular}
\end{table*}

\newpage
\clearpage

\newpage
\clearpage

\begin{figure*}[t]
   \centering
        \includegraphics[width=0.45\textwidth]{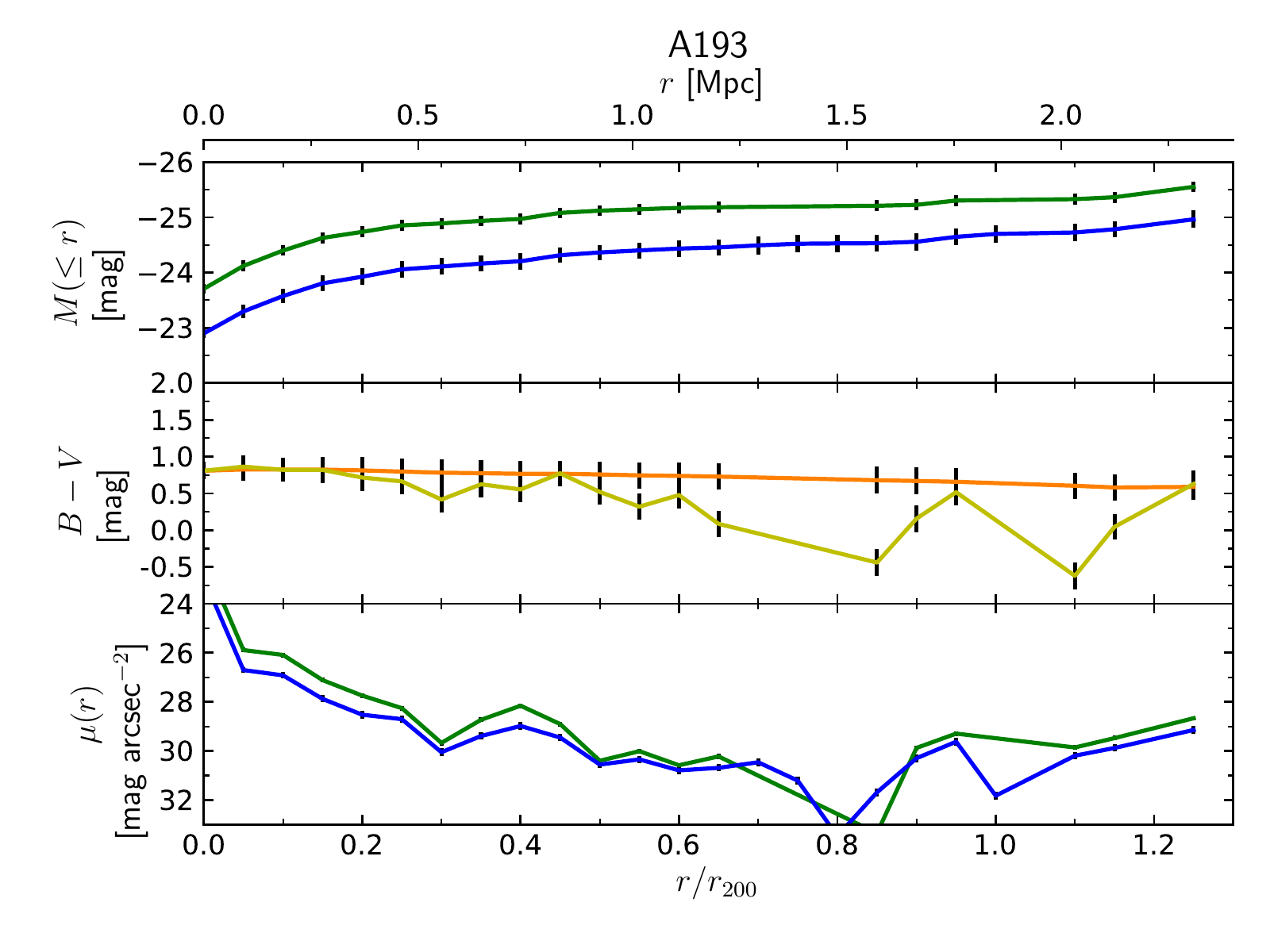}       \includegraphics[width=0.45\textwidth]{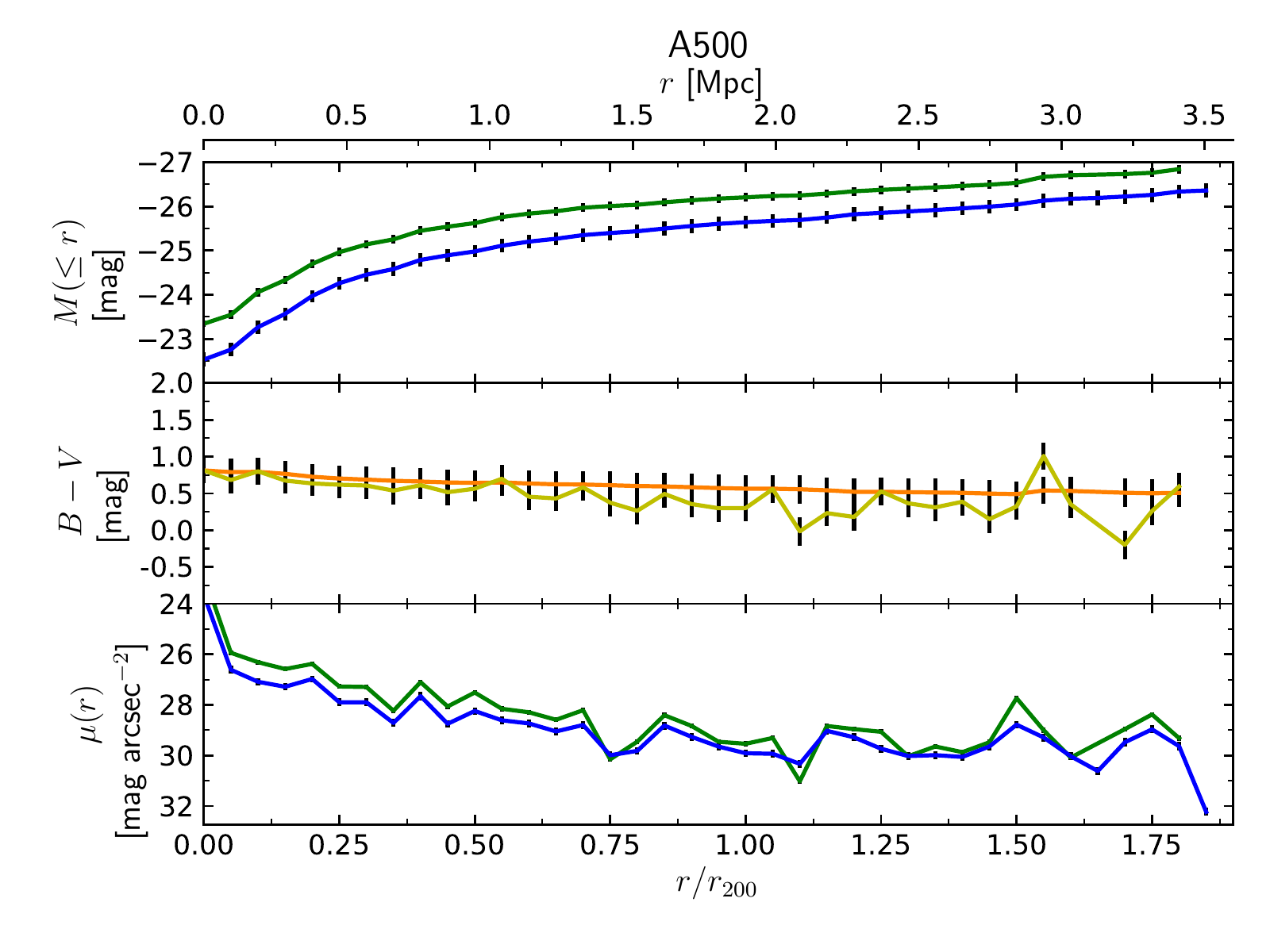}
        \includegraphics[width=0.45\textwidth]{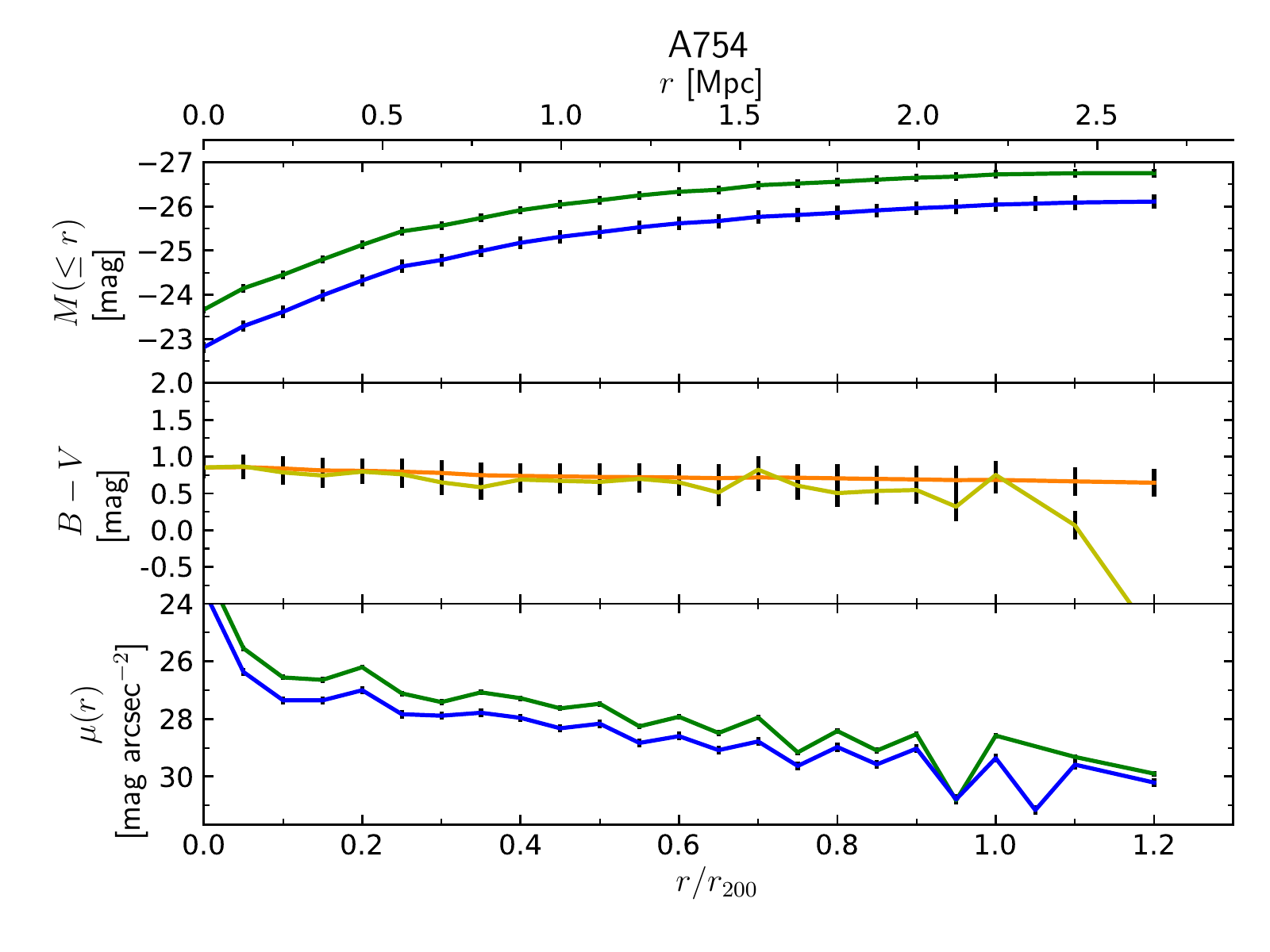}       \includegraphics[width=0.45\textwidth]{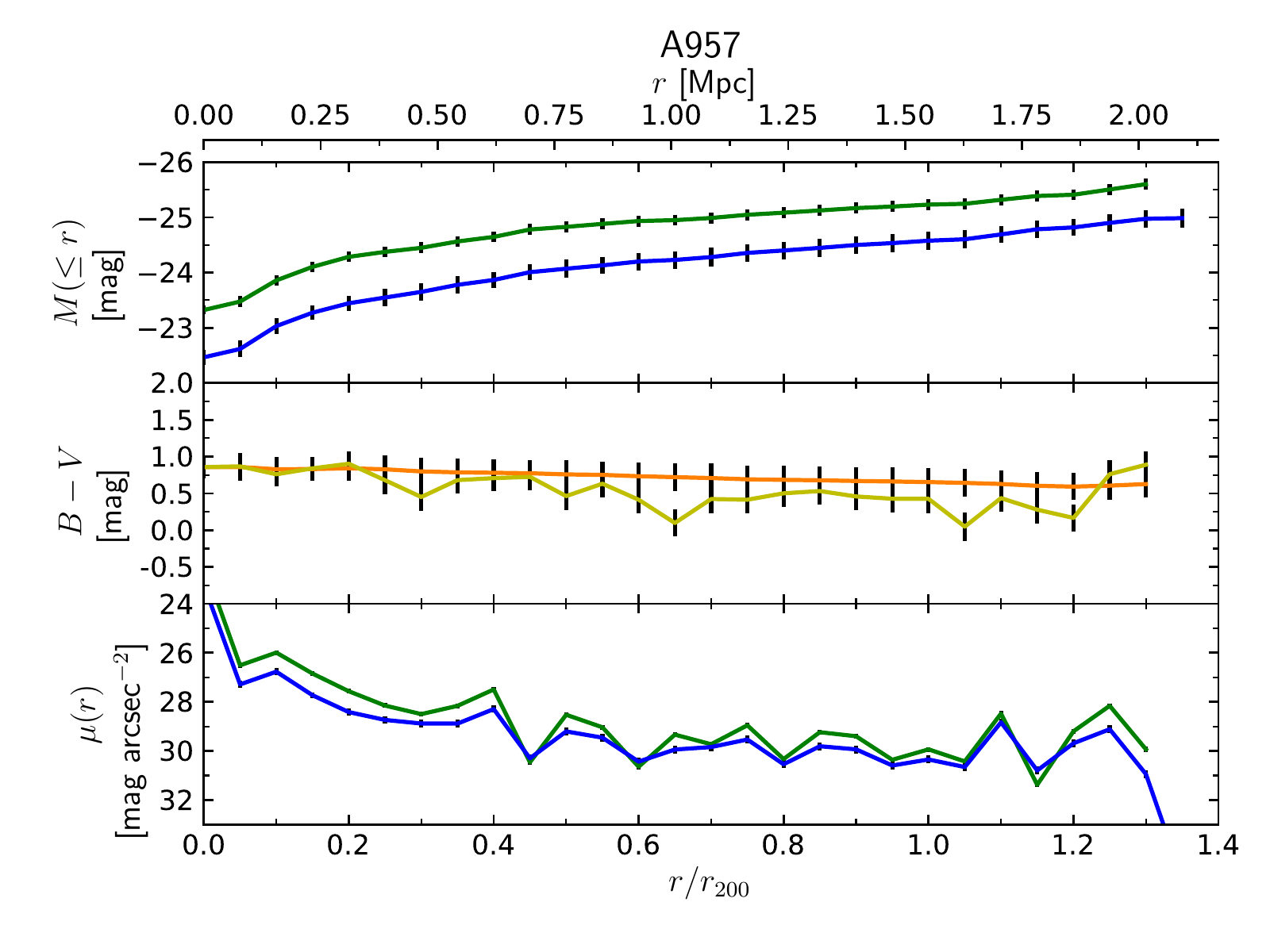}
        \includegraphics[width=0.45\textwidth]{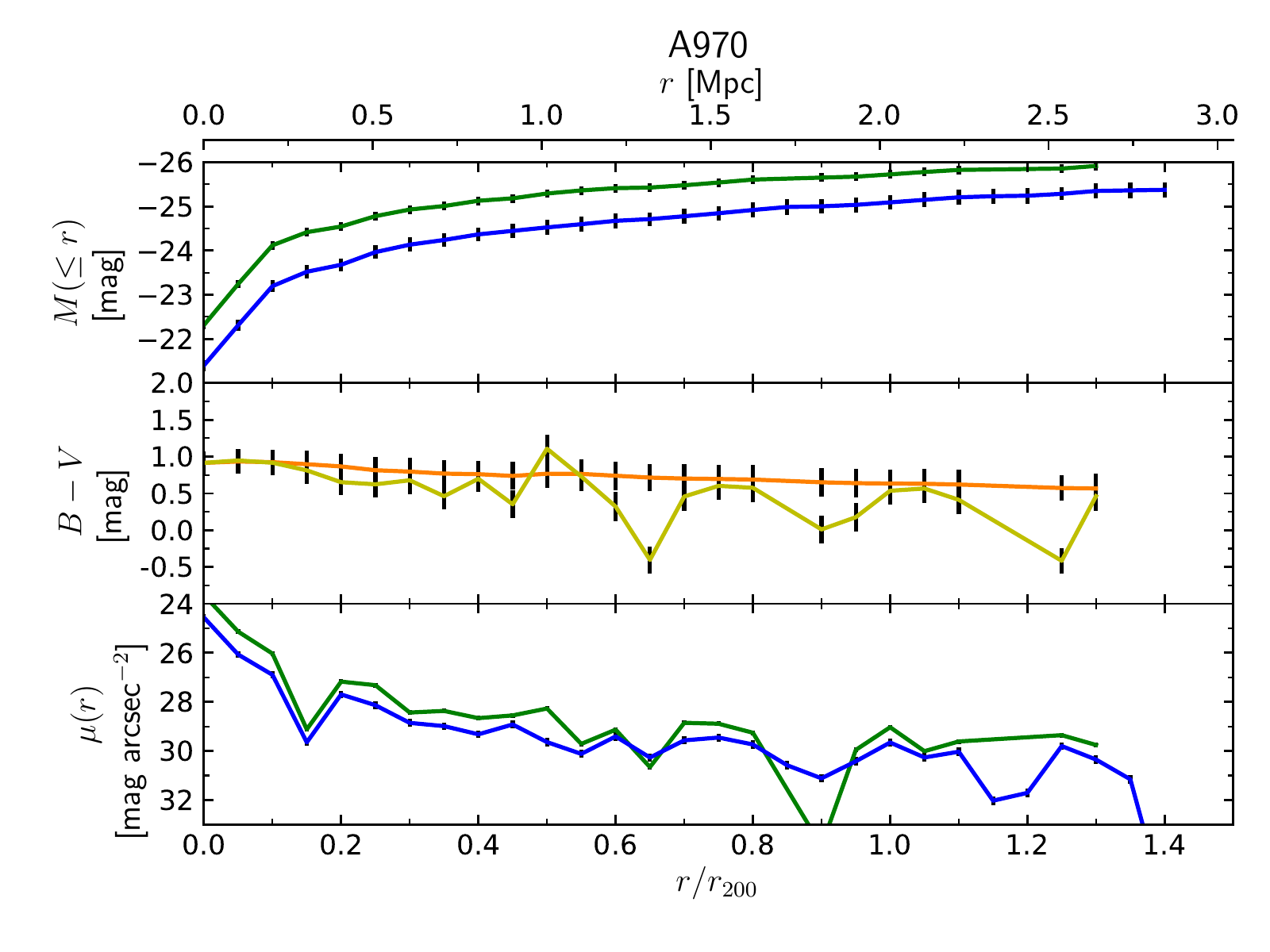}       \includegraphics[width=0.45\textwidth]{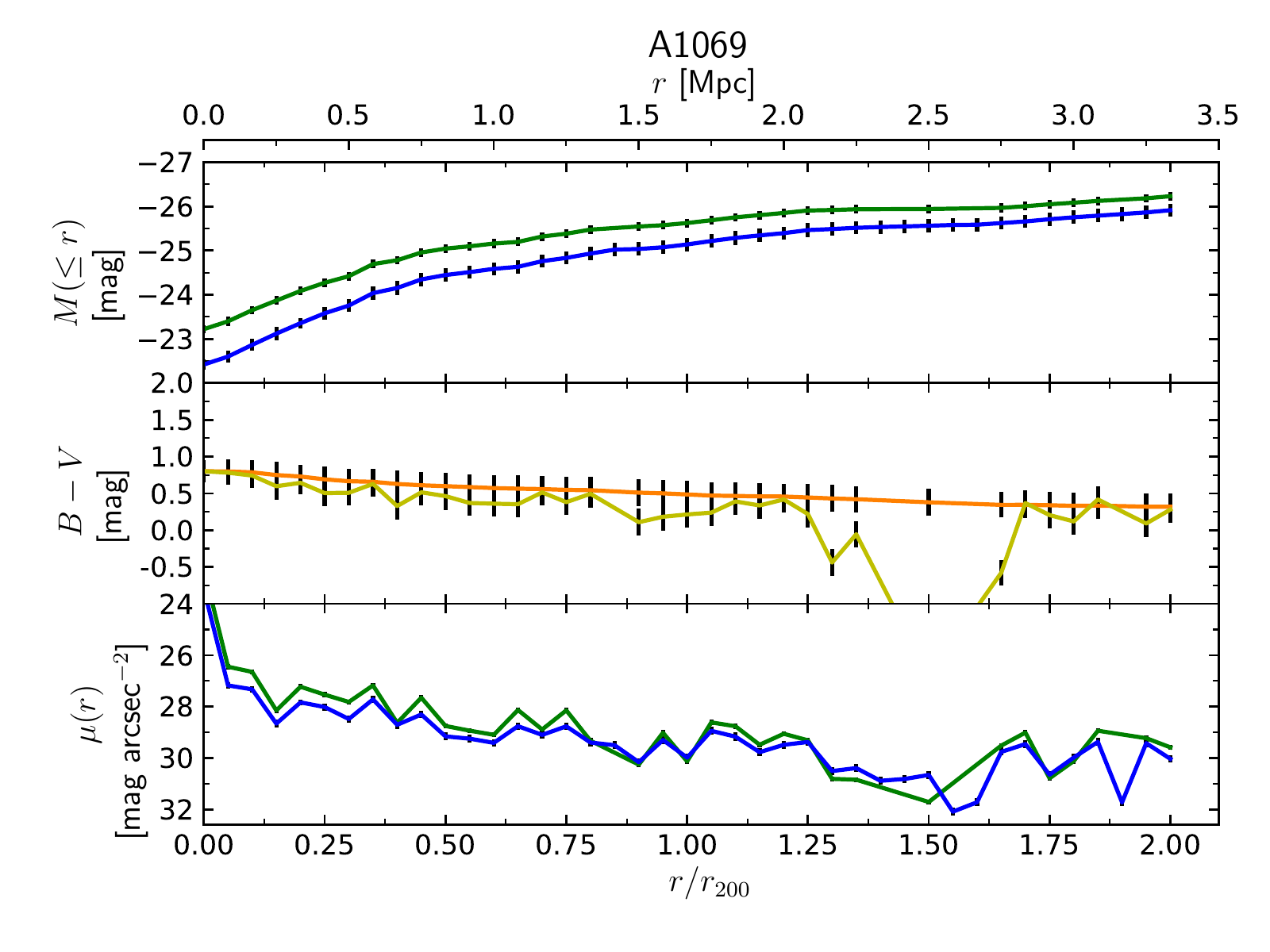}
    \caption{Photometric profiles of Omega-WINGS galaxy clusters, continued.}
\end{figure*}

\newpage
\clearpage

\begin{figure*}[t]
   \centering
        \includegraphics[width=0.45\textwidth]{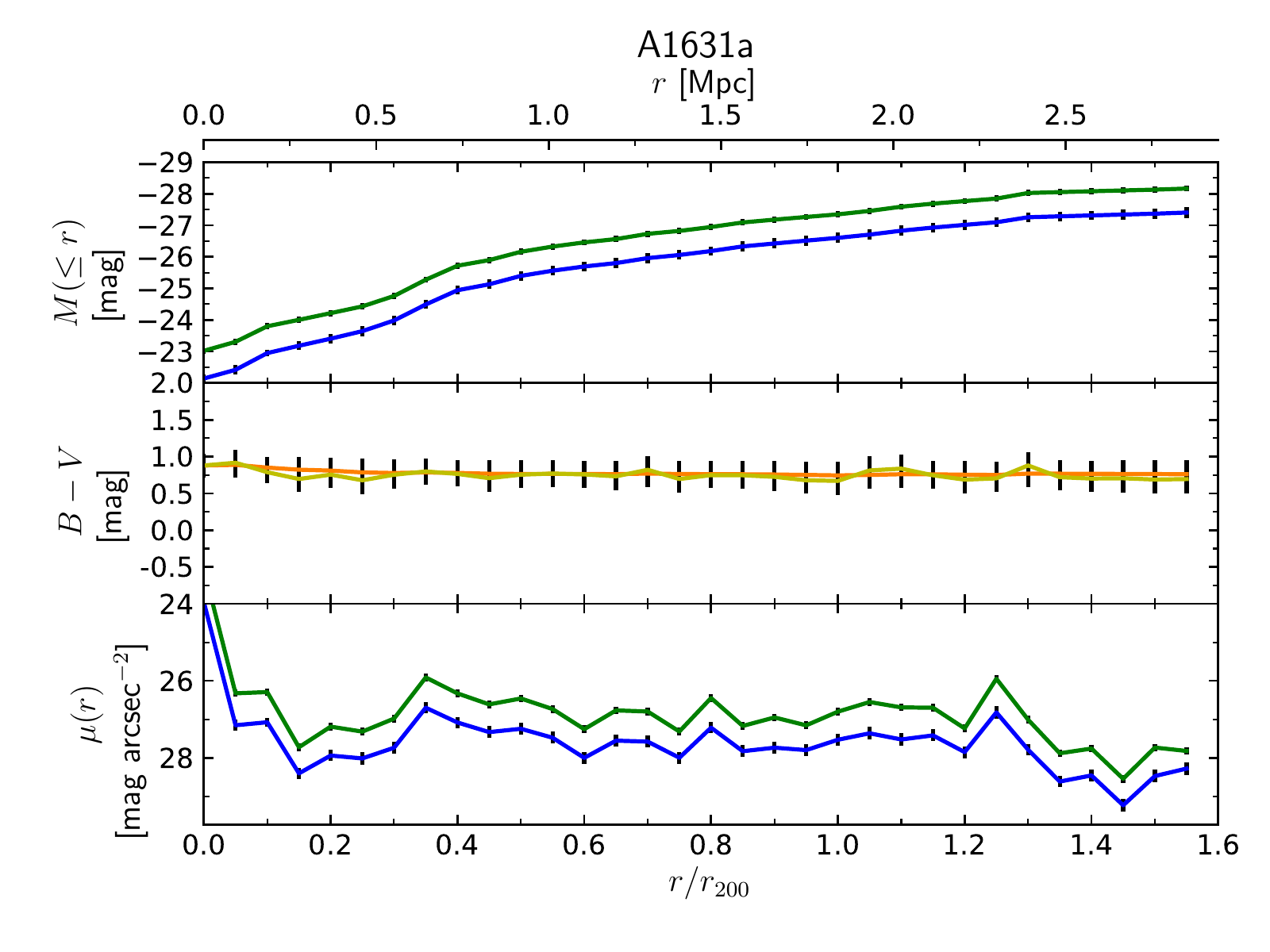}     \includegraphics[width=0.45\textwidth]{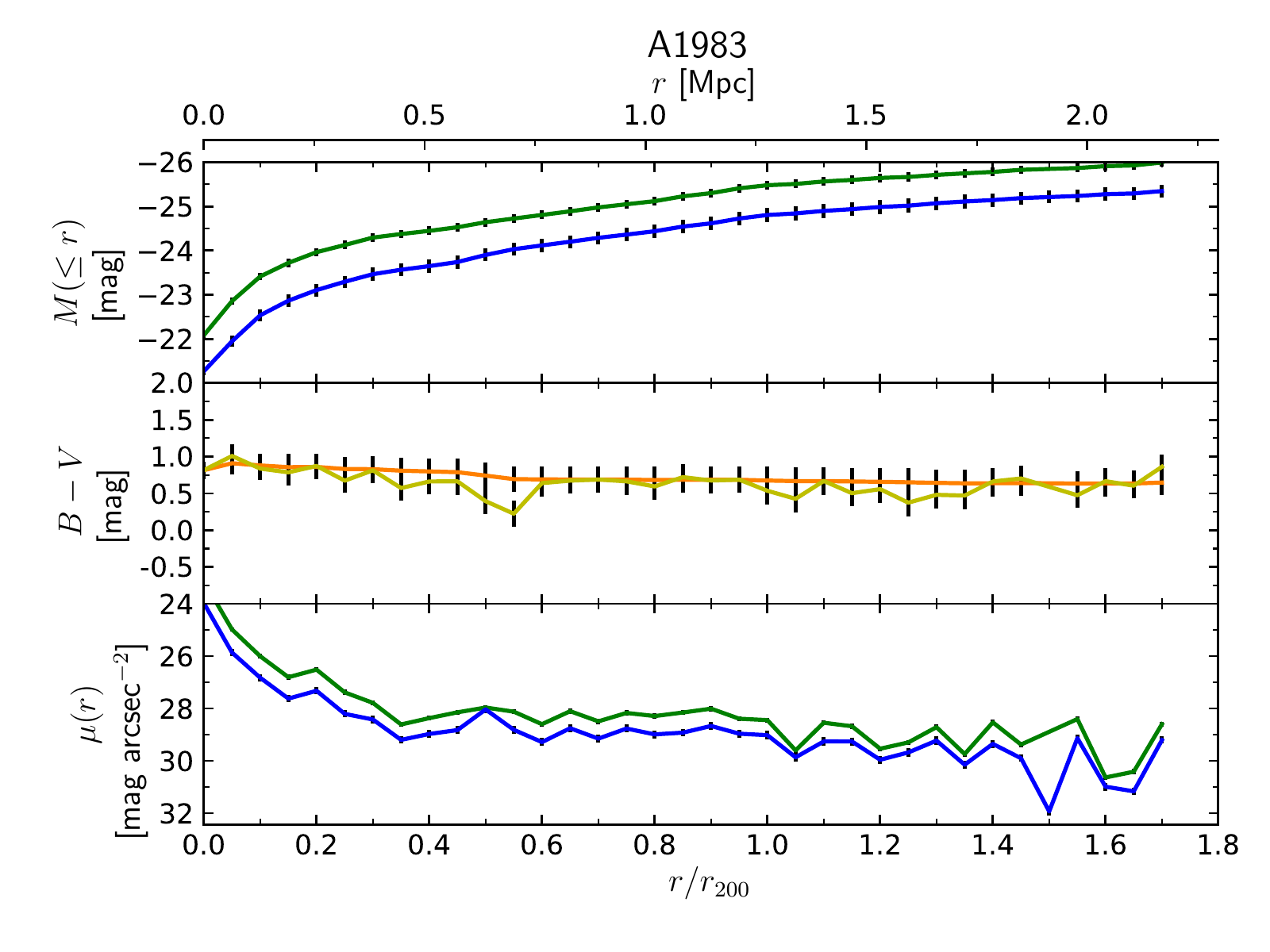}
        \includegraphics[width=0.45\textwidth]{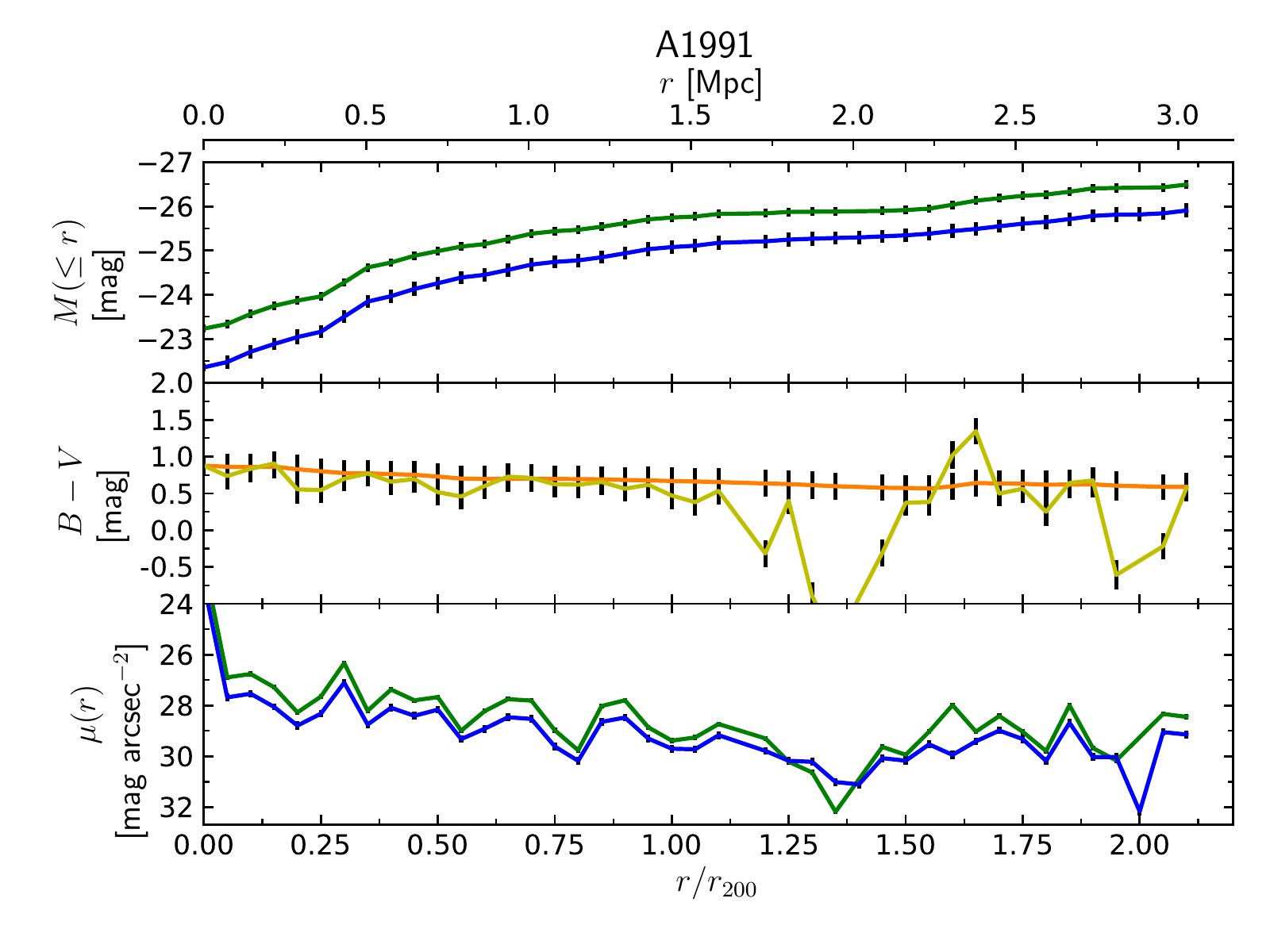}      \includegraphics[width=0.45\textwidth]{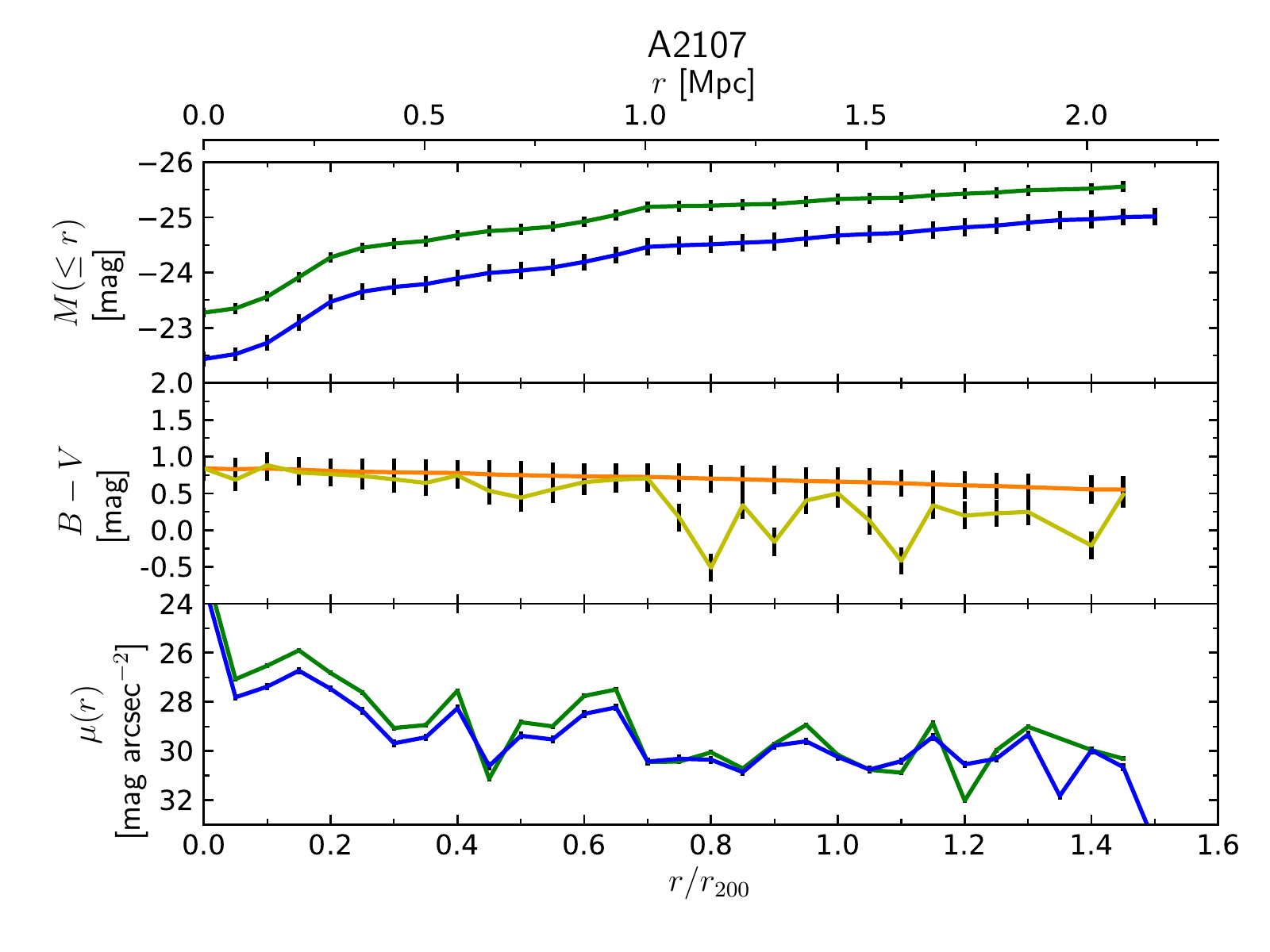}
        \includegraphics[width=0.45\textwidth]{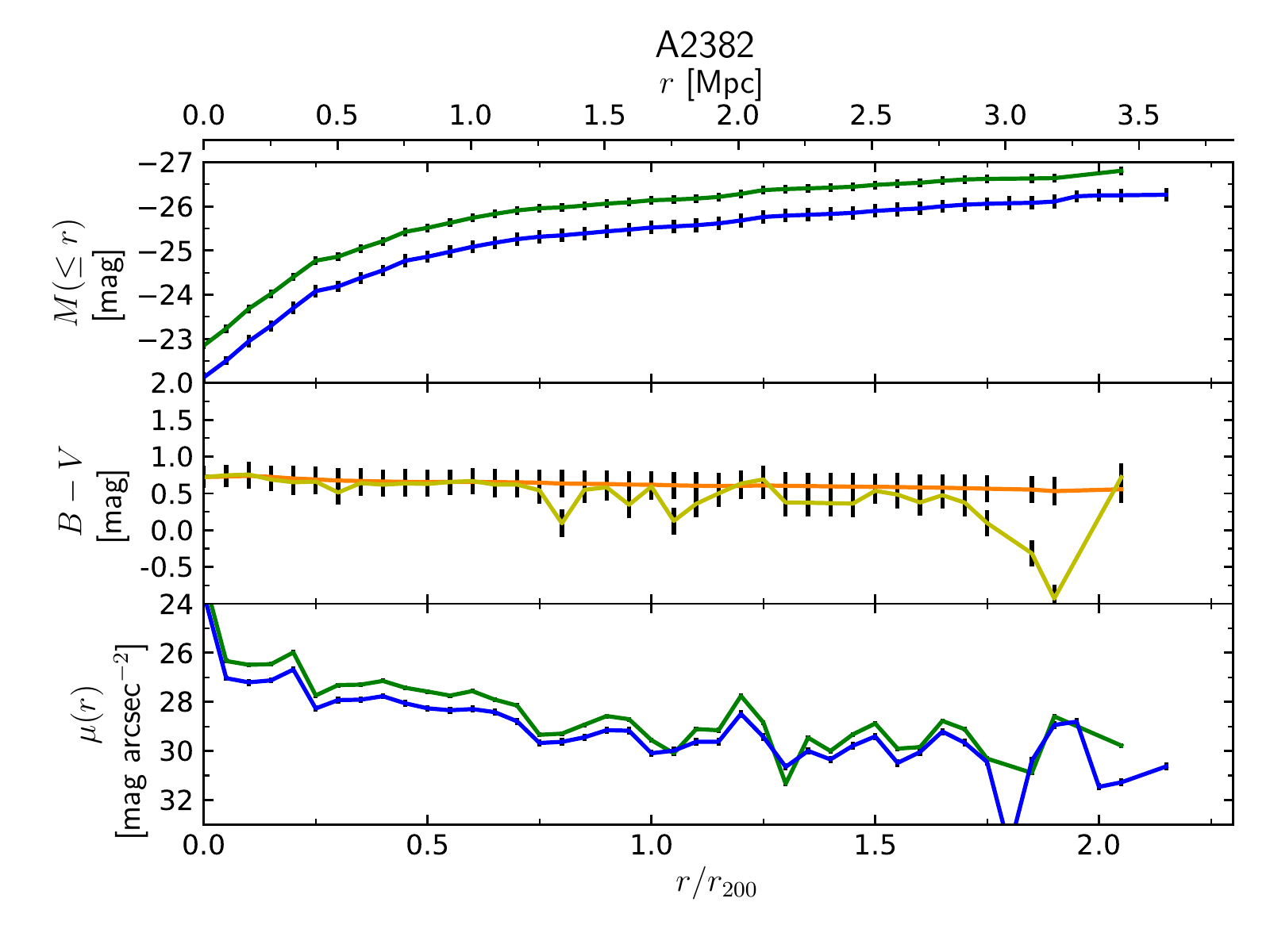}      \includegraphics[width=0.45\textwidth]{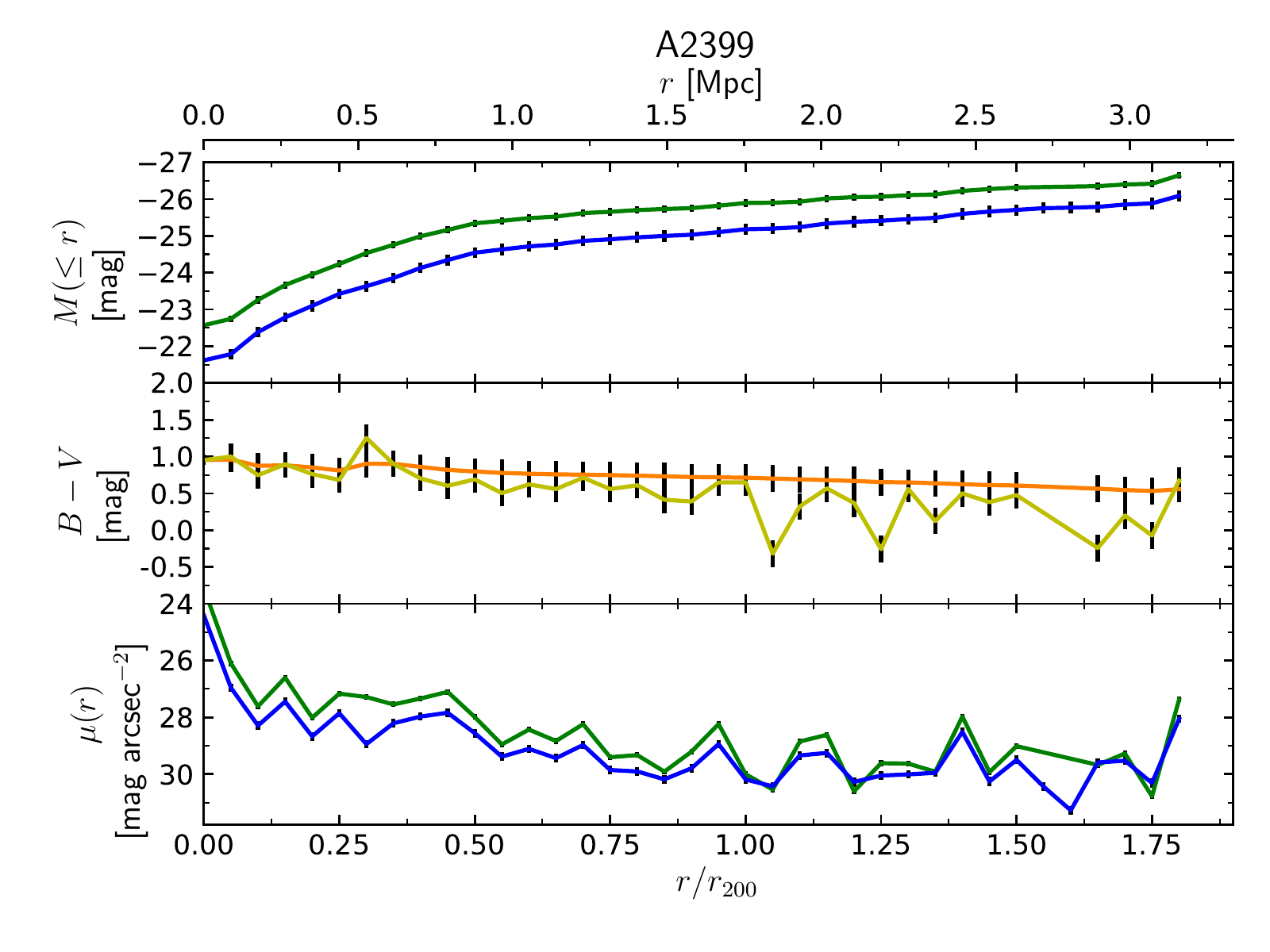}
    \caption{Photometric profiles of Omega-WINGS galaxy clusters, continued.}
\end{figure*}

\newpage
\clearpage

\begin{figure*}[t]
   \centering
        \includegraphics[width=0.45\textwidth]{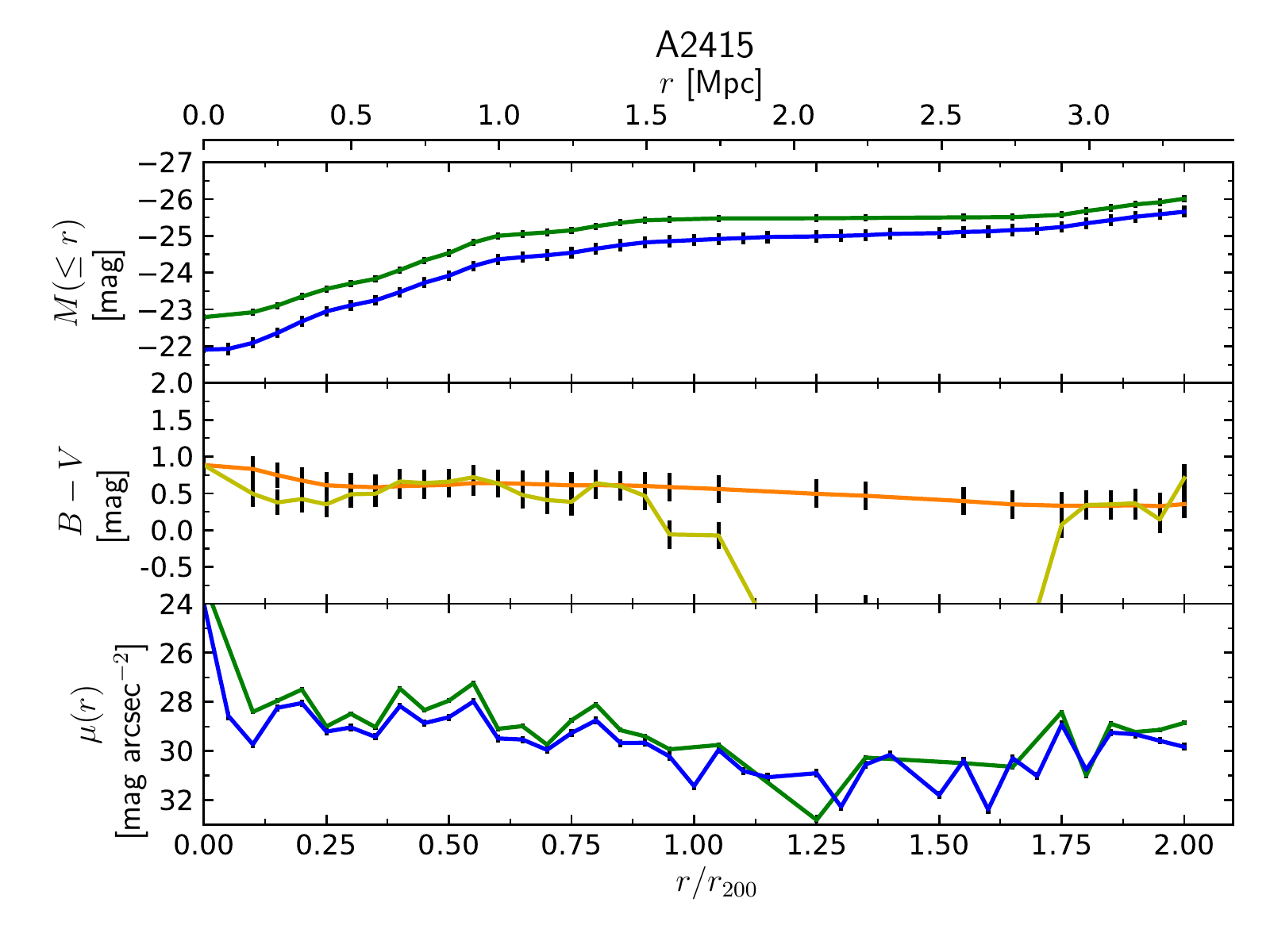}      \includegraphics[width=0.45\textwidth]{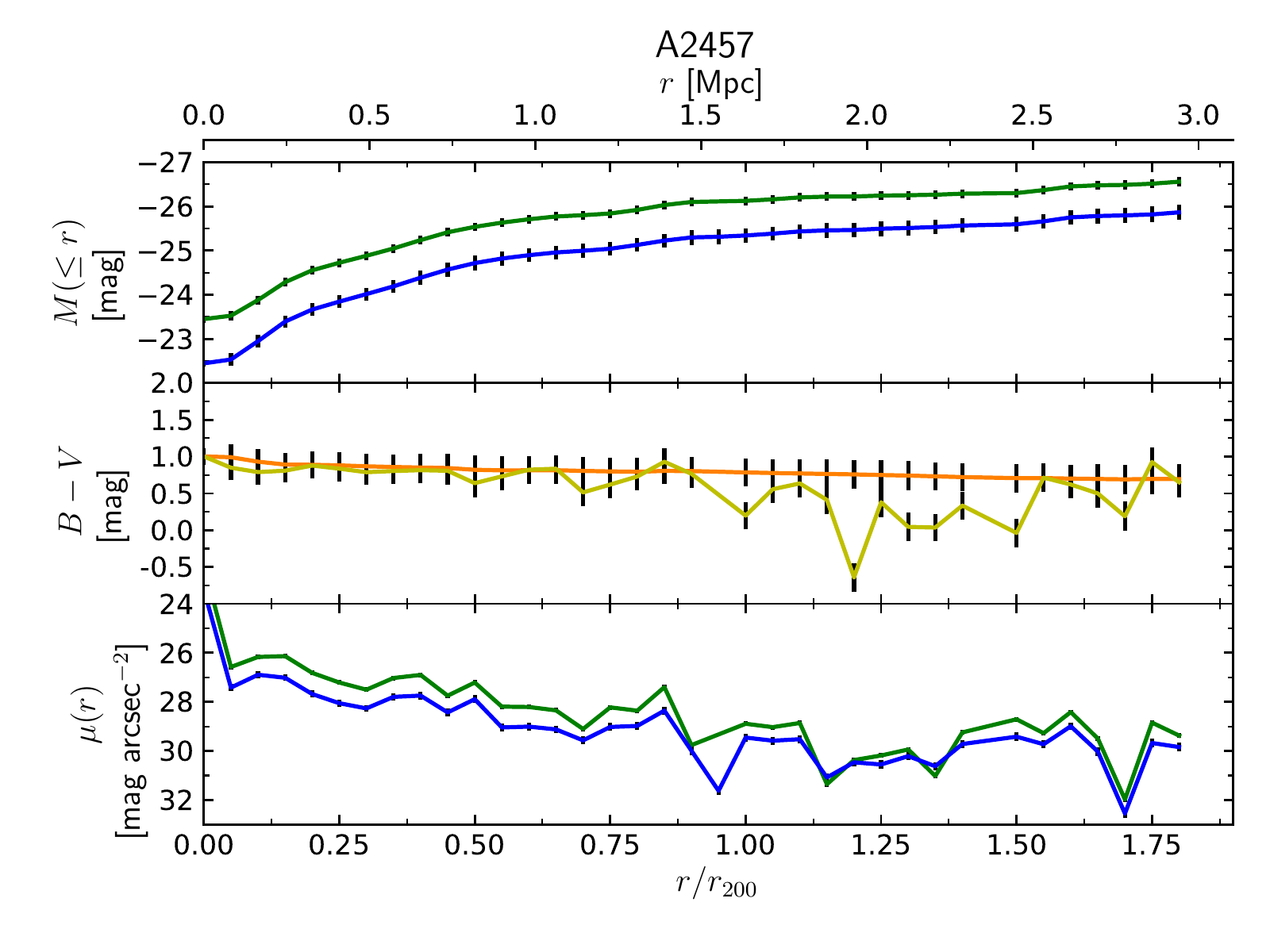}
        \includegraphics[width=0.45\textwidth]{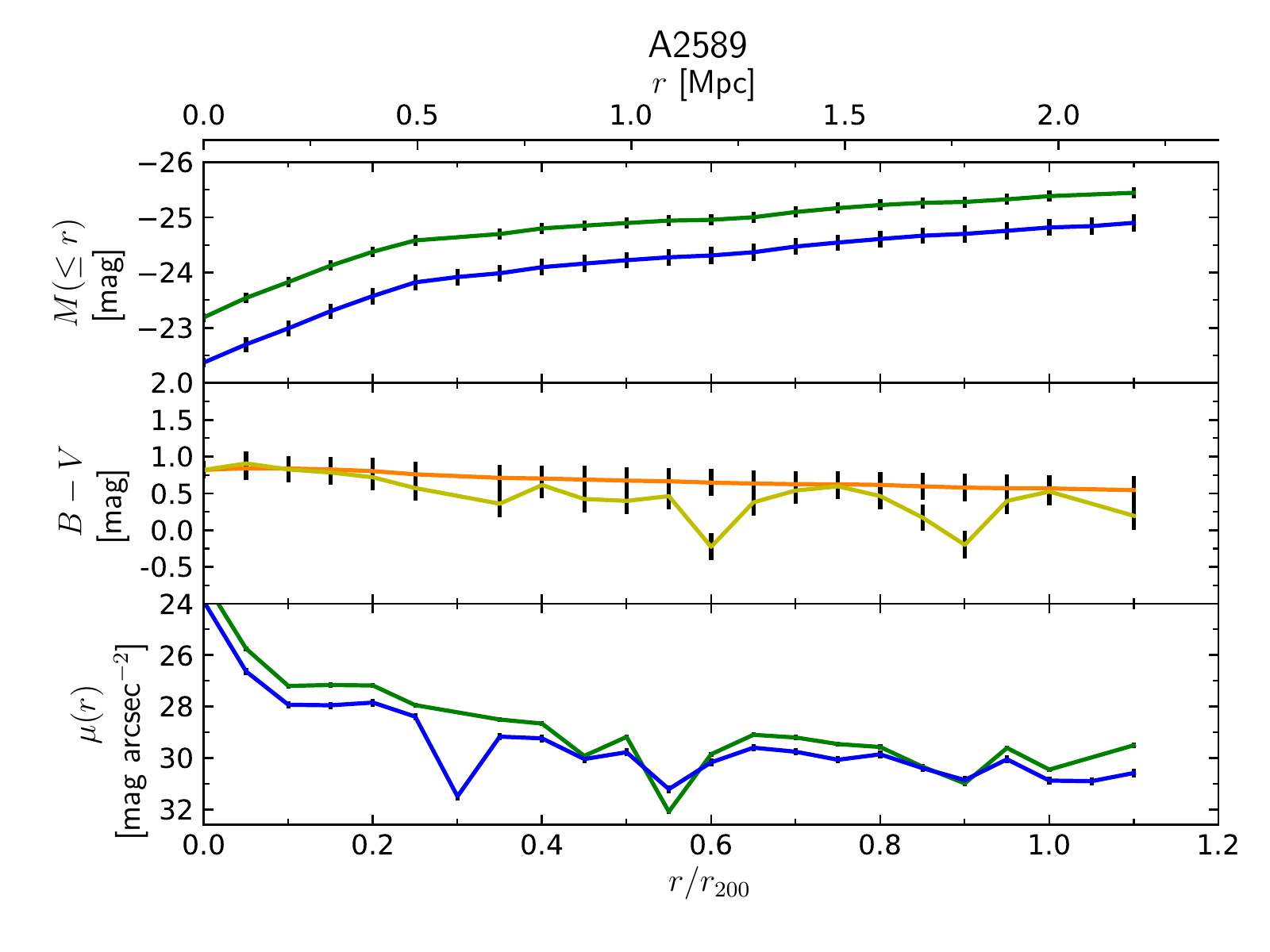}      \includegraphics[width=0.45\textwidth]{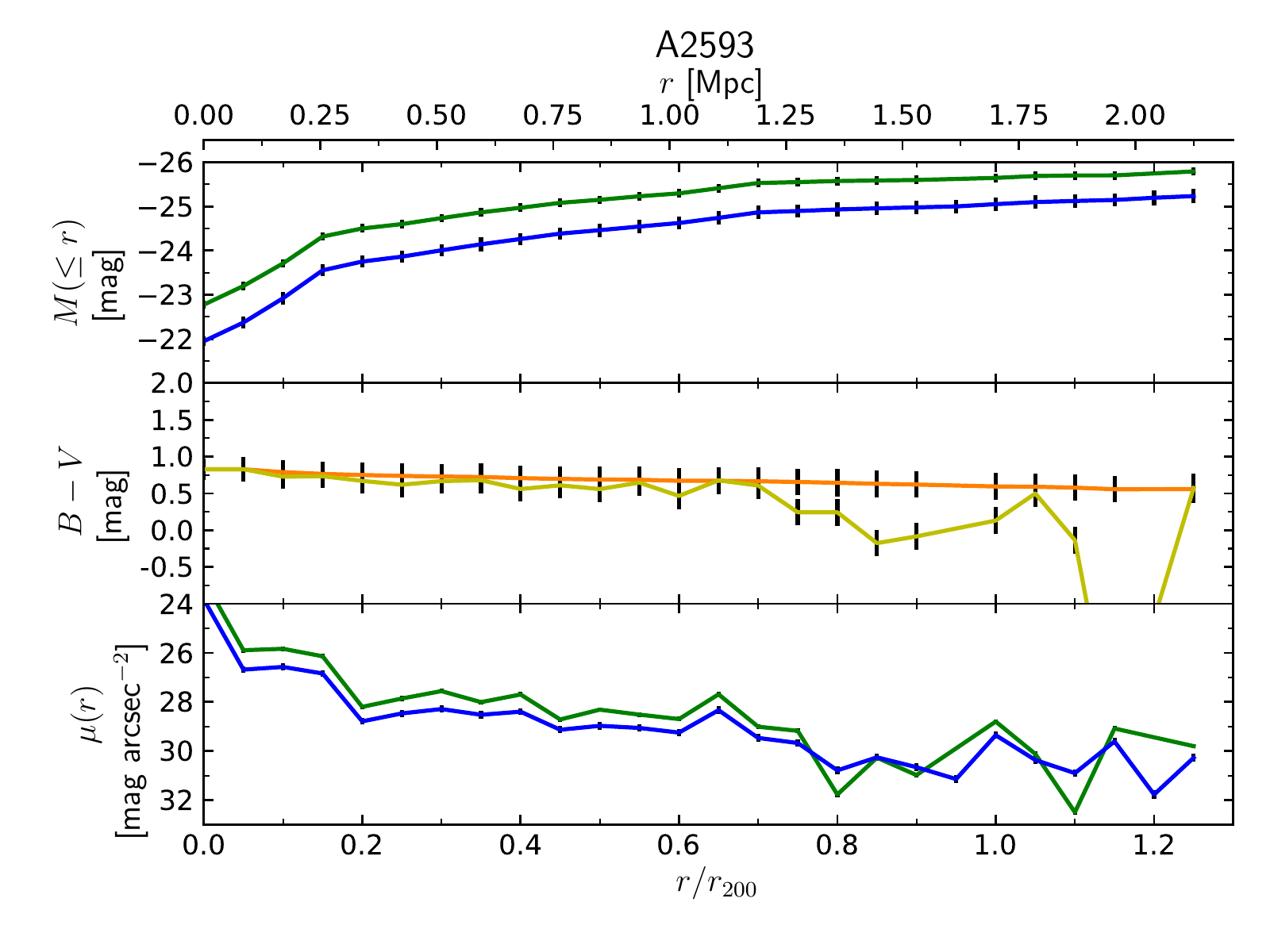}
        \includegraphics[width=0.45\textwidth]{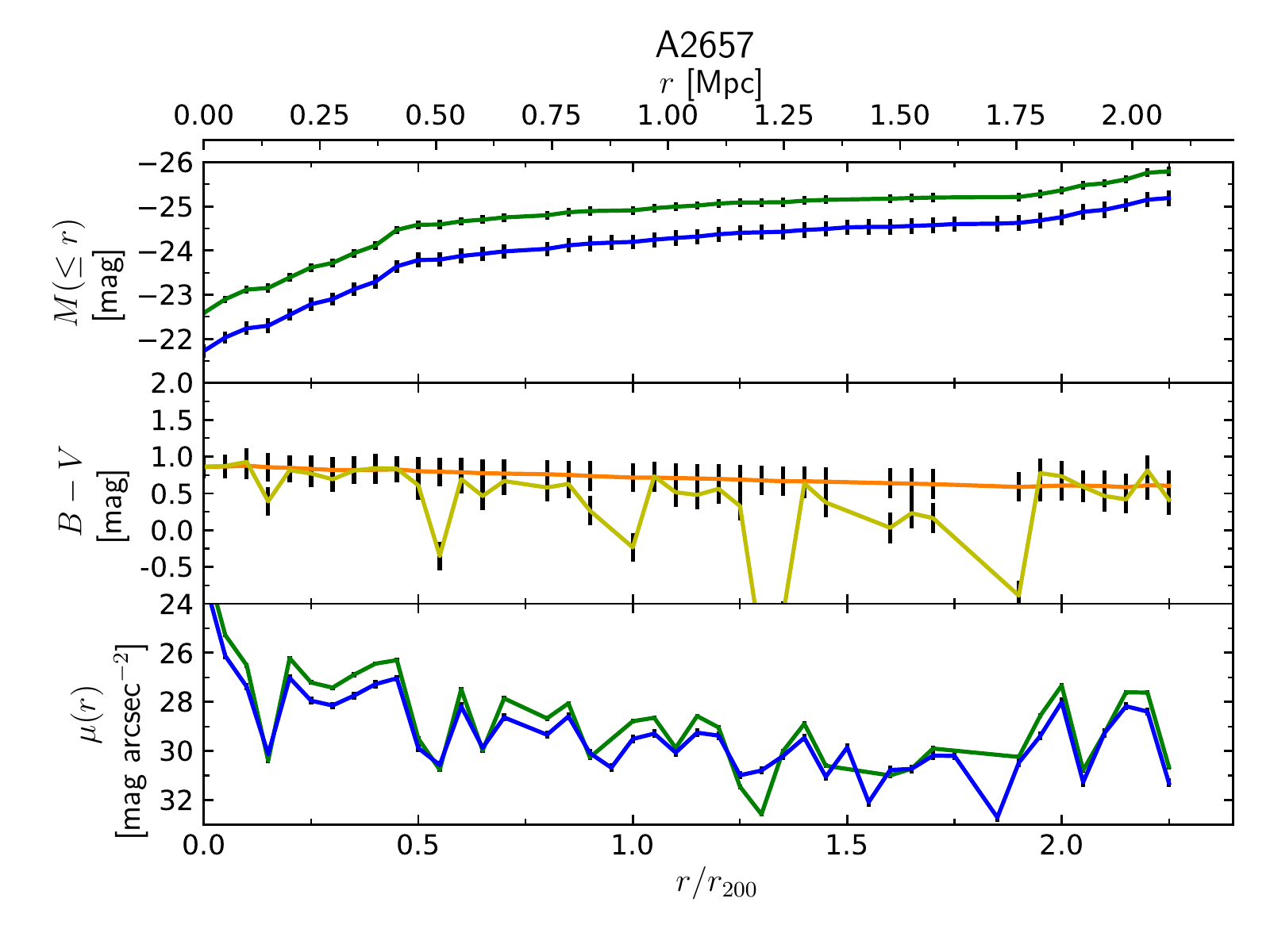}      \includegraphics[width=0.45\textwidth]{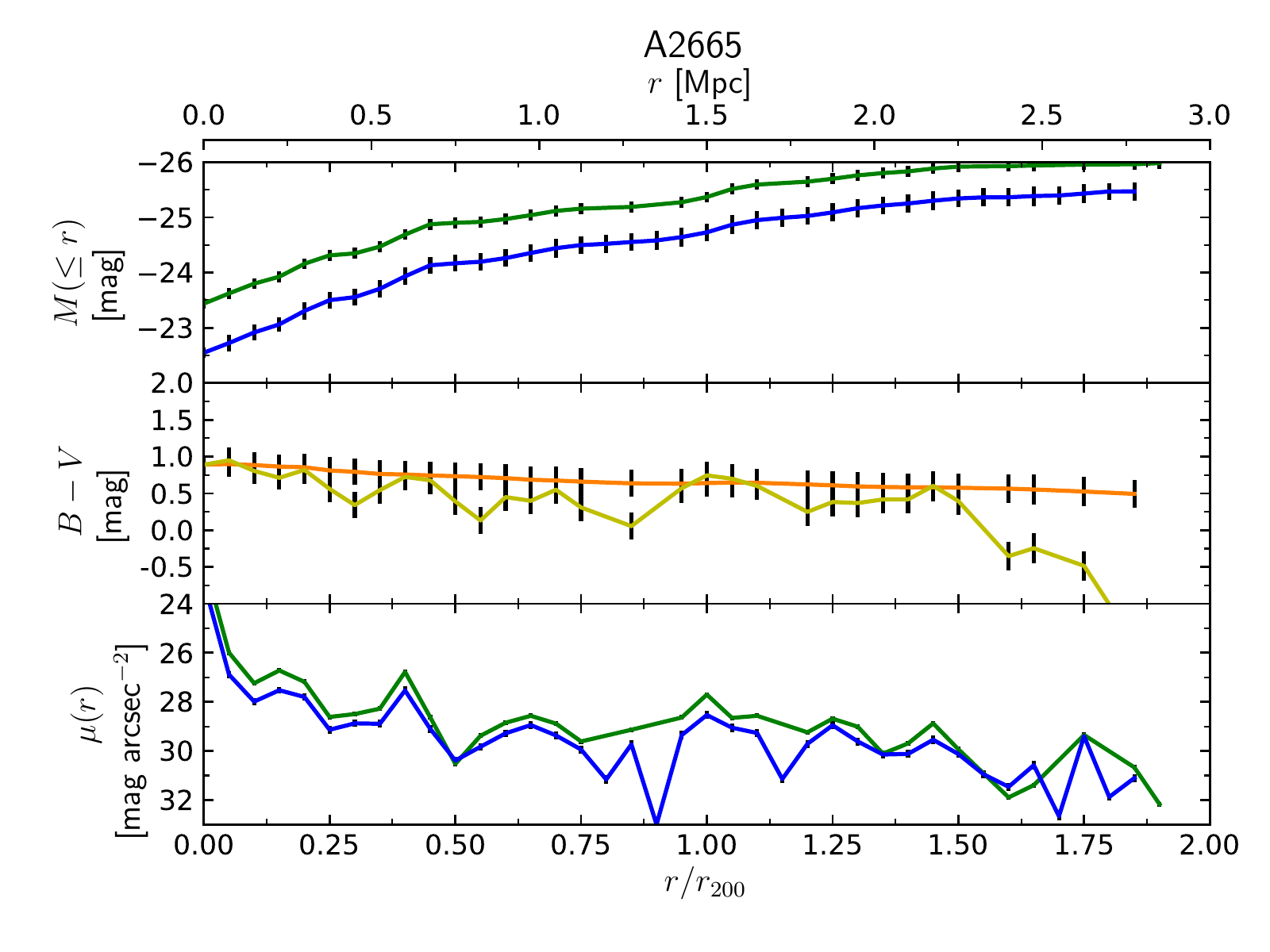}
    \caption{Photometric profiles of Omega-WINGS galaxy clusters, continued.}
\end{figure*}

\newpage
\clearpage

\begin{figure*}[t]
   \centering
        \includegraphics[width=0.45\textwidth]{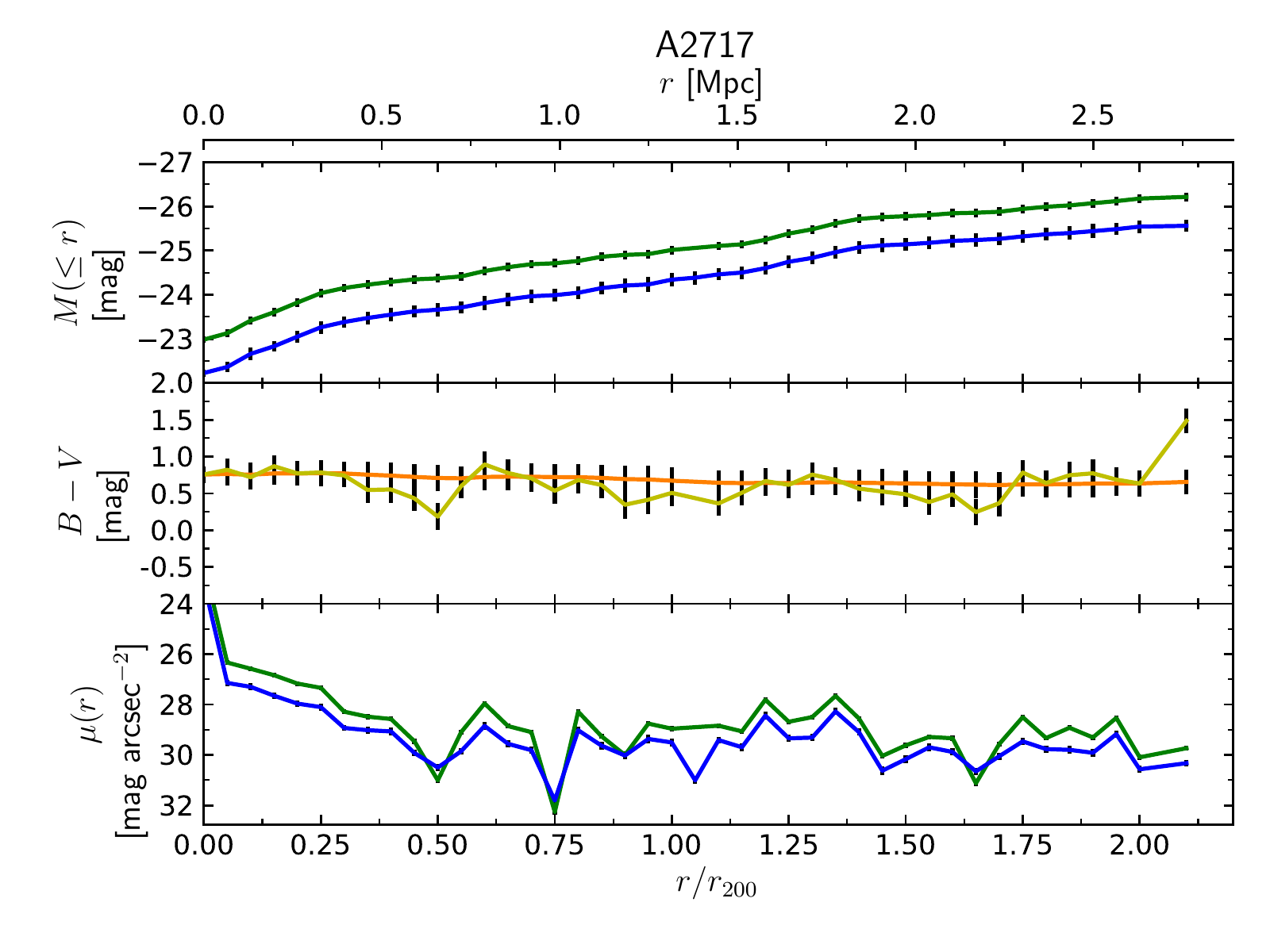}      \includegraphics[width=0.45\textwidth]{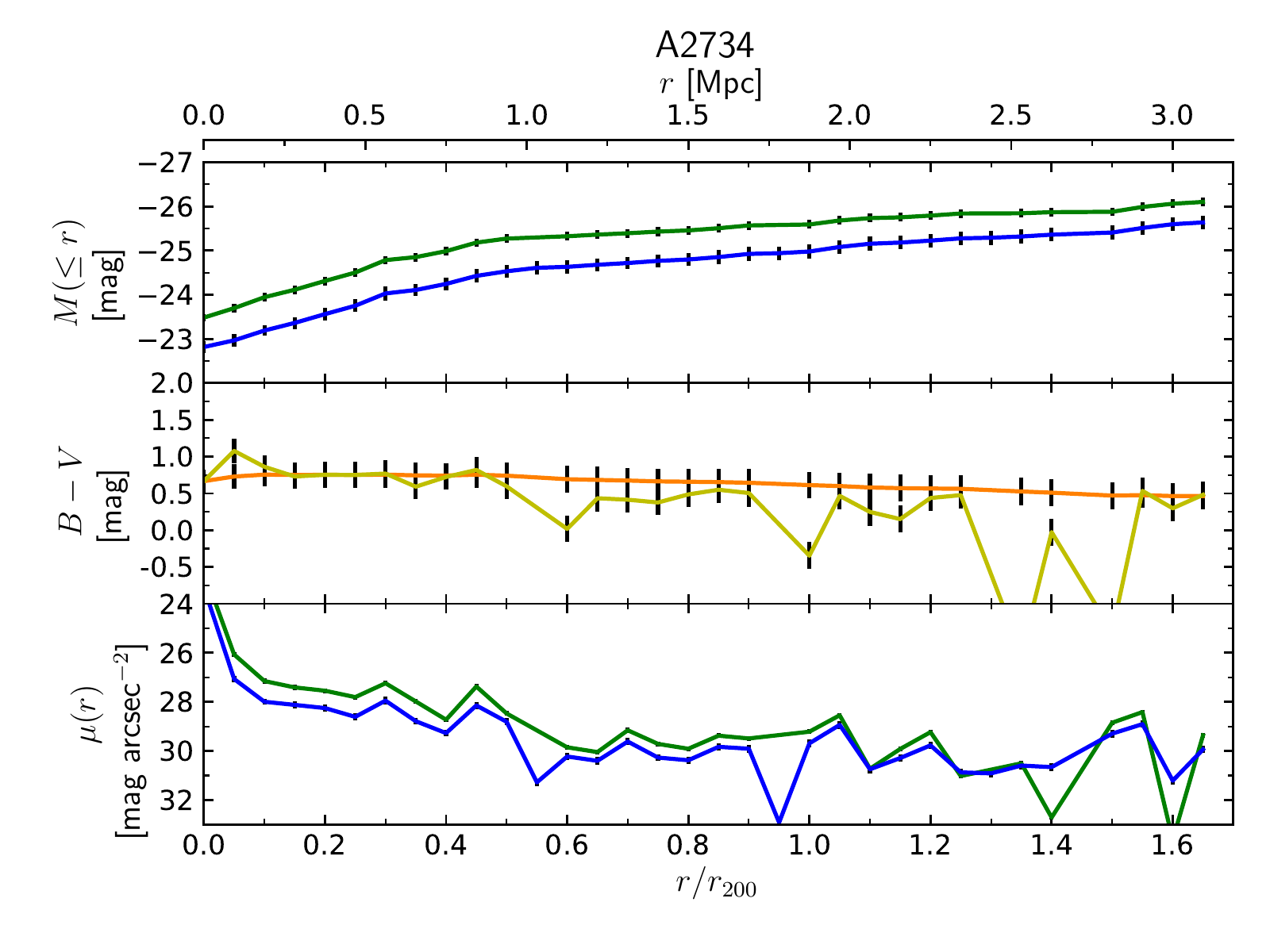}
        \includegraphics[width=0.45\textwidth]{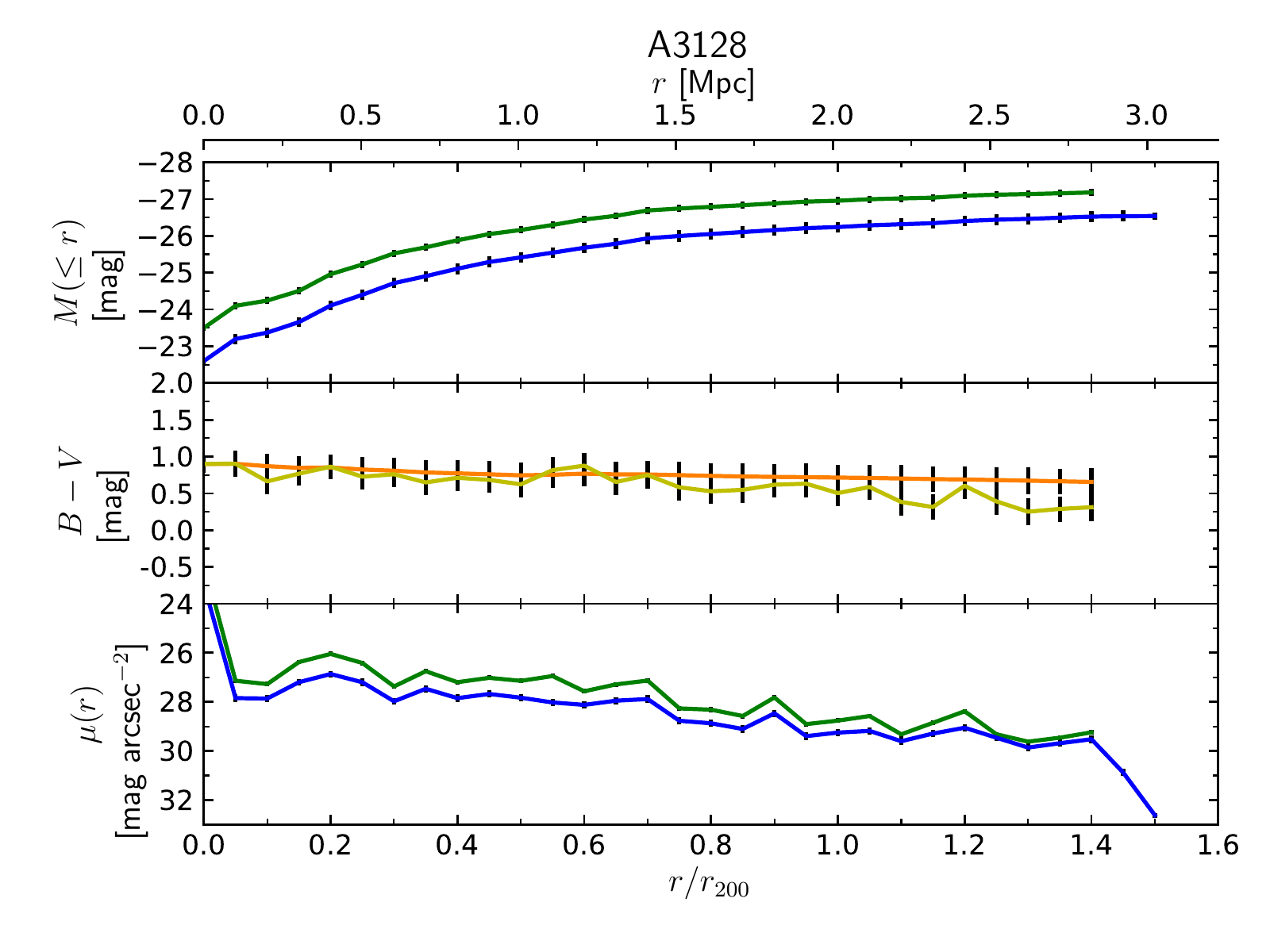}      \includegraphics[width=0.45\textwidth]{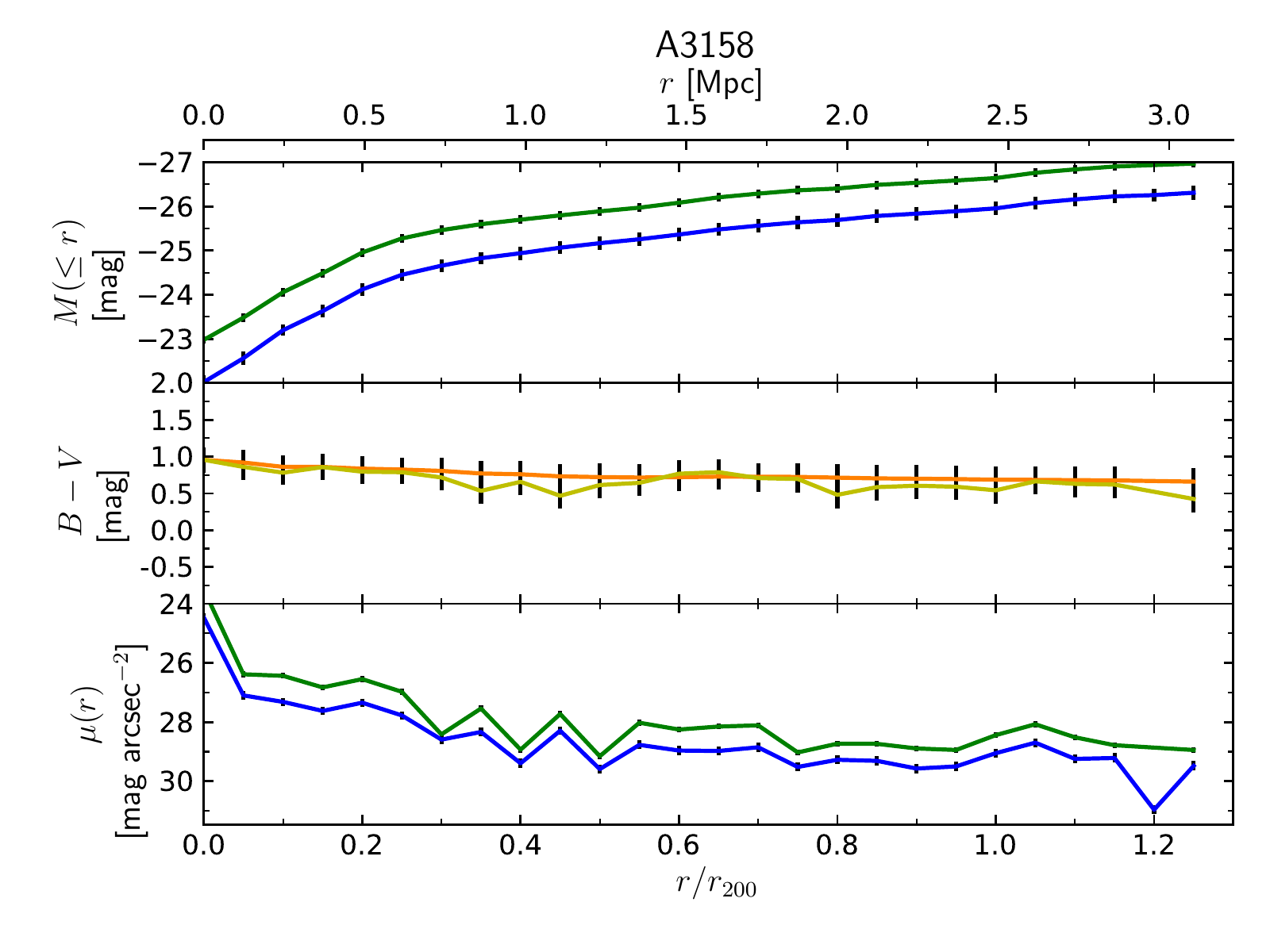}
        \includegraphics[width=0.45\textwidth]{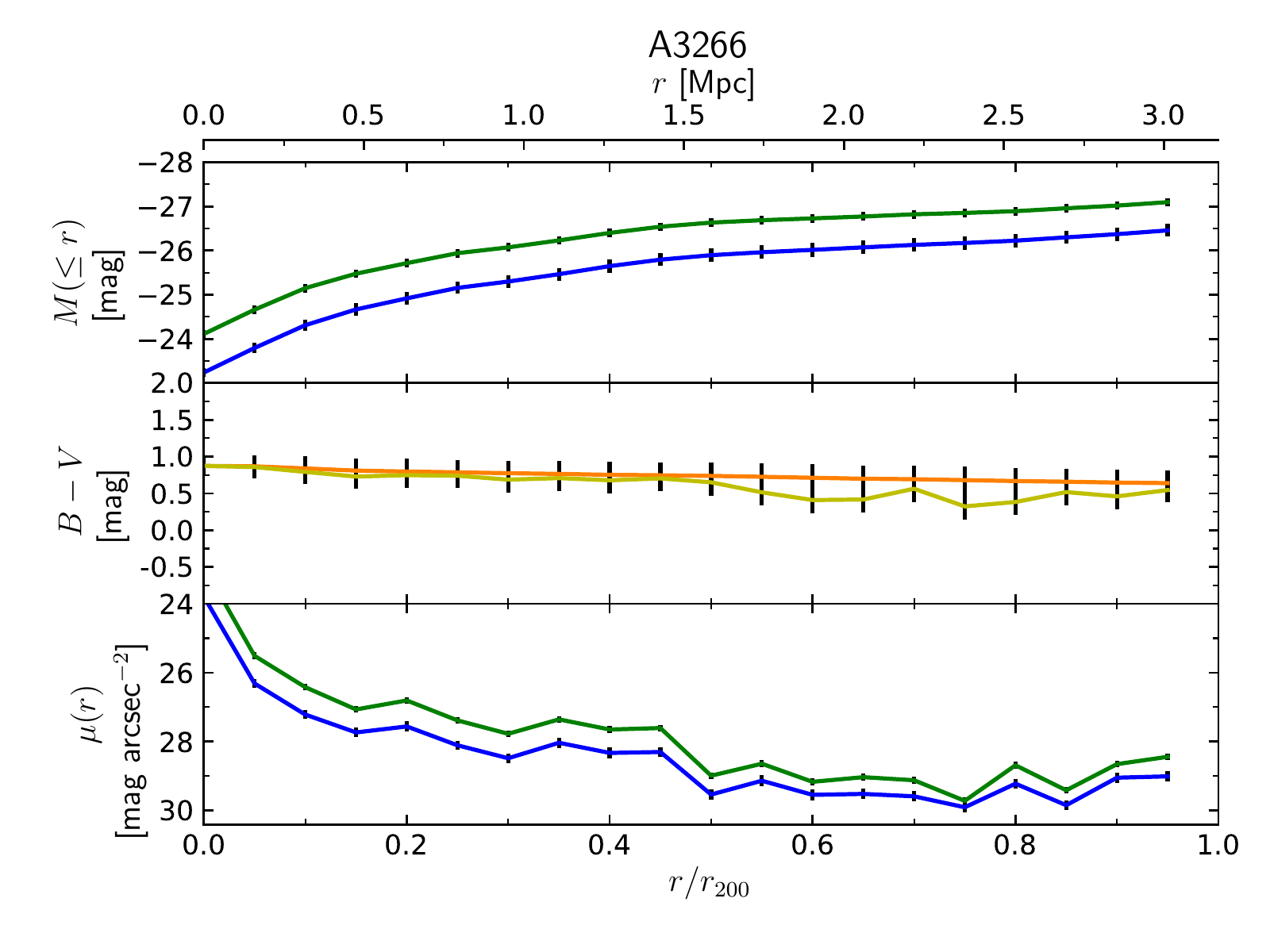}      \includegraphics[width=0.45\textwidth]{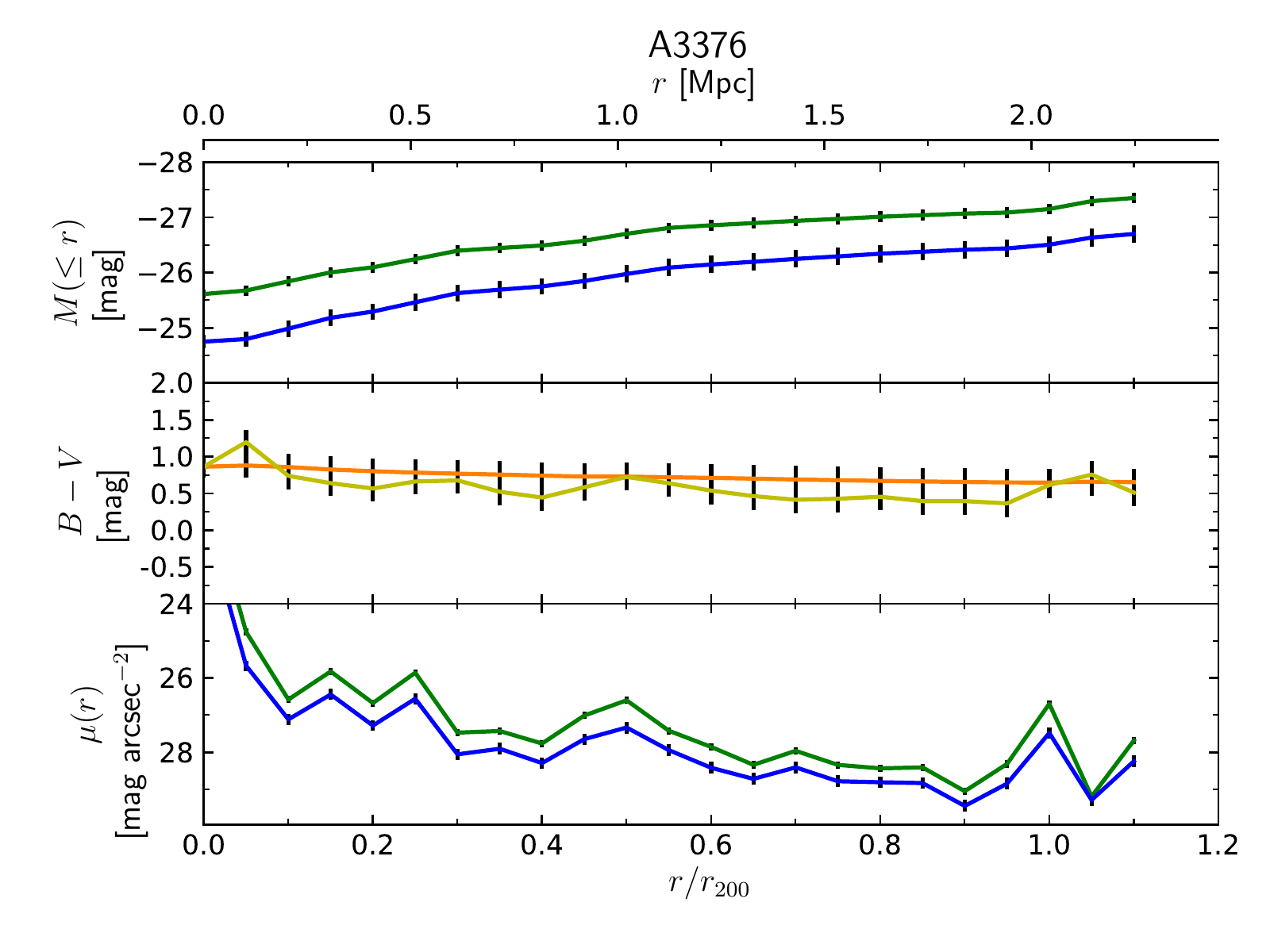}
    \caption{Photometric profiles of Omega-WINGS galaxy clusters, continued.}
\end{figure*}

\newpage
\clearpage

\begin{figure*}[t]
   \centering
        \includegraphics[width=0.45\textwidth]{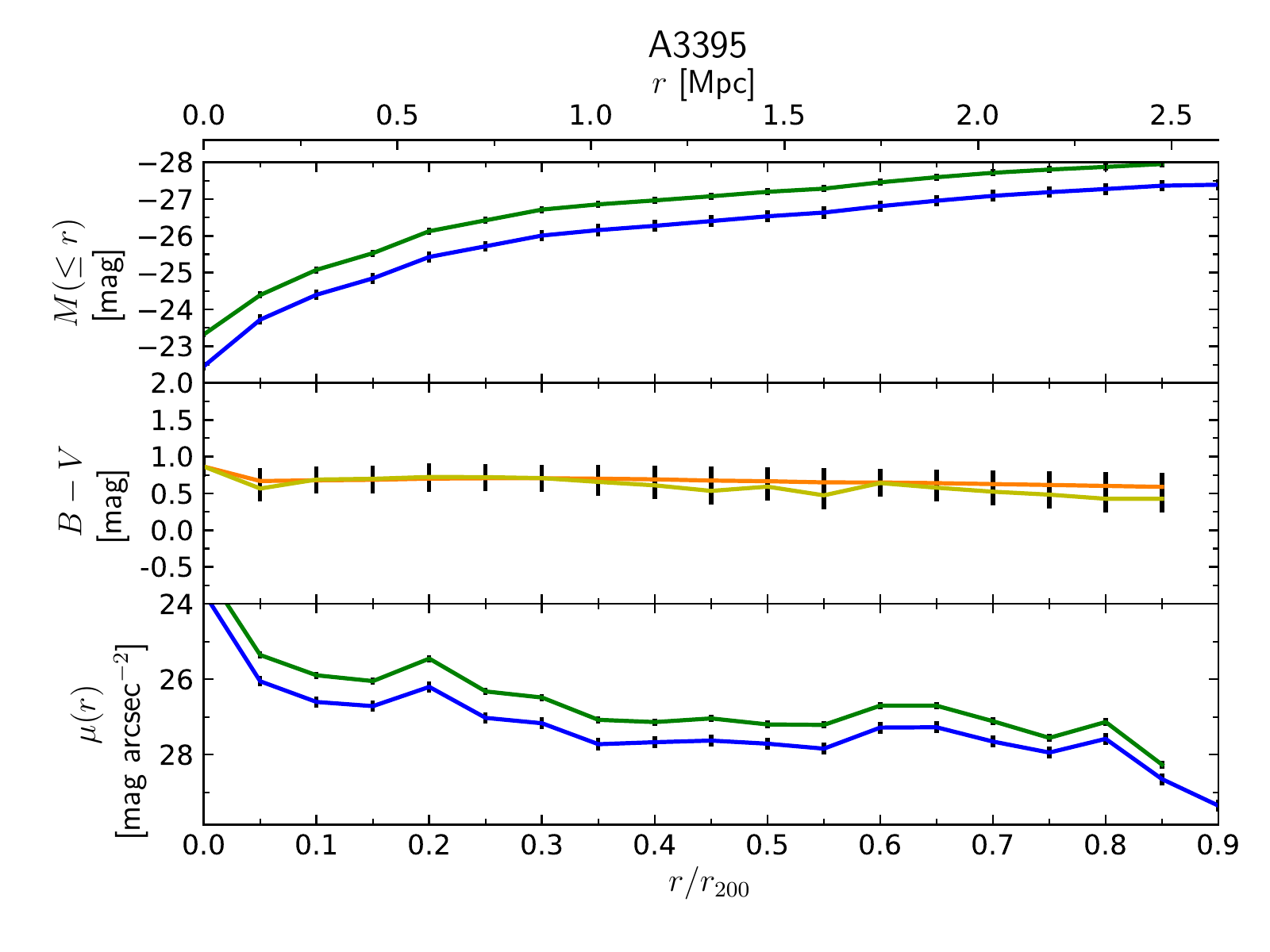}      \includegraphics[width=0.45\textwidth]{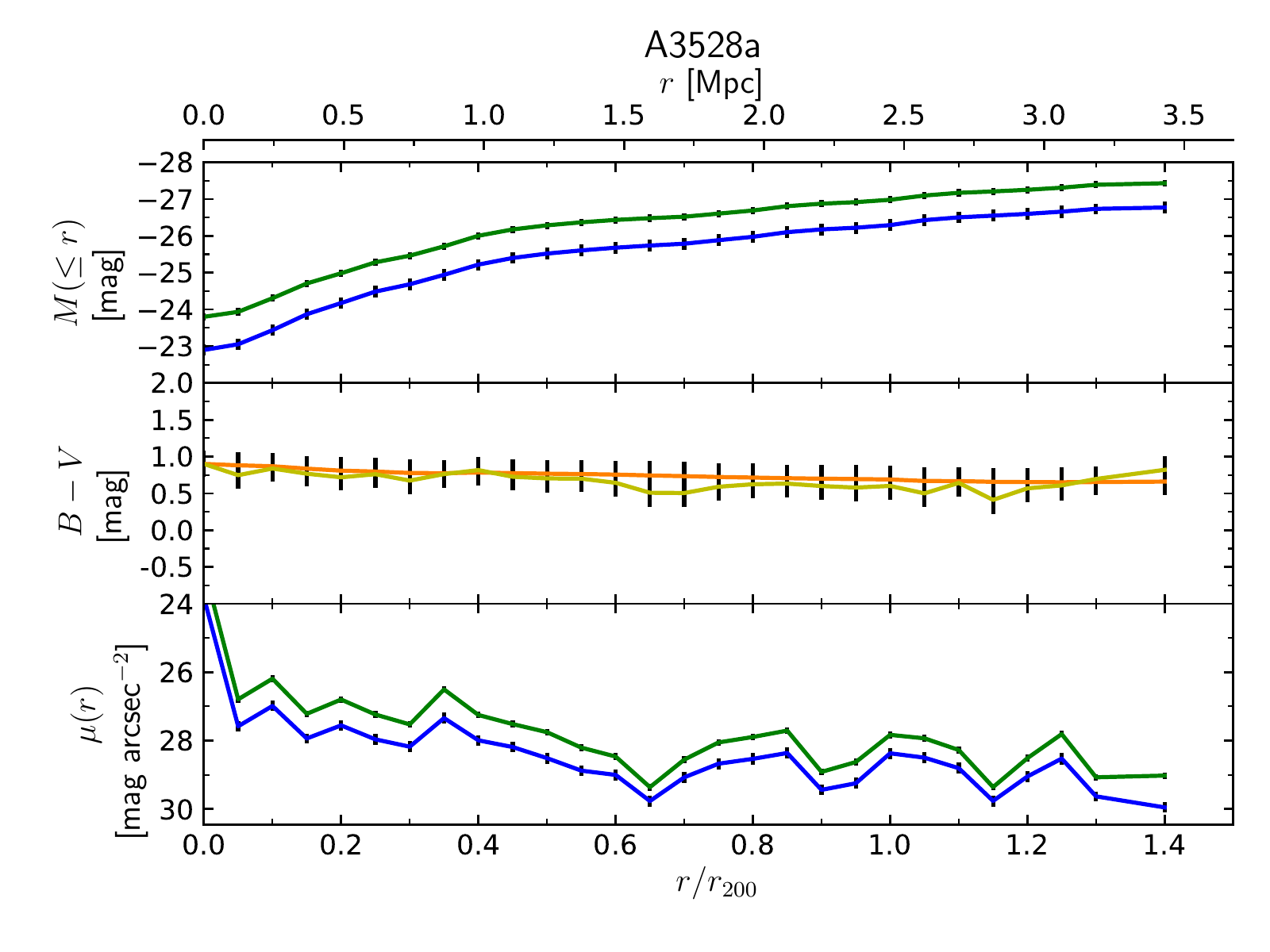}
        \includegraphics[width=0.45\textwidth]{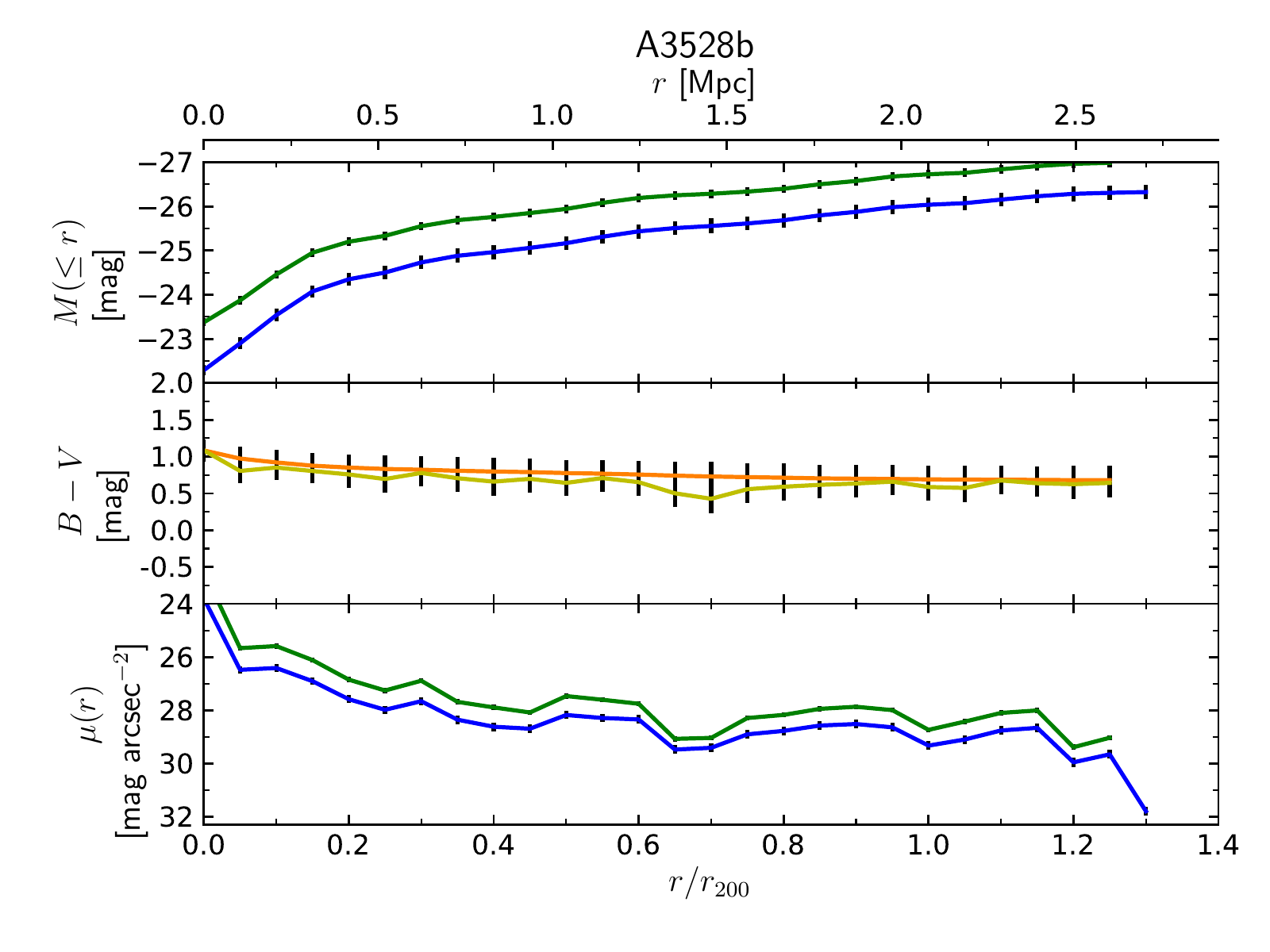}     \includegraphics[width=0.45\textwidth]{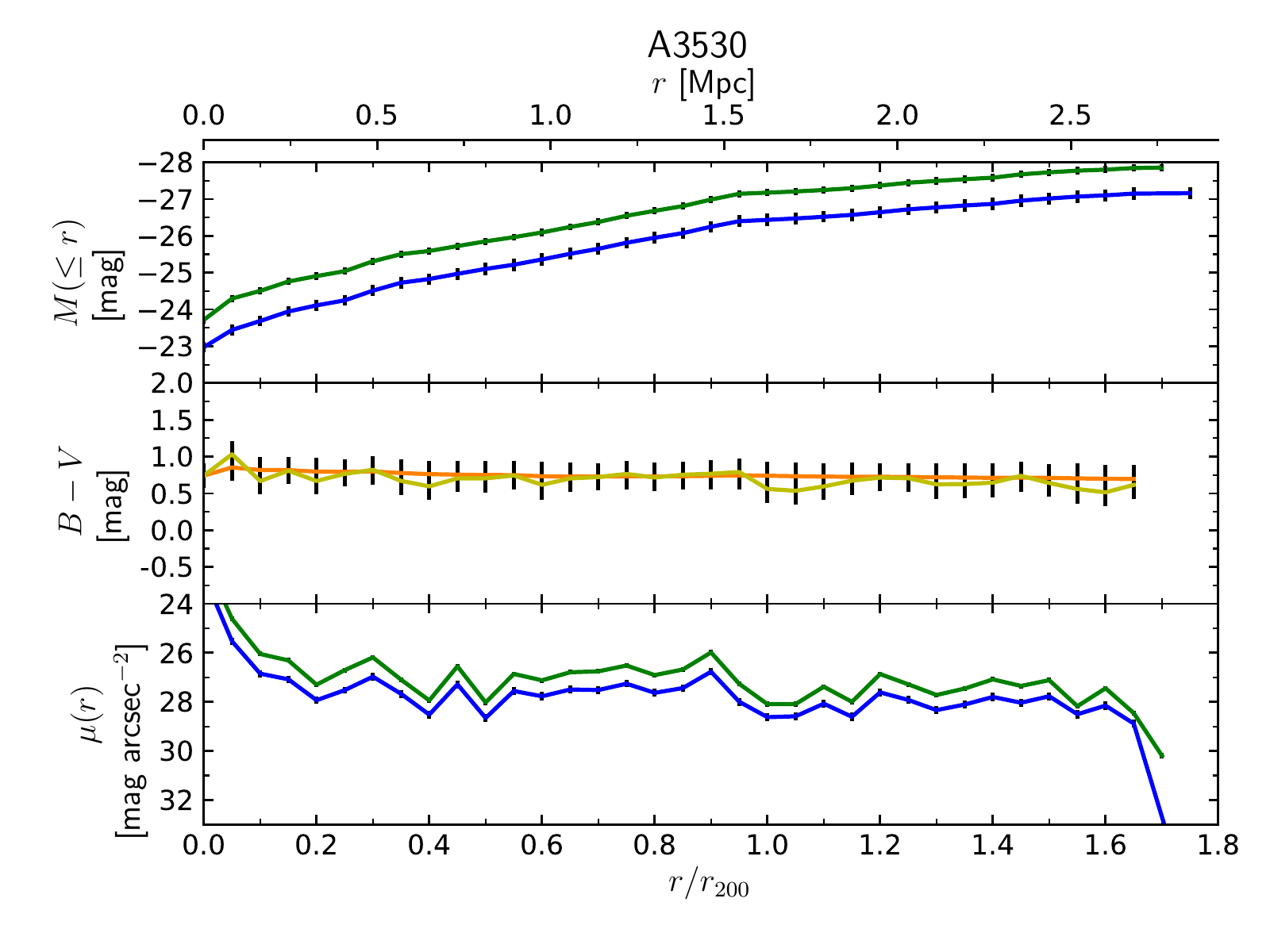}
        \includegraphics[width=0.45\textwidth]{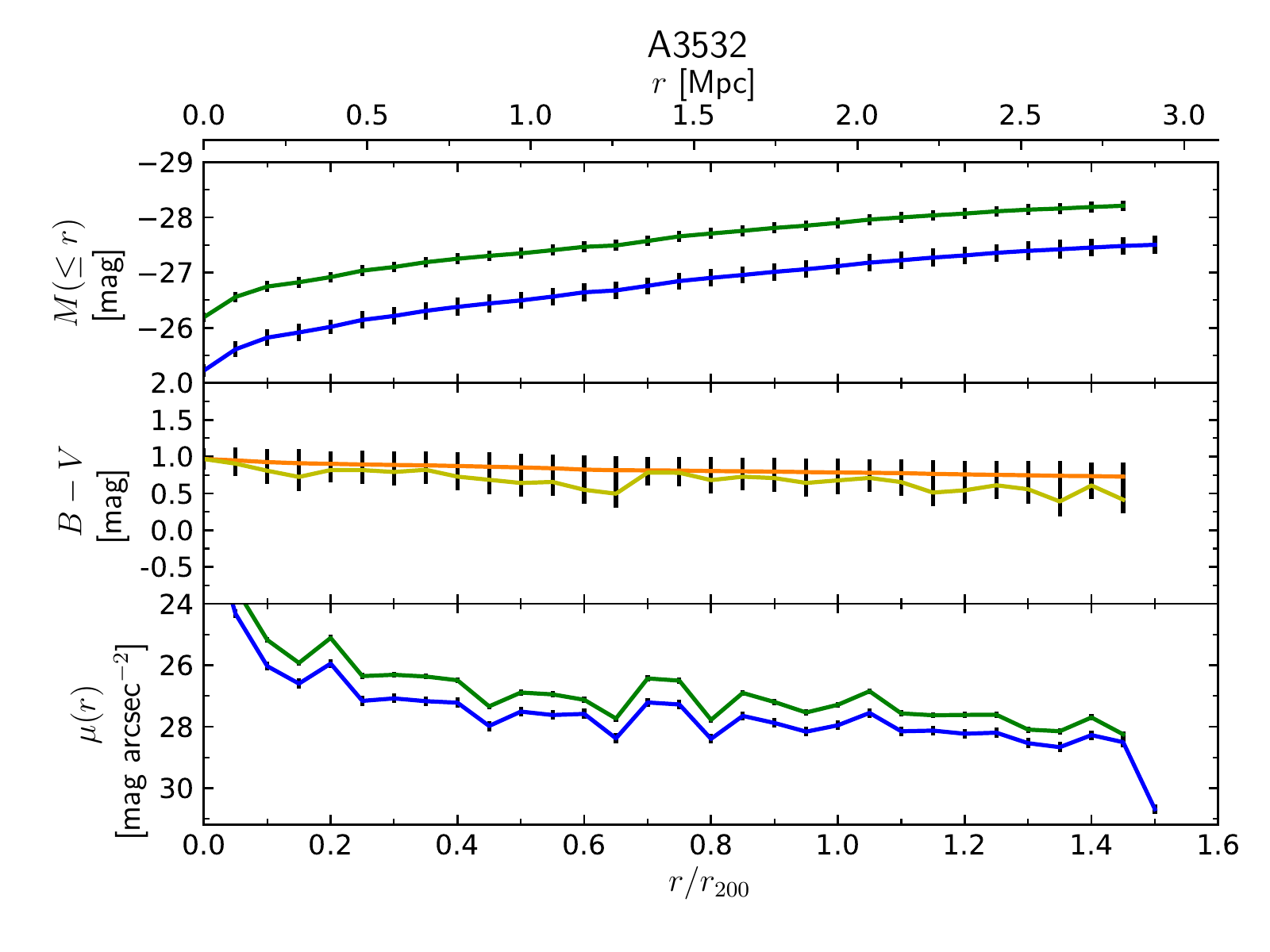}      \includegraphics[width=0.45\textwidth]{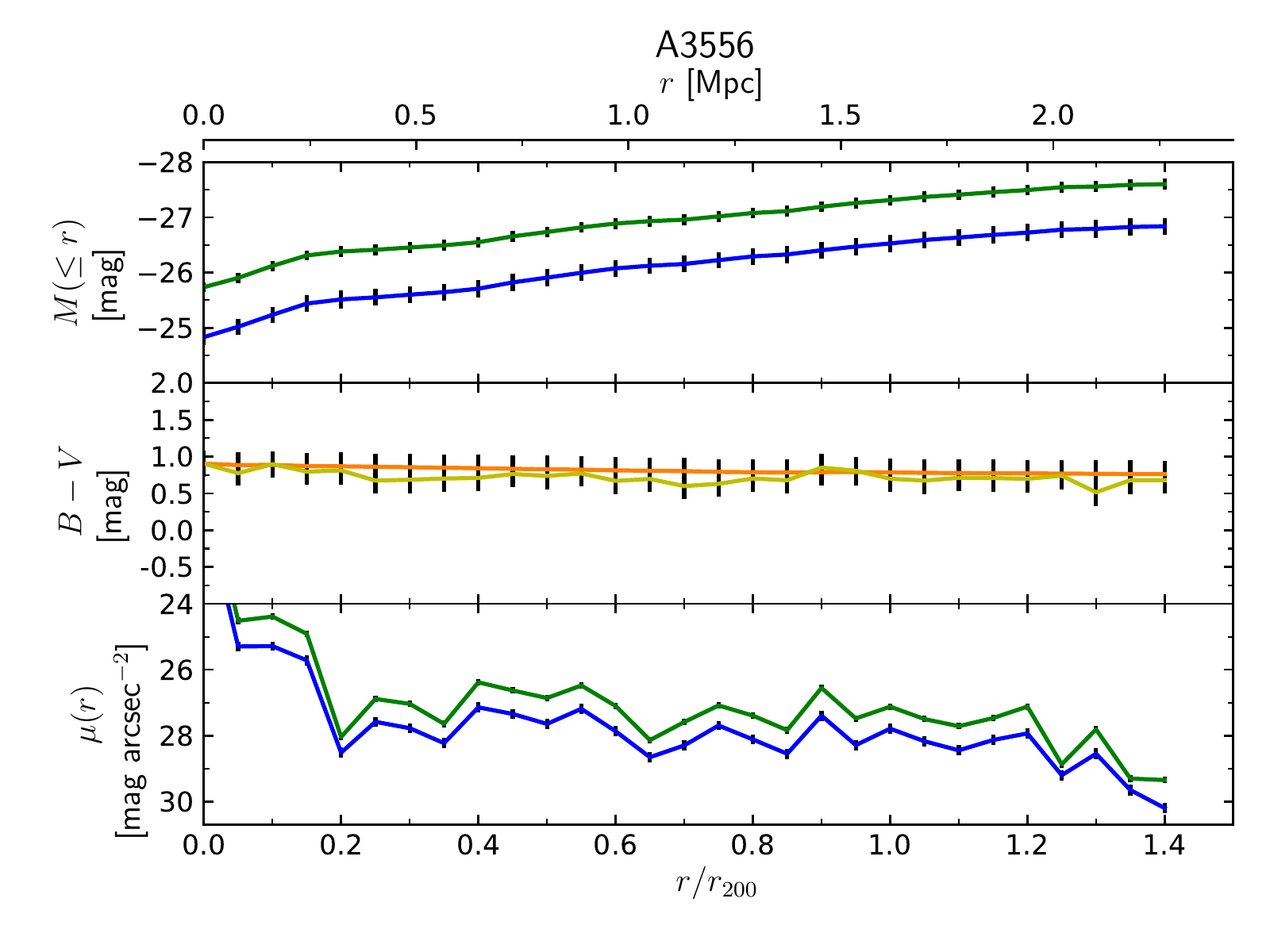}
    \caption{Photometric profiles of Omega-WINGS galaxy clusters, continued.}
\end{figure*}

\newpage
\clearpage

\begin{figure*}[t]
   \centering
        \includegraphics[width=0.45\textwidth]{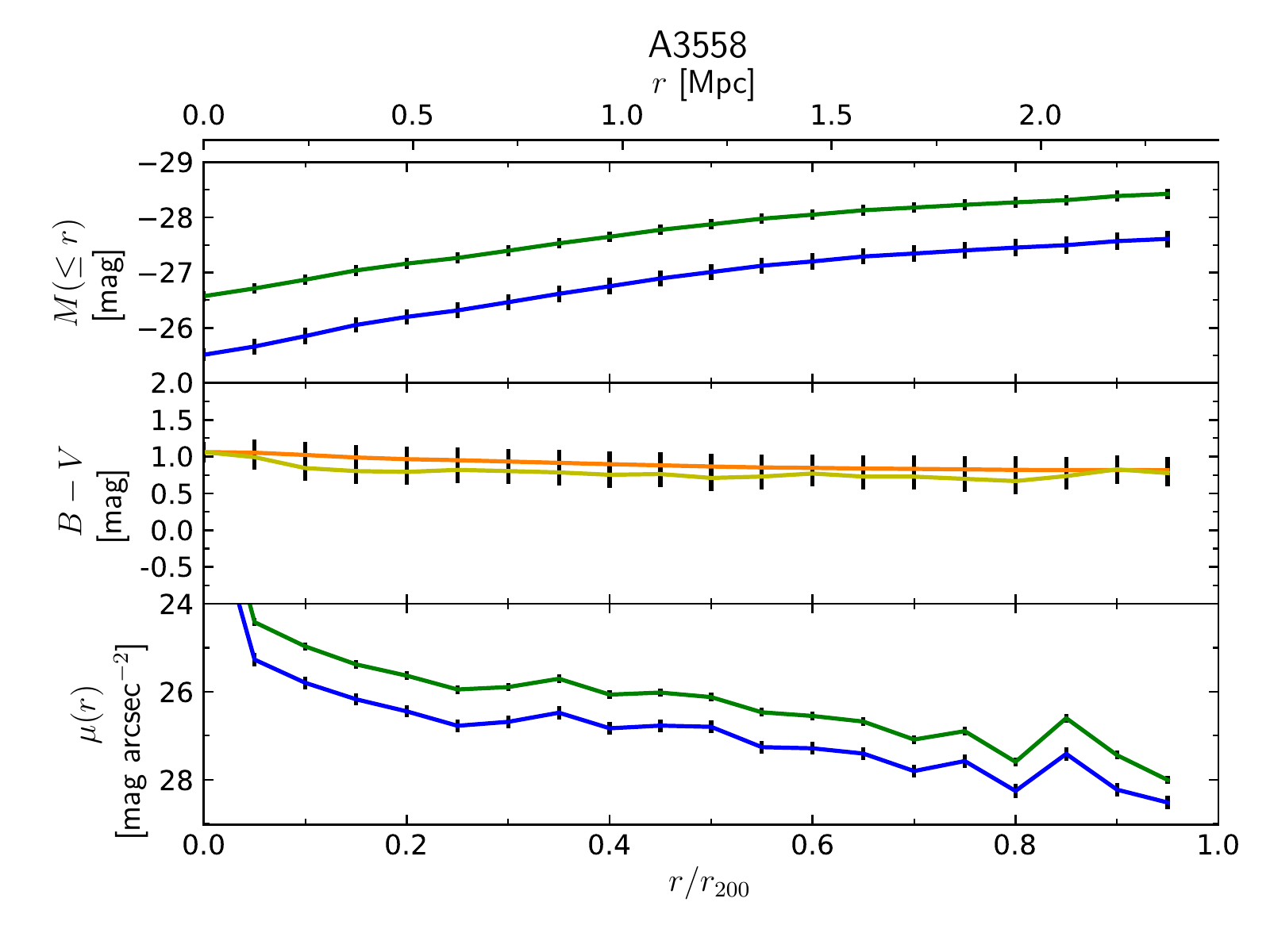}      \includegraphics[width=0.45\textwidth]{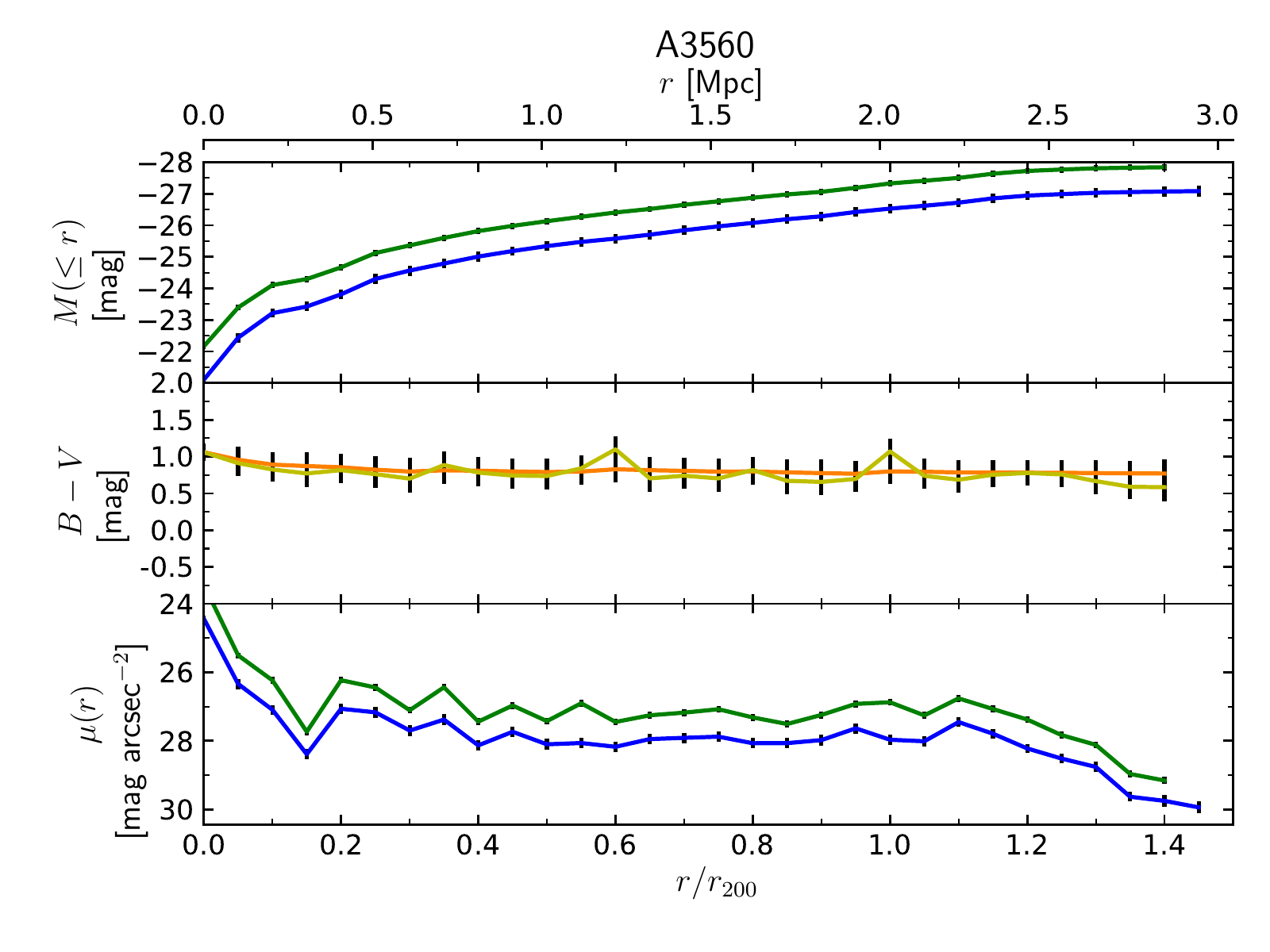}
        \includegraphics[width=0.45\textwidth]{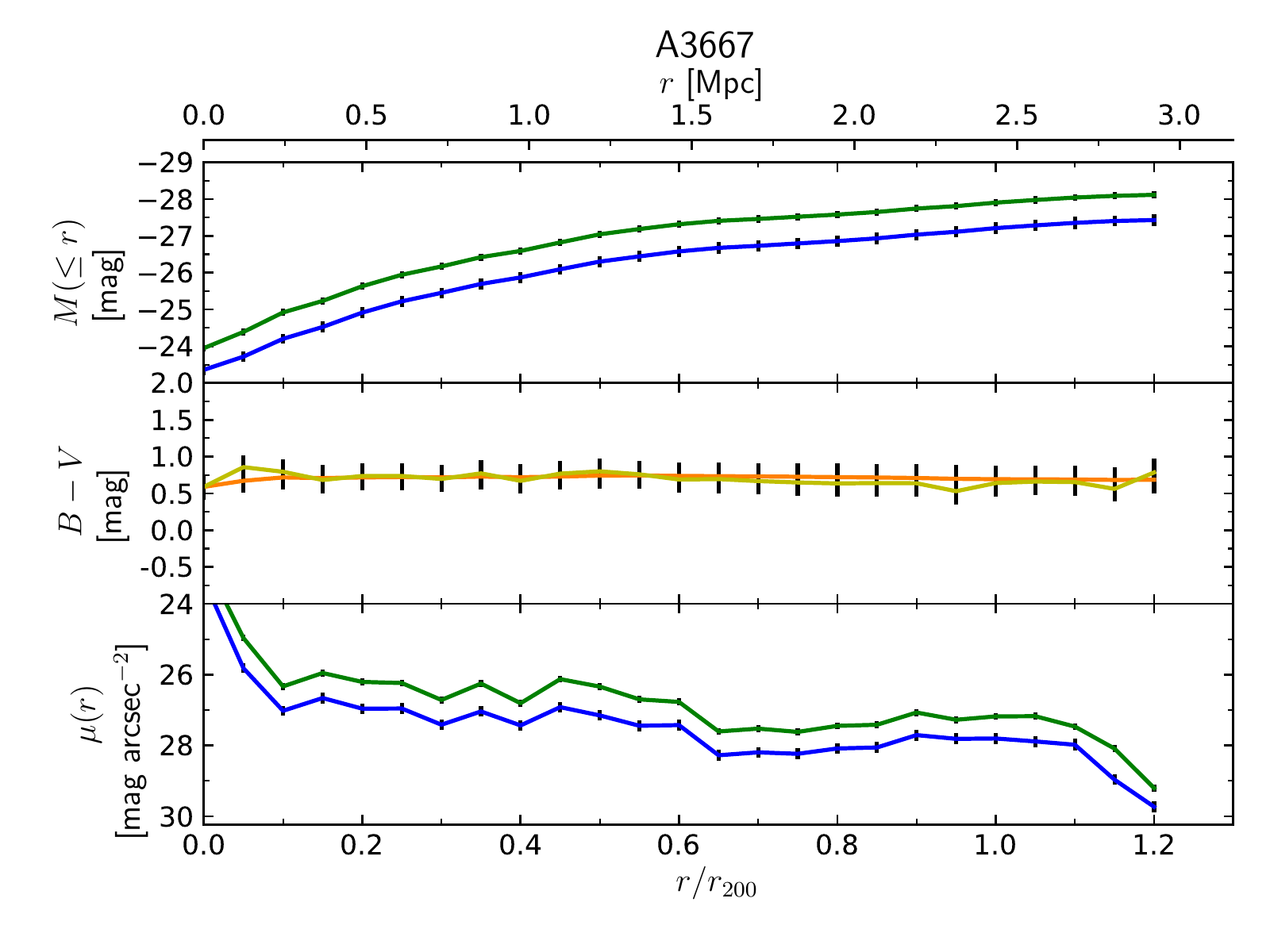}      \includegraphics[width=0.45\textwidth]{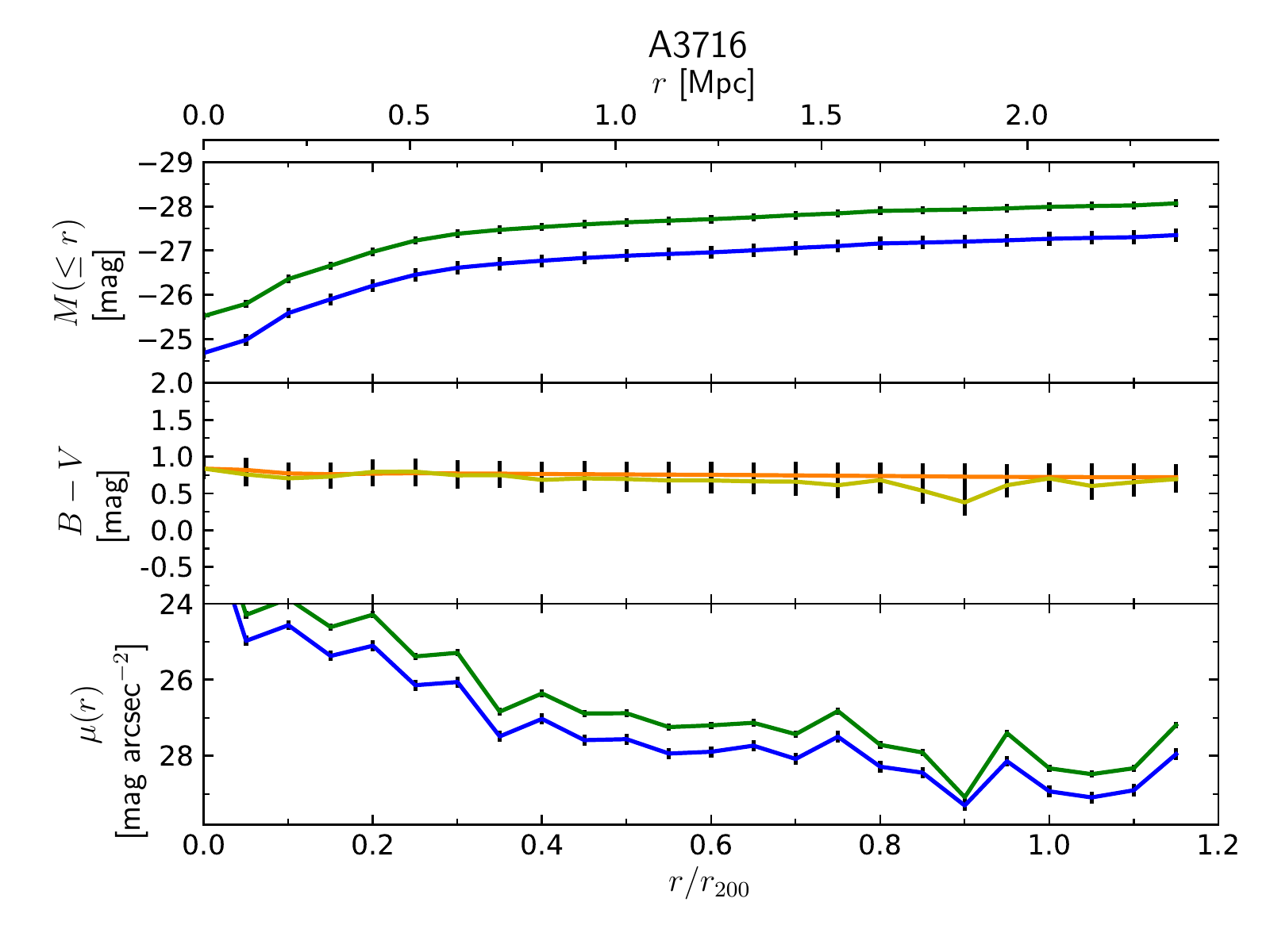}
        \includegraphics[width=0.45\textwidth]{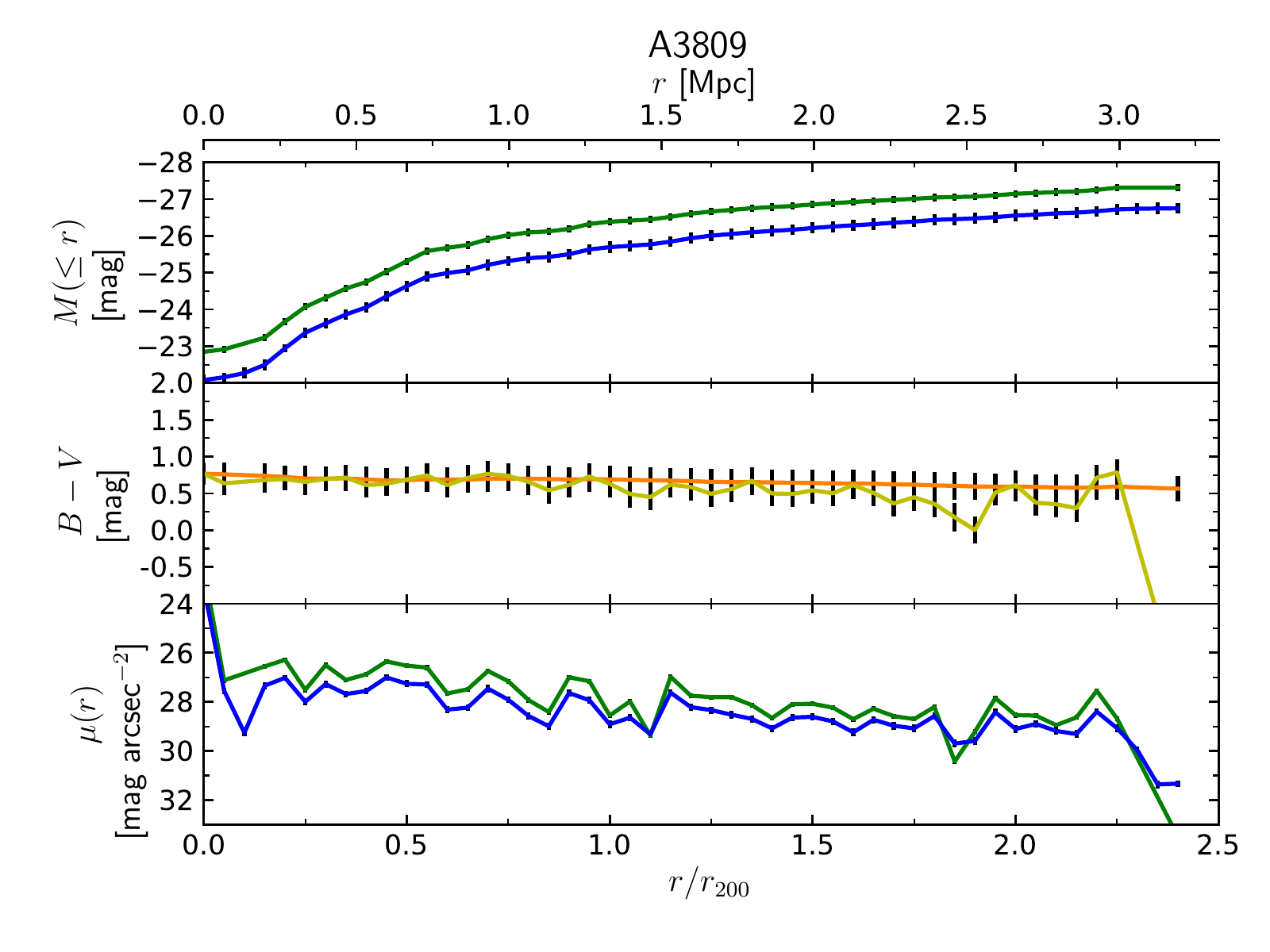}      \includegraphics[width=0.45\textwidth]{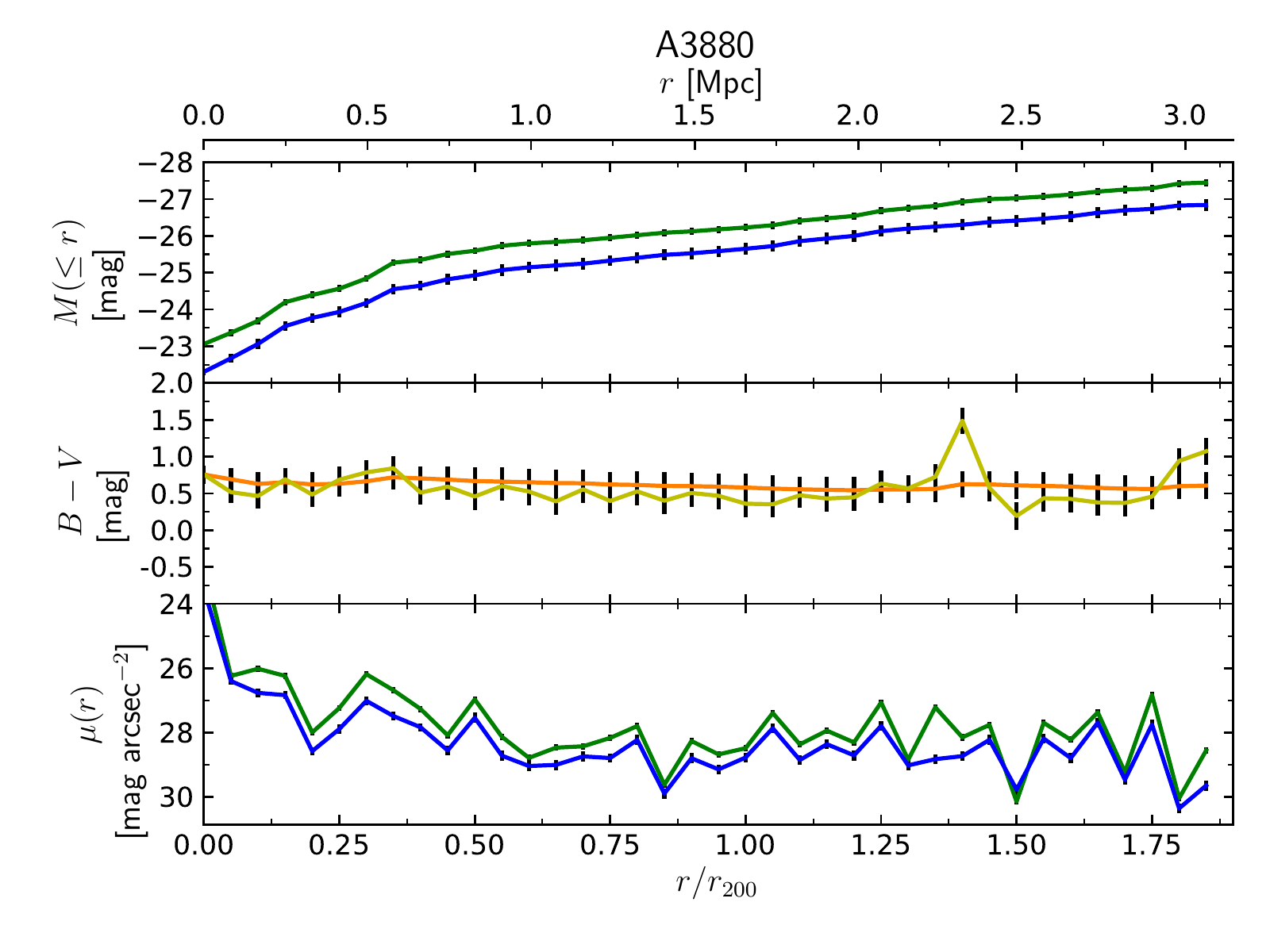}
    \caption{Photometric profiles of Omega-WINGS galaxy clusters, continued.}
\end{figure*}

\newpage
\clearpage

\newpage
\clearpage

\newpage
\clearpage

\begin{figure*}[t]
   \centering
        \includegraphics[width=0.45\textwidth]{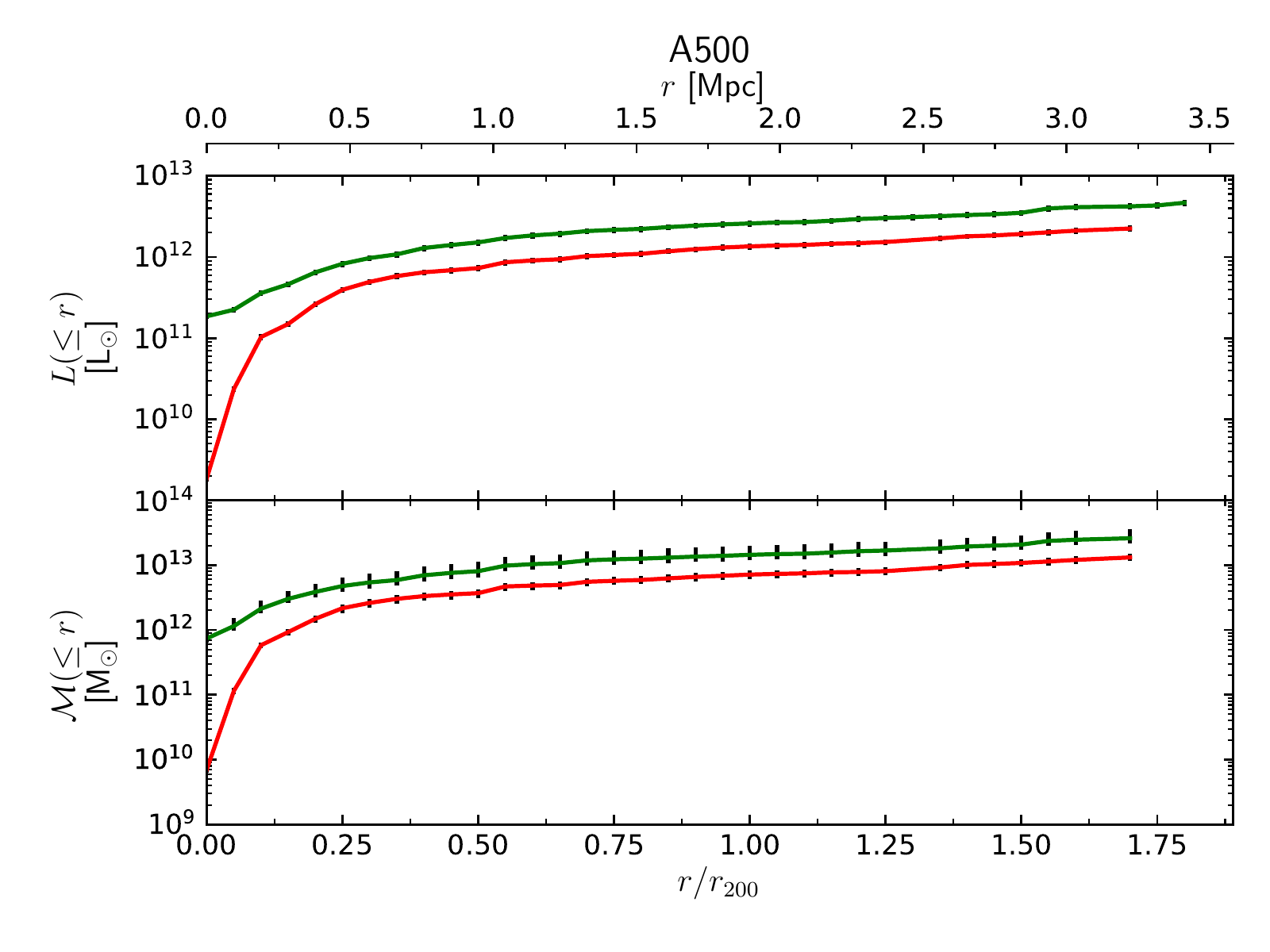}\includegraphics[width=0.45\textwidth]{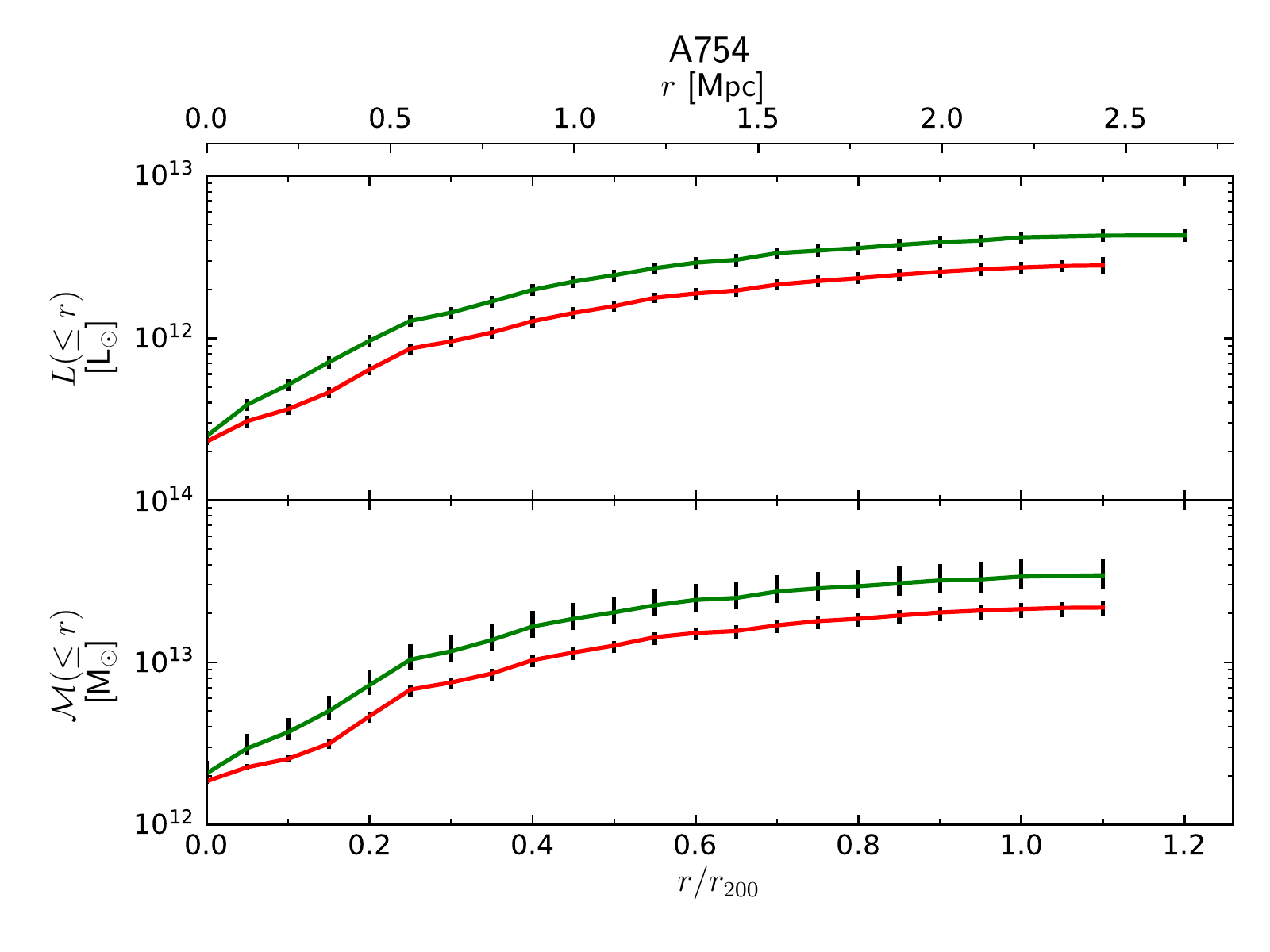}
        \includegraphics[width=0.45\textwidth]{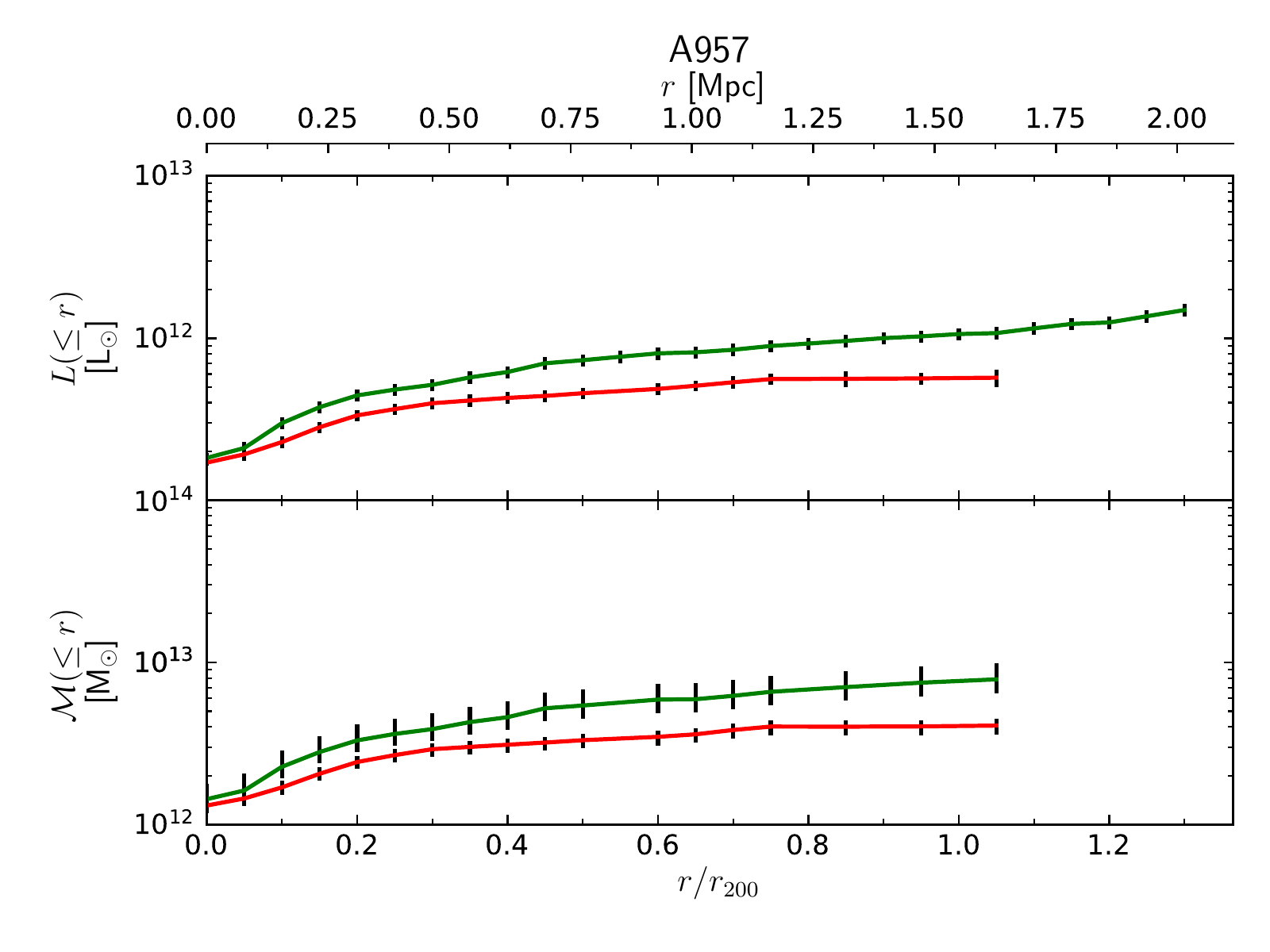}  \includegraphics[width=0.45\textwidth]{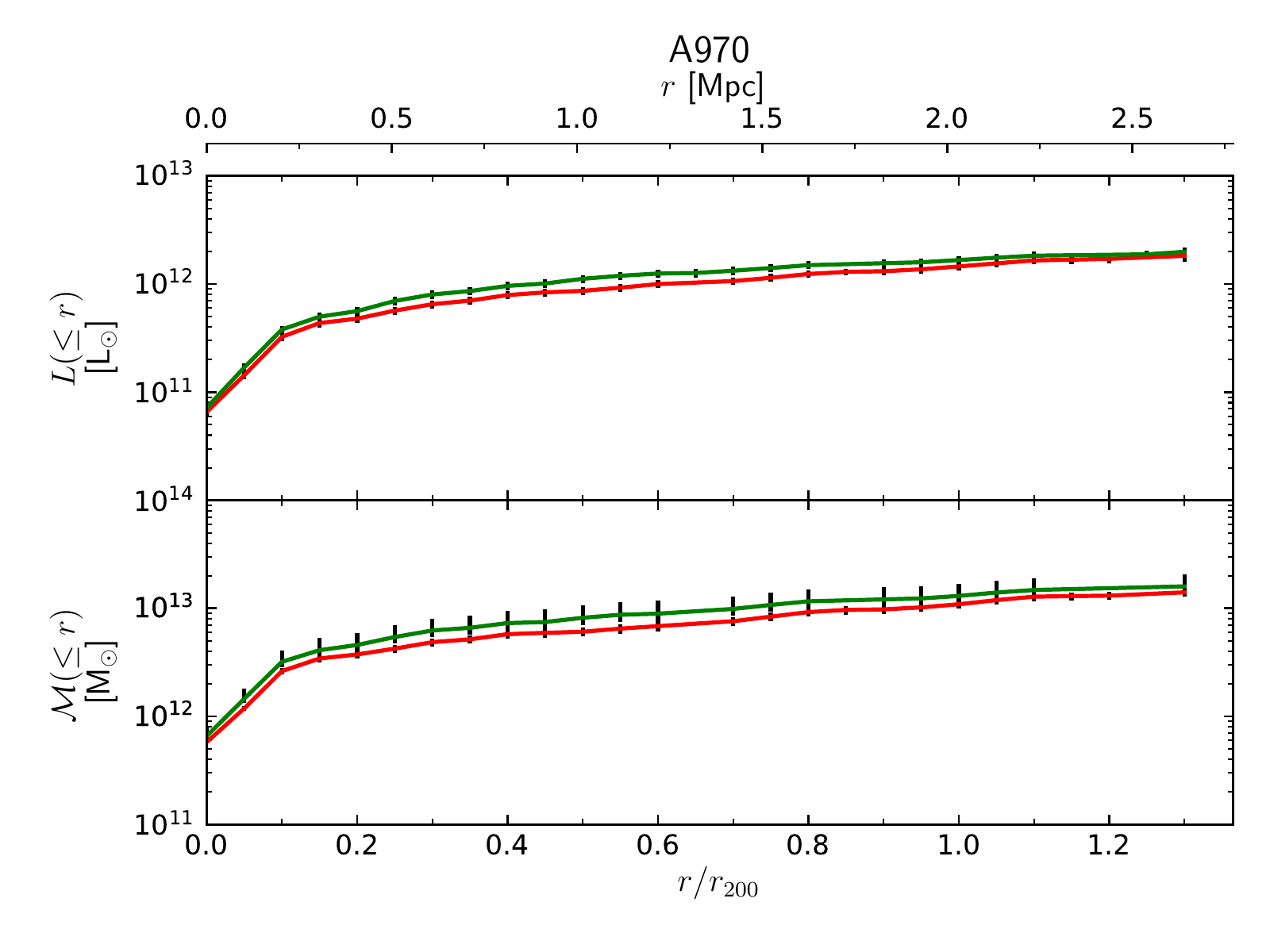}
        \includegraphics[width=0.45\textwidth]{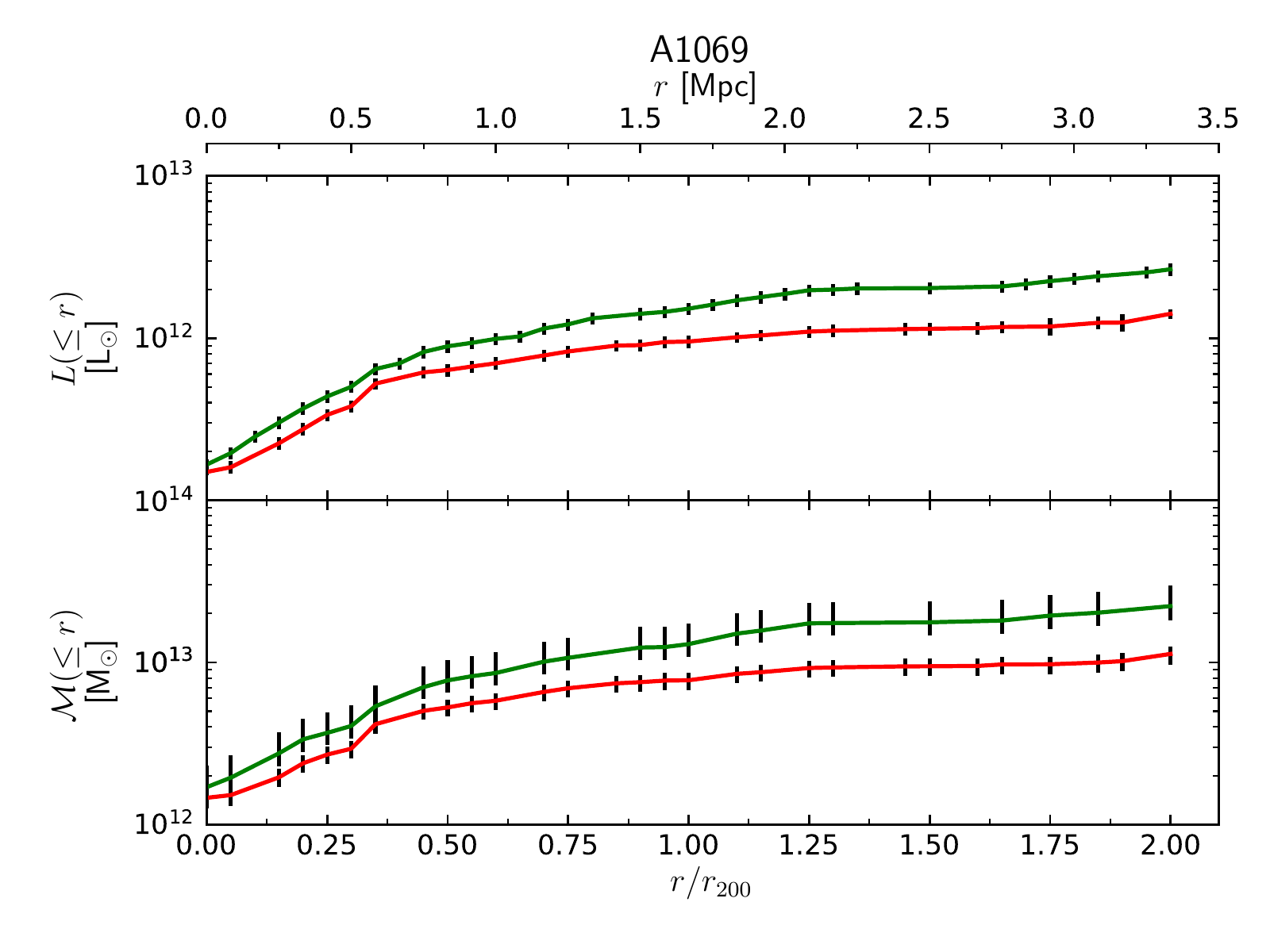}\includegraphics[width=0.45\textwidth]{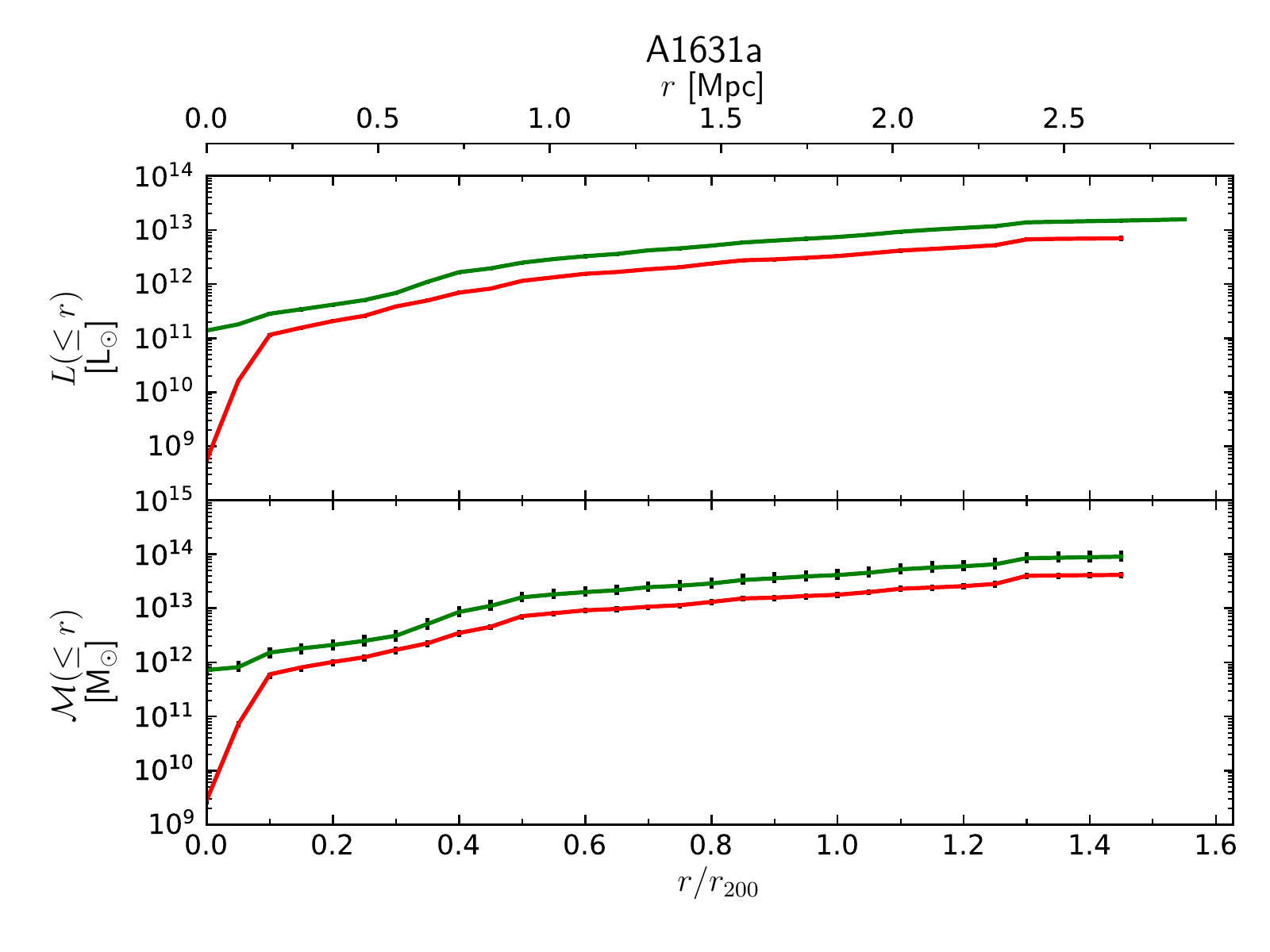}
    \caption{Mass profiles of Omega-WINGS galaxy clusters. Continued.}
\end{figure*}

\newpage
\clearpage

\begin{figure*}[t]
   \centering
        \includegraphics[width=0.45\textwidth]{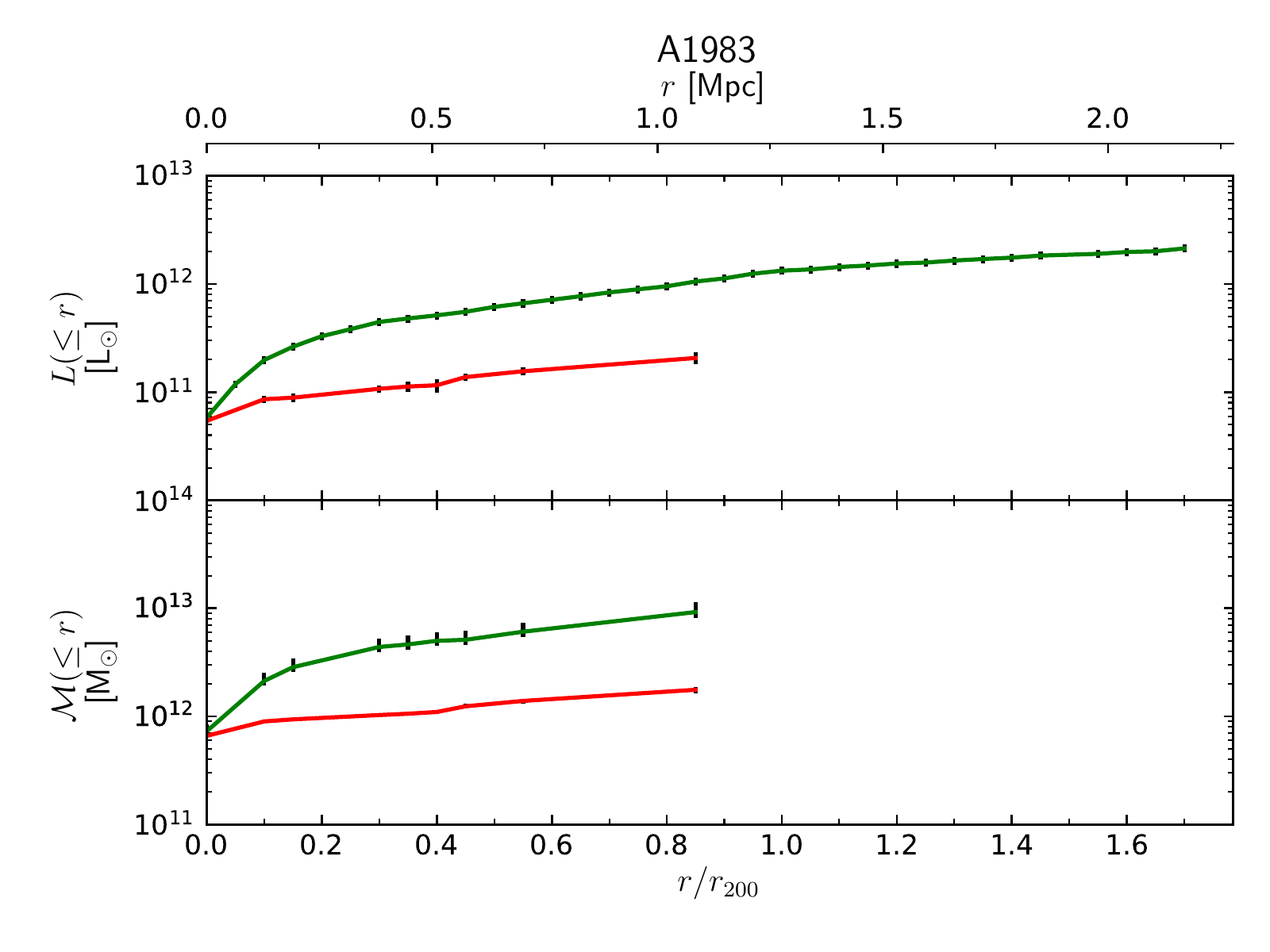}\includegraphics[width=0.45\textwidth]{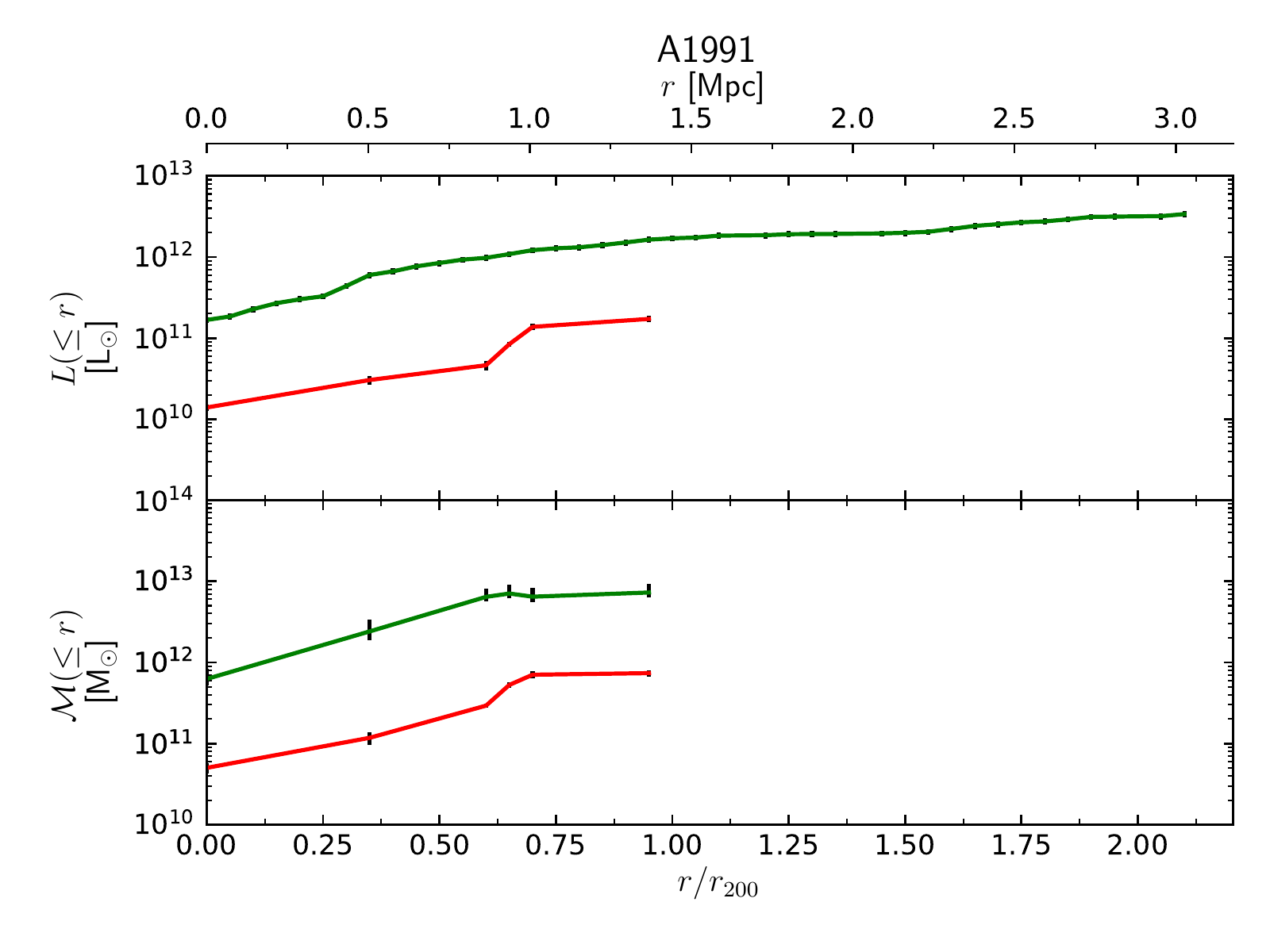}
        \includegraphics[width=0.45\textwidth]{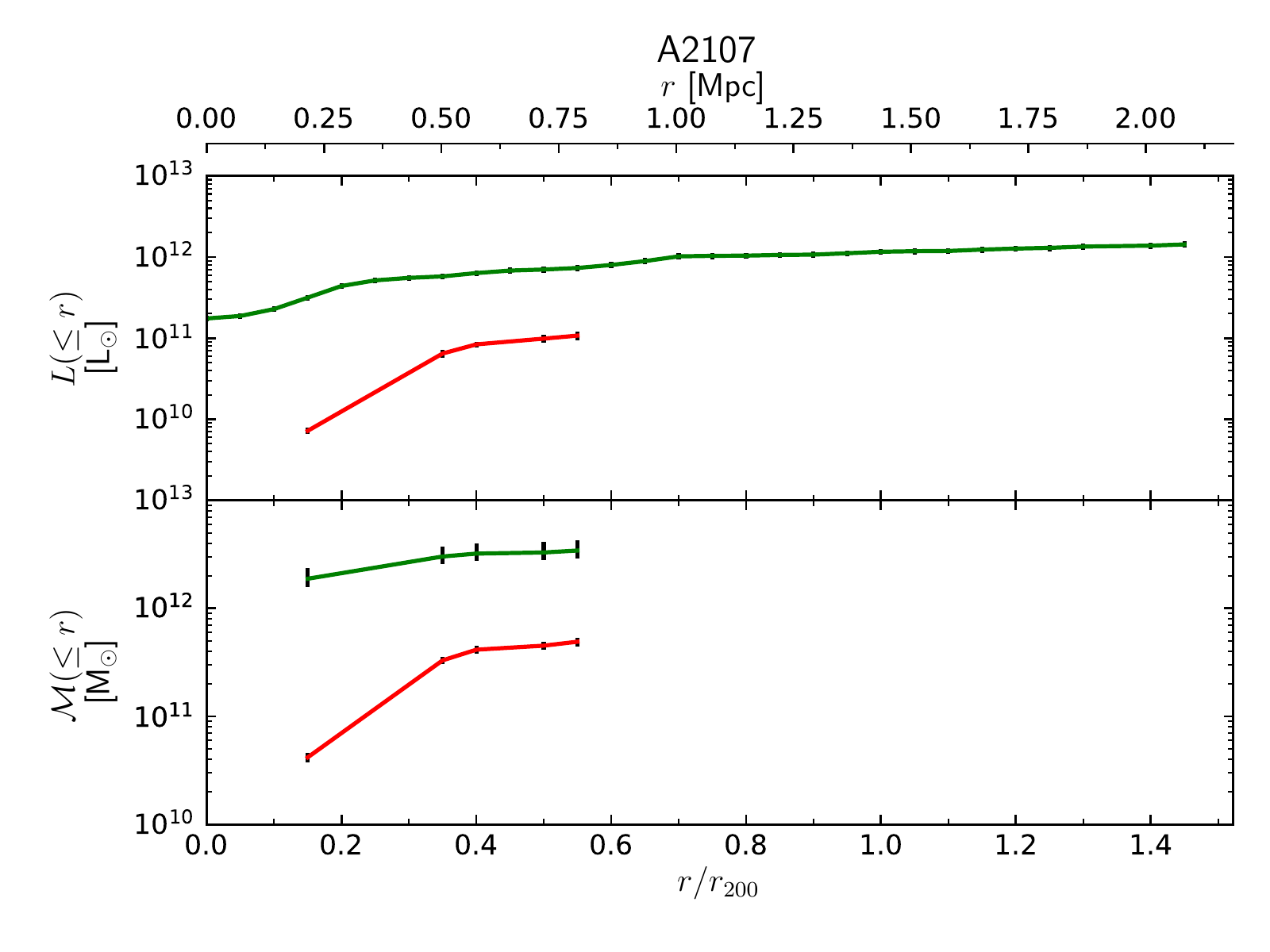}\includegraphics[width=0.45\textwidth]{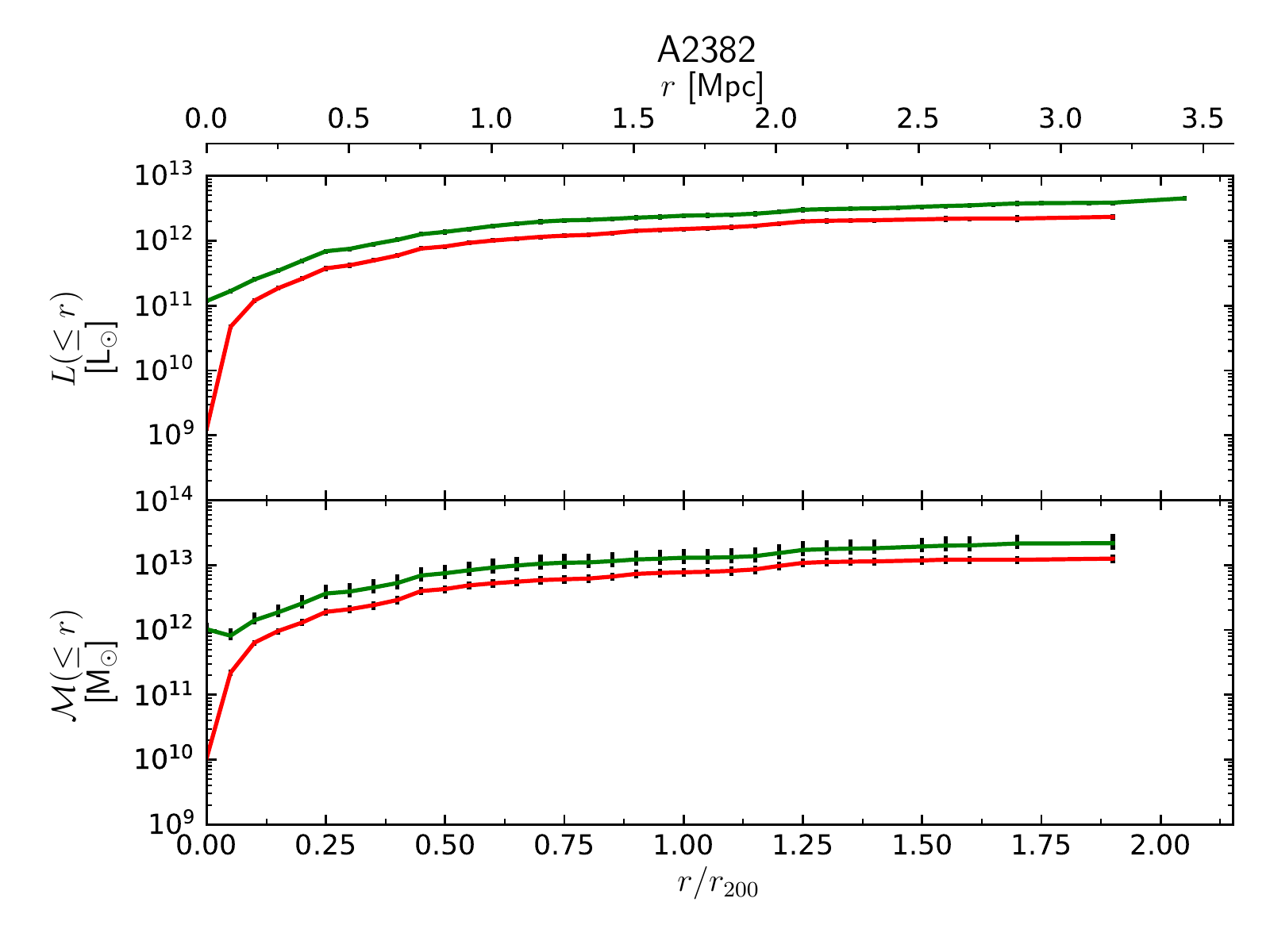}
        \includegraphics[width=0.45\textwidth]{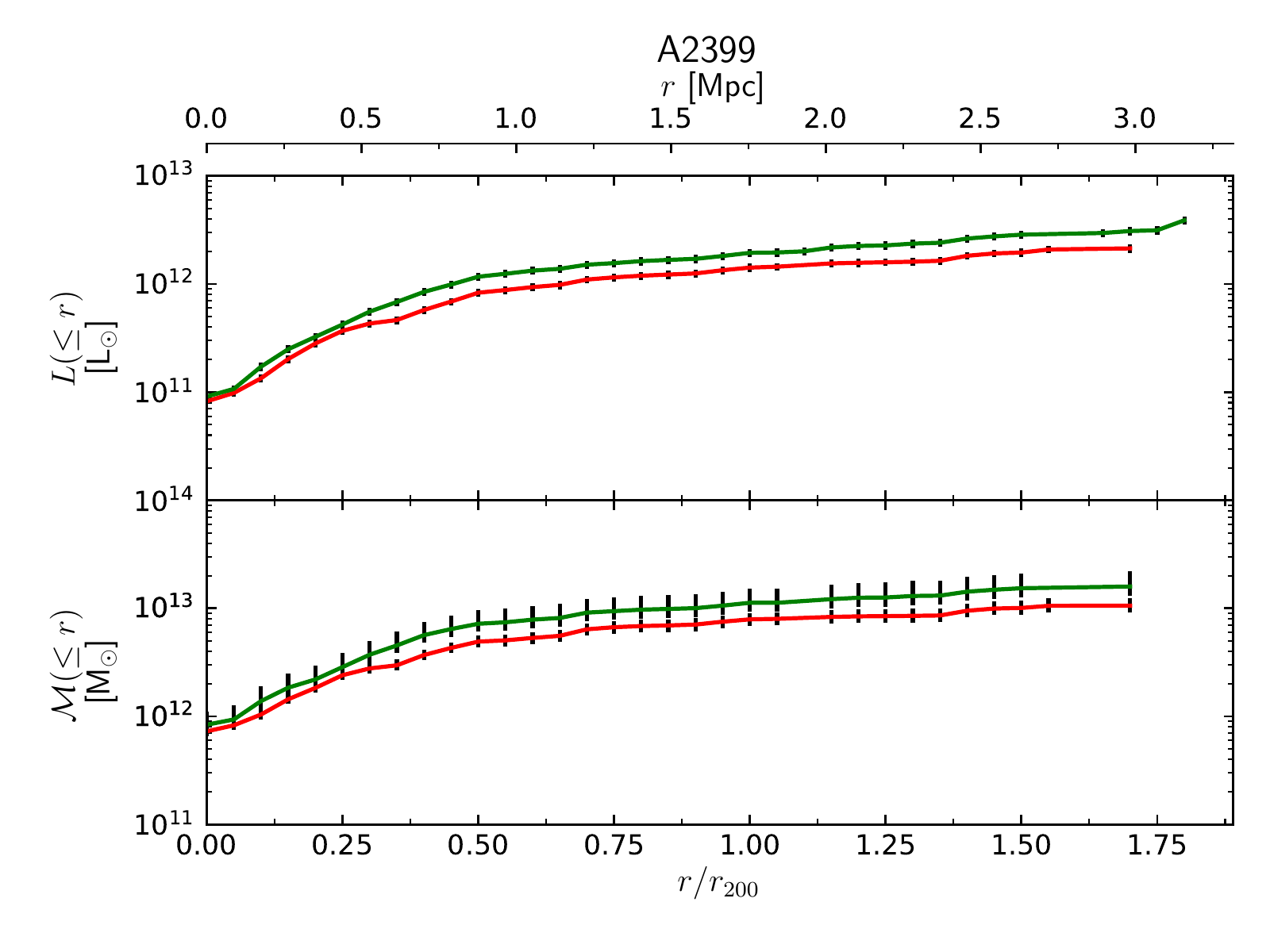}\includegraphics[width=0.45\textwidth]{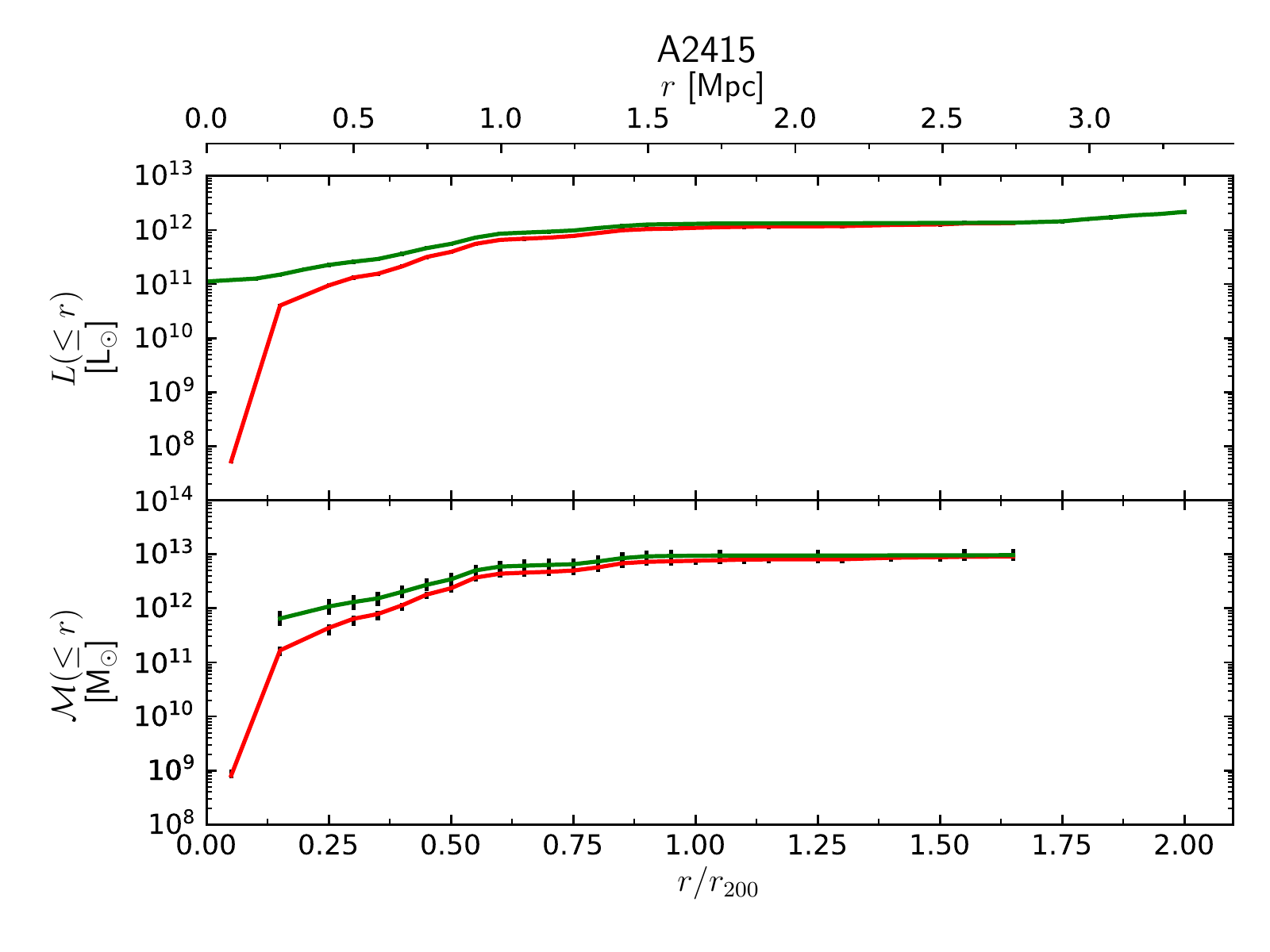}
    \caption{Mass profiles of Omega-WINGS galaxy clusters. Continued.}
\end{figure*}

\newpage
\clearpage

\begin{figure*}[t]
   \centering
        \includegraphics[width=0.45\textwidth]{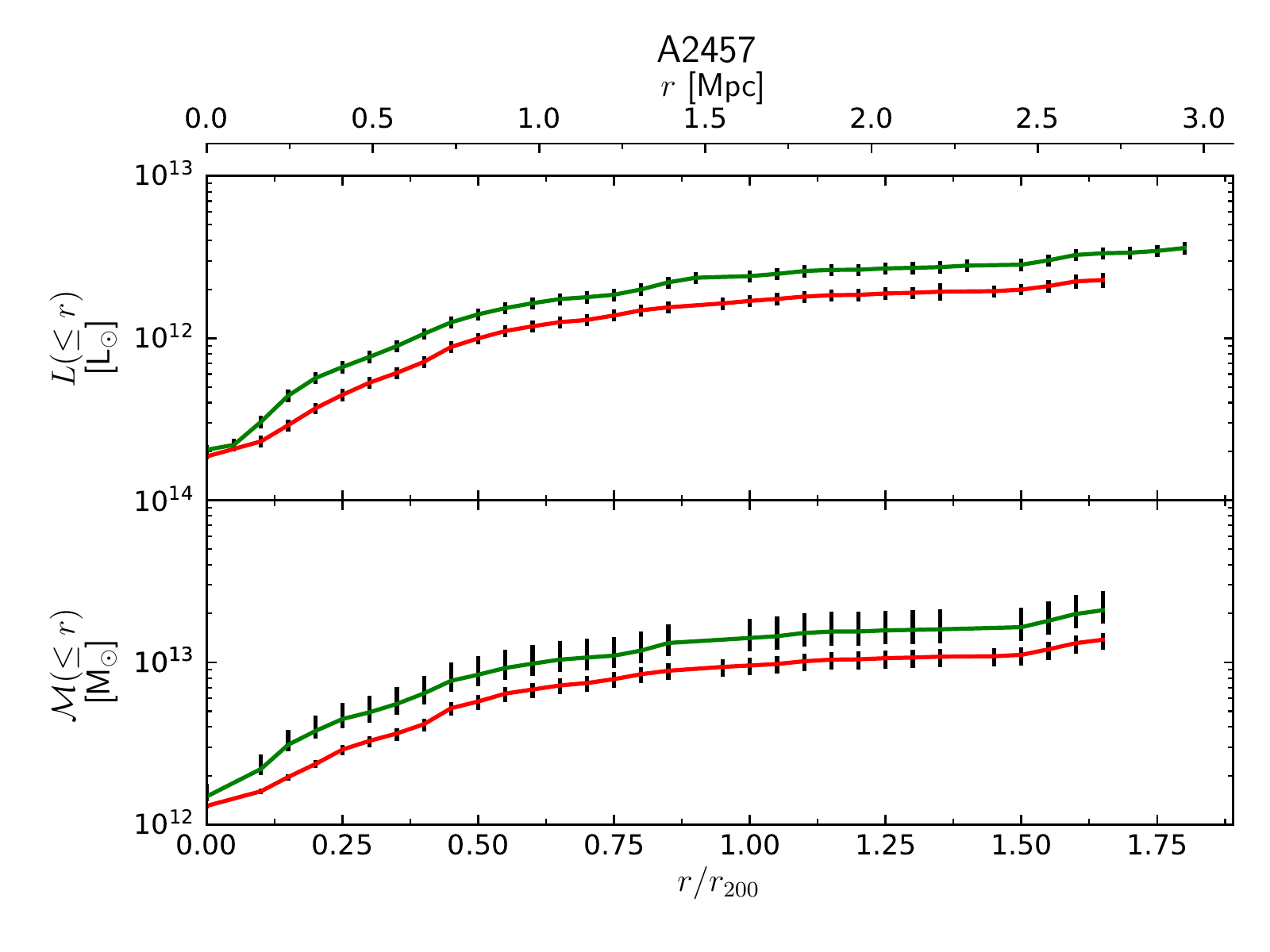}\includegraphics[width=0.45\textwidth]{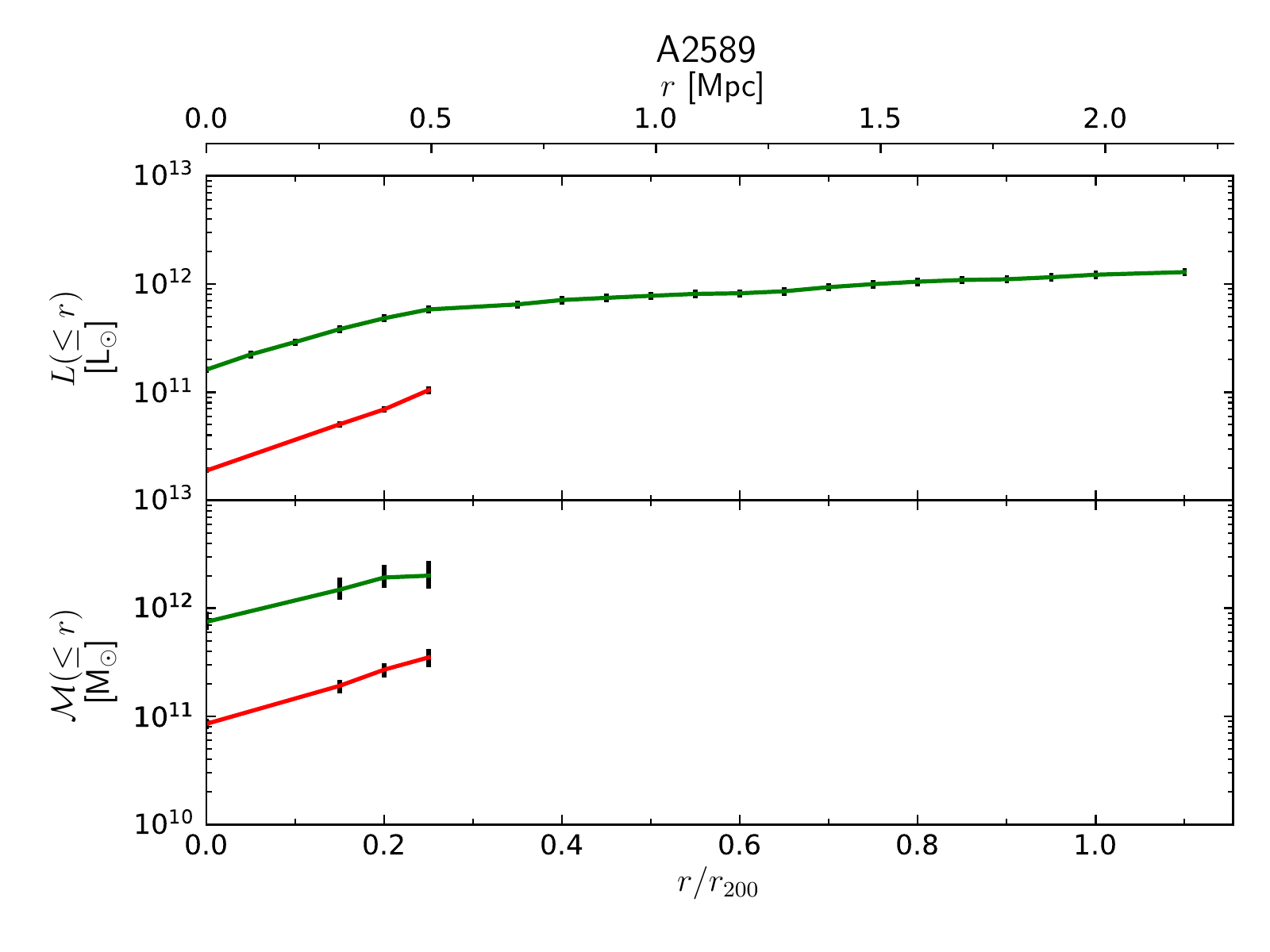}  
        \includegraphics[width=0.45\textwidth]{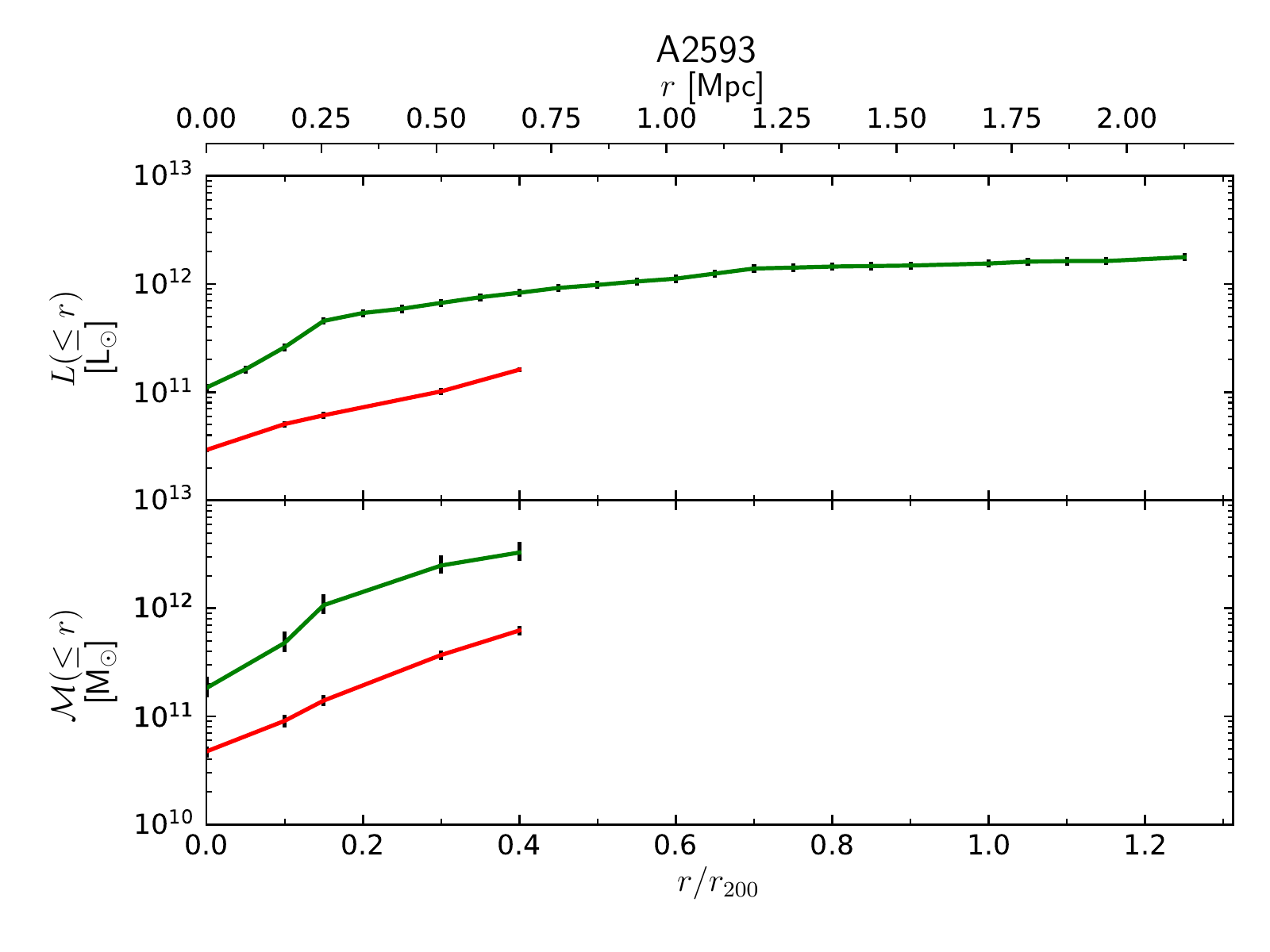}\includegraphics[width=0.45\textwidth]{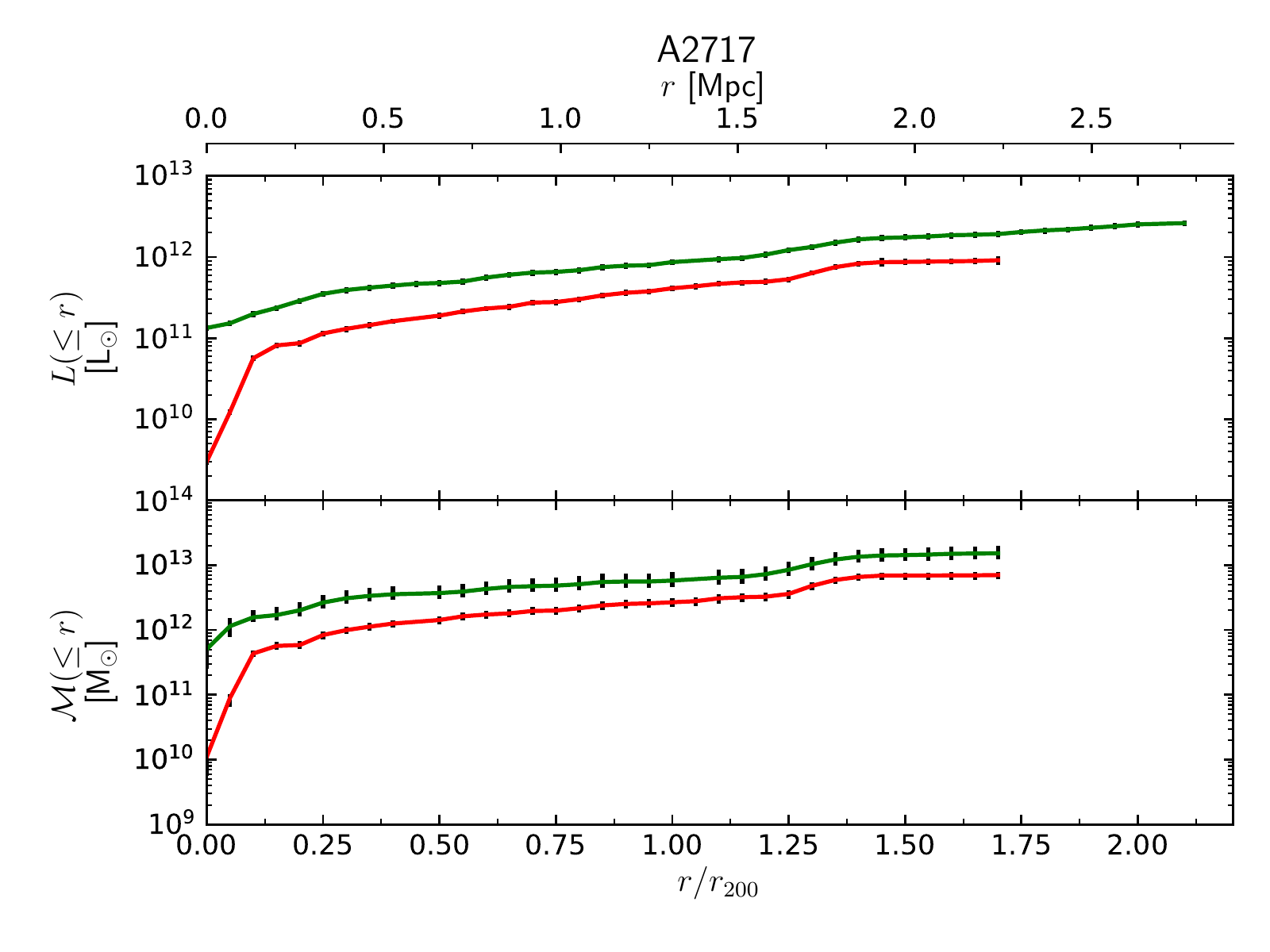}
        \includegraphics[width=0.45\textwidth]{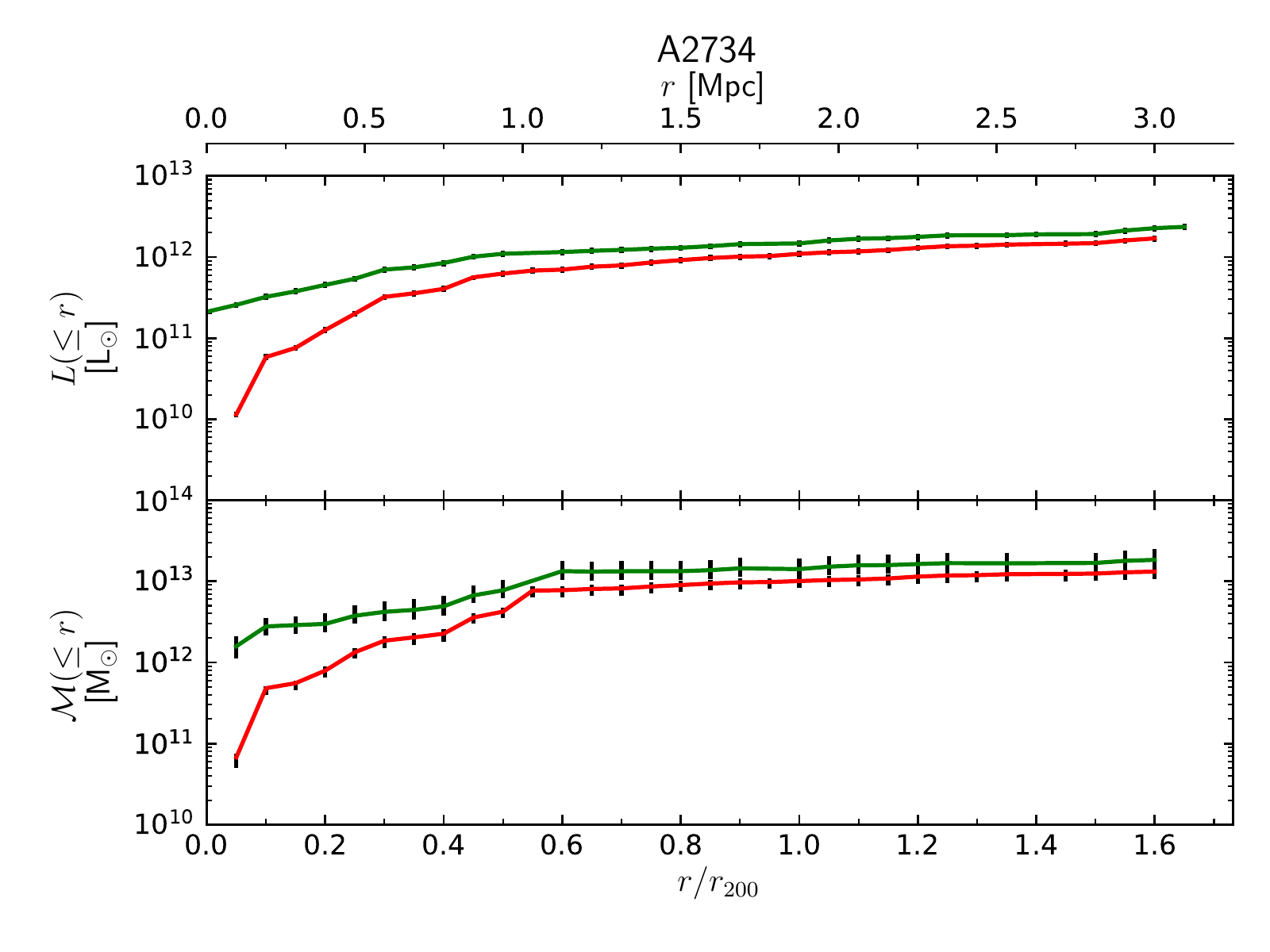}\includegraphics[width=0.45\textwidth]{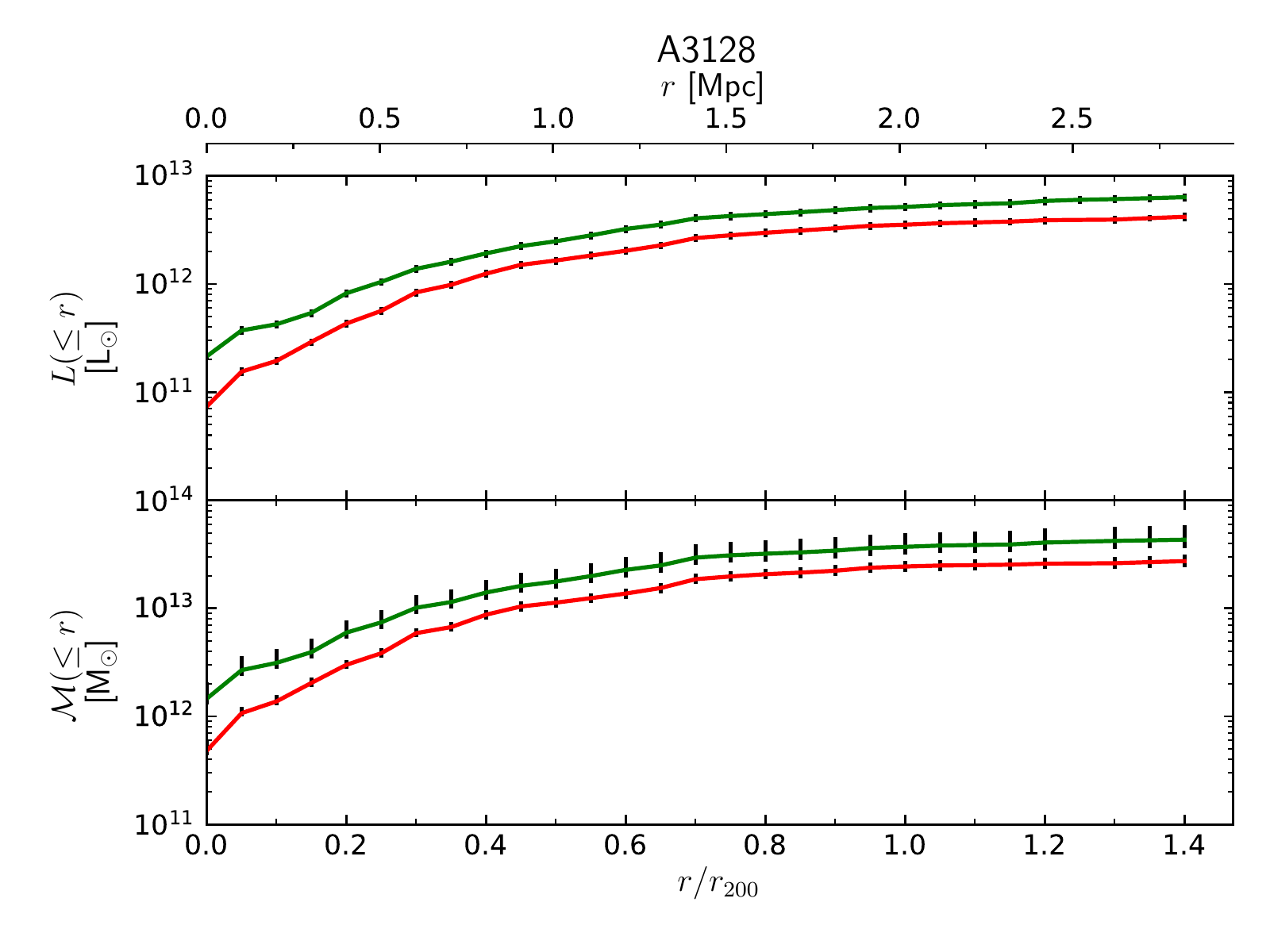}
    \caption{Mass profiles of Omega-WINGS galaxy clusters. Continued.}
\end{figure*}

\newpage
\clearpage

\begin{figure*}[t]
   \centering
        \includegraphics[width=0.45\textwidth]{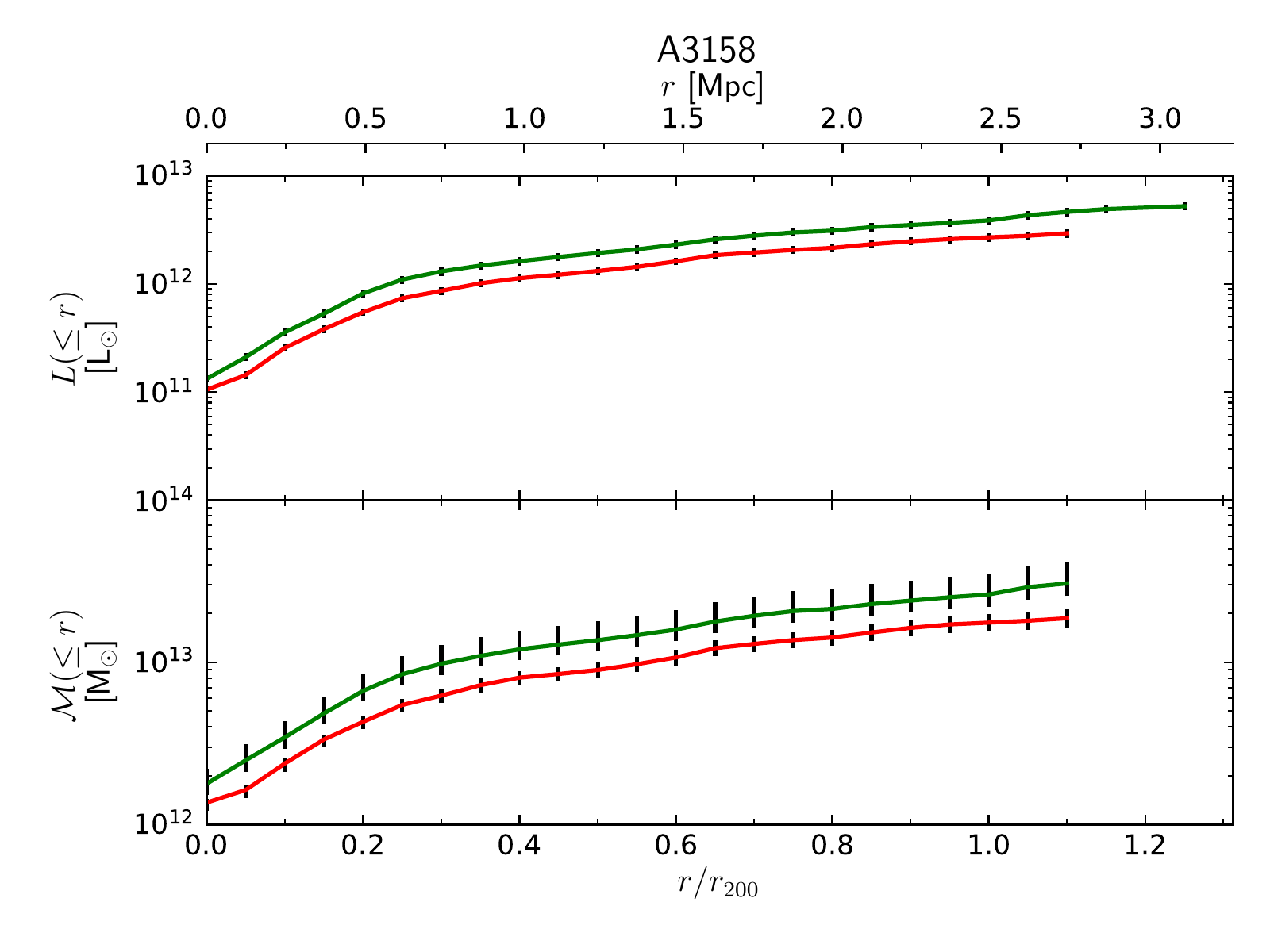}\includegraphics[width=0.45\textwidth]{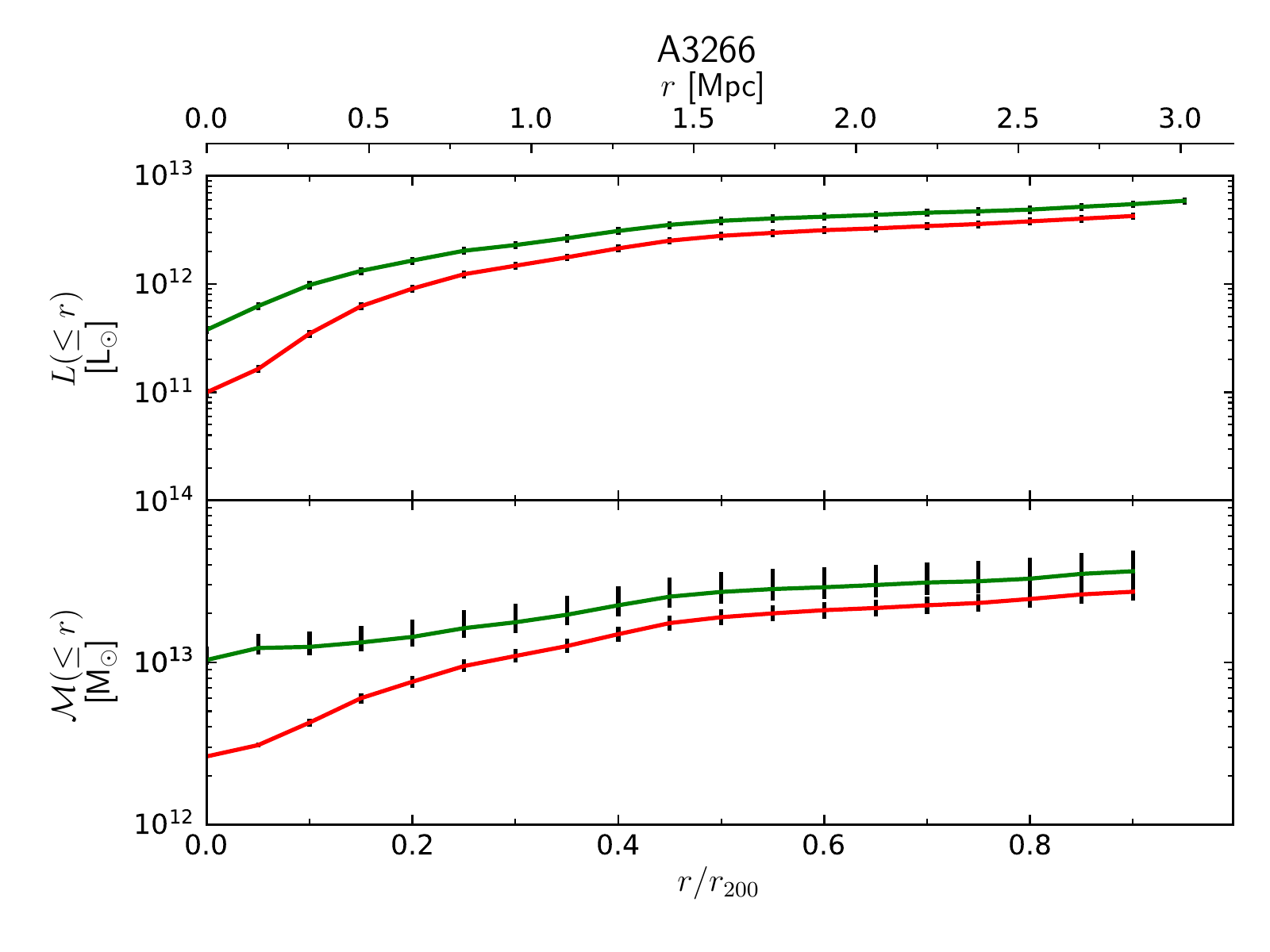}
        \includegraphics[width=0.45\textwidth]{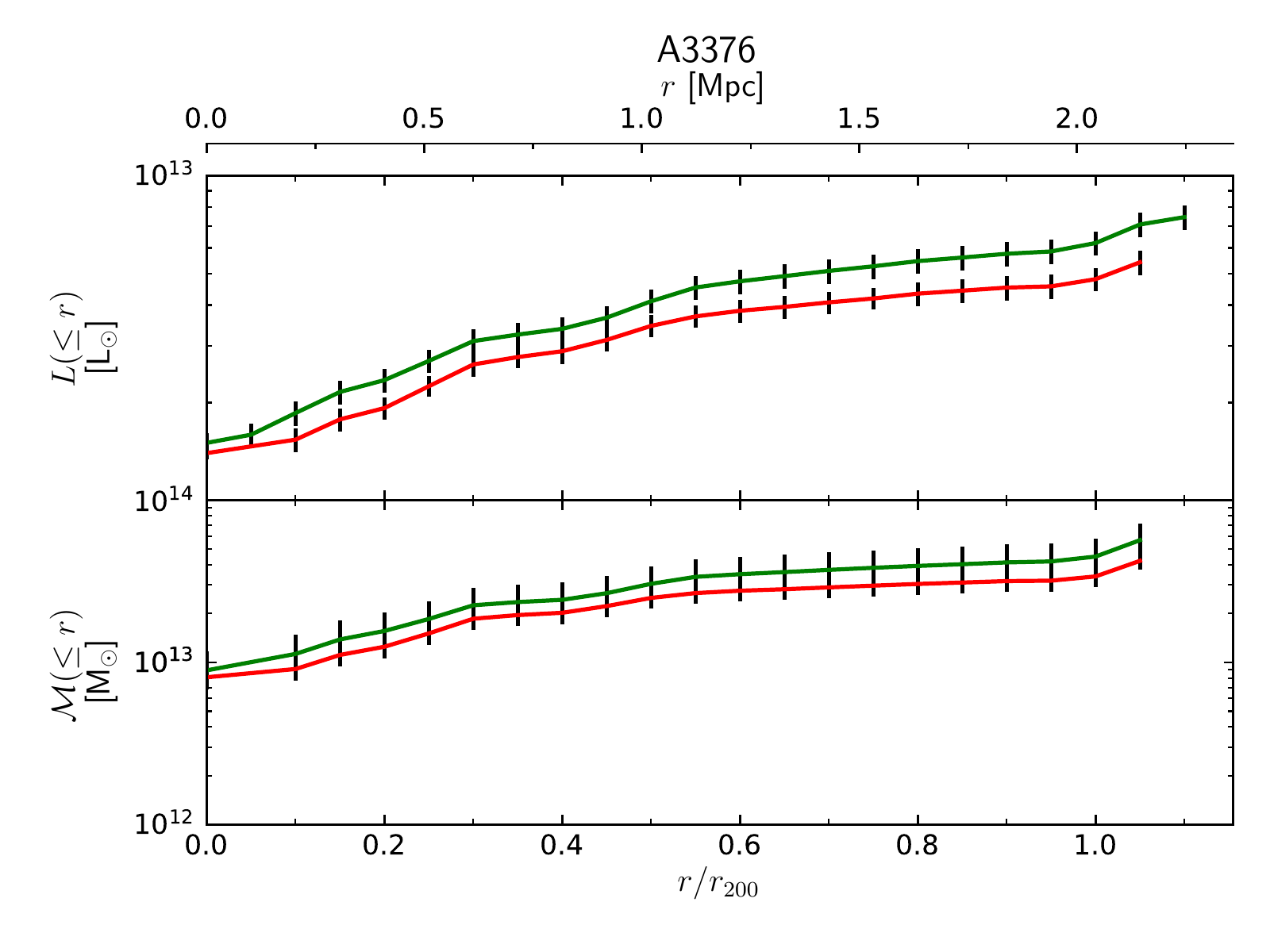}\includegraphics[width=0.45\textwidth]{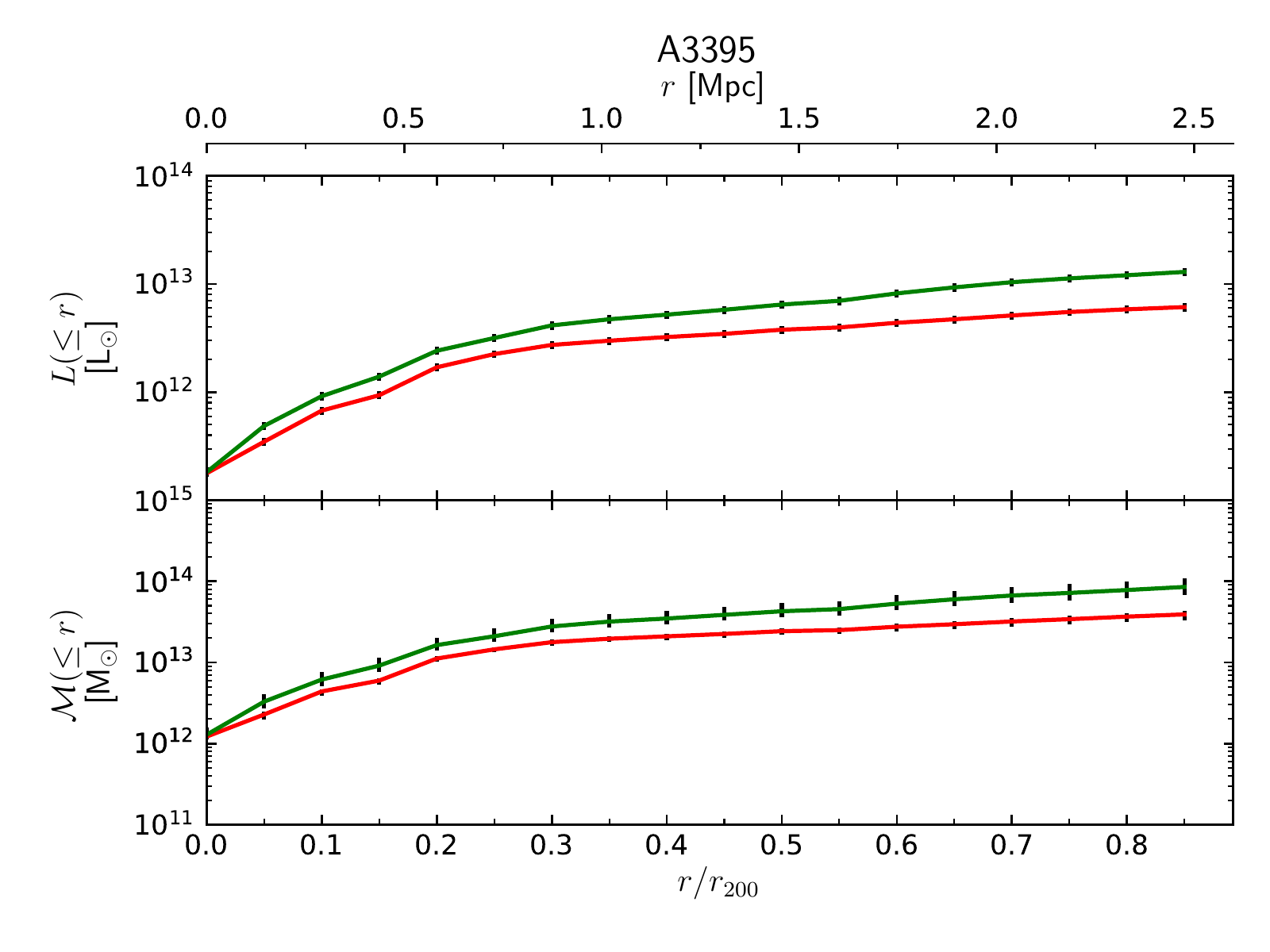}
        \includegraphics[width=0.45\textwidth]{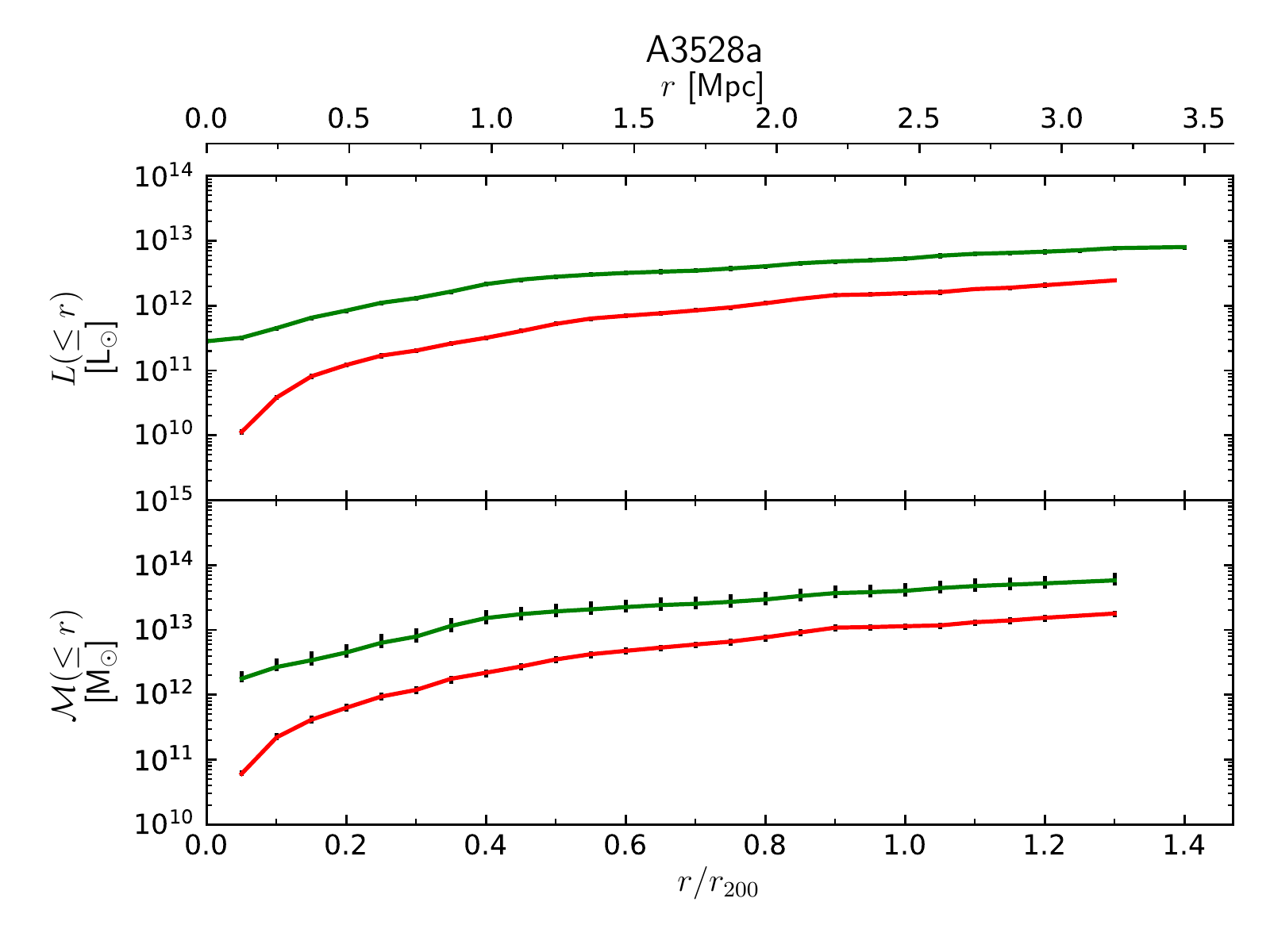}\includegraphics[width=0.45\textwidth]{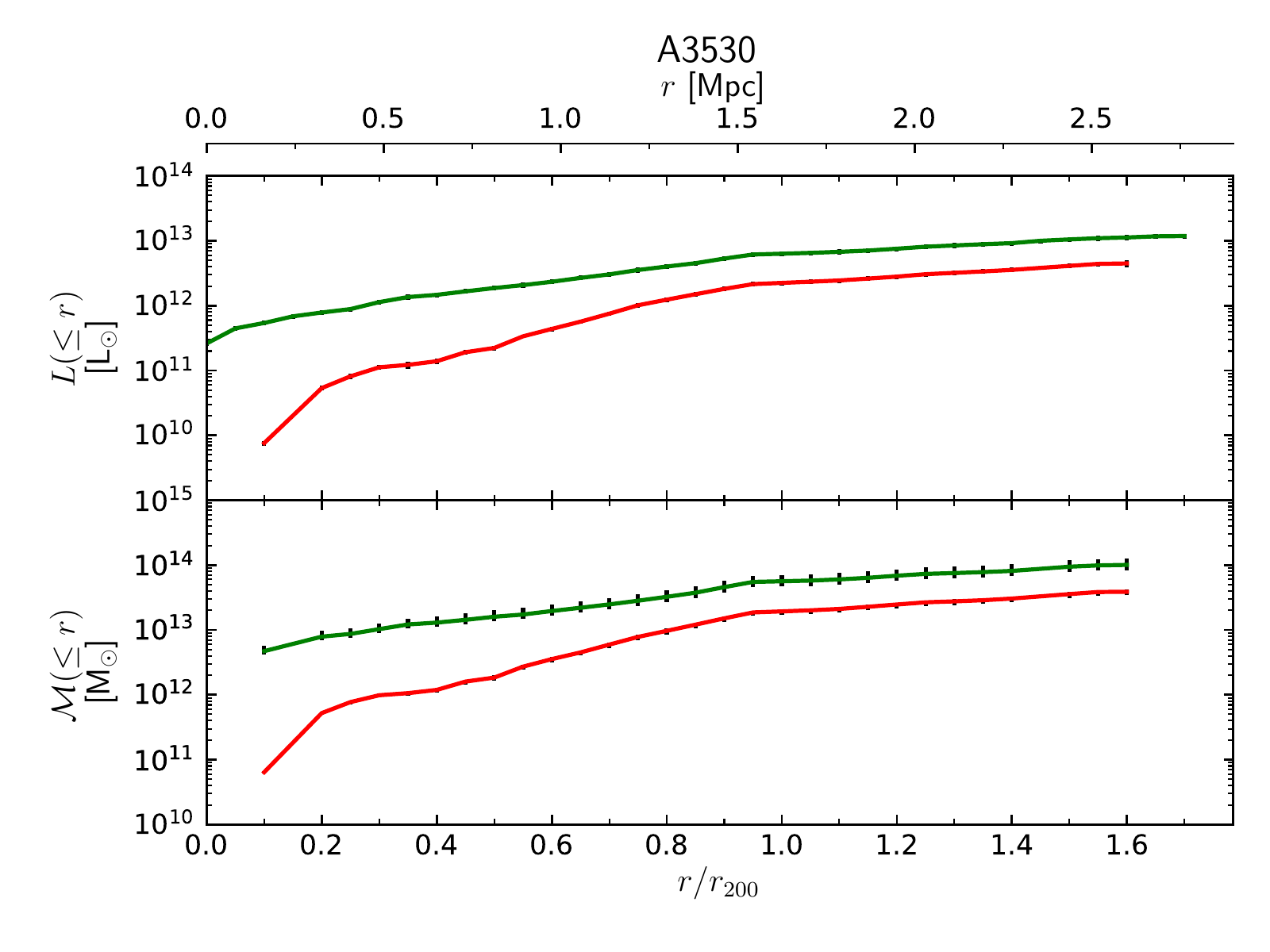}
    \caption{Mass profiles of Omega-WINGS galaxy clusters, continued.}
\end{figure*}

\newpage
\clearpage

\begin{figure*}[t]
   \centering
        \includegraphics[width=0.45\textwidth]{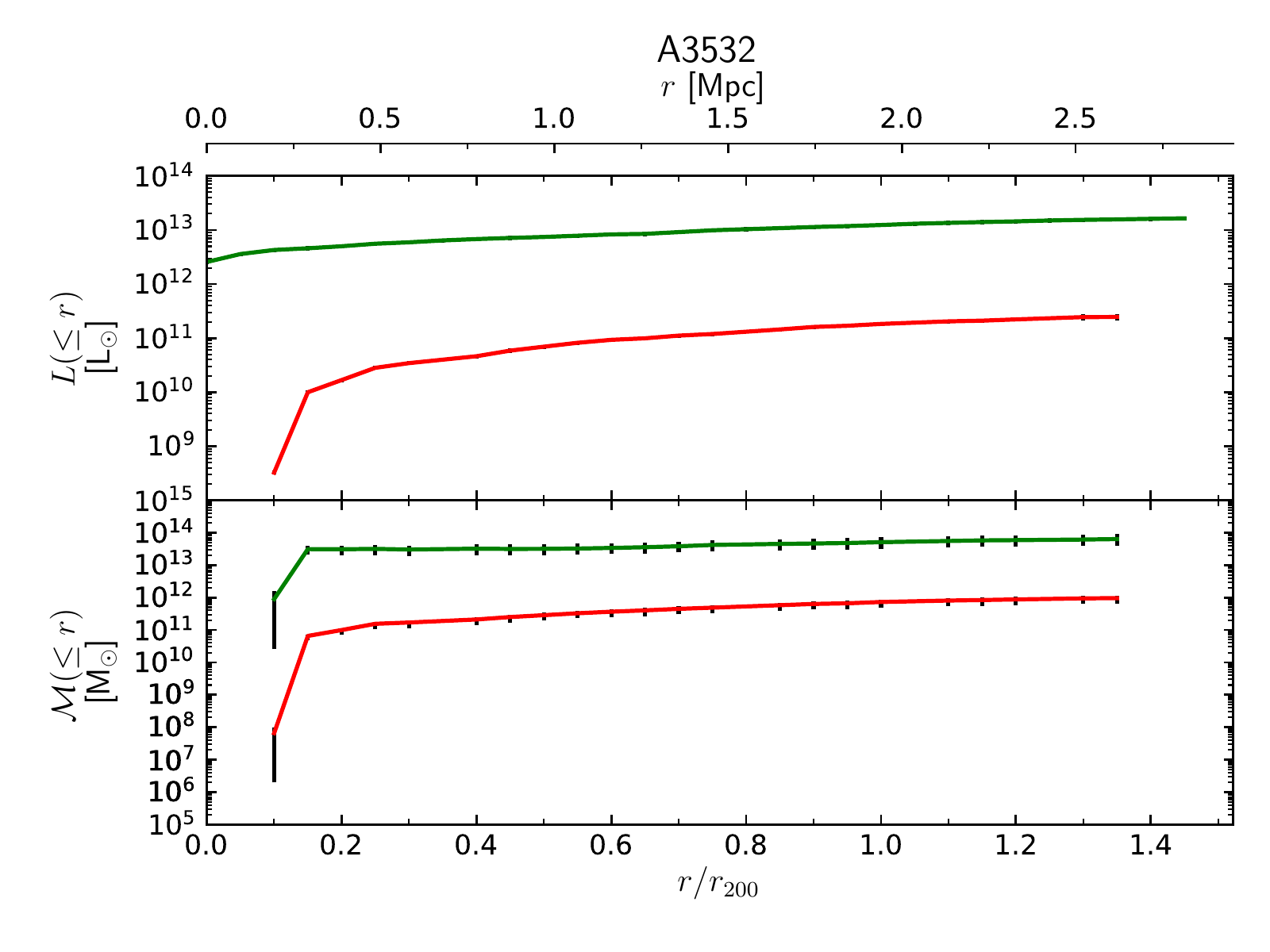} \includegraphics[width=0.45\textwidth]{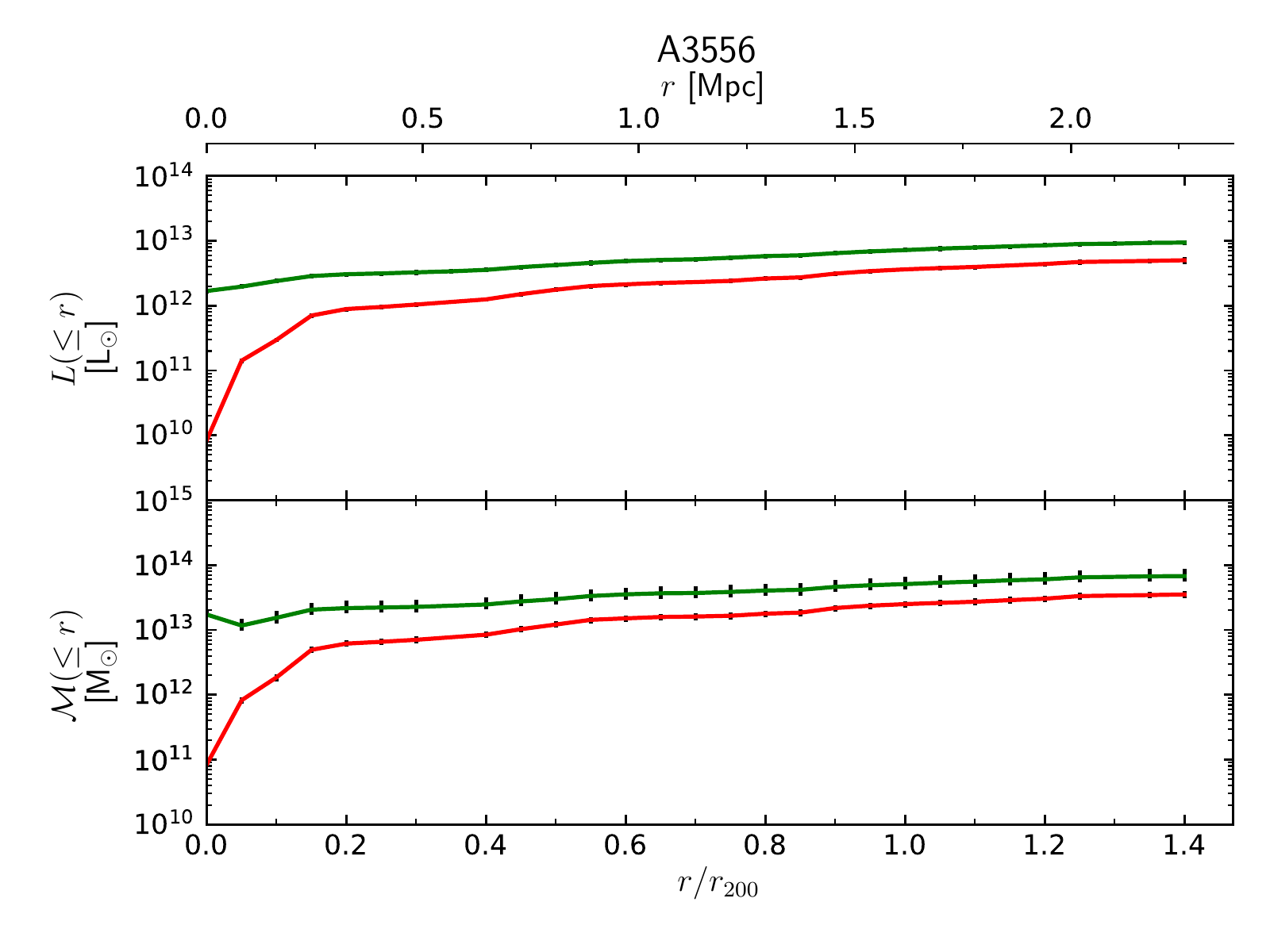}
        \includegraphics[width=0.45\textwidth]{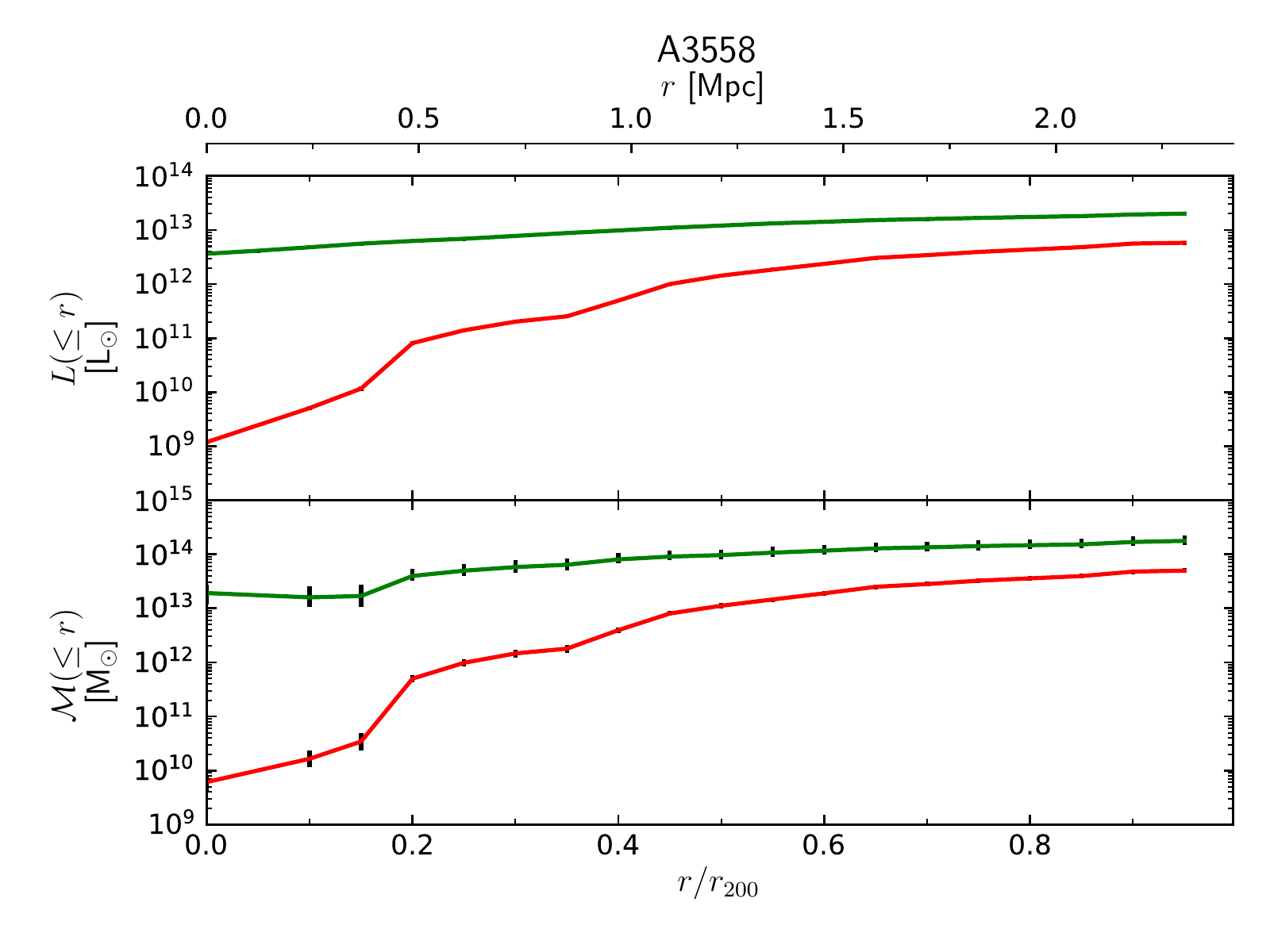} \includegraphics[width=0.45\textwidth]{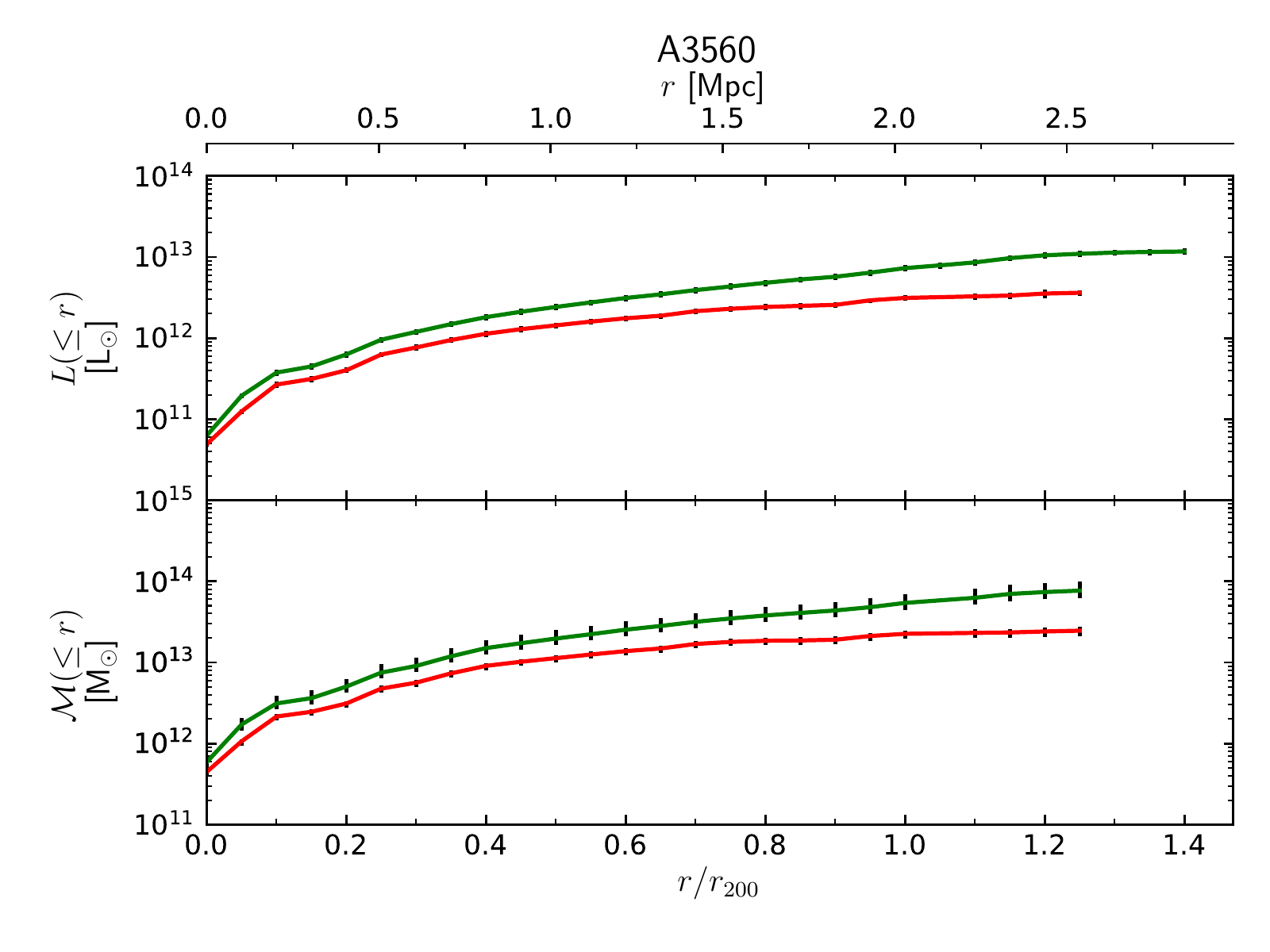}
        \includegraphics[width=0.45\textwidth]{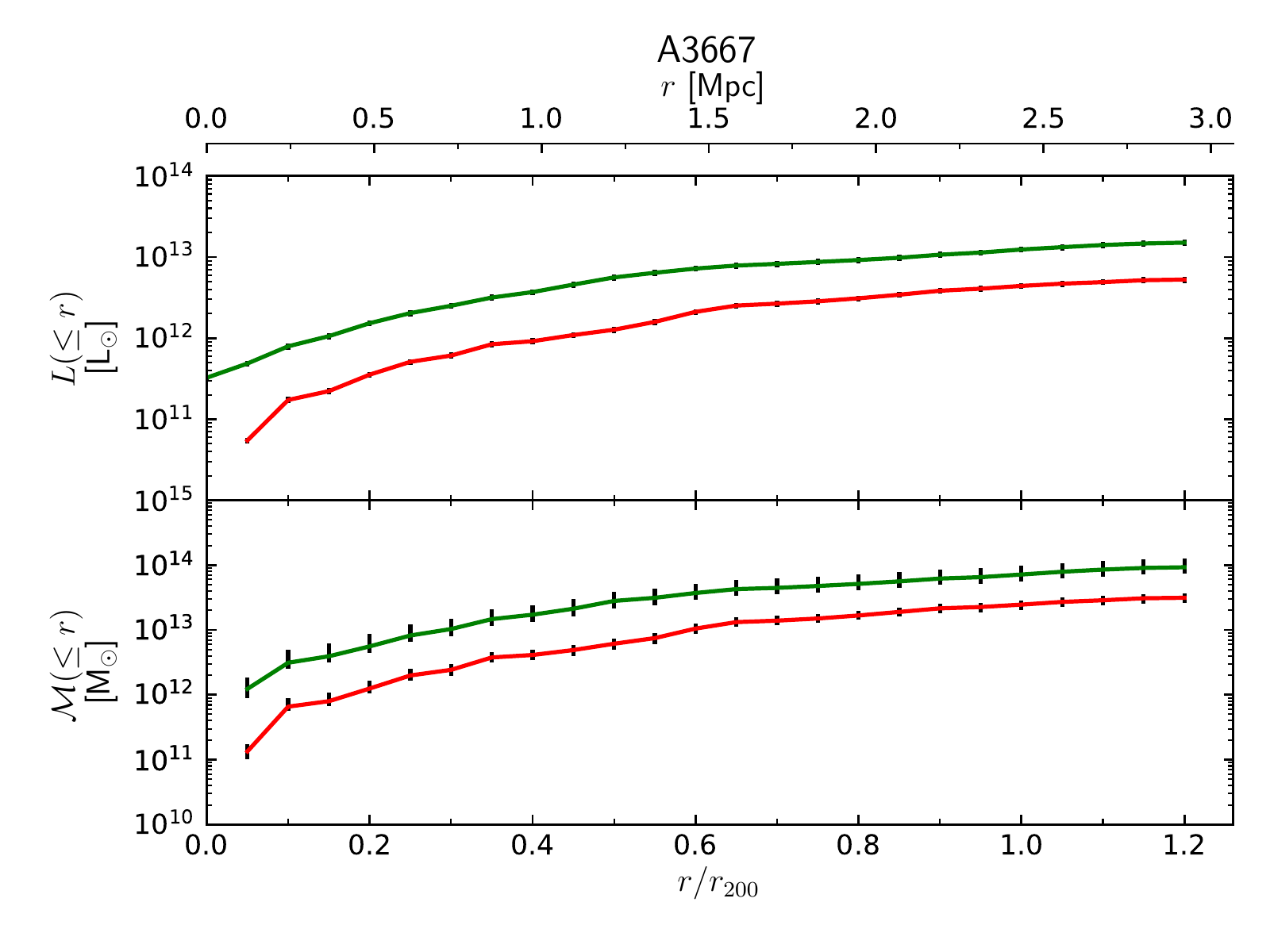} \includegraphics[width=0.45\textwidth]{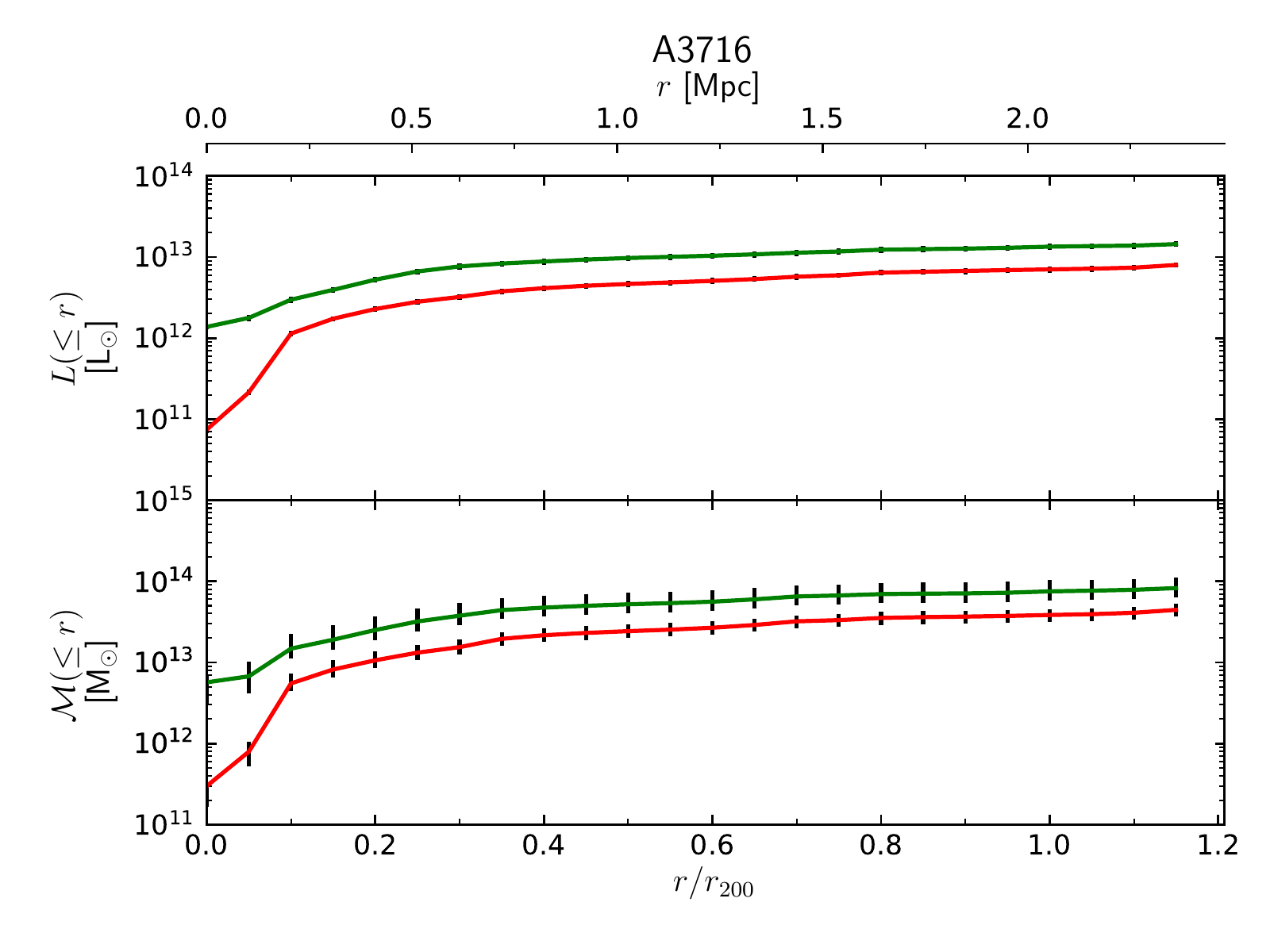}
    \caption{Mass profiles of Omega-WINGS galaxy clusters, continued.}
\end{figure*}

\newpage
\clearpage

\newpage
\clearpage

\newpage
\clearpage

\begin{figure*}[t]
   \centering
        \includegraphics[width=0.45\textwidth]{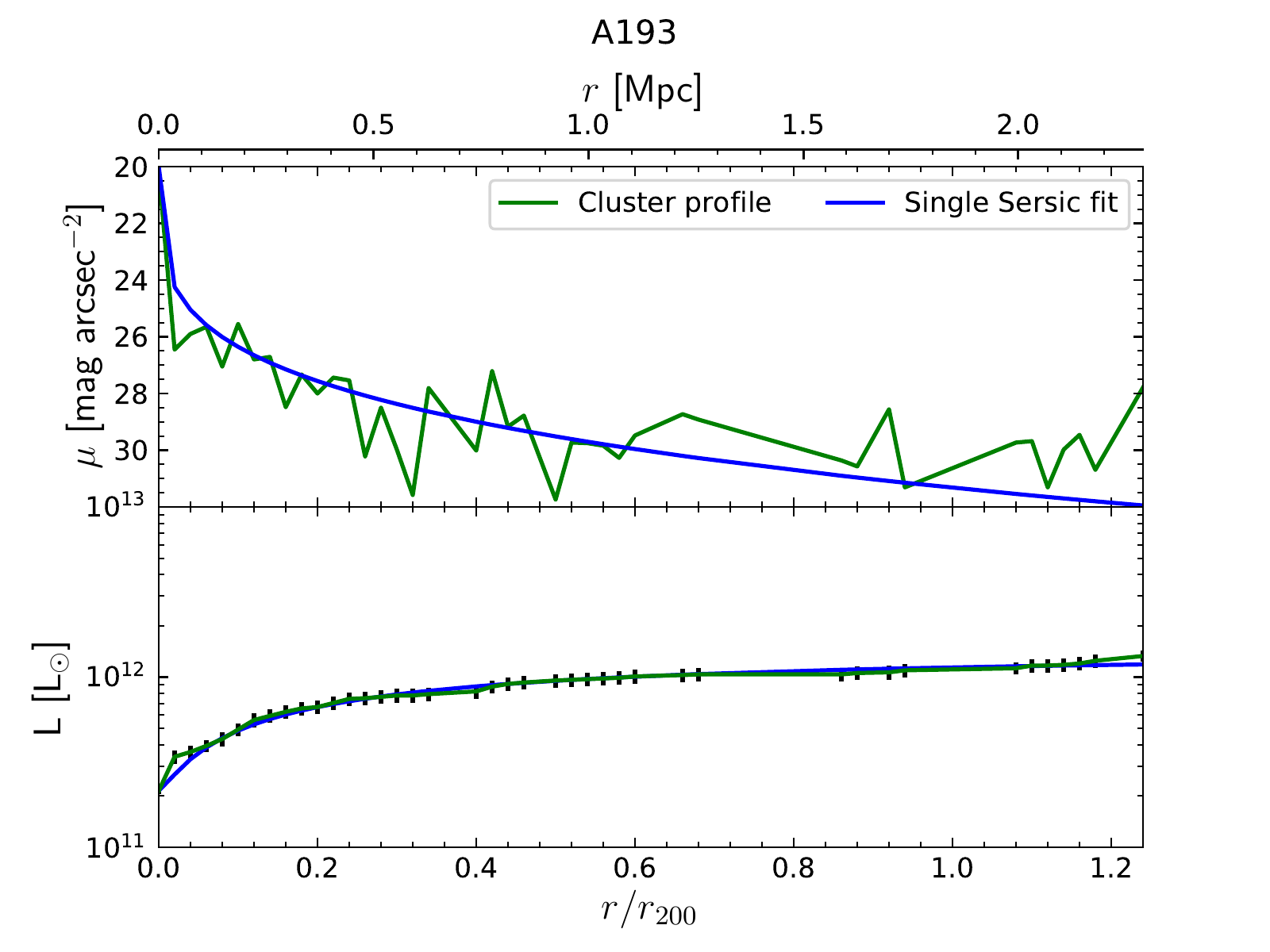} \includegraphics[width=0.45\textwidth]{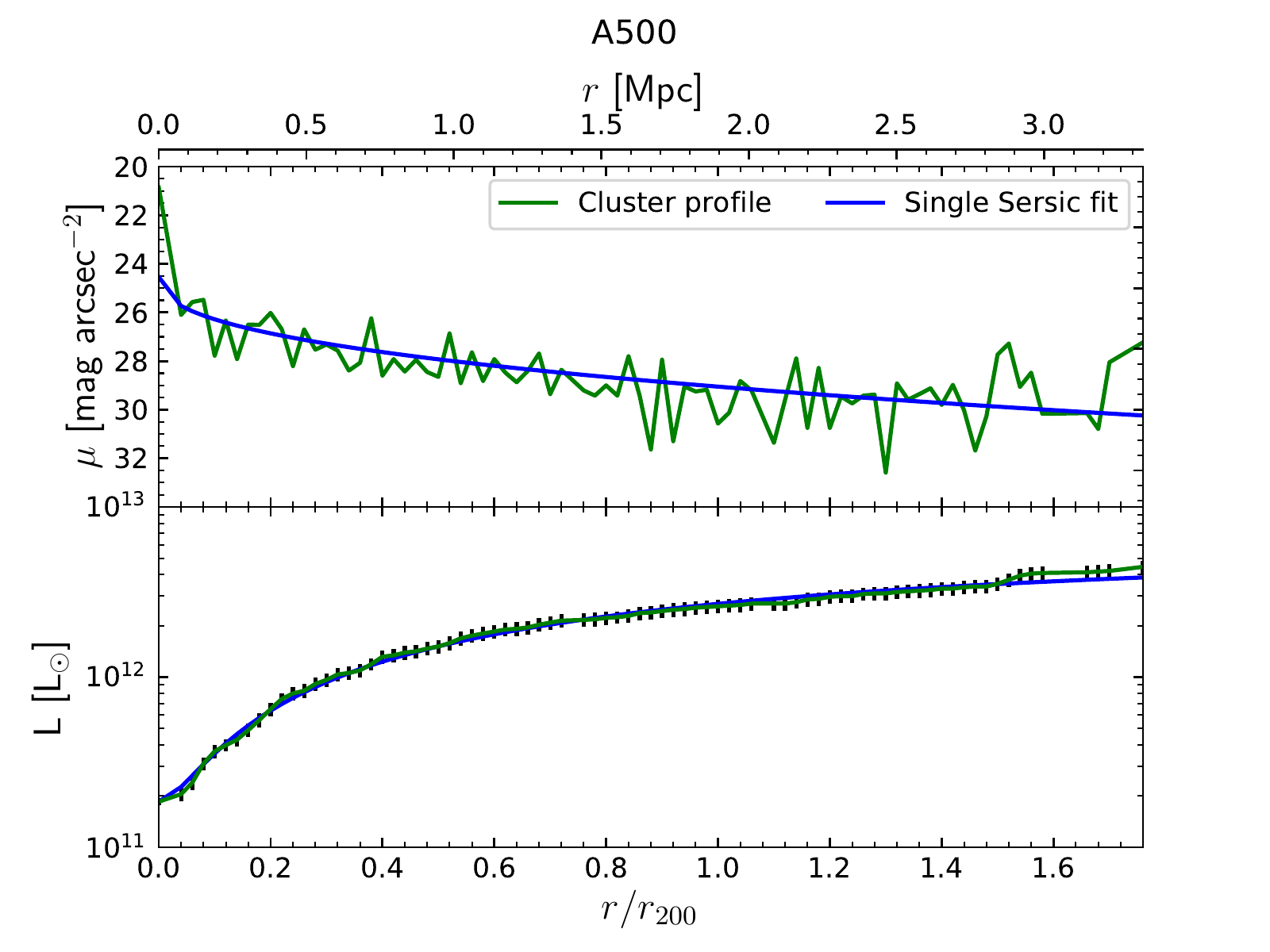}
        \includegraphics[width=0.45\textwidth]{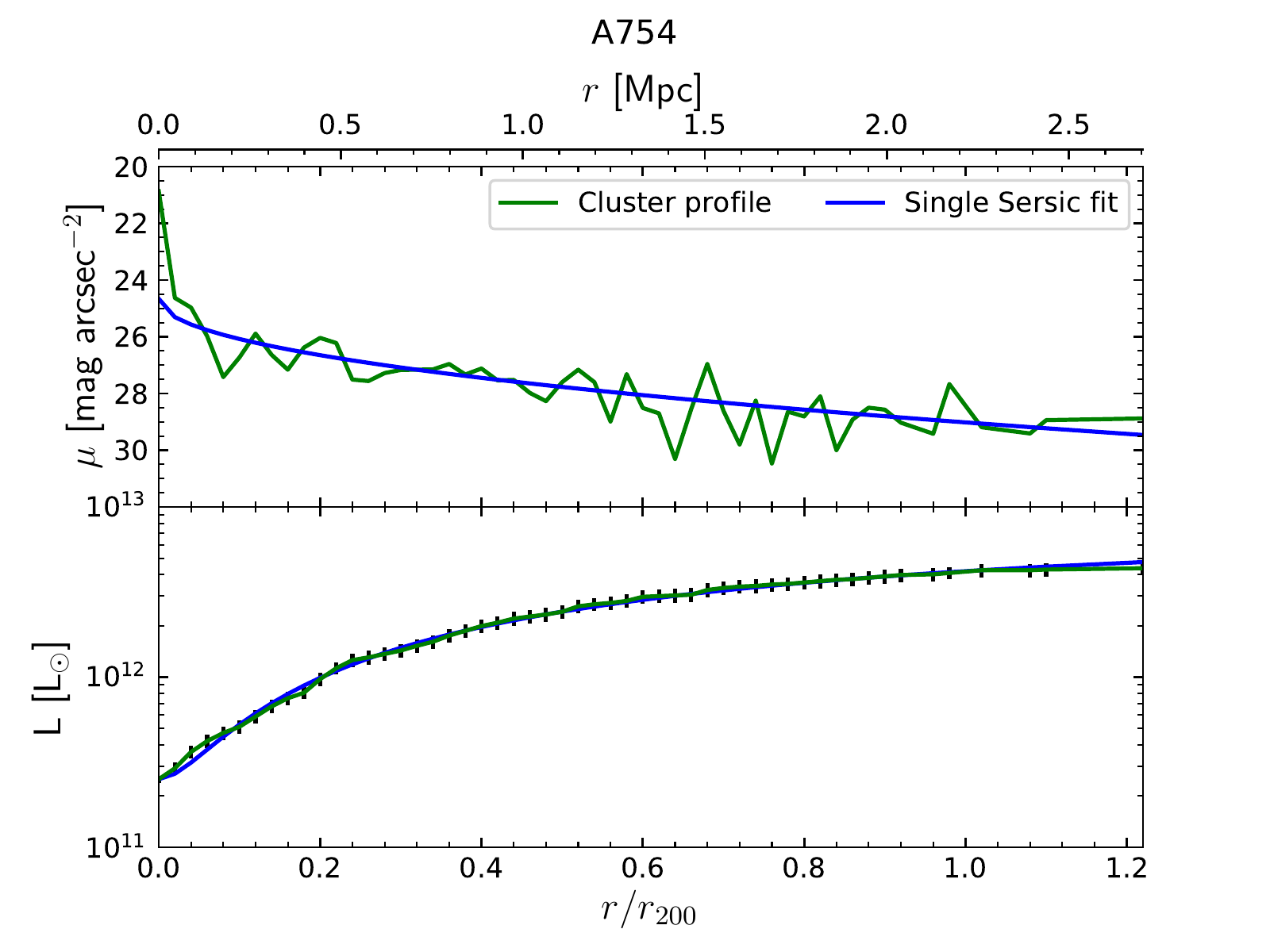} \includegraphics[width=0.45\textwidth]{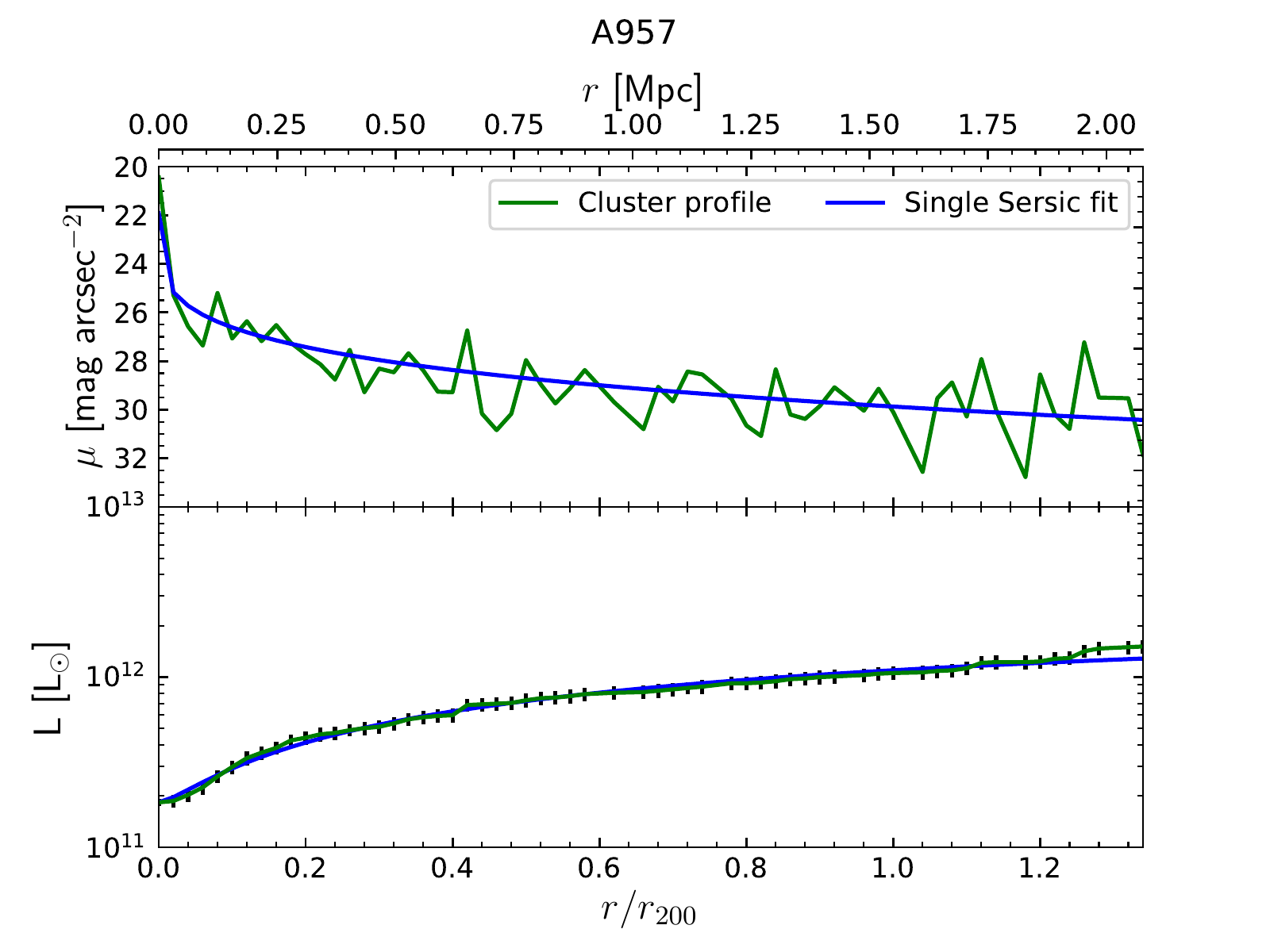}
        \includegraphics[width=0.45\textwidth]{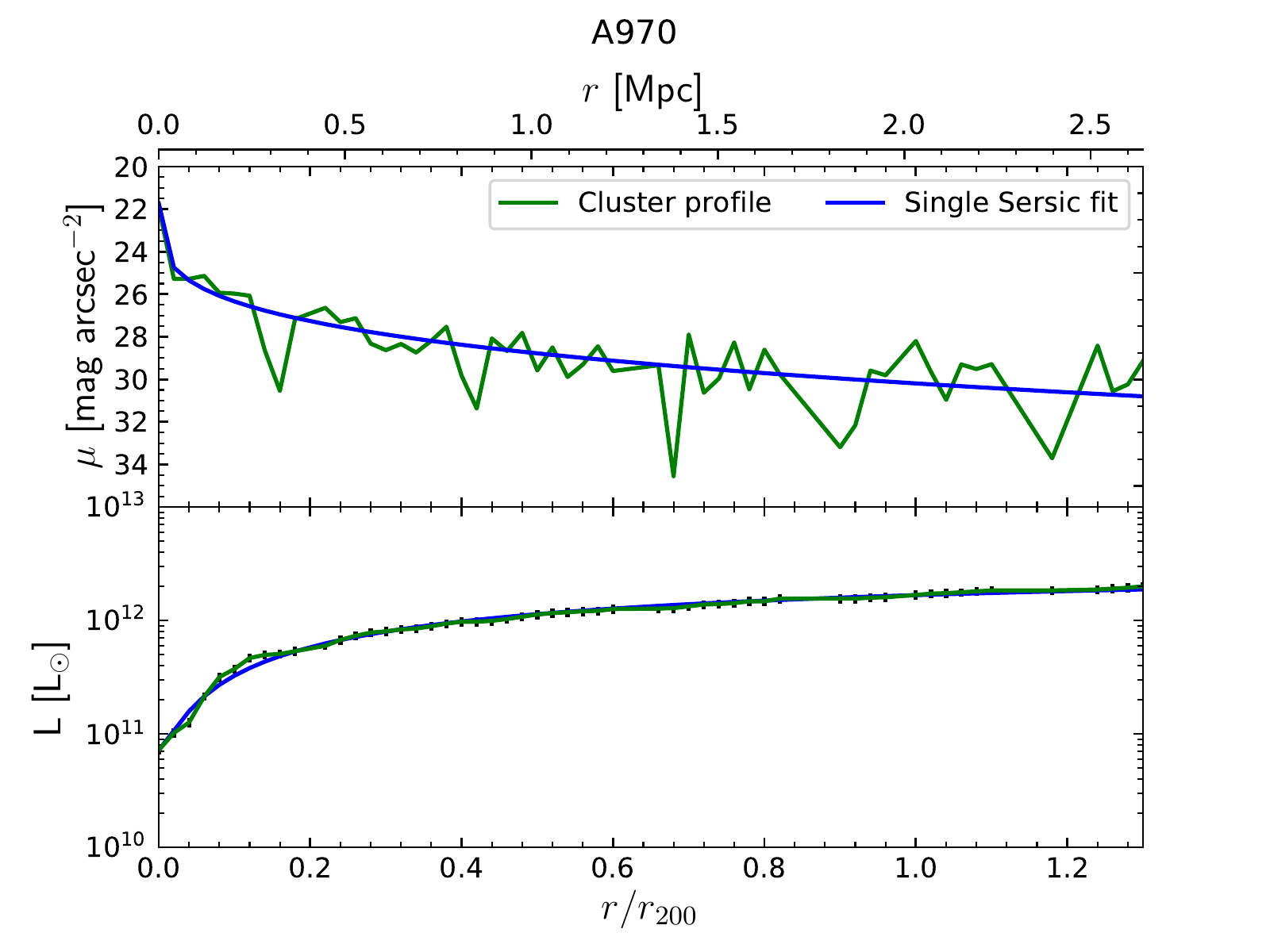} \includegraphics[width=0.45\textwidth]{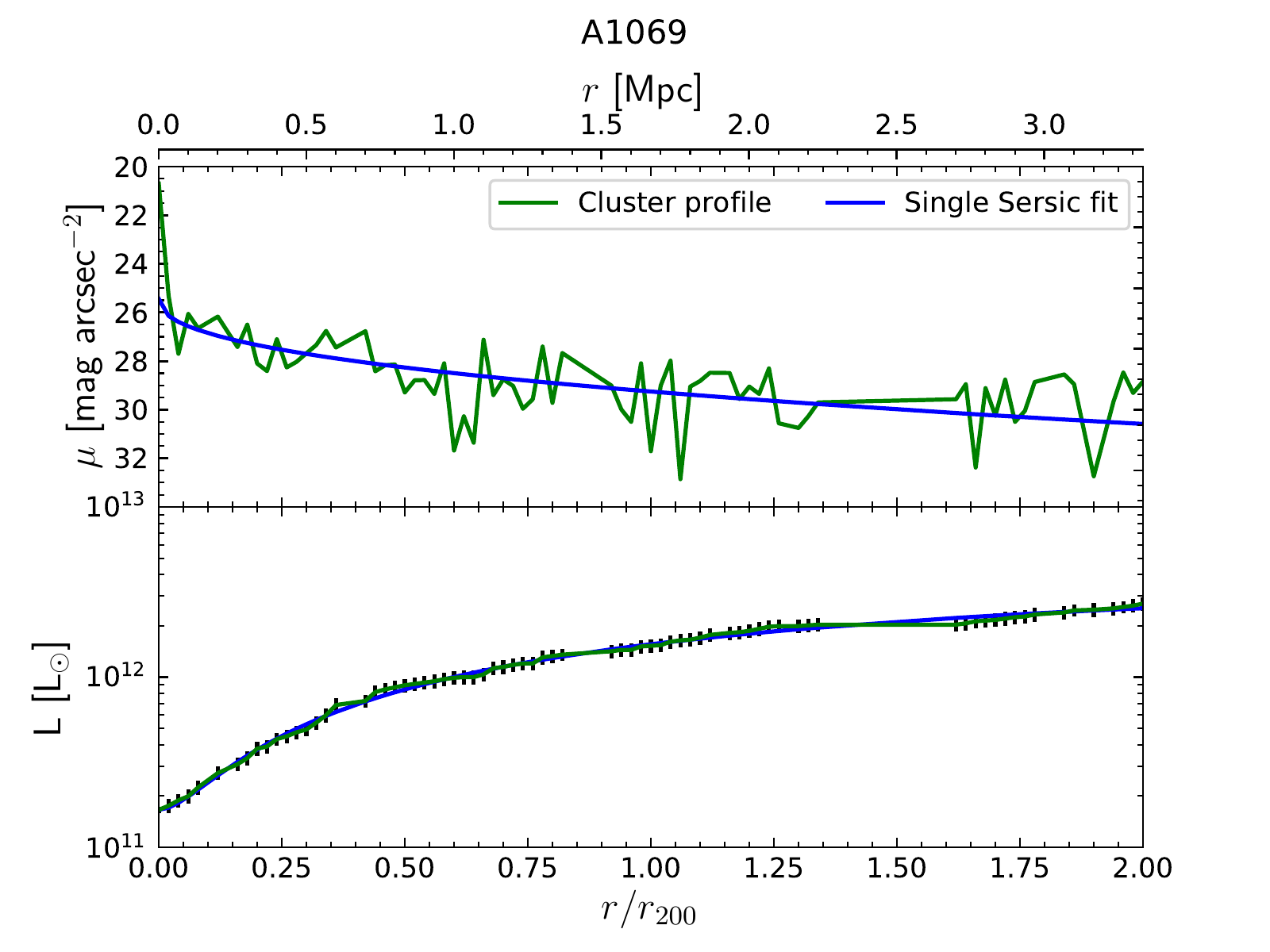}
    \caption{Photometric decomposition of Omega-WINGS galaxy clusters luminosity profiles, continued.}
\end{figure*}

\newpage
\clearpage

\begin{figure*}[t]
   \centering
        \includegraphics[width=0.45\textwidth]{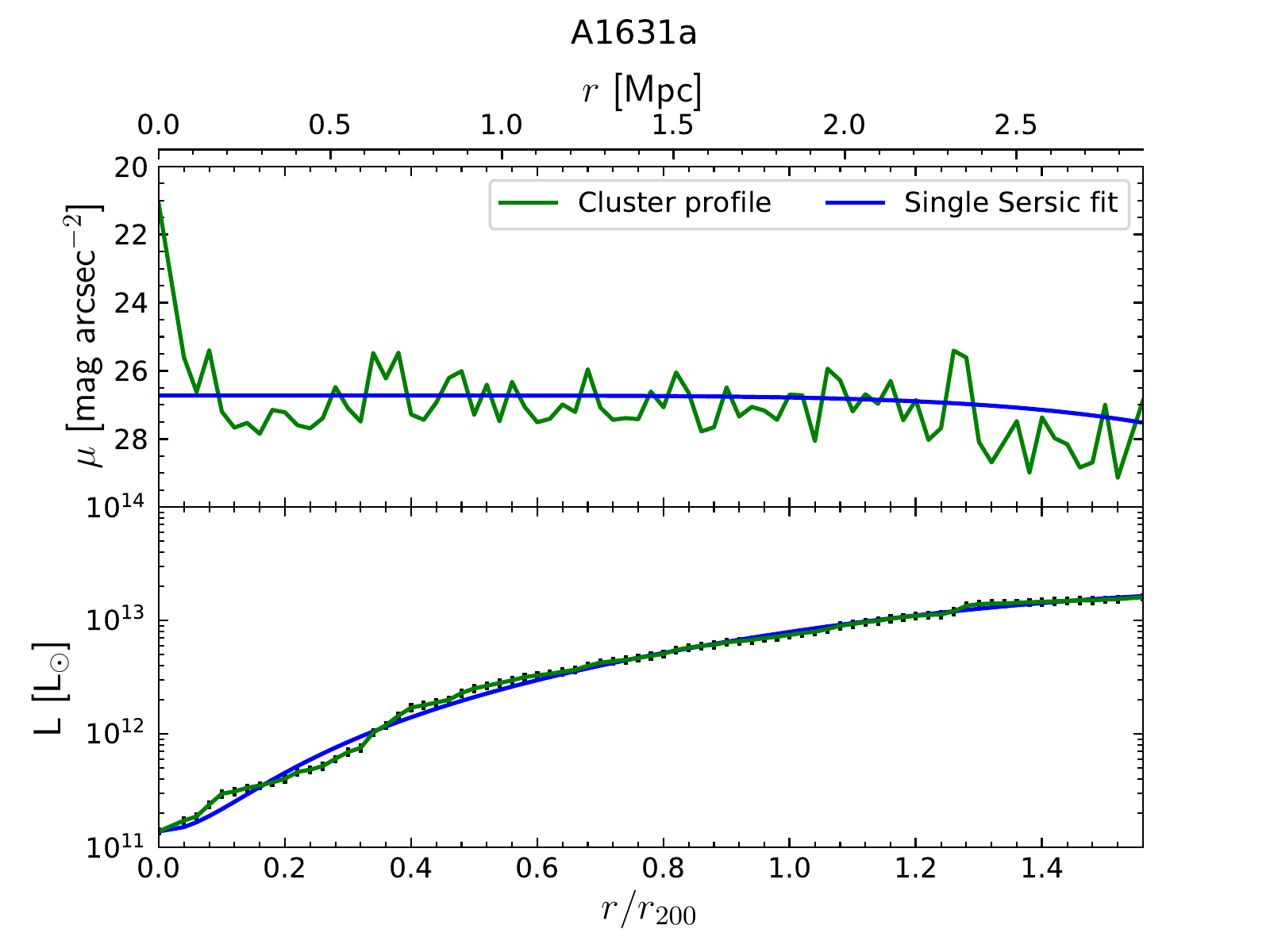}       \includegraphics[width=0.45\textwidth]{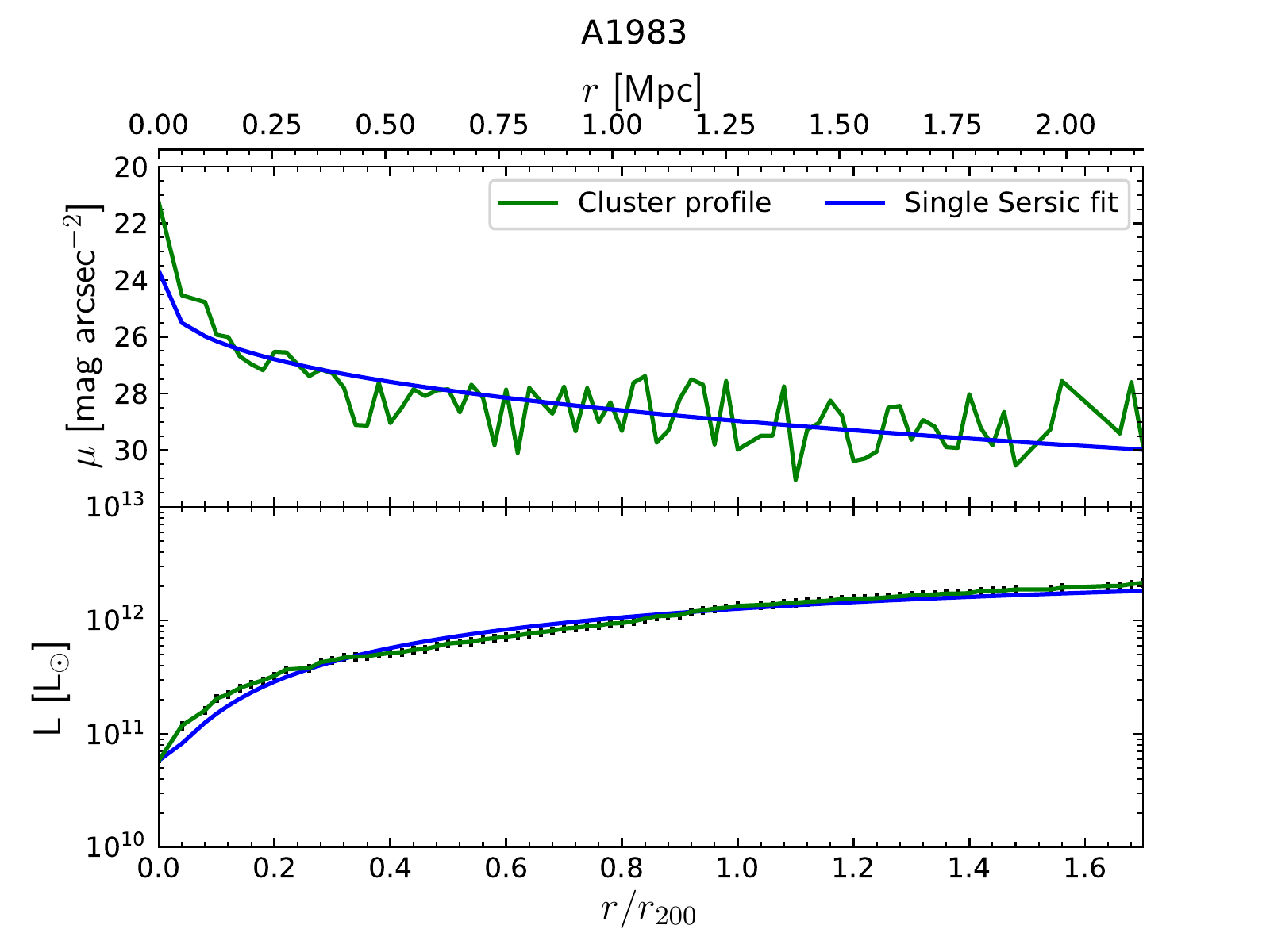}
        \includegraphics[width=0.45\textwidth]{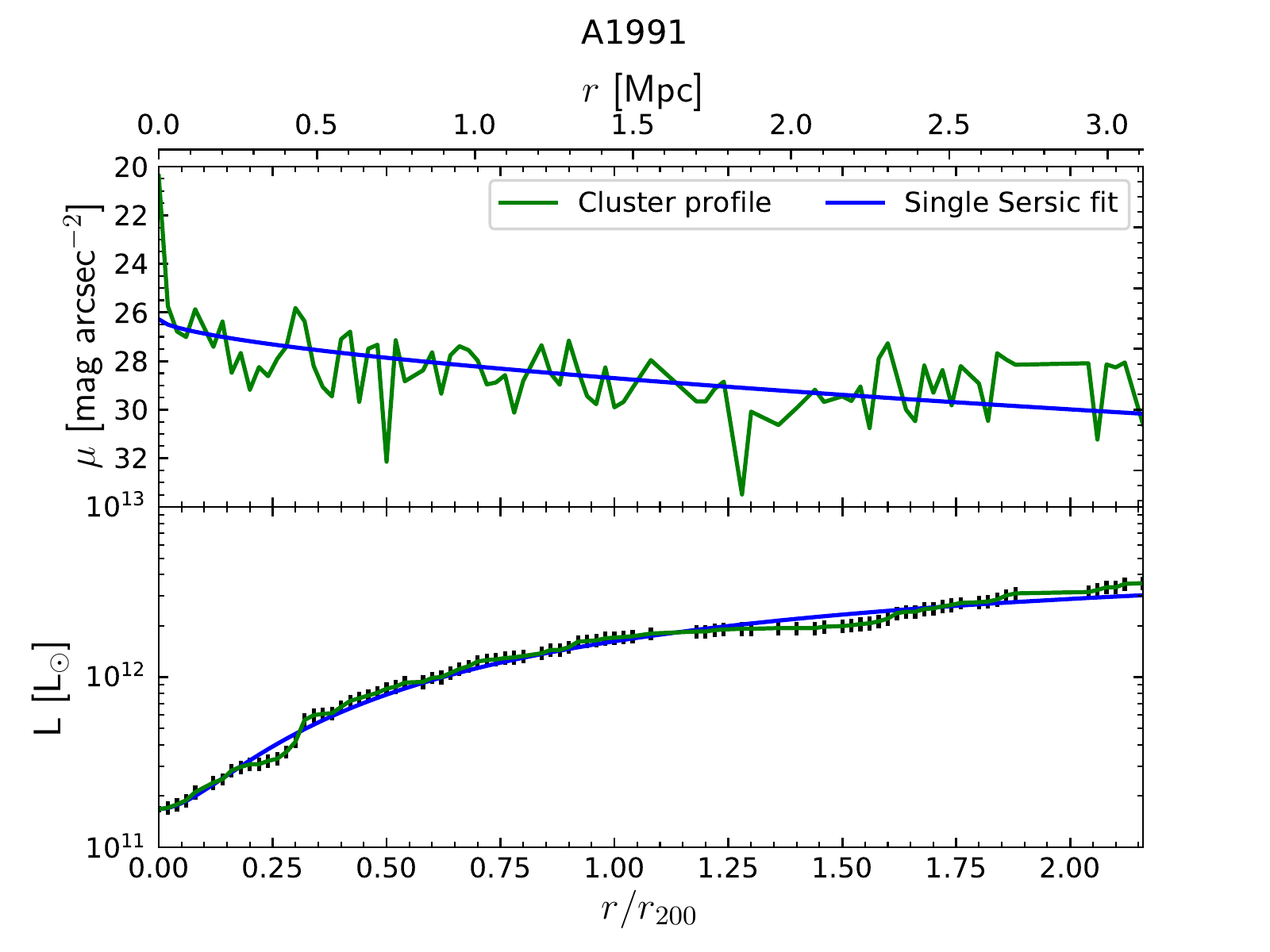}        \includegraphics[width=0.45\textwidth]{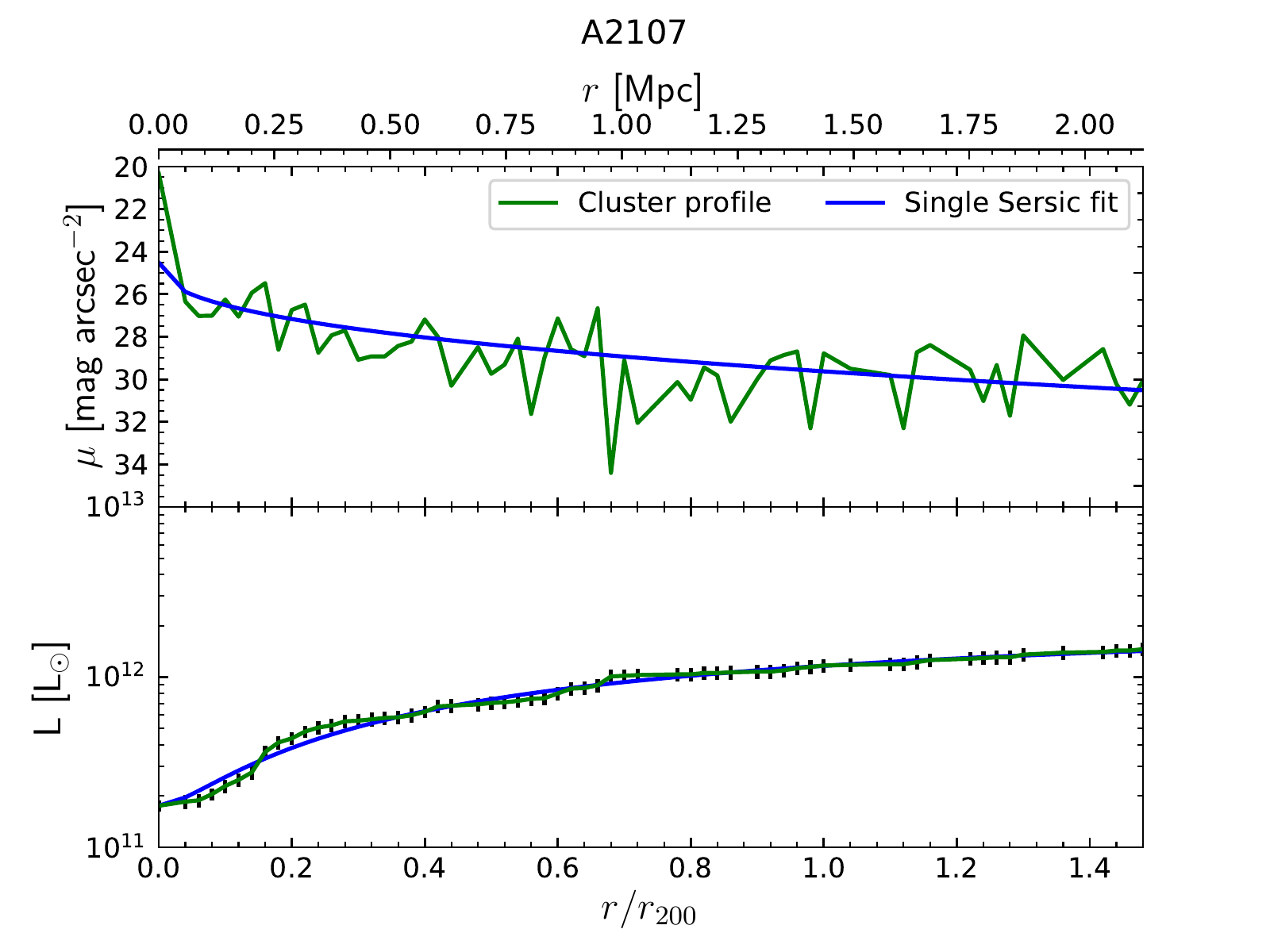}
        \includegraphics[width=0.45\textwidth]{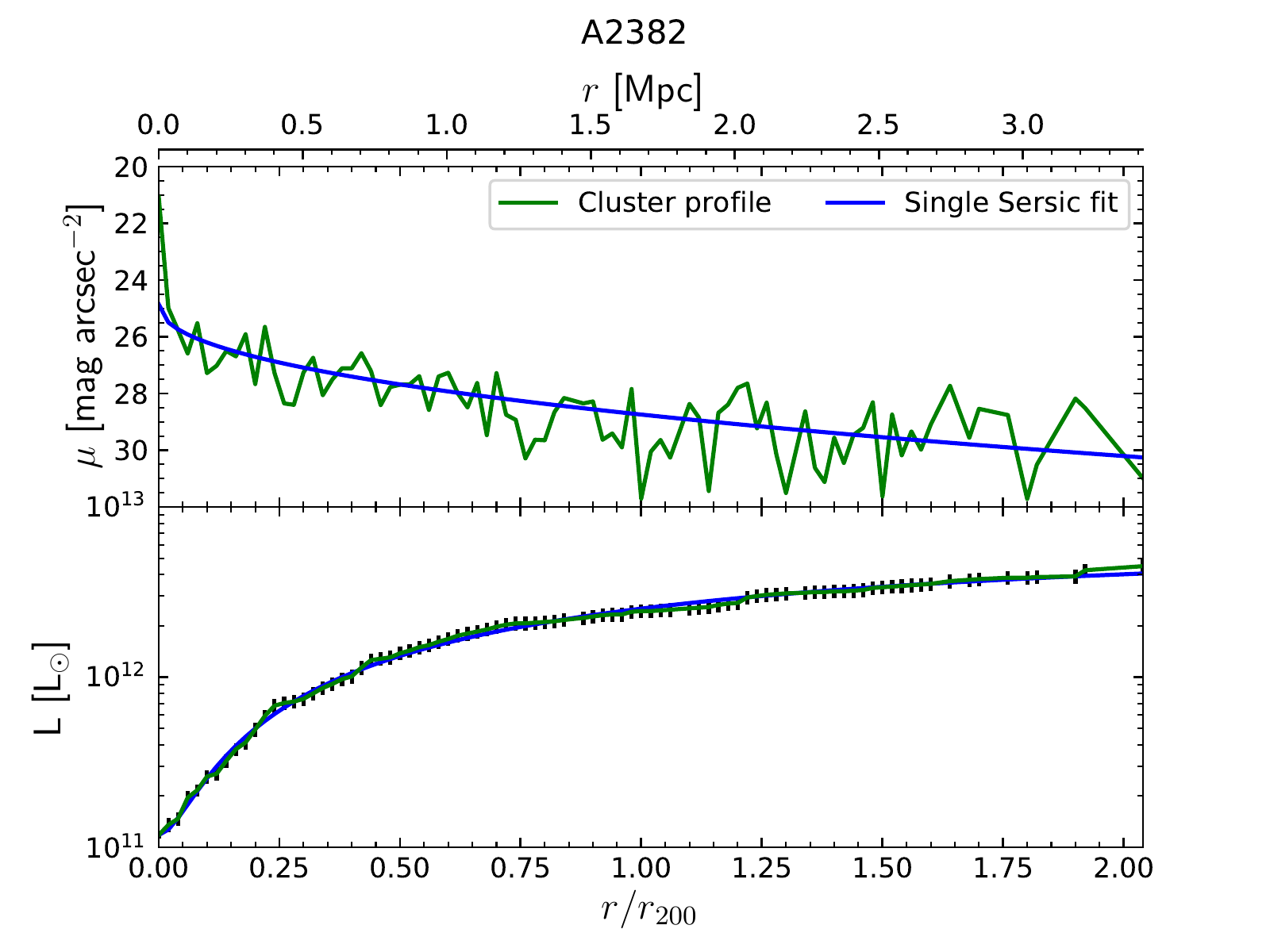}        \includegraphics[width=0.45\textwidth]{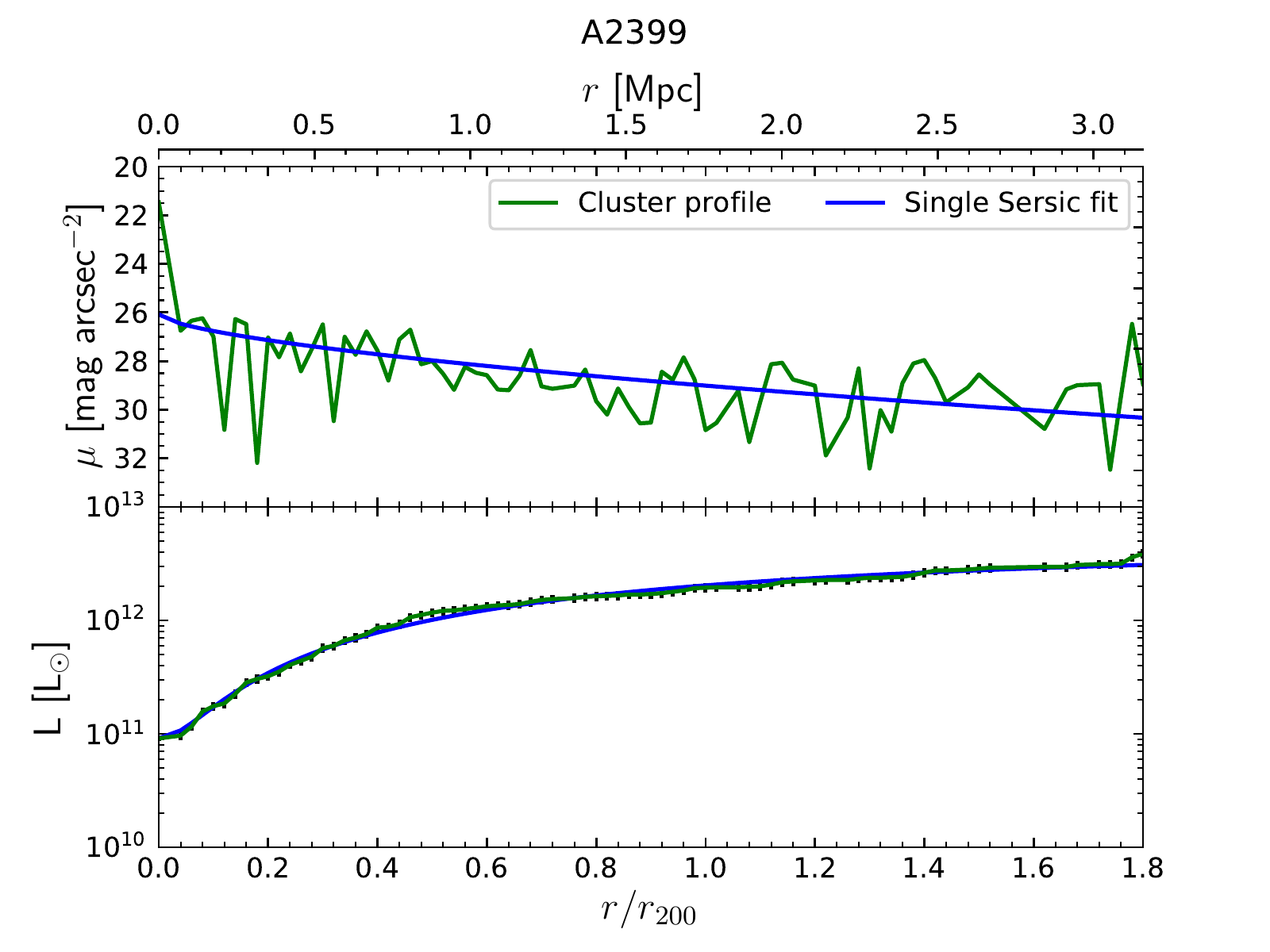}
    \caption{Photometric decomposition of Omega-WINGS galaxy clusters luminosity profiles, continued.}
\end{figure*}
    
\newpage
\clearpage

\begin{figure*}[t]
   \centering
        \includegraphics[width=0.45\textwidth]{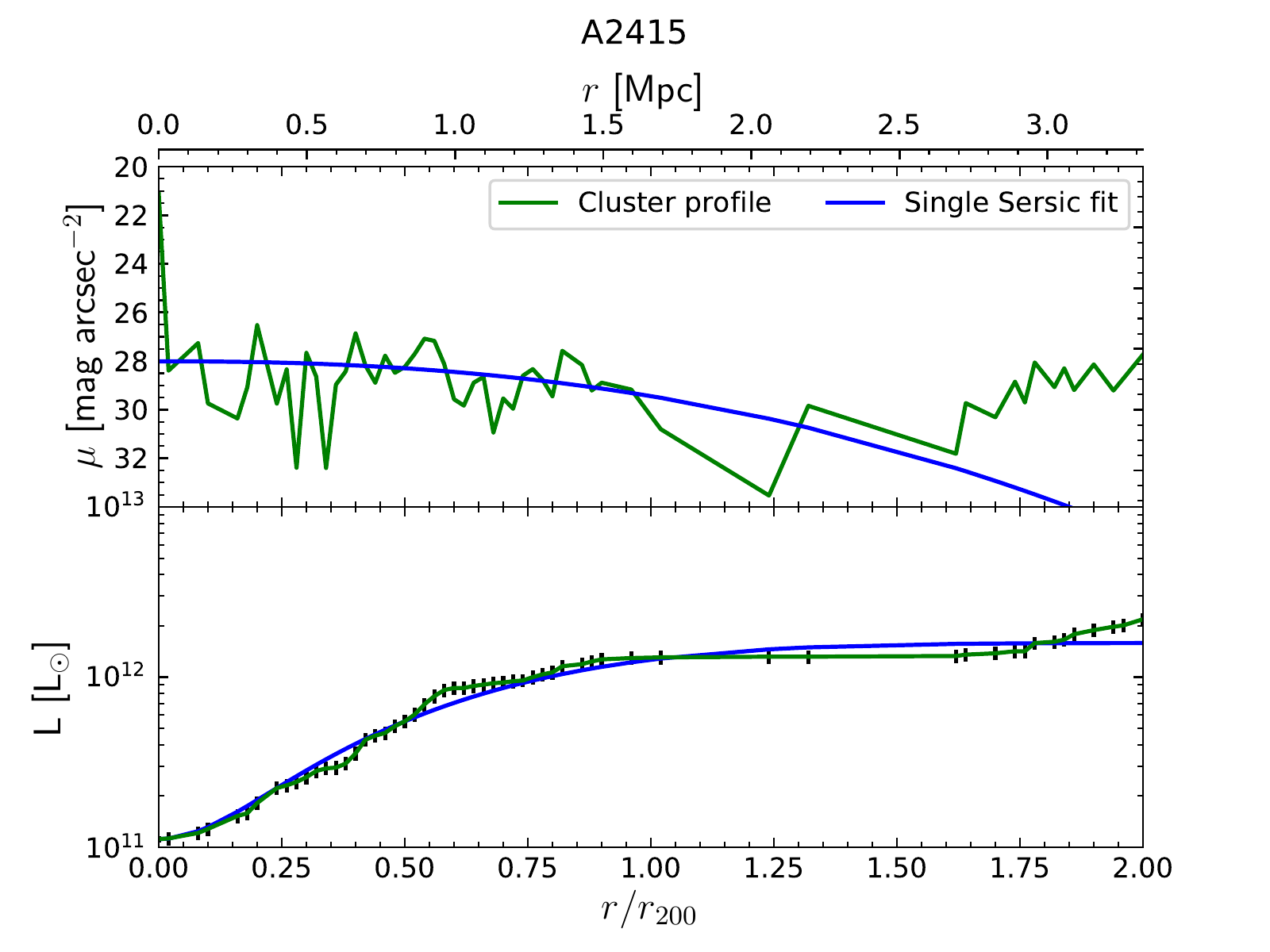}        \includegraphics[width=0.45\textwidth]{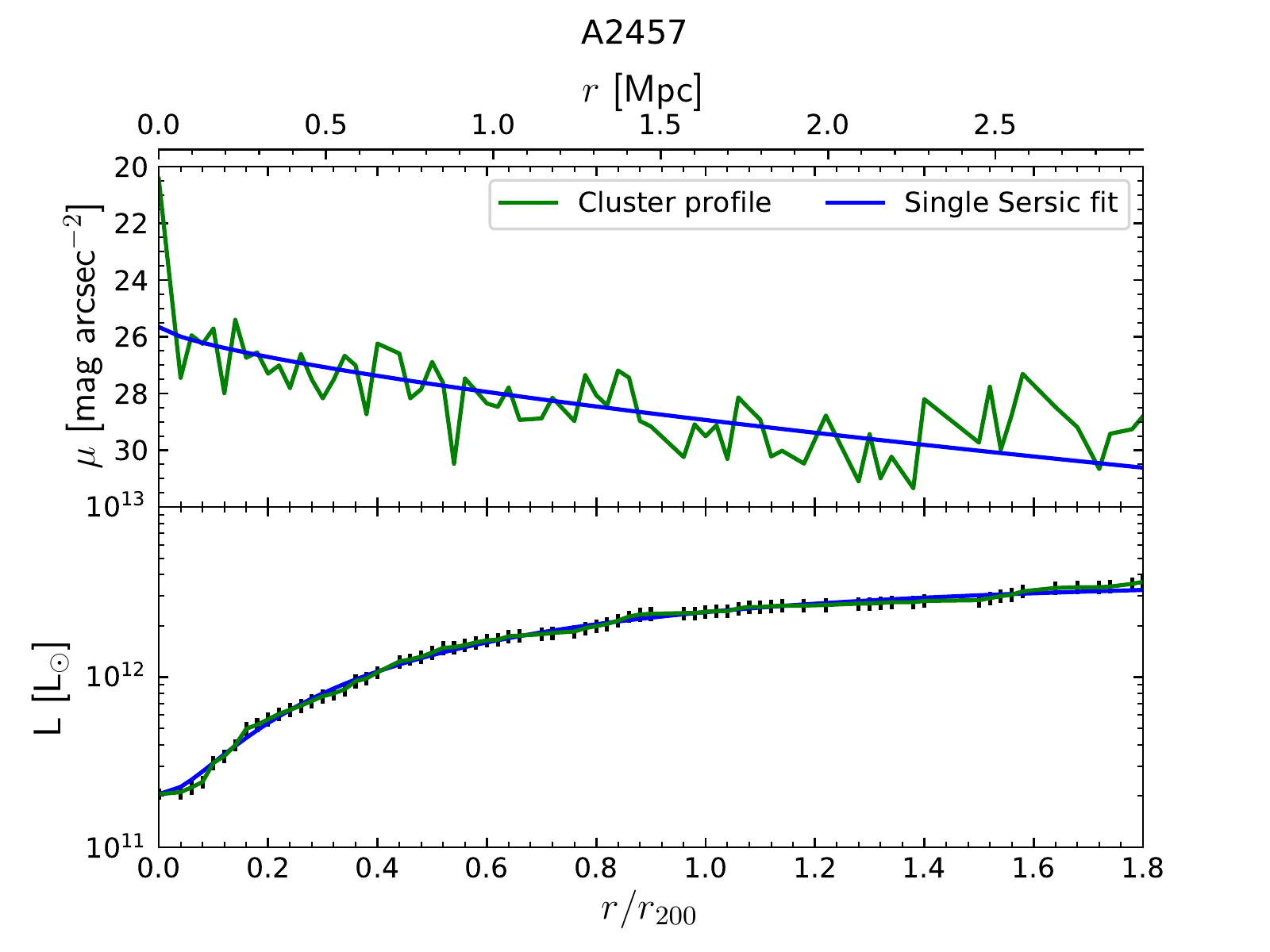}
        \includegraphics[width=0.45\textwidth]{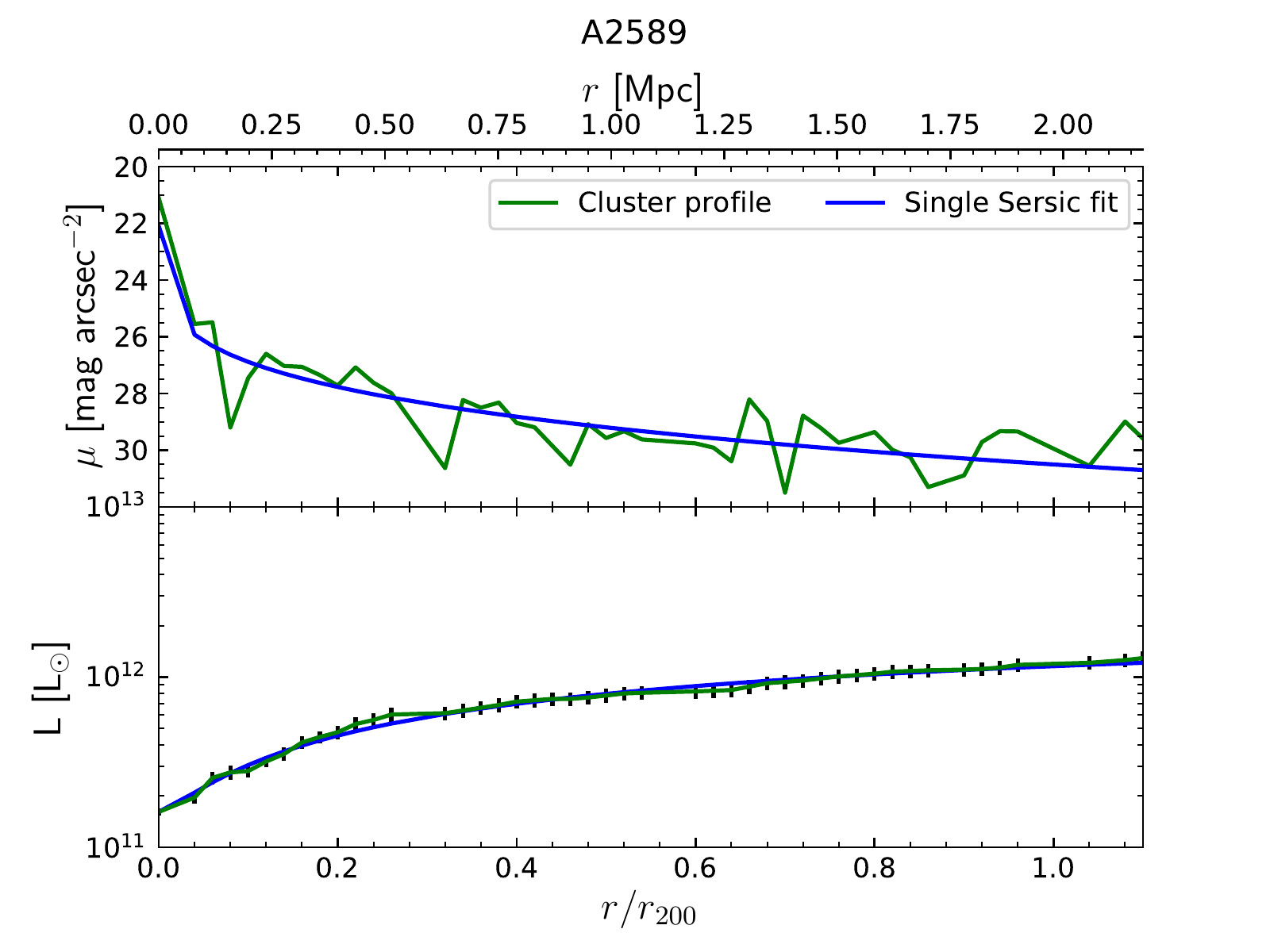}        \includegraphics[width=0.45\textwidth]{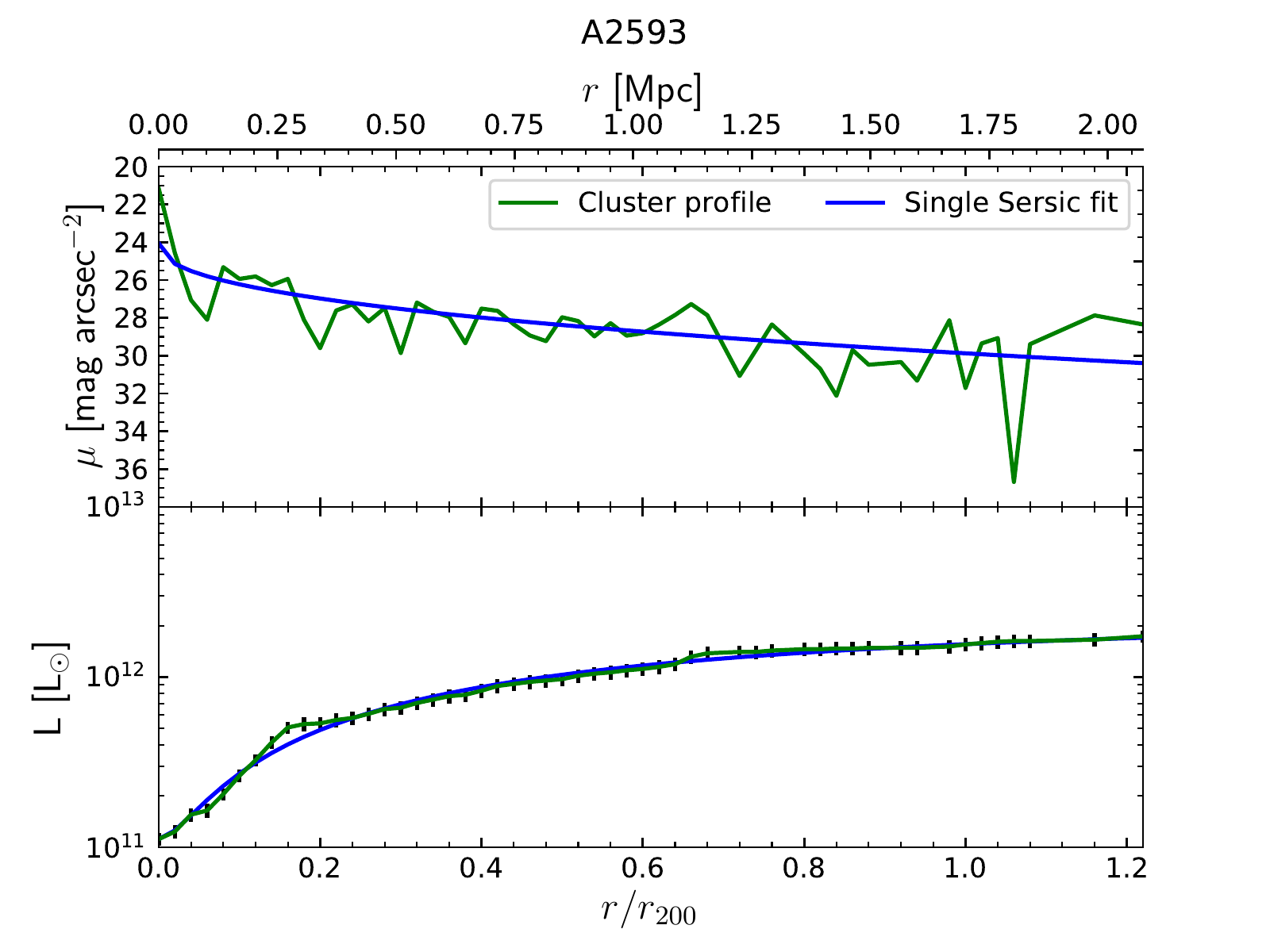}
        \includegraphics[width=0.45\textwidth]{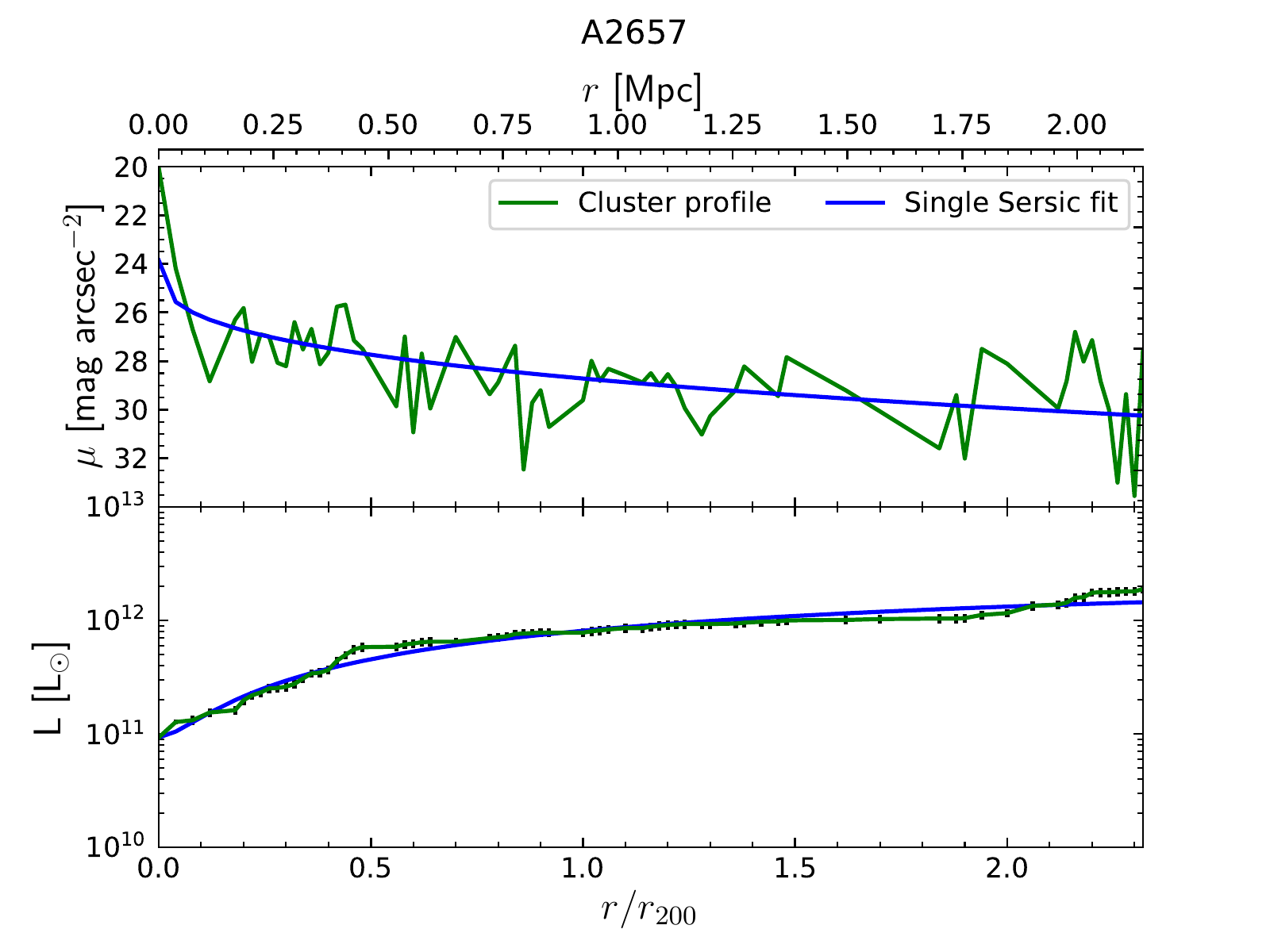}        \includegraphics[width=0.45\textwidth]{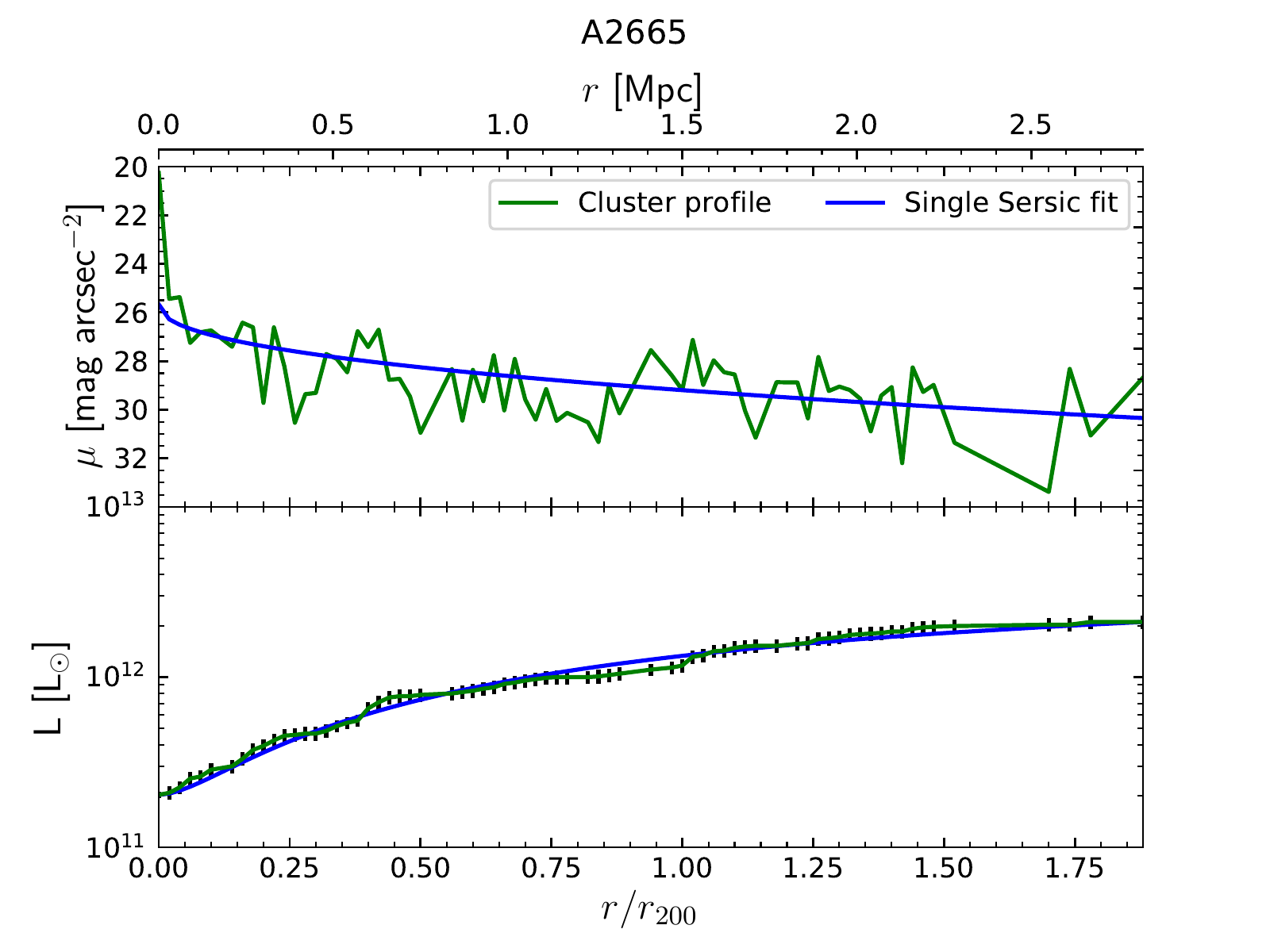}
    \caption{Photometric decomposition of Omega-WINGS galaxy clusters luminosity profiles, continued.}
\end{figure*}

\newpage
\clearpage

\begin{figure*}[t]
   \centering
        \includegraphics[width=0.45\textwidth]{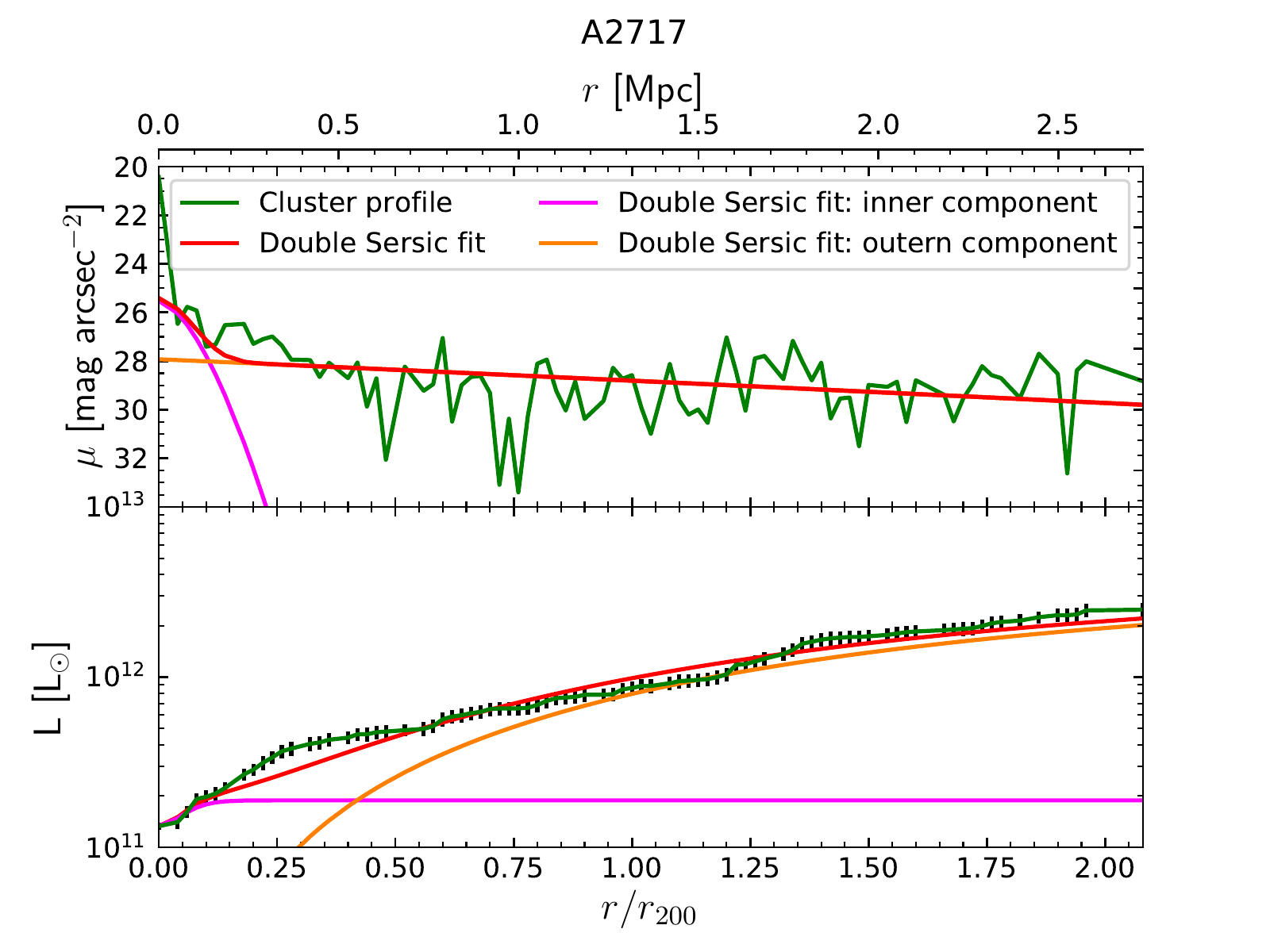}        \includegraphics[width=0.45\textwidth]{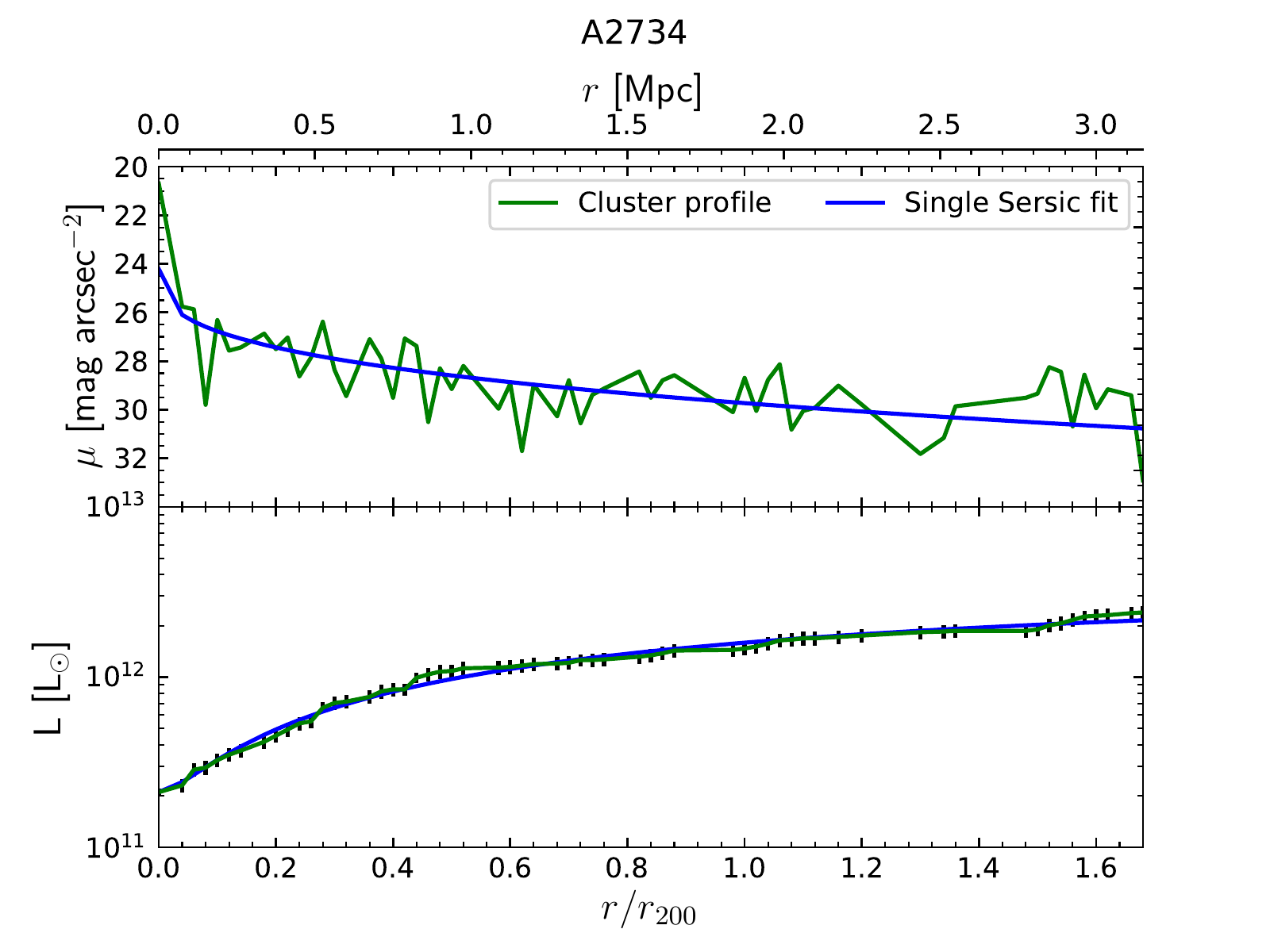}
        \includegraphics[width=0.45\textwidth]{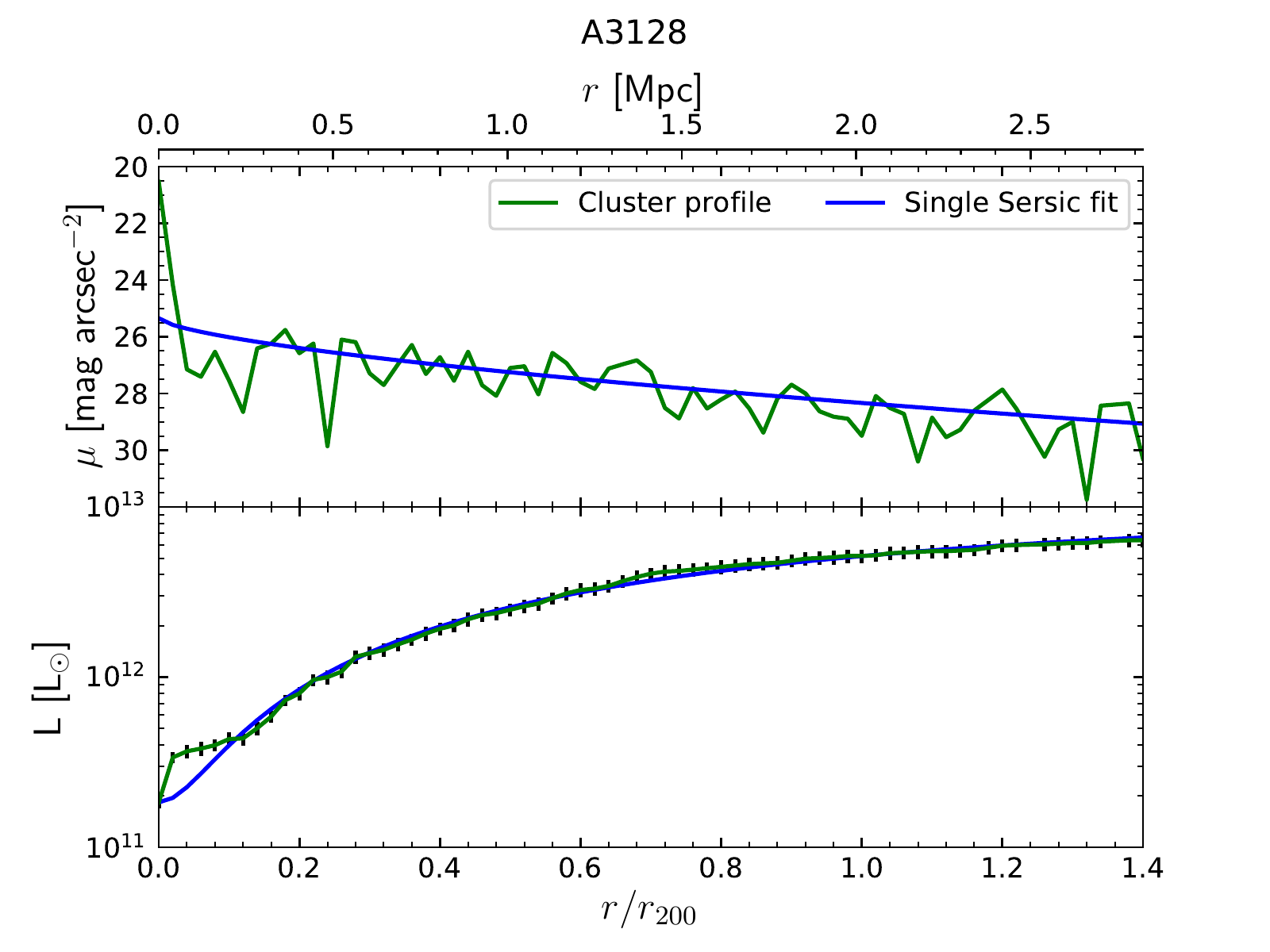}        \includegraphics[width=0.45\textwidth]{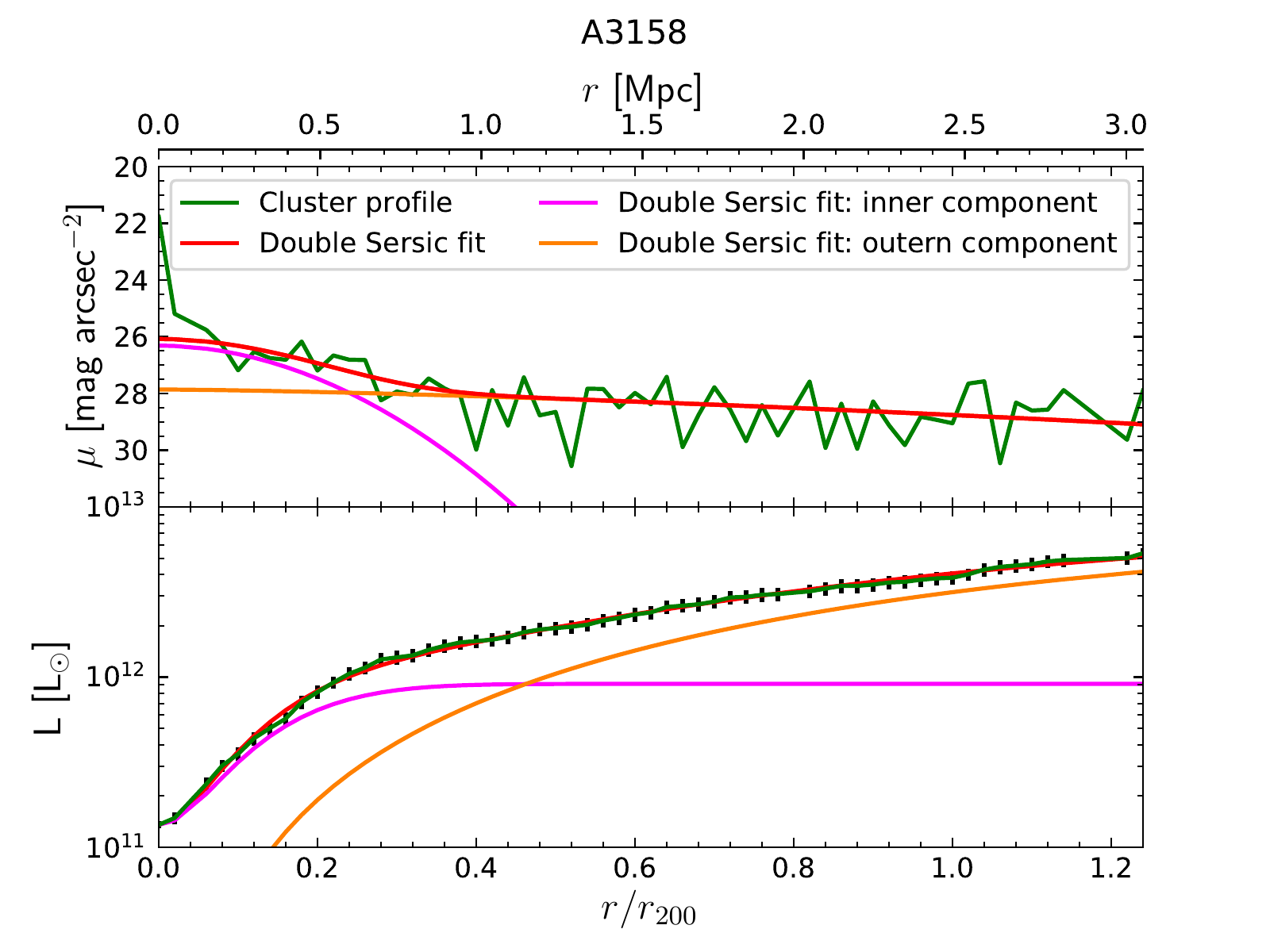}
        \includegraphics[width=0.45\textwidth]{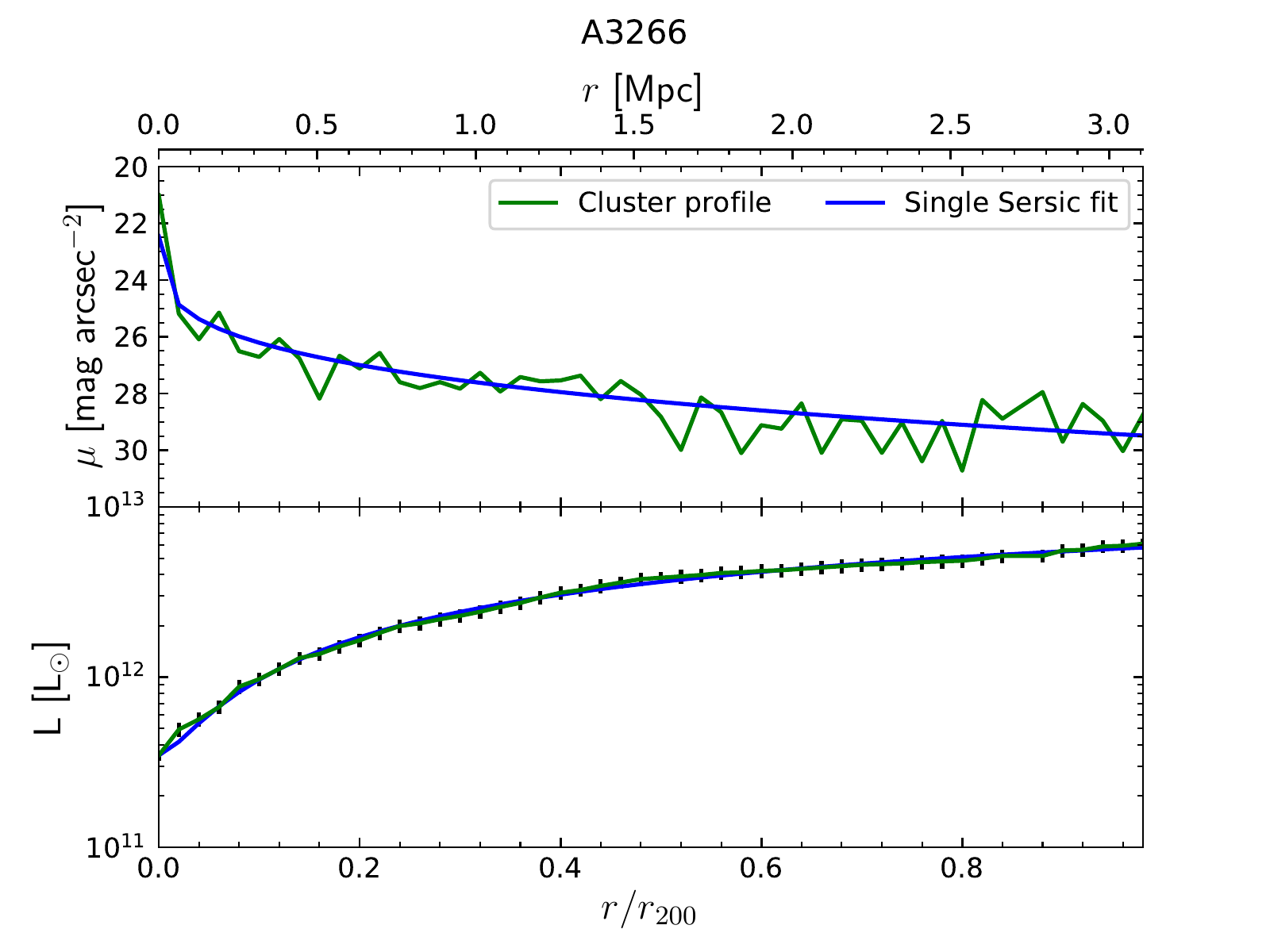}        \includegraphics[width=0.45\textwidth]{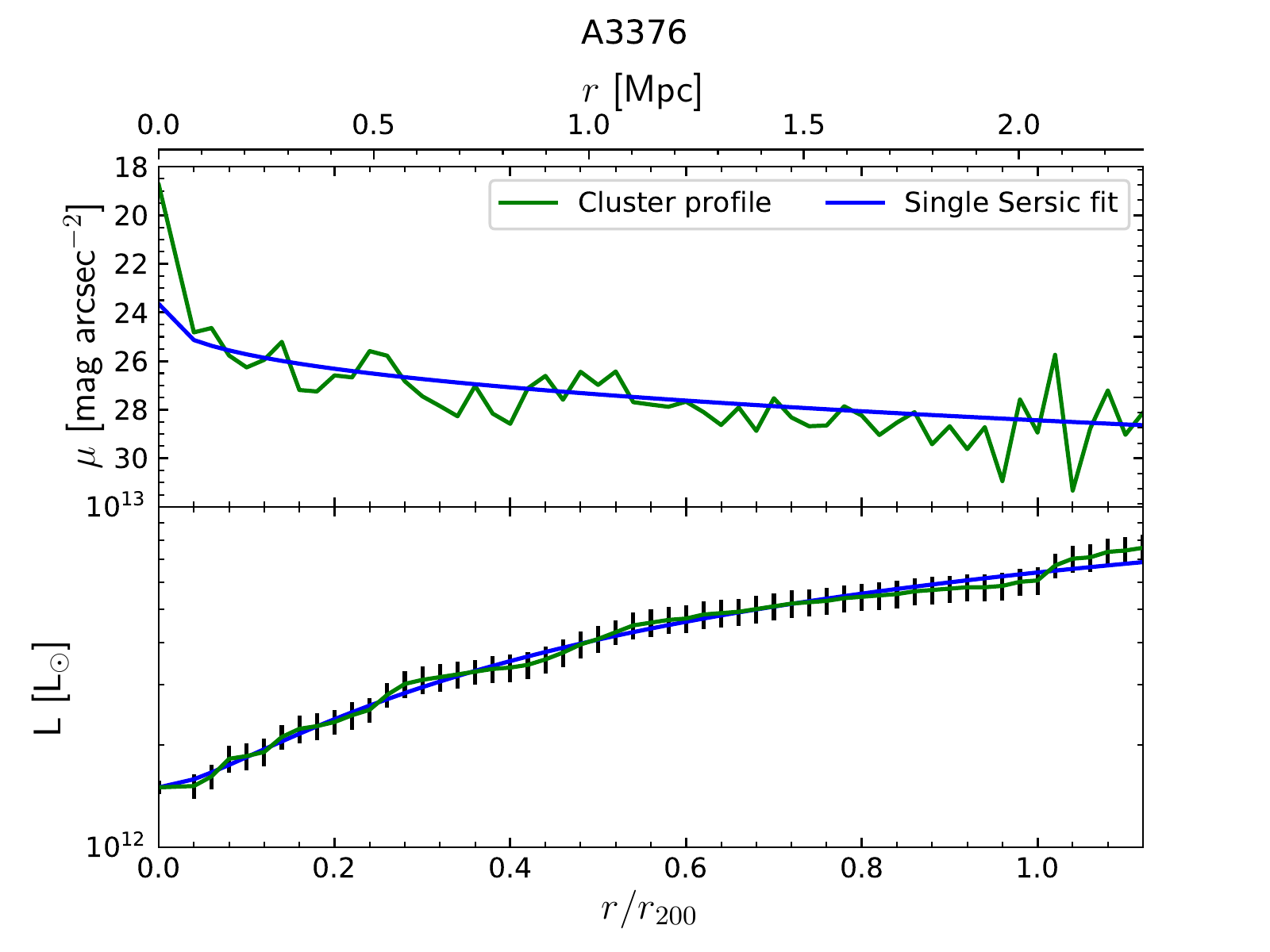}
    \caption{Photometric decomposition of Omega-WINGS galaxy clusters luminosity profiles, continued.}
\end{figure*}

\newpage
\clearpage

\begin{figure*}[t]
   \centering
        \includegraphics[width=0.45\textwidth]{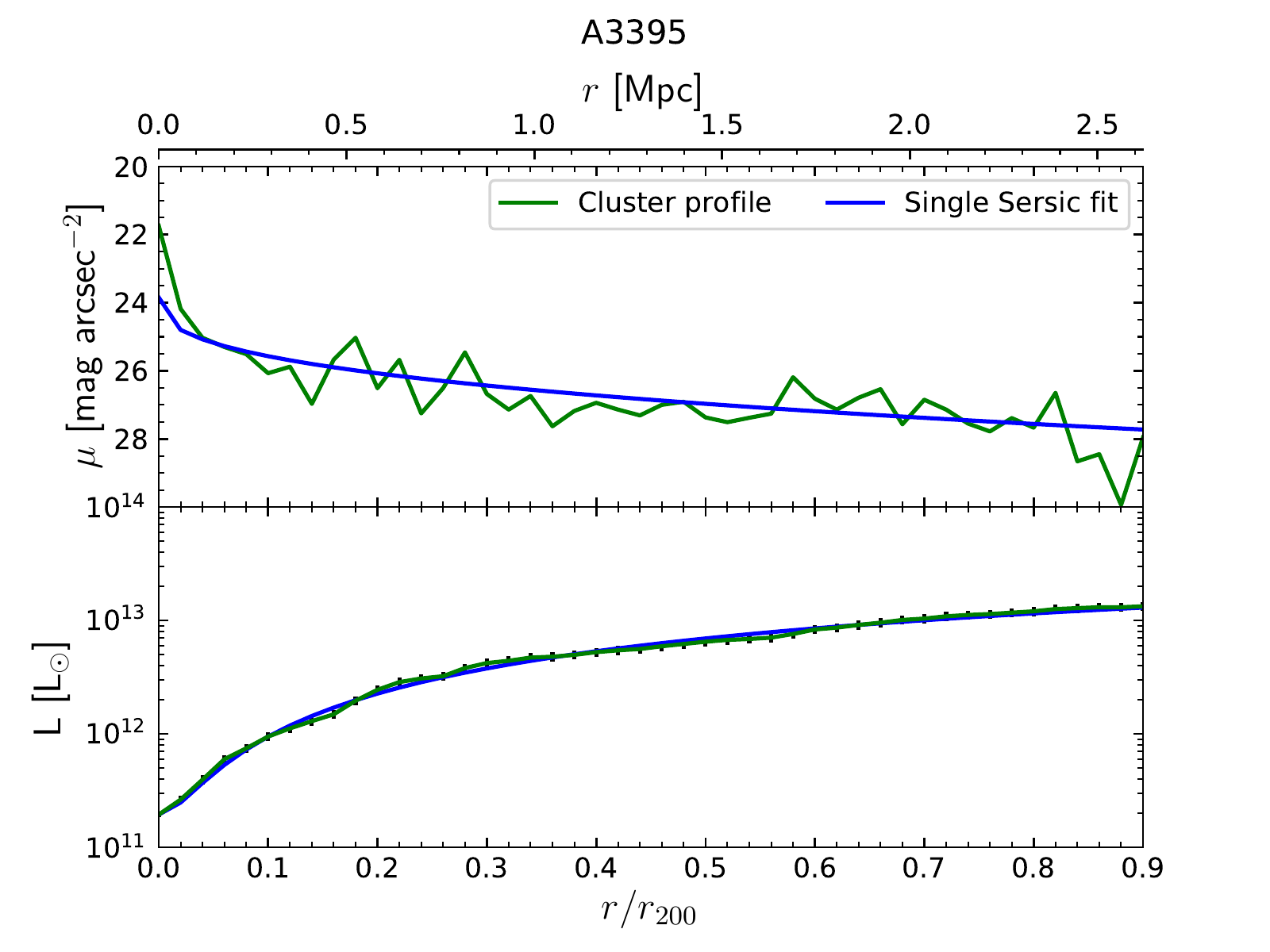}        \includegraphics[width=0.45\textwidth]{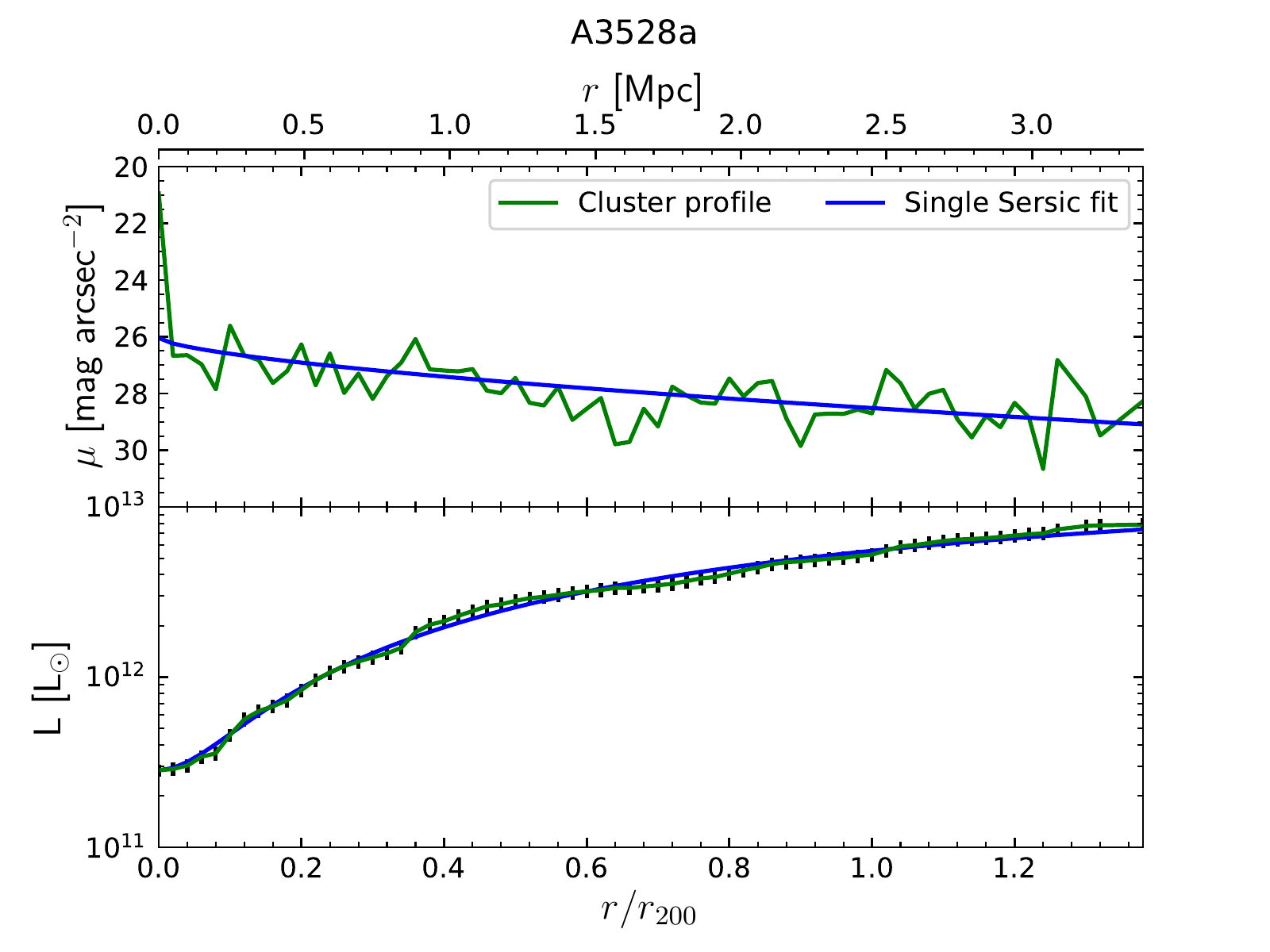}
        \includegraphics[width=0.45\textwidth]{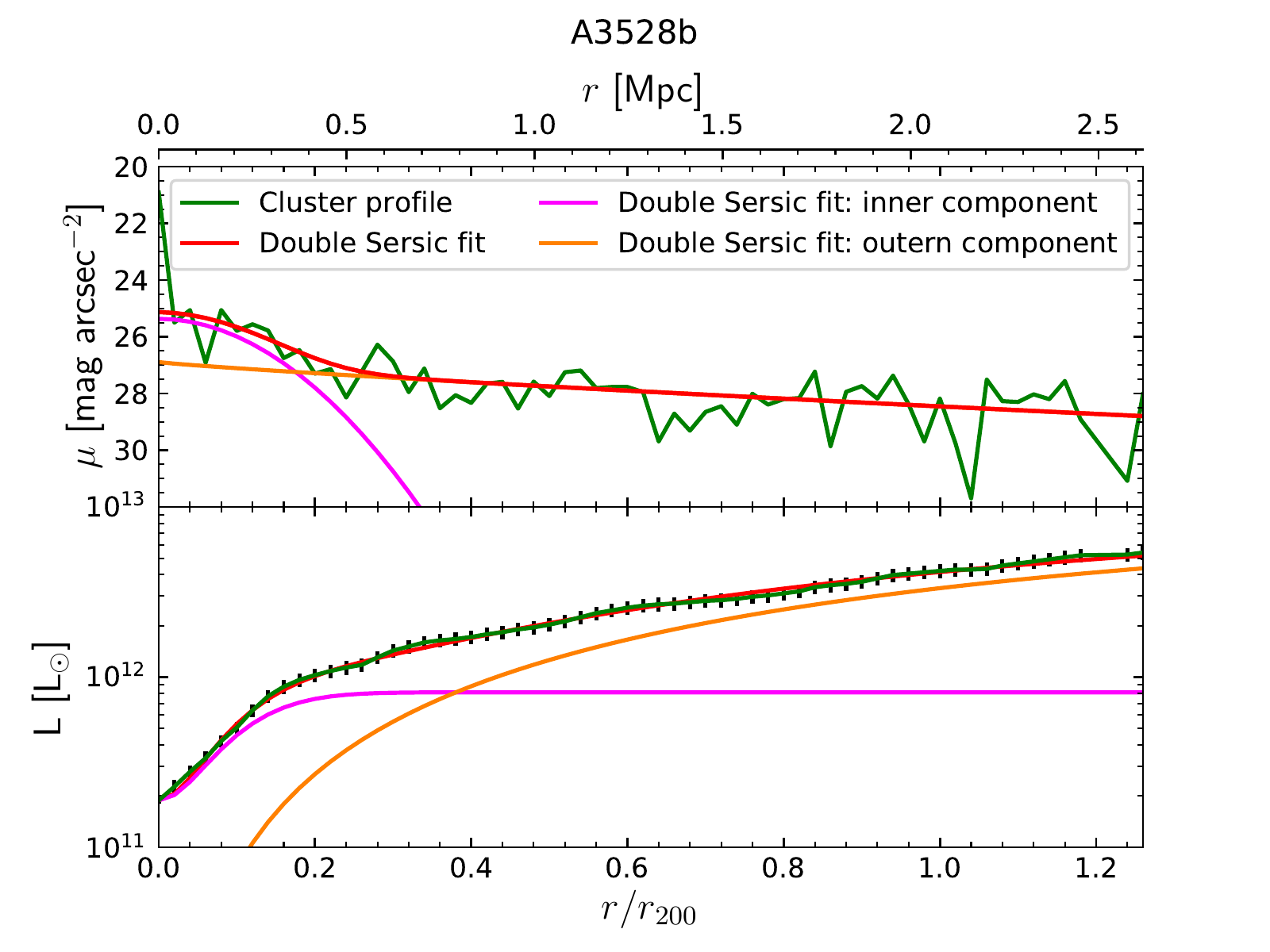}       \includegraphics[width=0.45\textwidth]{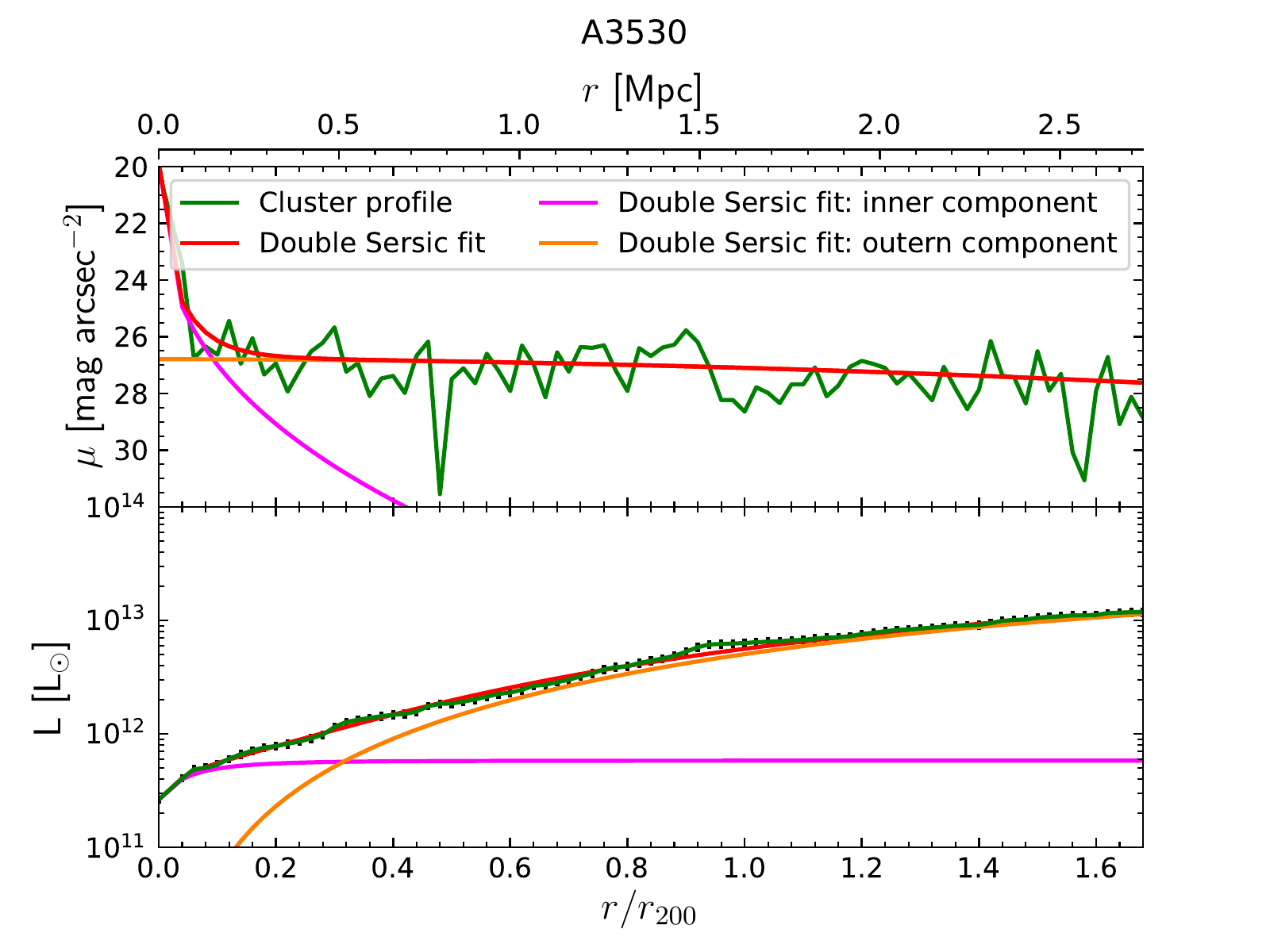}
        \includegraphics[width=0.45\textwidth]{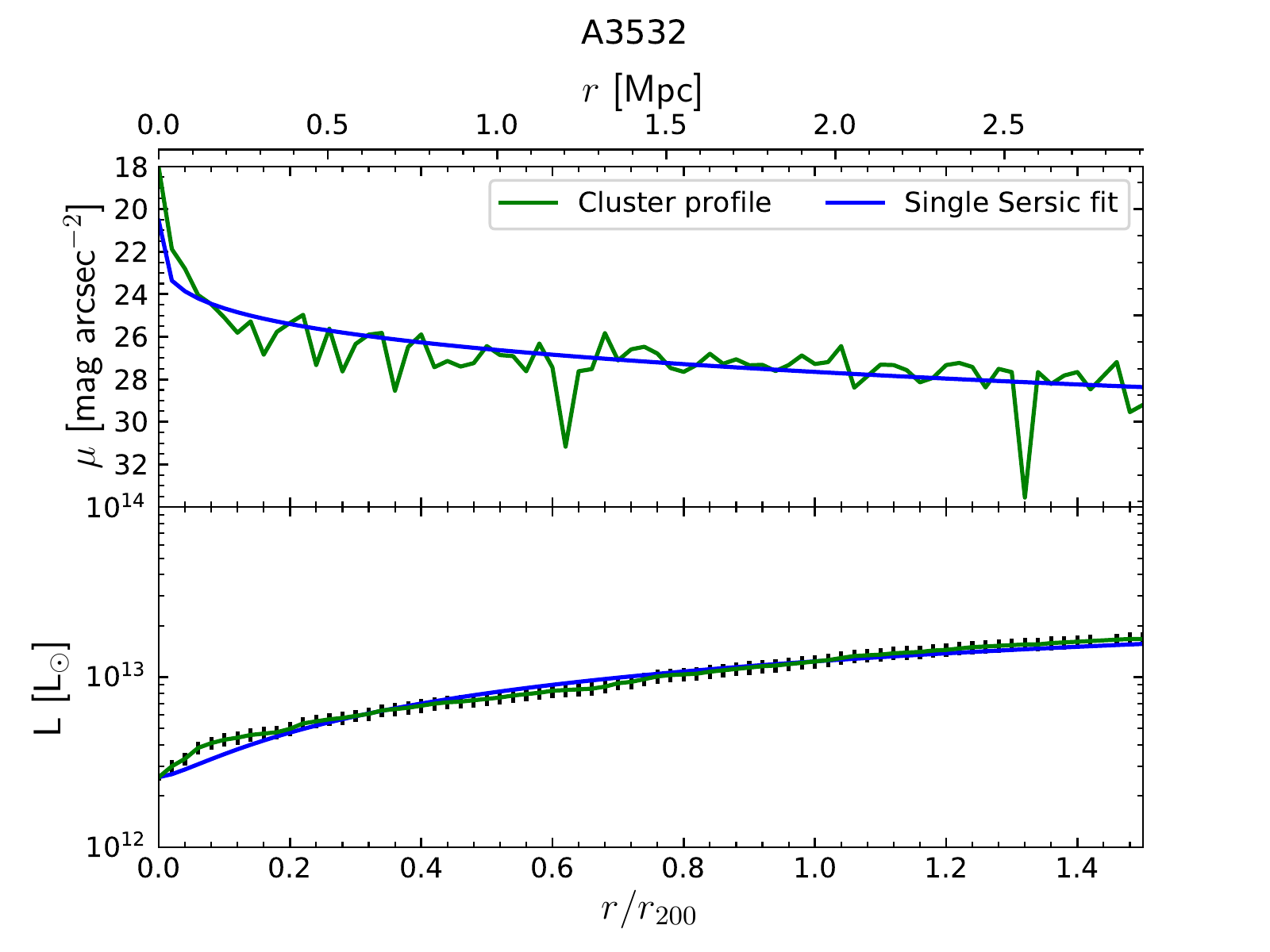}        \includegraphics[width=0.45\textwidth]{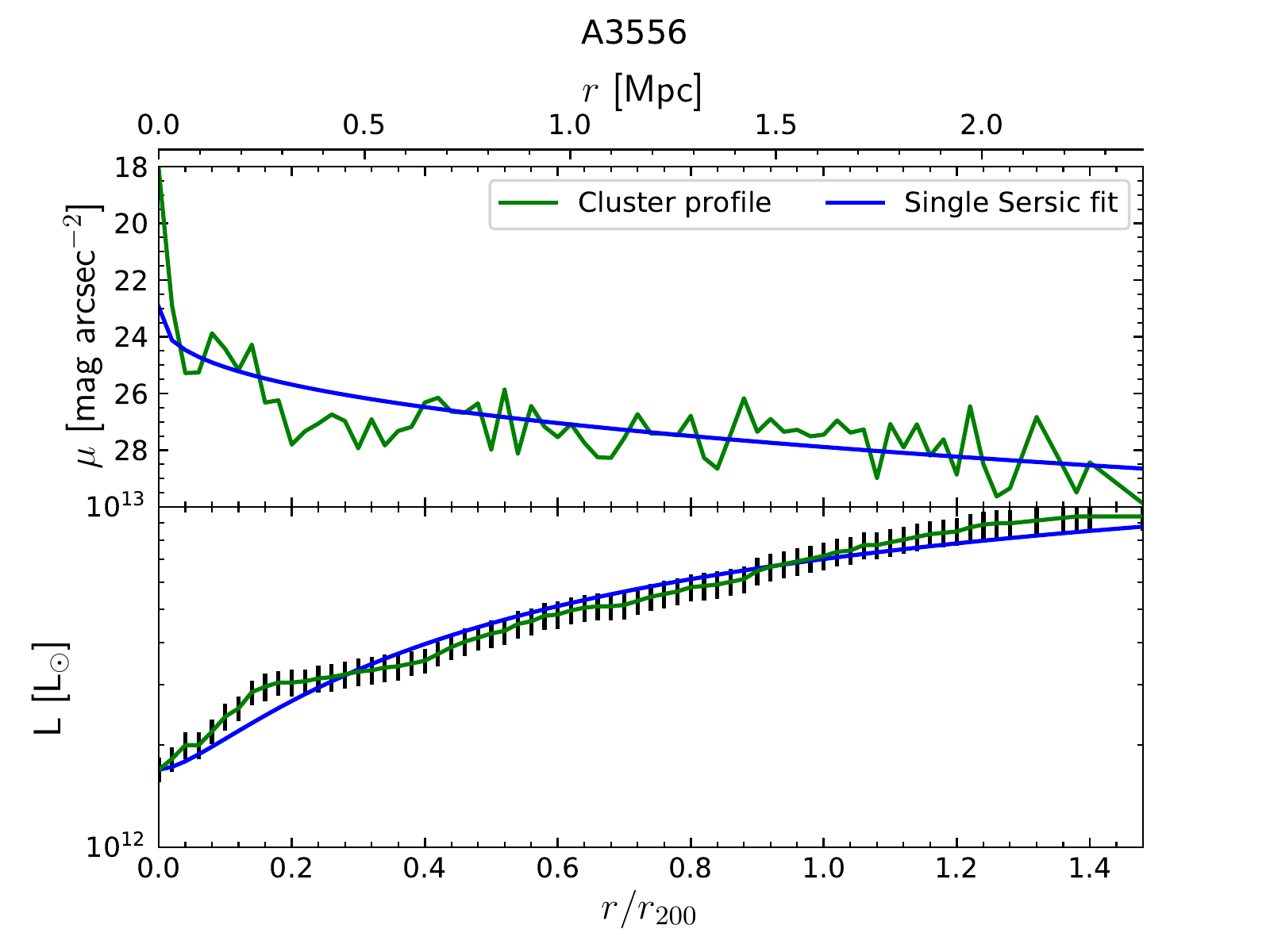}
    \caption{Photometric decomposition of Omega-WINGS galaxy clusters luminosity profiles, continued.}
\end{figure*}

\newpage
\clearpage

\begin{figure*}[t]
   \centering
        \includegraphics[width=0.45\textwidth]{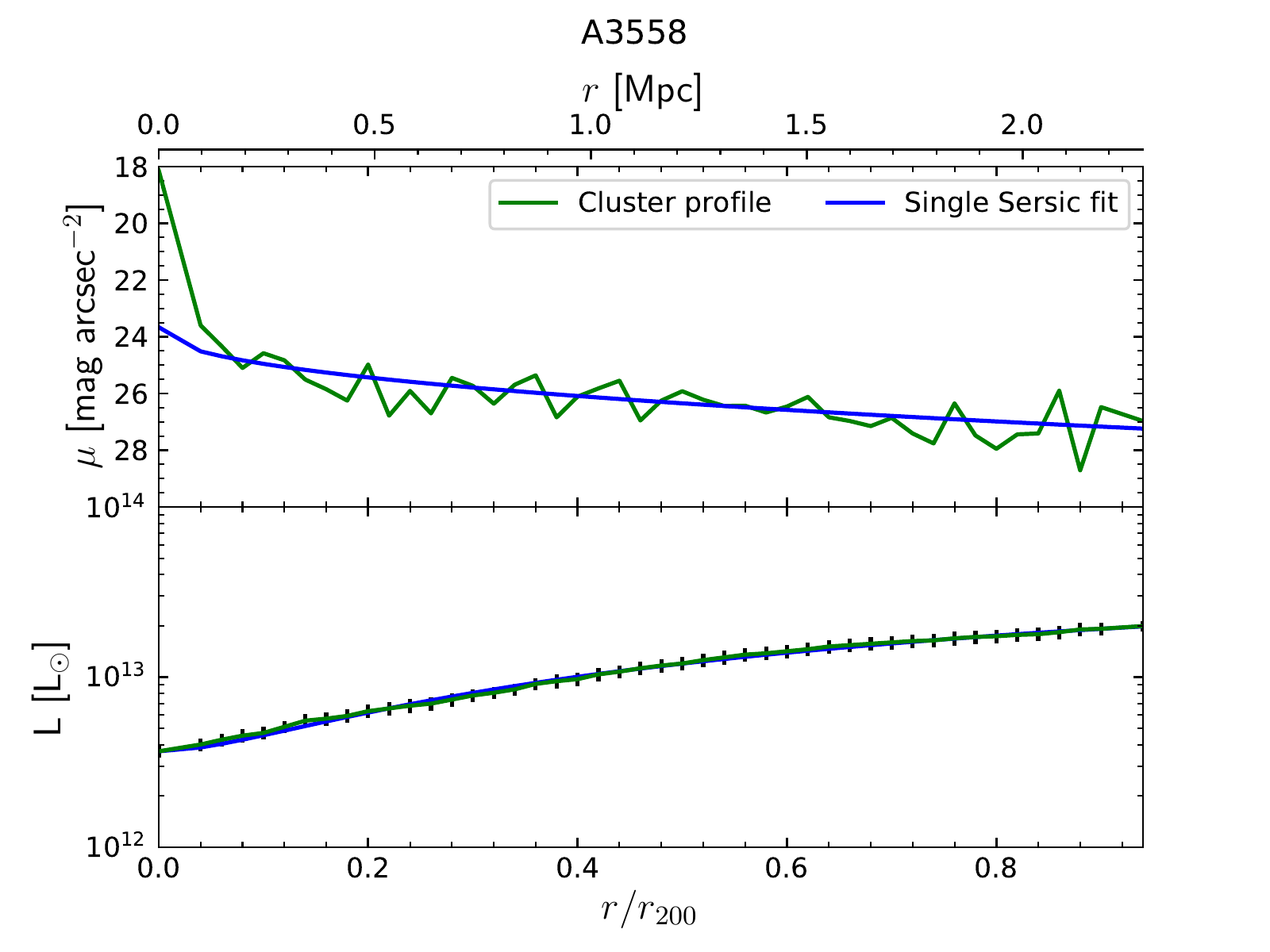}        \includegraphics[width=0.45\textwidth]{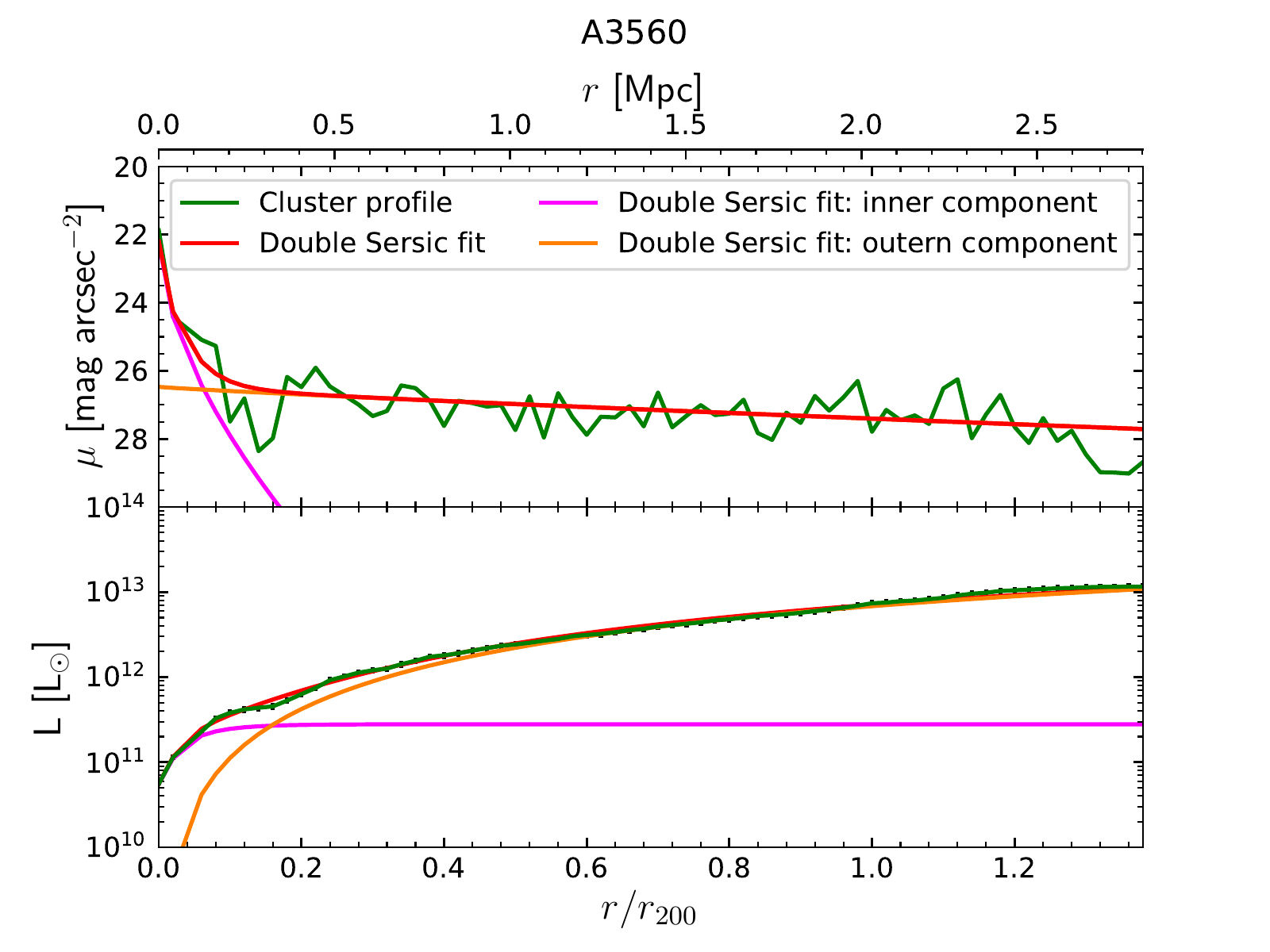}
        \includegraphics[width=0.45\textwidth]{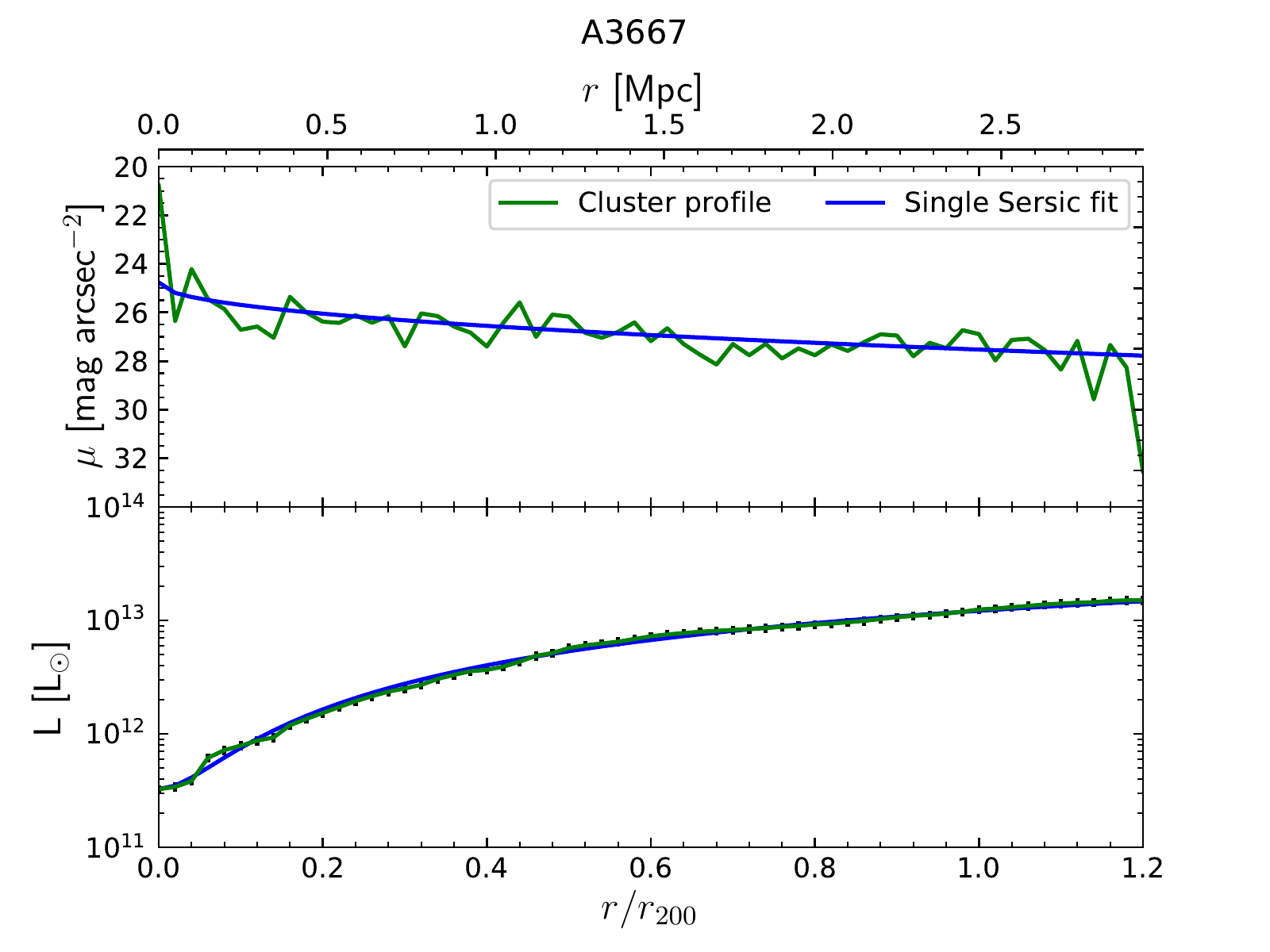}        \includegraphics[width=0.45\textwidth]{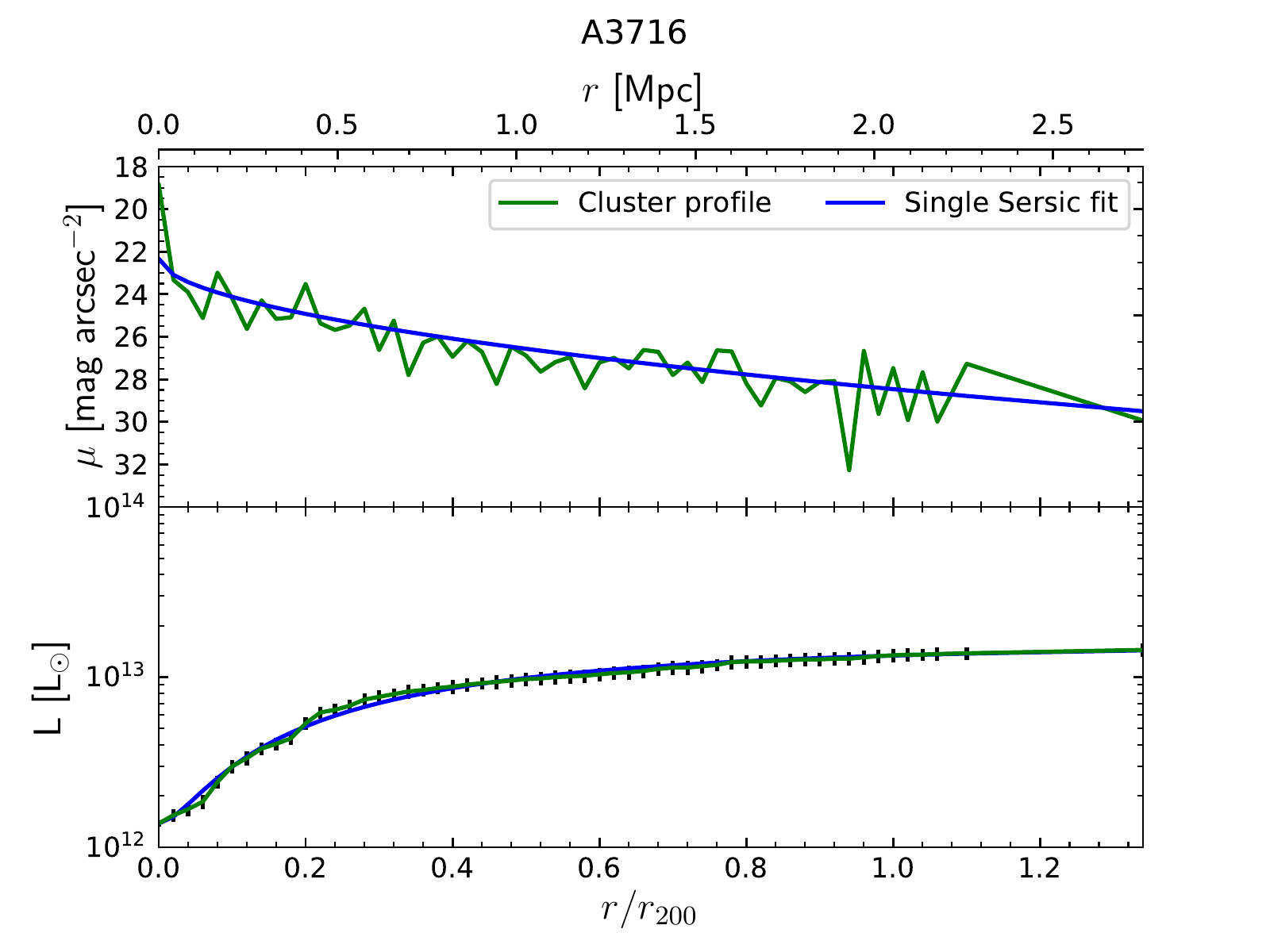}
        \includegraphics[width=0.45\textwidth]{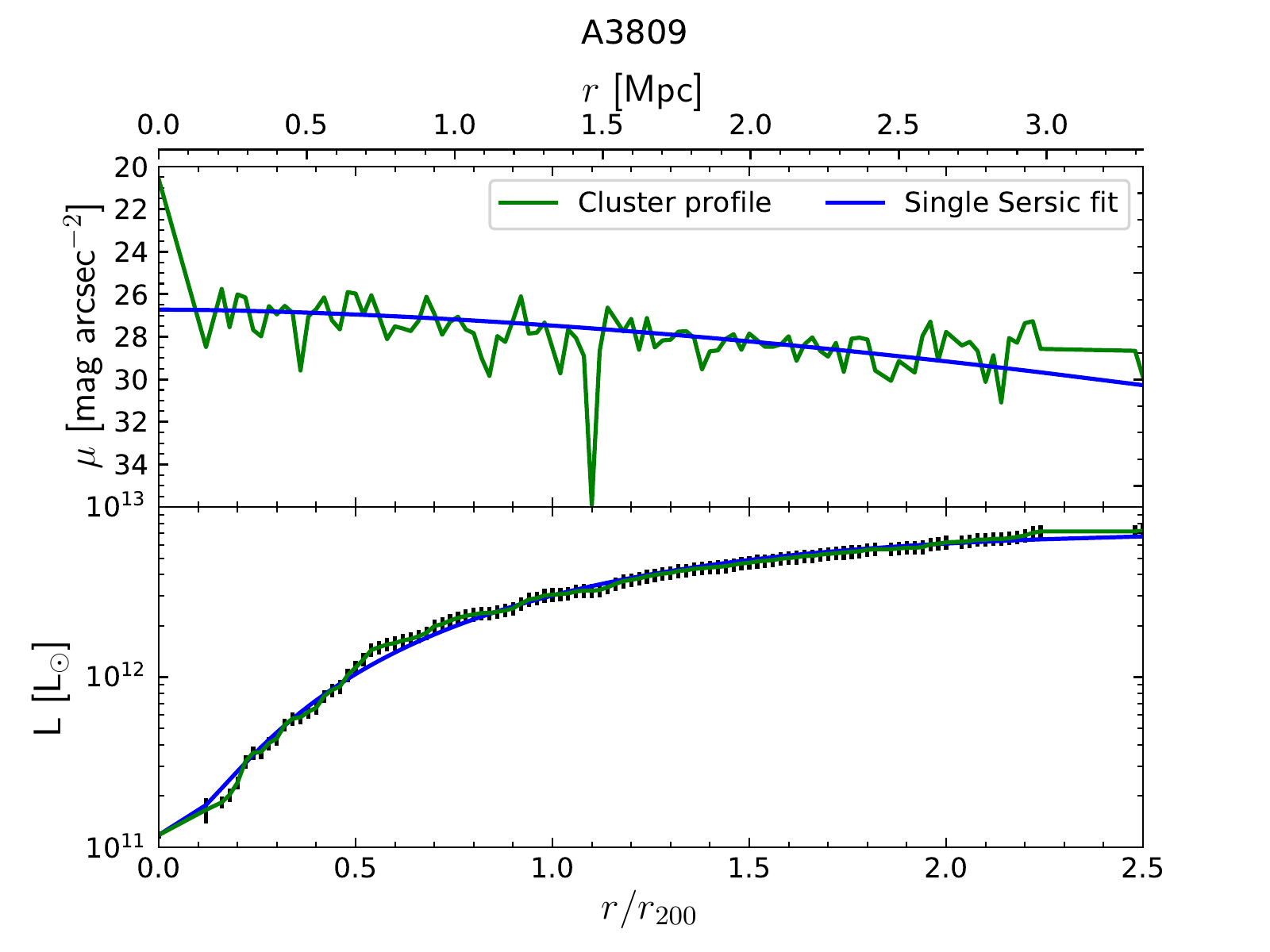}        \includegraphics[width=0.45\textwidth]{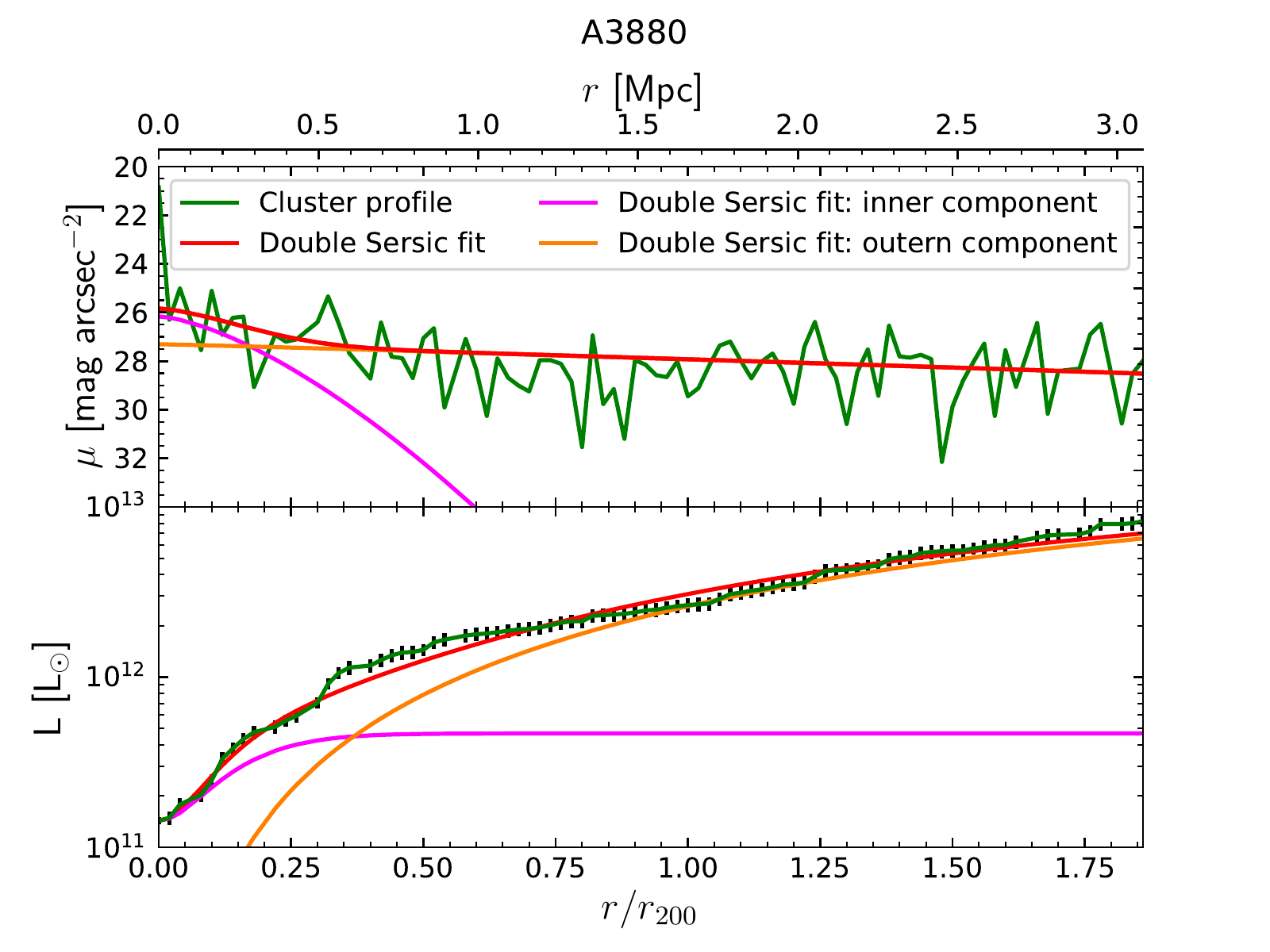}
    \caption{Photometric decomposition of Omega-WINGS galaxy clusters luminosity profiles, continued.}
\end{figure*}

\newpage
\clearpage

\end{appendix}
\end{document}